\documentclass[natbib]{svjour3}       % onecolumn (second format)
\smartqed  % flush right qed marks, e.g. at end of proof
\usepackage{graphicx}
\usepackage[lastpage,user]{zref}

\usepackage{amsmath}
\usepackage{amssymb}
\usepackage{lscape}
\usepackage{textcomp}
\usepackage[export]{adjustbox}
\usepackage{braket}
\usepackage{bm} %% for some math fonts, e.g. boldsymbol
\usepackage{footnote}
\usepackage{epstopdf}
\usepackage[linktocpage]{hyperref}
\usepackage{url}
\usepackage{xcolor}
\usepackage{nicefrac}
\usepackage{colortbl} 
\usepackage{pifont} 
\usepackage{soul}
\usepackage{doi}

\usepackage{rotating}
\usepackage{pdflscape}

\usepackage[11pt]{extsizes}

\hypersetup{
  colorlinks   = true, %Colours links instead of ugly boxes
  urlcolor     = blue, %Colour for external hyperlinks
  linkcolor    = blue, %Colour of internal links
  citecolor   = blue %Colour of citations
 }

\journalname{Living Reviews in Relativity}
\newcommand{\lcdm}{\Lambda\mathrm{CDM}}
\newcommand{\nn}{\nonumber}
\newcommand*{\mpl}{M_{\rm Pl}}
\newcommand{\be}{\begin{equation}}
\newcommand{\ee}{\end{equation}}
\newcommand{\bea}{\begin{eqnarray}}
\newcommand{\eea}{\end{eqnarray}}
\def\[{\begin{equation}}
\def\]{\end{equation}}

\def\hmpcinv{\,h\,{\rm Mpc^{-1}}}
\def\hinvmpc{\,h^{-1}{\rm Mpc}}

\newcommand{\Hu}{km $\rm s^{-1}$ Mpc$^{-1}$}

\providecommand{\Planck}{\textit{Planck}}

\providecommand{\planck}{\Planck}

\newcommand{\mksym}[1]{\ifmmode {\rm #1}\else #1\fi}

\newcommand{\vnhat}{{\hat{\mathbf{n}}}}
\newcommand{\tT}{\tilde{T}}
\newcommand{\valpha}{{\boldsymbol{\alpha}}}

\newcommand{\grad}{\nabla}
\newcommand{\ud}{{\text{d}}}
\newcommand{\clp}{\mathcal{P}}

\newcommand{\metE}{\ensuremath{\tilde{g}}}
\newcommand{\metM}{\ensuremath{g}}
\newcommand{\RiemE}{\ensuremath{\tilde{R}}}
\newcommand{\volE}{\ensuremath{\sqrt{-\metE}}}
\newcommand{\metS}{\ensuremath{\hat{g}}}
\newcommand{\connE}{\ensuremath{\tilde{\nabla}}}
\newcommand{\EinE}{\ensuremath{\tilde{G}}}

\newcommand{\mplr}{m_{\rm Pl}}

\newcommand{\kh}{k_H}

\def\R{{\cal R}}
\def\g{\gamma}

\def\lsim{\raise 0.4ex\hbox{$<$}\kern -0.8em\lower 0.62
ex\hbox{$\sim$}}

\def\gsim{\raise 0.4ex\hbox{$>$}\kern -0.7em\lower 0.62
ex\hbox{$\sim$}}

\def\lbar{{\hbox{$\lambda$}\kern -0.7em\raise 0.6ex
\hbox{$-$}}}

\newcommand{\Gmn}{G_{\mu\nu}}
\newcommand{\RMN}{R^{\mu\nu}}
\newcommand{\Tmn}{T_{\mu\nu}}
\newcommand{\iBox}{\square^{-1}}

\renewcommand\({\left(}
\renewcommand\){\right)}
\renewcommand\[{\left[}
\renewcommand\]{\right]}

\def\nucubic{{\nu} \rm{Galileon}}
\def\nulcdm{{\nu} \Lambda\rm{CDM}}

\def\[{\begin{equation}}
\def\]{\end{equation}}

\newcommand{\acmarklr}{\ding{51}}
\newcommand{\axmarklr}{\ding{55}}

\newcommand{\aTstrutlr}{\rule{0pt}{2.6ex}}       % "top" strut
\newcommand{\aBstrutlr}{\rule[-0.9ex]{0pt}{0pt}} % "bottom" strut
\newcommand{\aTBstrutlr}{\aTstrutlr\aBstrutlr} % top&bottom struts
\usepackage{geometry}
\begin{document}

\title{Testing general relativity in cosmology}

\author{Mustapha Ishak}

\institute{Mustapha\ Ishak%
\at
Department of Physics,\\
The University of Texas at Dallas,\\ 
Richardson, TX 75080, USA\\
\email{mishak@utdallas.edu}
}

\date{Received: 28 May 2018 / Accepted: 6 November 2018 / Published online: 18 December 2018}

\maketitle
%%%%%%%%%%%%%%%%%%%%%%%%%%%%%%%%%%%%%%%%%%%%%%%%%%%%%%%%%%%%%%%%%%%%%%%%%%%%%%%%%%%
%%%%%%%%%%%%%%%%%%%%%%%%%%%%%%%%%%%%%%%%%%%%%%%%%%%%%%%%%%%%%%%%%%%%%%%%%%%%%%%%%%%
%%%%%%%%%%%%%%%%%%%%%%%%%%%%%%%%%%%%%%%%%%%%%%%%%%%%%%%%%%%%%%%%%%%%%%%%%%%%%%%%%%%
%%%%%%%%%%%%%%%%%%%%%%%%%%%     ABSTRACT     %%%%%%%%%%%%%%%%%%%%%%%%%%%%%%%%%%%%%%
%%%%%%%%%%%%%%%%%%%%%%%%%%%%%%%%%%%%%%%%%%%%%%%%%%%%%%%%%%%%%%%%%%%%%%%%%%%%%%%%%%%
%%%%%%%%%%%%%%%%%%%%%%%%%%%%%%%%%%%%%%%%%%%%%%%%%%%%%%%%%%%%%%%%%%%%%%%%%%%%%%%%%%%
%%%%%%%%%%%%%%%%%%%%%%%%%%%%%%%%%%%%%%%%%%%%%%%%%%%%%%%%%%%%%%%%%%%%%%%%%%%%%%%%%%%

\begin{abstract}
We review recent developments and results in testing general relativity (GR) at cosmological scales. The subject has witnessed rapid growth during the last two decades with the aim of addressing the question of cosmic acceleration and the dark energy associated with it. However, with the advent of precision cosmology, it has also become a well-motivated endeavor by itself to test gravitational physics at cosmic scales. We overview cosmological probes of gravity, formalisms and parameterizations for testing deviations from GR at cosmological scales, selected modified gravity (MG) theories, gravitational screening mechanisms, and computer codes developed for these tests. We then provide summaries of recent cosmological constraints on MG parameters and selected MG models. We supplement these cosmological constraints with a summary of implications from the recent binary neutron star merger event. Next, we summarize some results on MG parameter forecasts with and without astrophysical systematics that will dominate the uncertainties. The review aims at providing an overall picture of the subject and an entry point to students and researchers interested in joining the field. It can also serve as a quick reference to recent results and constraints on testing gravity at cosmological scales. 
\end{abstract}

\newpage

\setcounter{tocdepth}{3}
\tableofcontents

%%%%%%%%%%%%%%%%%%%%%%%%%%%%%%%%%%%%%%%%%%%%%%%%%%%%%%%%%%%%%%%%%%%%%%%%%%%%%%%%%%%
%%%%%%%%%%%%%%%%%%%%%%%%%%%%%%%%%%%%%%%%%%%%%%%%%%%%%%%%%%%%%%%%%%%%%%%%%%%%%%%%%%%
%%%%%%%%%%%%%%%%%%%%%%%%%%%%%%%%%%%%%%%%%%%%%%%%%%%%%%%%%%%%%%%%%%%%%%%%%%%%%%%%%%%
%%%%%%%%%%%%%%%%%%%%%%%%%%%%%%%%%%%%%%%%%%%%%%%%%%%%%%%%%%%%%%%%%%%%%%%%%%%%%%%%%%%
%%%%%%%%%%%%%%%%%%%%%%%%                        %%%%%%%%%%%%%%%%%%%%%%%%%%%%%%%%%%%
%%%%%%%%%%%%%%%%%%%%%%%%   INTRODUCTION         %%%%%%%%%%%%%%%%%%%%%%%%%%%%%%%%%%%
%%%%%%%%%%%%%%%%%%%%%%%%                        %%%%%%%%%%%%%%%%%%%%%%%%%%%%%%%%%%%
%%%%%%%%%%%%%%%%%%%%%%%%%%%%%%%%%%%%%%%%%%%%%%%%%%%%%%%%%%%%%%%%%%%%%%%%%%%%%%%%%%%
%%%%%%%%%%%%%%%%%%%%%%%%%%%%%%%%%%%%%%%%%%%%%%%%%%%%%%%%%%%%%%%%%%%%%%%%%%%%%%%%%%%
%%%%%%%%%%%%%%%%%%%%%%%%%%%%%%%%%%%%%%%%%%%%%%%%%%%%%%%%%%%%%%%%%%%%%%%%%%%%%%%%%%%
%%%%%%%%%%%%%%%%%%%%%%%%%%%%%%%%%%%%%%%%%%%%%%%%%%%%%%%%%%%%%%%%%%%%%%%%%%%%%%%%%%%

\newpage 

\section{Introduction}
\label{sec:introduction}
For over a century, Einstein's General Relativity (GR) has continued to be an impressive theory of gravity that fits observations from our solar system to the entire cosmological model of the universe. Guided by some key principles, Einstein came to the important realization of a very close relationship between the curvature of spacetime and gravity. Taking into account further requirements, such as coordinate invariance, conservation laws, and limits that must be consistent with Newtonian gravity, he proposed his gravitational field equations \citep{Einstein1915}. Astonishingly, the same simple but powerful equations remain to date the most accurate description of gravitional physics at all scales.  

Shortly after that, GR gave birth to the current standard model of cosmology predicting exact solutions with expanding or contracting universes. It allowed the combination of ideas from Friedmann and Lema\^{i}tre about expanding universes \citep{Friedmann1922,Lemaitre1931} along with the geometry of homogeneous and isotropic spacetimes of Robertson and Walker \citep{Robertson1935,Walker1937} in order to produce the so-called Friedmann--Lema\^{i}tre--Robertson--Walker models (FLRW). These models describing the background cosmological evolution were completed by the addition of cosmological perturbation theory to populate them with cosmic structures. Over the years and decades to follow, the FLRW models plus cosmological perturbations benefited from a number of theoretical developments and observational techniques that allowed us to map the whole history of cosmic evolution from very early times to the current stages of the universe as we observe it today.

However, this scientific triumph in cosmology came with two conundrums: dark matter and cosmic acceleration (or dark energy). Indeed, in order for the FLRW model to fit current  observations, we first need $\sim$25\% of the mass-energy content in the universe to be in the form of a pressureless dark matter component that interacts only gravitationally with baryons and light (possibly weakly with baryons as well). The requirement for the presence of such a dark  matter component does not come only from cosmology but also from rotation curves of galaxies, gravitational lensing observations, and the requirement of deep initial potential wells that would have allowed the formation of the largest structures that we observe today; see for example  
\cite{Trimble1987,Bertone2005,DAmico2009,Einasto2014,Freese2017} and references therein. The dark matter problem motivated  the introduction of modified gravity (MG) theories that would explain such observations by a small modification to Kepler laws such as Modified Newtonian Dynamics (MOND) \citep{MOND1}, its relativistic generalization known as TeVeS (tensor-vector-scalar) theory \citep{Bekenstein2004}, or other vector-tensor theories. While Dark Matter motivated proposals of some MG models, the main focus of this review is rather on models that attempt to address the question of cosmic acceleration that we describe next. 

The second problem in standard cosmology is indeed that of cosmic acceleration and the dark energy associated with it. Two decades ago, two independent groups using supernova measurements found that the universe's expansion is speeding up rather than slowing down \citep{RiessEtAl1998,PerlmutterEtAl1999}. A plethora of complementary  cosmological observations have continued since to confirm this result and require that an FLRW model fitting observations must have a genuine or effective dark energy component that would account for $\sim$70\% of the total energy budget in the universe. In such a universe, the baryons constitute only $\sim$5\% of this budget. This picture has become the concordance model in cosmology referred to as the Lambda-Cold-Dark-Matter ($\Lambda$CDM) model. This best fit model is spatially flat. $\Lambda$ is the cosmological constant, and its addition to the Einstein's equations can produce an accelerated expansion of the universe. 

The cosmological constant can be cast into the model as an effective cosmic fluid with an equation of state of minus one. This coincides exactly with the equation of state of the vacuum energy associated with zero-level quantum fluctuations. Interestingly, this connects the problem of cosmic acceleration to the problems of the cosmological constant/vacuum energy problems \citep{Weinberg1988CC,Carroll1992,Sahni2000,Carroll2001CC,Peebles2003,
Padmanabhan2003,Copeland2006,2007-Ishak-Remarks-DE}.
Namely, why is the value measured from cosmology so small compared to the one predicted from quantum field calculations? This is known as the old cosmological constant problem. A second question (the new problem) is why the energy density associated with the cosmological constant/vacuum energy is of the same order of magnitude as the matter density at present cosmic time? (If it were any larger it would have prevented cosmic structure from forming.) Other types of dark energy have been proposed with an equation of state that is very close to minus one and would be not connected to the cosmological constant/vacuum energy. These are for example quintessence models based on a scalar field with kinetic energy and potential terms that can be cast as well into an effective dark energy model with a negative equation of state also close to minus one \citep{Peebles1988,Ratra1988CC,Caldwell1998}. It is worth noting that most of these dark energy models do not address the cosmological constant problem and may suffer from some form of fine-tuning as well.  

Relevant to our review, the question of cosmic acceleration motivated a number of proposals for modified gravity models that could produce such an acceleration without the need for a cosmological constant. Such models are said to be self-accelerating. 
In most cases, these models do not address the cosmological constant problems and it is hoped that by some mechanism, for example degravitation or some given cancellation, vacuum energy does not contribute to gravitational and cosmological dynamics. However, in some cases, modified gravity models do provide some degravitation mechanism, although not successfully so far. We discuss these further in this review. 

Finally, there are also modified gravity models at high energies that have been  motivated by the search for quantum gravity and other unified theories of physics which may or may not have any consequences at cosmological scales.

While the rapid growth of the subject of deviations from GR and MG models has been motivated by cosmic acceleration/dark energy and to some extent by dark matter, the subject of MG models is an old one. Indeed, just a few years after GR was introduced, Weyl gravity was proposed by \citep{Weyl1918}, and so were the theories of \cite{Eddington1924TM}, \cite{Cartan1922b}, \cite{BD1961}, and many others. Testing GR and gravity theories within the solar system and using other astrophysical objects have been the subject of intense work with a number of important results over the last five decades or so; see for a review \cite{WillReview2014}. 
Impressively, GR fits all these local tests of gravity. In fact, it fits them so well that these tests are commonly referred to as GR local tests. 
This is very useful to the current cosmological developments, because it has established very stringent constraints at the level of the solar system that any gravity theory must pass. Nevertheless, to address these requirements, some MG models have some gravitational screening mechanisms that allow them to deviate from GR at cosmological scales but then become indistinguishable from it at small scales. 

Further motivation for testing GR at cosmological scales is the increasing quantity and quality of available cosmological data. These are indeed good times for cosmology where a plethora of complementary observational data from ongoing and planned surveys will continue to flow for the many decades to come. 
These include the cosmic microwave background radiation, weak gravitational lensing, galaxy surveys, distances to supernovae, baryon acoustic oscillations, and gravitational waves. A good piece of news is that one can not only combine these  data sets to increase their constraining power, but one can also apply consistency tests between such complementary data sets. This would allow one to identify any problems with systematic effects in the data or any problems with the underlying model. Furthermore, nature has also given us a break in cosmology as we have two types of data sets. Indeed, some data sets are sensitive to the geometry and expansion of the universe and some other sets are sensitive to the growth of large-scale structure (i.e., the rate at which structures cluster in the universe). These two sets of observations must be consistent with one another. 
For testing deviations from GR and constraining MG models, it was realized that MG models can mimic an expansion history of the universe that is identical to that of the concordance $\Lambda$CDM model while they can still have a structure formation history that is different and distinguishable from that of $\Lambda$CDM. It has become common practice that the background expansion is modeled with an effective dark energy with an equation of state close to the minus one value of $\Lambda$CDM. Meanwhile, any departure from GR is constrained by using the growth data from large-scale structure observables.    

There are two general approaches that have been developed to test departures from GR at cosmological scales. The first one is where the deviation is parameterized in a phenomenological way with no necessary exact knowledge of the specific alternative theory. The growth equations are modified by the addition of MG parameters that represent the departure from GR. These MG parameters are expected to take values of unity for GR but depart from it for MG models. 
It is worth noting that such an effective description may not necessarily remain valid at all scales constrained by observations and so must be used with some caution when compared to various observations. The second approach is to choose a specific class of MG models (like the popular $f(R)$ or DGP models (see Sect.~\ref{sec:f(R)} and Sect.~\ref{sec:DGP})) and derive cosmological perturbations and observables for these models. These are then implemented in cosmological analysis software and used to compare to the data. We cover both approaches in this review. A related question is what one could call a modified gravity model versus a dark energy model. There are some guiding helpful prescriptions that we discuss in the review but the spectrum of models has a grey zone where such a distinction is not unambiguous. We characterize various types of deviations from General Relativity and organize MG models accordingly with some illustrative examples.

In this review, we aim at providing an overall current picture of the field of testing gravity at cosmological scales including a selection of recent important results on the subject. The review is meant to provide an entry point for students and researchers interested in the field where they can find summaries and references to further readings. This review can also serve for experienced researchers or other readers to find quickly recent developments or results in the field. As required for the Living Review guidelines, this review is written with the depth and style of a plenary review talk on the subject. It is not meant to replace thorough comprehensive reviews on various parts of this topic and we refer the reader constantly to such specialized reviews as we discuss each sub-topic.

%%%%%%%%%%%%%%%%%%%%%%%%%%%%%%%%%%%%%%%%%%%%%%%%%%%%%%%%%%%%%%%%%%%%%%%%%%%%%%%%%%%
%%%%%%%%%%%%%%%%%%%%%%%%%%%%%%%%%%%%%%%%%%%%%%%%%%%%%%%%%%%%%%%%%%%%%%%%%%%%%%%%%%%
%%%%%%%%%%%%%%%%%%%%%%%%%%%%%%%%%%%%%%%%%%%%%%%%%%%%%%%%%%%%%%%%%%%%%%%%%%%%%%%%%%%
%%%%%%%%%%%%%%%%%%%%%%%%%%%%%%%%%%%%%%%%%%%%%%%%%%%%%%%%%%%%%%%%%%%%%%%%%%%%%%%%%%%
%%%%%%%%%%%%%%%%%%%%%%%%%%%%%%%%%%%%%%%%%%%%%%%%%%%%%%%%%%%%%%%%%%%%%%%%%%%%%%%%%%%
%%%%%%%%%%%%%%%%%%%%%%%%                       %%%%%%%%%%%%%%%%%%%%%%%%%%%%%%%%%%%%
%%%%%%%%%%%%%%%%%%%%%%%%  GENERAL RELATIVITY   %%%%%%%%%%%%%%%%%%%%%%%%%%%%%%%%%%%%
%%%%%%%%%%%%%%%%%%%%%%%%                       %%%%%%%%%%%%%%%%%%%%%%%%%%%%%%%%%%%%
%%%%%%%%%%%%%%%%%%%%%%%%%%%%%%%%%%%%%%%%%%%%%%%%%%%%%%%%%%%%%%%%%%%%%%%%%%%%%%%%%%%
%%%%%%%%%%%%%%%%%%%%%%%%%%%%%%%%%%%%%%%%%%%%%%%%%%%%%%%%%%%%%%%%%%%%%%%%%%%%%%%%%%%
%%%%%%%%%%%%%%%%%%%%%%%%%%%%%%%%%%%%%%%%%%%%%%%%%%%%%%%%%%%%%%%%%%%%%%%%%%%%%%%%%%%
%%%%%%%%%%%%%%%%%%%%%%%%%%%%%%%%%%%%%%%%%%%%%%%%%%%%%%%%%%%%%%%%%%%%%%%%%%%%%%%%%%%
%%%%%%%%%%%%%%%%%%%%%%%%%%%%%%%%%%%%%%%%%%%%%%%%%%%%%%%%%%%%%%%%%%%%%%%%%%%%%%%%%%%

\section{General Relativity (GR)}
\label{sec:General_Relativity}

%%%%%%%%%%%%%%%%%%%%%%%%%%%%%%%%%%%%%%%%%%%%%%%%%%%%%%%%%%%%%%%%%%%%%%%%%%%%%%%%%%%
%%%%%%%%%%%%%%%%%%%%%%%                                         %%%%%%%%%%%%%%%%%%%
%%%%%%%%%%%%%%%%%%%%%%%          SUB-SECTION                    %%%%%%%%%%%%%%%%%%%
%%%%%                         Basic principles                                 %%%%
%%%%%                                                                          %%%%
%%%%%%%%%%%%%%%%%%%%%%%%%%%%%%%%%%%%%%%%%%%%%%%%%%%%%%%%%%%%%%%%%%%%%%%%%%%%%%%%%%%
%%%%%%%%%%%%%%%%%%%%%%%%%%%%%%%%%%%%%%%%%%%%%%%%%%%%%%%%%%%%%%%%%%%%%%%%%%%%%%%%%%%

\subsection{Basic principles}
\label{sec:Basic_principles}

Einstein considered some key guiding principles and well-known limits that a successful theory of gravity must obey. At the forefront is the principle of covariance -- that is the laws of physics must be independent of any coordinate system. So the right language must be that of tensors or another coordinate independent formulation. 
Such a successful theory should locally be consistent with special relativity and must inherit its principles including the equivalence of local inertial frames of reference, the universal constancy of the speed of light in vacuum, and the Lorentz-invariance of the theory. 

An important part of Einstein's reflections when he proposed special relativity and then continued to work toward general relativity was about the principles of equivalence. He found guidance in Mach's ideas about relativity and the nature of inertia \citep{Mach1988}, although, he had to abandon some of them later on. 

From the principle of equivalence between gravity and inertia that we provide below, Einstein developed the important insight that gravity seems to have a privileged status compared to other interactions. That is gravity is equivalent to inertia. The principle of universality of free-fall and gravitational interaction as expressed below in the equivalence principles combined with some insight that gravity is omnipresent in spacetime, led Einstein to formulate gravity as the curvature of spacetime. See various discussions and perspectives in reviews and books, e.g., \cite{WillReview2014,Will2018,DInvero1992,Rindler2006,Weinberg1972,MTW1973,Carroll2003}. 

\begin{itemize}
 
\item{
\textbf{Weak equivalence principle (WEP):}  WEP is stated in a variety of formulations. One of them is usually stated as the equivalence between the inertial mass and the gravitational mass which has been been tested to a few parts in $10^{13}$ \citep{Adelberger2001,Wagner2012EP}  
 and a few parts in
$10^{14}$ (Touboul2017), 
see \cite{WillReview2014} for WEP test timeline. Einstein then advocated that inertia and gravity must be the same and that an observer inside a ``cabin'' (with no windows) at rest in the presence of a gravitational acceleration will not be able to distinguish that situation from one where the ``cabin'' is on a rocket moving up with the exact opposite acceleration. The WEP is expressed as the universality of the gravitational interaction and free-fall for all particles. For our review, we focus on the notions of universality of free fall and the matter coupling in the context of GR+dark energy versus modified gravity (MG) models following for example \cite{2016-Joyce-Lombriser-Schmidt-DEvsMG}. WEP is satisfied if there exists some spacetime metric (in the Jordan frame) to which all species of matter are universally coupled. Test particles fall then along geodesics of this metric.  
}
\item{
\textbf{Einstein equivalence principle (EEP):}   
The EEP requires the validity of the WEP, and that in all local freely falling frames, the laws of physics reduce to those of special relativity (assuming tidal gravitational forces are absent). It is also customary to add here that the EEP contains the statement that the outcome of any local non-gravitational experiment is independent of where and when it is performed \citep{WillReview2014}.
}
\item{
\textbf{Strong equivalence principle (SEP):} 
The SEP extends the universality of free fall of the WEP to massive gravitating objects so it is completely independent of the composition of the objects as well as their gravitational binding energy. Compact objects like Black Holes will also fall along geodesics  like test particles \citep{Will1993,WillReview2014}. The SEP extends also the EEP to include all of the laws of physics, gravitational or otherwise. 
}

\end{itemize}

One more remark is worth mentioning about the relationship between the equivalence prinicples and the spacatime metric. Let us recall that metric theories of gravity satisfy the following properties, see for example \cite{WillReview2014}: (i) a symmetric metric exists, (ii) test particles follow geodesics of such a metric, and (iii) in local reference frames, the non-gravitational laws of physics are those of special relativity. From this definition, it follows that metric theories obey the EEP. It also encourages one to imply that theories that obey the EEP are metric theories, e.g., \cite{WillReview2014}.

We conclude this subsection by commenting on a few other notions that guided Einstein in formulating his equations of the gravitational field. The geometrical nature of GR and the principles it is built upon are certainly far from the Newtonian gravity based on forces and potentials, not to mention the notions of absolute space and other shortcomings that had to be abolished. However, it is interesting to remark that the notion of spacetime and its metric to explain gravity can be compared to the notion of the gravitational potential field in space created by massive objects. However, there is a major difference, in GR there is no gravitational potential or gravity that is added on the top of spacetime, but gravity is curvature of spacetime itself. This was a major insight that Einstein got from his EEP principle. In fact, he knew well that GR must have Newtonian gravity as a limit in the weak regime and that provided to him many hints on how to formulate the field equations that we provide in the next section. 

%%%%%%%%%%%%%%%%%%%%%%%%%%%%%%%%%%%%%%%%%%%%%%%%%%%%%%%%%%%%%%%%%%%%%%%%%%%%%%%%%%%
%%%%%%%%%%%%%%%%%%%%%%%                                         %%%%%%%%%%%%%%%%%%%
%%%%%%%%%%%%%%%%%%%%%%%          SUB-SECTION                    %%%%%%%%%%%%%%%%%%%
%%%%%    Einstein field equations (EFEs) and their exact solutions             %%%%
%%%%%                                                                          %%%%
%%%%%%%%%%%%%%%%%%%%%%%%%%%%%%%%%%%%%%%%%%%%%%%%%%%%%%%%%%%%%%%%%%%%%%%%%%%%%%%%%%%
%%%%%%%%%%%%%%%%%%%%%%%%%%%%%%%%%%%%%%%%%%%%%%%%%%%%%%%%%%%%%%%%%%%%%%%%%%%%%%%%%%%

\subsection{Einstein field equations (EFEs) and their exact solutions}
\label{sec: EFE}

In addition to the principles above, Einstein used the fact that, in the weak field limit, the gravitational field equations must locally reduce to those of Newtonian gravity where the metric tensor components would be related to the gravitational potential and the field equations must reduce to Poisson equations. From the latter, he imposed that the curvature side of the equations must contain only up to second order derivatives of the metric and must also be of the same tensor rank as the energy-momentum tensor. This naturally led Einstein to consider the Ricci tensor, derived from contracting twice the Riemann curvature tensor, but there was a little bit more into it. Indeed, he knew that the equations must satisfy conservation laws and thus must be divergence-free. While the vanishing of the divergence of the matter-energy source side of the equations is assured by energy conservation laws and continuity equations, on the curvature side, the Ricci tensor is not divergence-free so more work was required. For that, Einstein built precisely the tensor that holds his name which, by the Bianchi identity, is divergence-free hence complies with conservation laws, as it should. Some technical or historical entire books or articles have been devoted to what led Einstein to derive his equations and we refer the reader to the extended study by \cite{GRGN} and references therein. 
  
With no further discussion, the Einstein's Field Equations (EFEs) read
\begin{equation}
G_{\mu\nu}+\Lambda g_{\mu\nu}=8\pi G T_{\mu\nu},\label{eq:efe}
\end{equation}
where $G_{\mu\nu}\equiv R_{\mu\nu}-\frac{1}{2}g_{\mu\nu}R$ is the Einstein tensor representing the curvature of spacetime, $R_{\mu\nu}$ is the Ricci tensor, $R$ the Ricci scalar, $g_{\mu\nu}$ is the metric tensor, and  $\Lambda$ is the cosmological constant. For brevity we use units such that $c=1$ throughout. On the RHS, the source (content) of spacetime is represented by the energy momentum tensor   
\begin{equation}
T_{\mu\nu}=(\rho+p)u_{\mu}u_{\nu}+pg_{\mu\nu}+q_{\mu}u_{\nu}+u_{\mu}q_{\nu}+\pi_{\mu \nu},\label{eq:stress}
\end{equation}
where $u^{\mu}$ is the tangent velocity 4-vector (e.g., the tangent field to the cosmic fluid particle world-lines) normalized by $u_{\mu}u^{\mu}=-1$, $\rho$ is the relativistic mass-energy density, $p$ is the isotropic pressure, $q^{\mu}$ the energy flux, and $\pi_{\mu \nu}$ is the trace-free anisotropic pressure or stress, all relative to $u^{\mu}$. The quantities $\rho$, $p$, $q_{\mu}$, and $\pi_{\mu \nu}$ are functions of time and space. We use the signature $(-,+,+,+)$ and a $3+1$ decomposition of spacetime unless stated otherwise. 

In standard cosmology, it is assumed that the cosmic fluid is well-described by a perfect fluid (i.e., $q_{\mu}=0$ and $\pi_{\mu \nu}=0$) at the cosmic background level which accounts for baryons, dark matter, radiation and a cosmological constant or another dark energy component. The energy-momentum tensor then reduces to

\begin{equation}
T_{\mu\nu}=(\bar{\rho}+\bar{p})u_{\mu}u_{\nu}+\bar{p}g_{\mu\nu},\label{eq:perfect}
\end{equation}
where the last three terms of \eqref{eq:stress} are set to zero, and the over bar means average over space of quantities and are now functions of time only. However, at the perturbation level, the velocity field contributes to the heat flux and neutrinos, for example, generate anisotropic shear at early times in the universe.   

It is not widely known that the EFEs have over 1300 exact solutions that have been derived over the last century, see for example the classical compilation book by \cite{StephaniEtAl2003} and also Online Interactive Geometric Databases equipped with a live tensor component calculator \citep{GRDB}. These solutions are based on symmetries of the spacetime and defined forms of the energy momentum source. 

While the large number of exact solutions exhibit the richness and mathematical beauty of the field, a number of solutions still lack any physical interpretation \citep{StephaniEtAl2003,Delgaty1998,Ishak2001}. 
Some of these solutions have found direct applications to real astrophysical systems. 
These include the popular Schwarzschild static spherically symmetric vacuum solution around a concentric mass \citep{Schwarzschild1916}. The solution is often used to model space around Earth, Sun, or other slowly rotating objects where it leads to more accurate predictions than Newtonian gravity, see e.g., \cite{WillReview2014}. The solution is also used to represent the exterior spacetime around a static spherically symmetric black hole. A second well-know exact solution is that of \cite{Kerr1963} representing the vacuum space around an axially symmetric rotating compact object or black hole. Next, several other static spherically symmetric non-vacuum solutions such as the Tolman family of solutions \citep{Tolman1939} and the Buchdahl solutions \citep{Buchdahl1967} have been used to model the interior of compact astrophysical objects such as Neutron stars \citep{LP2007}. Finally, some solutions have found applications in cosmology. These include, for example, the isotropic and homogeneous Friedmann--Lema\^{i}tre--Robertson--Walker (FLRW) solutions \citep{Friedmann1922,Lemaitre1931,Robertson1935,Walker1937}, the inhomogeneous Lema\^{i}tre--Tolman--Bondi solutions \citep{Lemaitre1933,Tolman1934,Bondi1947}, the inhomogeneous Szekeres models \citep{Szekeres1975}, the anisotropic Bianchi models \citep{Ellis1969}, and others \citep{EllisElst1999}.

Einstein's Equations of general relativity connected naturally the isotropic and homogeneous geometry of space given by the Robertson--Walker metric to the cosmic fluid substratum described by a perfect fluid, giving birth to the standard model of cosmology that we describe in the next section. 

It is important to note, and in particular in the context of this review, that while Einstein derived his equations from the principles and approach discuss above, the field equations also derive immediately from a variational principle where the action for the curvature sector is simply the Ricci scalar. This was derived simultaneously by Einstein and Hilbert and the curvature part of the action bears their names. The GR action with a cosmological constant term reads 
\begin{equation}
S_{GR}=\int d^{4}x \sqrt{-g} \left[\frac{R- 2 \Lambda}{16 \pi G}+{\cal L}_{M}\right],
\label{eq:GRaction}
\end{equation}
where $g$ is the determinant of the metric tensor and ${\cal L}_{M}$ is the Lagrangian for the matter fields. Variations of Eq.~\eqref{eq:GRaction} with respect to the metric, $g_{\mu\nu}$, gives the field equations \eqref{eq:efe} above. Modified gravity models are often introduced at the level of the action.

Finally, with regards to this review, it is worth clarifying that modifications to GR mean also that the above exact solutions are not anymore valid and need to be replaced by their homologous solutions in the modified theory. For cosmology, an FLRW metric is often used but then leads to modified dynamical equations often referred to as modified Friedmann's equations. 

%
%%%%%%%%%%%%%%%%%%%%%%%%%%%%%%%%%%%%%%%%%%%%%%%%%%%%%%%%%%%%%%%%%%%%%%%%%%%%%%%%%%%
%%%%%%%%%%%%%%%%%%%%%%%%%%%%%%%%%%%%%%%%%%%%%%%%%%%%%%%%%%%%%%%%%%%%%%%%%%%%%%%%%%%
%%%%%%%%%%%%%%%%%%%%%%%%%%%%%%%%%%%%%%%%%%%%%%%%%%%%%%%%%%%%%%%%%%%%%%%%%%%%%%%%%%%
%%%%%%%%%%%%%%%%%%%%%%%%%%%%%%%%%%%%%%%%%%%%%%%%%%%%%%%%%%%%%%%%%%%%%%%%%%%%%%%%%%%
%%%%%%%%%%%%%%%%%%%%%%%%%%%%%%%%%%%%%%%%%%%%%%%%%%%%%%%%%%%%%%%%%%%%%%%%%%%%%%%%%%
%%%%%%%%%%%%%%%%%%%%%%%%%%%%%%%%%%%%%%%%%%%%%%%%%%%%%%%%%%%%%%%%%%%%%%%%%%%%%%%%%%%
%%%%%%%%%%%%%%%%%%%%%%%%                                  %%%%%%%%%%%%%%%%%%%%%%%%%
%%%%%%%%%%%%%%%%%%%%%%%%  STANDARD MODEL OF COSMOLOGY     %%%%%%%%%%%%%%%%%%%%%%%%%
%%%%%%%%%%%%%%%%%%%%%%%%                                  %%%%%%%%%%%%%%%%%%%%%%%%%
%%%%%%%%%%%%%%%%%%%%%%%%%%%%%%%%%%%%%%%%%%%%%%%%%%%%%%%%%%%%%%%%%%%%%%%%%%%%%%%%%%%
%%%%%%%%%%%%%%%%%%%%%%%%%%%%%%%%%%%%%%%%%%%%%%%%%%%%%%%%%%%%%%%%%%%%%%%%%%%%%%%%%%%
%%%%%%%%%%%%%%%%%%%%%%%%%%%%%%%%%%%%%%%%%%%%%%%%%%%%%%%%%%%%%%%%%%%%%%%%%%%%%%%%%%%
%%%%%%%%%%%%%%%%%%%%%%%%%%%%%%%%%%%%%%%%%%%%%%%%%%%%%%%%%%%%%%%%%%%%%%%%%%%%%%%%%%%
%%%%%%%%%%%%%%%%%%%%%%%%%%%%%%%%%%%%%%%%%%%%%%%%%%%%%%%%%%%%%%%%%%%%%%%%%%%%%%%%%%%

\section{The standard model of cosmology}
\label{sec:standardModel}

%%%%%%%%%%%%%%%%%%%%%%%%%%%%%%%%%%%%%%%%%%%%%%%%%%%%%%%%%%%%%%%%%%%%%%%%%%%%%%%%%%%
%%%%%%%%%%%%%%%%%%%%%%%                                         %%%%%%%%%%%%%%%%%%%
%%%%%%%%%%%%%%%%%%%%%%%          SUB-SECTION                    %%%%%%%%%%%%%%%%%%%
%%%%%                    The homogeneous cosmological background               %%%%
%%%%%                                                                          %%%%
%%%%%%%%%%%%%%%%%%%%%%%%%%%%%%%%%%%%%%%%%%%%%%%%%%%%%%%%%%%%%%%%%%%%%%%%%%%%%%%%%%%
%%%%%%%%%%%%%%%%%%%%%%%%%%%%%%%%%%%%%%%%%%%%%%%%%%%%%%%%%%%%%%%%%%%%%%%%%%%%%%%%%%%

\subsection{The homogeneous cosmological background}
\label{sec: Homogeneous}

%%%%%%%%%%%%%%%%%%%%%%%%%%%%%%%%%%%%%%%%%%%%%%%%%%%%%%%%%%%%%%%%%%%%%%%%%%%%%%%%%%%
%%%%%%%%%%%%%%%%%%%%%%%%%%%%%%%%%%%%%%%%%%%%%%%%%%%%%%%%%%%%%%%%%%%%%%%%%%%%%%%%%%%
%%%%%%%%%%%%%%%%%%%%%%%%%%%%%%%%%%%%%%%%%%%%%%%%%%%%%%%%%%%%%%%%%%%%%%%%%%%%%%%%%%%
\subsubsection{FLRW metric and Friedmann's equations}

From the nearly isotropic large scale observations around us and the assumption that it should not look any different from another point in the universe (i.e., the cosmological principle), one can infer that the universe can be described by a spacetime that is globally isotropic and thus homogeneous. The geometry is then described by the metric of Friedmann-Lema\^{i}tre-Robertson-Walker (FLRW) with line element
\begin{equation}
ds^2=-dt^2+a^2(t)\left(\frac{dr^2}{1-kr^2}+r^2(d\theta^2+\sin^2\theta d\phi^2)\right),\label{eq:flrw}
\end{equation}
where $a(t)$ is the expansion scale factor representing the time-dependent evolution of the spatial part of the metric (surfaces of constant $t$), and $k\in\{-1,0,+1\}$ determines the geometry of these spatial sections: negatively curved, flat, or positively curved, respectively.

The EFEs \eqref{eq:efe} solved for the FLRW metric \eqref{eq:flrw} and a perfect fluid source energy momentum tensor \eqref{eq:perfect} give the dynamical Friedmann equations.  The first equation derives from time-time components of the EFEs as
\begin{equation}
\frac{\dot{a}^2}{a^2}=H(t)^2=\frac{8\pi G}{3}\bar{\rho} +\frac{\Lambda}{3}-\frac{k}{a^2},
\label{eq:FriedmannEq1}
\end{equation}
where an \textit{overdot} denotes the derivative with respect to the cosmic time $t$, and we isolated on the LHS the Hubble parameter defined as,   
\be
H(t)\equiv \frac{\dot{a}(t)}{a(t)}.
\ee
This allows us to define a first cosmological parameter, the Hubble constant as $H_0=H(t_0)$  where $t_0$ is the present time. It is common to use instead the normalized parameter $h \equiv H_0/$(100 \Hu).  As usual, in the spatially flat case, the scale factor can be normalized such that its present value $a_0=a(t_0)\equiv 1$. We recall that in spatially curved space, one cannot normalize simultaneously the spatial curvature and the scale factor. The cosmological
redshift is related to the scale factor by $1+z=a_0/a$.

The second Friedmann equation derives from the combination of the space-space component and the time-time component of the EFEs, and can be written as an acceleration/deceleration equation  as follows
\begin{equation}
\frac{\ddot{a}}{a}\, = -\frac{4\pi G}{3}\left(\bar{\rho}\, +\, 3\bar{p} \right)\, +\, \frac{\Lambda}{3}.
\label{eq:FriedmannEq2}
\end{equation}

It is sometimes more convenient to replace the radial coordinate, $r$, by the comoving coordinate $\chi$ using $d\chi \equiv {dr}/{\sqrt{1-kr^2}}$  so that the line element reads 
\begin{equation}
ds^2=-dt^2+a^2(t)\left(d\chi^2+f_{K}^2(\chi) (d\theta^2+\sin^2\theta d\phi^2)\right),\label{eq:flrw2}
\end{equation}
where 
\begin{equation}
f_K(\chi)=\left\{ \begin{array}{lcl}
\sin (\chi) & & k=+1 \\
\chi & & k=0 \\
\sinh (\chi) & & k=-1 \end{array} \right. .
\label{f_K}
\end{equation}

Finally, it is also sometimes convenient to change the coordinate (cosmic) time to the conformal time defined as $d\tau \equiv {dt}/{a(t)}$ so the line element now reads 
\begin{equation}
ds^2=a^2 (\tau) \left[ -d\tau^2 + d\chi^2+f_{K}^2(\chi) (d\theta^2+\sin^2\theta d\phi^2)\right].
\label{eq:flrw3}
\end{equation}

The Friedmann equations and the FLRW metric provide a description of the homogeneous universe and its dynamics serving as a basis to study the propagation of light, the expansion history, distance measures, and the energy budget of the universe. 

Again, with regards to modifications to GR, the Friedmann's equations above, i.e.,  \eqref{eq:FriedmannEq1} and \eqref{eq:FriedmannEq2}, are modified and so are all the observables and distance measurements described below that build on these equations. For example, in relation to cosmic acceleration, the cosmological constant term can be replaced by extra terms coming from the modification and that could play a similar role to it. However, as we already mentioned in the introduction, some of these models are able to fit well the expansion and background observations so any further distinction will have to come from the growth of structure constraints and observables.  

%%%%%%%%%%%%%%%%%%%%%%%%%%%%%%%%%%%%%%%%%%%%%%%%%%%%%%%%%%%%%%%%%%%%%%%%%%%%%%%%%%%
%%%%%%%%%%%%%%%%%%%%%%%%%%%%%%%%%%%%%%%%%%%%%%%%%%%%%%%%%%%%%%%%%%%%%%%%%%%%%%%%%%%
%%%%%%%%%%%%%%%%%%%%%%%%%%%%%%%%%%%%%%%%%%%%%%%%%%%%%%%%%%%%%%%%%%%%%%%%%%%%%%%%%%%
\subsubsection{Cosmic mass-energy budget, dark energy and cosmic acceleration}

In general relativity, conservation laws are given by the vanishing of the covariant derivative of the energy momentum tensor, i.e., $T^{\mu\nu}_{\;\;\;\;\; ;\nu}=0$. This provides the continuity equation
\be
\dot{\bar{\rho}}+3 \frac{\dot{a}}{a}(\bar{\rho}+\bar{p})=\dot{\bar{\rho}}+3 \frac{\dot{a}}{a}\bar{\rho}(1+w)=0,
\label{eq:continuity}
\ee
where in the last step we used the equation of state variable, $w$, defined as 
\be
\bar{p}=w\bar{\rho}.
\label{eq:EOS}
\ee 
It follows from the continuity Eq. \eqref{eq:continuity}, that for a matter (baryon and dark matter) dominated epoch (i.e., $w=0$) $\bar{\rho}_m \propto a^{-3}$, for a radiation dominated epoch  (i.e., $w=1/3$) $\bar{\rho}_r \propto a^{-4}$, and for a cosmological constant (i.e., $w=-1$) $\rho_{\Lambda}$ is a constant,  while for a dynamical dark energy with $w_{de}$, 
\be
\bar{\rho}_{de} = \bar{\rho}_{de}^{0} a^{-3(1+w_{de})}.
\ee
In models of dynamical dark energy, $w_{de}$  is another cosmological parameter that is allowed to be different from $-1$ in cosmological analyses. It can also be allowed to vary in redshift (or scale factor) in which case it can, for example, take the form $w(a)=w_0+w_a(1-a)$ known as CPL parameterization \citep{Chevallier2001IAU,Linder2003ET}. Other parametrizations for $w$ have been introduce in order to fit other dark energy or modified gravity models. Alternatively, the equation of state can also be binned in the redshift. 

It is trivial to observe from the second Friedmann equation \eqref{eq:FriedmannEq2} that a cosmic effective dark energy fluid with an equation of state $w_{de}=p_{de}/\rho_{de}<-1/3$ gives an accelerated expansion. This is the case for a cosmological constant. The field equations of GR have no difficulty in mathematically producing an accelerated expansion, but the real  challenge is to figure out what is the physical nature of such an  effective dark energy fluid. 

So far, most analyses are consistent with the value of $w=-1$ of a cosmological constant with shrinking error bars around it; see for example DES Year-1 cosmological parameter paper \citep{DES2017} where combining most available data sets gave $w_{de}= -1.00{\,}^{+0.04}_{-0.05}$. Although the latest data from Planck and Planck combined with other data sets was found to slightly favor $w_{de}$ values slightly smaller than $-1$ \citep{Planck2015MG}. 
However, current data do not yet significantly constrain the $w_0$ and $w_a$ parameters for a time-varying equation of state of DE. 

In order to describe the energy budget in the universe as measured from observations, we first need to describe the critical density of the universe evaluated today, noted as  $\rho_{crit}^0$. This will serve as a reference density and is determined from the first Friedmann equation \eqref{eq:FriedmannEq1} in a spatially flat universe with no cosmological constant. That is:
\bea
\rho_{crit}^0 & = & \frac{3H^2_0}{8\pi G} \\ \nonumber
            & = & 1.9\times 10^{-29}h^2 \rm grams \,\,\rm cm^{-3} \\ \nonumber
            & = & 2.8\times 10^{11}h^2 \rm M_{\odot} \rm Mpc^{-3}.
\label{eq:criticalDensity}
\eea
The last line is given in solar masses, $\rm M_{\odot}$, per megaparsec cubed. We can now use this reference density to express the density parameters today for different species as the ratio
\be
\Omega_i^0=\frac{\bar{\rho}_i^0}{\rho_{crit}^0}.
\ee
This defines 3 other cosmological parameters with their values today as for example estimated from Planck and other data sets \citep{Planck2015}: $\Omega_b^0\approx 0.05$ for baryonic matter, $\Omega_dm^0 \approx 0.26$ for cold dark matter, $\Omega_{\Lambda}^0 \approx 0.69$ for a cosmological constant, and a tiny curvature ``density'' parameter $|\Omega_k^0\equiv -k/H_0^2| < 0.01$. These numbers characterize the standard spatially flat Lambda-Cold-Dark-Matter ($\Lambda$CDM) concordance model. 

The Friedmann equation \eqref{eq:FriedmannEq1} can be re-written in terms of these density parameters and the scale factor as 
\be
H^2(a)=H^2_0 \left[ \Omega_m^0 a^{-3}+ \Omega_r^0 a^{-4}+ \Omega_k^0 a^{-2}+\Omega_{de}^0 a^{-3(1+w)}  \right],
\label{eq:FriedmannOmegas}
\ee
where we use $\Omega_m^0 \equiv \Omega_b^0+\Omega_c^0$ and recall that $\Omega_r^0\approx 10^{-4}$ and is so negligible at the present time. So when evaluated today for a spatially  flat  universe with a cosmological constant, $\Lambda$, Eq.~\eqref{eq:FriedmannOmegas} reduces to simply $\Omega_m^0+\Omega_{\Lambda}^0=1$.

%%%%%%%%%%%%%%%%%%%%%%%%%%%%%%%%%%%%%%%%%%%%%%%%%%%%%%%%%%%%%%%%%%%%%%%%%%%%%%%%%%%
%%%%%%%%%%%%%%%%%%%%%%%%%%%%%%%%%%%%%%%%%%%%%%%%%%%%%%%%%%%%%%%%%%%%%%%%%%%%%%%%%%%
%%%%%%%%%%%%%%%%%%%%%%%%%%%%%%%%%%%%%%%%%%%%%%%%%%%%%%%%%%%%%%%%%%%%%%%%%%%%%%%%%%%
\subsubsection{Cosmological distances}

Another useful background information to cover is that of distances in cosmology. We start with the physical distance or proper distance (e.g., \citealt{Weinberg1972}), defined for example by integrating the line element \eqref{eq:flrw2} at a given instant along a radial direction so that $dt=d\theta=d\phi=0$ 
\be 
d_{phys}(t)={a(t)} \int^{\chi}_0 d\chi'={a(t)}\chi.
\label{eq:D_phys}
\ee

This is the distance that would be instantaneously measured if we used a gigantic ruler from us to a remote object. In \cite{Weinberg1972}, this is equivalently defined from \eqref{eq:flrw2} as  
\be 
d_{prop}(t)=\int^{r}_0 \sqrt{g_{rr}}dr'={a(t)}\int^{r}_0 \frac{dr'}{\sqrt{1-kr'^2}}={a(t)}\chi.
\label{eq:D_phys}
\ee 

This distance is time dependent so a radial comoving distance is often used as 

\be
\chi=\frac{d_{phys}}{{a(t)}}.
\ee

In the spatially flat case, with the normalization of $a\equiv 1$ today, the comoving distance is normalized to be equal to the proper distance today. Also, the normalized comoving distance to a galaxy with redshift $z$ (or $a=1/(1+z)$) is thus given from Eq.~\eqref{eq:flrw2} as 

\be 
\chi=\int^{t_{today}}_{t}\frac{dt'}{a(t')}= \int^{1}_{a}\frac{da'}{a'^2 H(a')}. 
\label{eq:chi}
\ee

Now, astronomers define other distances that can be measured by different methods. First, the angular diameter distance is defined for an object that has a typical diameter size, $\mathcal{D}$, and an angular observed size, $\delta \theta$  as \citep{Ellis1973,EllisElst1999}
\bea 
d_A&\equiv&\frac{\mathcal{D}}{\delta \theta}=\frac{\sqrt{g_{\theta\theta}}d\theta}{\delta\theta}\\ \nonumber \\ \nonumber
&=&{a(t)}f_K(\chi),
\label{eq:AngularDiameter}
\eea 
where we have used the metric \eqref{eq:flrw2} and $f_K(\chi)$ is given by \eqref{f_K}. Furthermore, the comoving angular diameter distance is defined as 
\be 
d_{AC}\equiv\frac{d_A}{{a(t)}}=f_K(\chi),
\label{eq:ComovingAngularDiameter}
\ee
so in a spatially flat cosmology, $\chi$ is also referred to as the comoving angular diameter distance.
 
Finally, for an object with luminosity, $L$, and flux, $F$, measured here at the observer (for example on a Charged-Coupled Device (CCD)), the luminosity distance, $d_L$, is defined from the relation  
\be
F \equiv\frac{L}{4\pi d_L^2}.
\label{eq:flux1}
\ee
From photon conservation, the flux measured at observer can be written in terms of the metric functions of  \eqref{eq:flrw2} and the source redshift as \citep{EllisElst1999}
\be
F=\frac{L}{4\pi (1+z)^2 r_G^2},
\label{eq:flux2}
\ee
where $r_G\equiv a(t_0) f_K(\chi)$ is called the galaxy area distance.  
Furthermore, two effects need to be considered. The first is that photons are redshifted by a factor $(1+z)$, and the second effect is that there is a time dilation due to cosmic expansion providing a second factor $(1+z)$. 

Now, comparing Eq.~\eqref{eq:flux1} and Eq.~\eqref{eq:flux2}, and using $r_G$, the luminosity distance is given by  
%with $a(t_0)$ normalized to unity,
\be 
d_L(z)=f_K(\chi)(1+z).
\label{eq:d_L}
\ee
$d_L$ is thus related to the angular diameter distance, $d_A$, by 
\be 
d_L=d_A(1+z)^2.
\ee
This is Etherington's reciprocity theorem (or distance-duality relation), which is true when the number of photons traveling on null geodesics is conserved.

%%%%%%%%%%%%%%%%%%%%%%%%%%%%%%%%%%%%%%%%%%%%%%%%%%%%%%%%%%%%%%%%%%%%%%%%%%%%%%%%%%%
%%%%%%%%%%%%%%%%%%%%%%%                                         %%%%%%%%%%%%%%%%%%%
%%%%%%%%%%%%%%%%%%%%%%%          SUB-SECTION                    %%%%%%%%%%%%%%%%%%%
%%%%%   The inhomogeneous lumpy universe and the growth of large-scale structure %%%
%%%%%                                                                          %%%%
%%%%%%%%%%%%%%%%%%%%%%%%%%%%%%%%%%%%%%%%%%%%%%%%%%%%%%%%%%%%%%%%%%%%%%%%%%%%%%%%%%%
%%%%%%%%%%%%%%%%%%%%%%%%%%%%%%%%%%%%%%%%%%%%%%%%%%%%%%%%%%%%%%%%%%%%%%%%%%%%%%%%%%%

\subsection{The inhomogeneous lumpy universe and the growth of large-scale structure}
\label{sec:perturbations}

%%%%%%%%%%%%%%%%%%%%%%%%%%%%%%%%%%%%%%%%%%%%%%%%%%%%%%%%%%%%%%%%%%%%%%%%%%%%%%%%%%%
%%%%%%%%%%%%%%%%%%%%%%%%%%%%%%%%%%%%%%%%%%%%%%%%%%%%%%%%%%%%%%%%%%%%%%%%%%%%%%%%%%%
%%%%%%%%%%%%%%%%%%%%%%%%%%%%%%%%%%%%%%%%%%%%%%%%%%%%%%%%%%%%%%%%%%%%%%%%%%%%%%%%%%%
\subsubsection{Large-scale structure and cosmological perturbations}

The universe we observe at large scales is rather full of clusters and superclusters of galaxies. Such a picture is mathematically realized by applying linear perturbations to Einstein's equations in an FLRW background. Sufficiently large scales are considered so linear perturbations are a valid description. 

This is done by adding to the metric tensor a small perturbation tensor. Then computing the Einstein tensor to the first order. At the same time, the energy momentum tensor is also linearly perturbed. The Einstein equations then give the usual background  Friedmann equations \eqref{eq.Friedmann} plus additional equations governing the evolution of the perturbations (see, e.g., \citealt{Carroll2003,Peter2013} for a pedagogical introductions and also some of the seminal references \citealt{Bardeen1980,Kodama1984}). An insightful approach to these linear perturbations is to decompose the components of the symmetric metric tensor perturbations according to how they transform under spatial rotations. The 00-component of the metric perturbation tensor is a scalar, the three 0i-components (or equally the three i0-components) constitute a vector, and the remaining nine ij components form a symmetric spatial tensor of rank two. This is known as the SVT decomposition of linear perturbations. The three parts transform only into components of the same type under spatial rotations. In GR, the scalar modes are, for example, associated with matter density fluctuations and used for large scale structure studies, tensor modes are associated with gravitational radiation used, for example, for primordial gravitational waves, while vector modes decay in    and are  usually ignored. Last, in addition to this decomposition, one needs to specify a gauge choice where the components of the perturbations can be different in the corresponding coordinate system, see e.g., \cite{Carroll2003,Peter2013} for pedagogical discussions. Modification to gravity can be implemented at the level of scalar mode perturbations as we discuss further below or at the level of tensor modes as in, e.g., \cite{Saltas2014,friction,MassiveGinCMB,Raveri2015MT,Amendola2014EO,Lin2016}.   

In this review, we will focus scalar perturbations. The perturbed spatially flat FLRW metric reads in, for example, the conformal Newtonian gauge as 
\be
ds^2=a(\tau)^2[-(1+2\Psi)d\tau^2+(1-2{\Phi})dx^idx_i],
\label{eq:PFLRW}
\ee
where $x_i$'s are the comoving coordinates, and $\tau$ the conformal time defined further above.  $\Phi$ and $\Psi$ are the gravitational scalar potentials describing the scalar mode of the metric perturbations.  

We consider subhorizon scales with $k \gg aH$. In many analyses and papers on testing gravity at cosmological scales, the perturbed equations are often specialized to the quasi-static limit or approximation. This means that the time evolution of the gravitational potentials is assumed to be small compared to the Hubble time so one can assume the derivatives of the potentials to be zero for sub- Hubble-horizon scales. For scalar-tensor theories, this approximation also means that one neglects the time derivatives of the fluctuations in the scalar field at scales below the scalar perturbation sound horizon. More on this approximation or its limits can be found in, e.g., \cite{Noller2014,Sawicki2015,PogosianEtAl2016}. 

The first-order perturbed Einstein equations in Fourier space give two equations that describe the evolution of the two scalar gravitational potentials, e.g., \cite{Ma-Bertschinger1995}. The combination of the time-time and time-space perturbed equations provides a Poisson equation for the potential $\Phi$.  The second equation includes the two potentials and comes from the traceless space-space components. The two equations read  (in the quasi-static approximation for the potentials)
\bea
k^2{\Phi}  &=&-4\pi G a^2\sum_i \bar{\rho}_i \delta_i
\label{eq:poisson_1}\\
k^2(\Psi-\Phi) &=& -12 \pi G a^2\sum_i \bar{\rho}_i(1+w_i)\sigma_i,
\label{eq:poisson_2}
\eea
where $\bar{\rho}_i$ and $\sigma_i$ are the density and the shear stress, respectively, for matter species denoted by the index $i$.  
$\delta_i$ is the gauge-invariant, rest-frame overdensity for matter species, $i$. Its evolution describes the growth of inhomogeneities. It is defined by
\be
\delta_i = \delta_i +3\mathcal{H}\frac{q_i}{k},
\label{eq:Deltadef}
\ee 
where $\mathcal{H} ={a}'/a$ is the Hubble factor in conformal time (where $'$ is for differentiation with respect to conformal time), and for species $i$, 
\be 
\delta_i=\frac{\rho_i-\bar{\rho}_i}{\bar{\rho}_i}
\ee
is the fractional overdensity; ${\bar{\rho}_i}$ is the background average density; $q_i$ is the heat flux related to the divergence of the peculiar velocity, $\theta_i$, by 
\be 
\theta_i=\frac{k\ q_i}{1+w_i}.
\ee
From conservation of the energy-momentum in the perturbed matter fluid, these quantities for uncoupled fluid species or the mass-averaged quantities for all the fluids evolve as, e.g., \cite{Ma-Bertschinger1995}:
\bea
{\delta}' & = & -k q +3(1+w){\Phi}'+3\mathcal{H}(w-\frac{\delta P}{\delta\rho})\delta
\label{eq:deltaevolution}\\
\frac{{q}'}{k}&= &-\mathcal{H}(1-3w)\frac{q}{k}+\frac{\delta P}{\delta\rho}\delta+(1+w)\left(\Psi-\sigma\right).
\label{eq:qevolution}
\eea

Combining these two equations, one obtains the evolution equation of $\delta$ as 
\be
{\delta}' = 3(1+w)\left({\Phi}'+\mathcal{H}\Psi\right)+3\mathcal{H}w\delta -\left[k^2+3\left(\mathcal{H}^2-{\mathcal{H}}'\right)\right]\frac{q}{k}-3\mathcal{H}(1+w)\sigma.
\label{eq:Deltaevolution}
\ee

Equations \eqref{eq:poisson_1}, \eqref{eq:poisson_2}, \eqref{eq:deltaevolution}, and \eqref{eq:qevolution} above are coupled to one another; their combinations, along with the evolution equations for the scale factor $a(\tau)$, can provide a full description of the growth history of structures in the universe.

%%%%%%%%%%%%%%%%%%%%%%%%%%%%%%%%%%%%%%%%%%%%%%%%%%%%%%%%%%%%%%%%%%%%%%%%%%%%%%%%%%%
%%%%%%%%%%%%%%%%%%%%%%%%%%%%%%%%%%%%%%%%%%%%%%%%%%%%%%%%%%%%%%%%%%%%%%%%%%%%%%%%%%%
%%%%%%%%%%%%%%%%%%%%%%%%%%%%%%%%%%%%%%%%%%%%%%%%%%%%%%%%%%%%%%%%%%%%%%%%%%%%%%%%%%%
\subsubsection{Growth factor and growth rate of large-scale structure}
\label{sec:growth}

Now, specializing the above equations to the case of matter (baryons plus cold dark matter) at late time, we can set $w=\delta P/\delta \rho=\sigma=0$. Also using the quasi-static approximation (i.e., ${\Phi}'=0$), Eq. \eqref{eq:deltaevolution} reduces to 
\be 
\delta'_m=-kq=-\theta.
\ee
Next, taking its derivative and using Eq. \eqref{eq:qevolution} as well as the two Poisson equations \eqref{eq:poisson_1} and \eqref{eq:poisson_2}, we write 
\be 
{\delta}_m''+\mathcal{H}{\delta}_m'-4 \pi G a^2 \bar\rho \delta_m=0. 
\label{eq:growthTau}
\ee
In cosmic time, this reads,
\be 
\ddot{\delta}_m+2{H}\dot {\delta}_m-4 \pi G \bar\rho \delta_m=0. 
\label{eq:growthCosmic}
\ee

\begin{figure*}[!ht]
\begin{center}
	\includegraphics[width=0.7\textwidth]{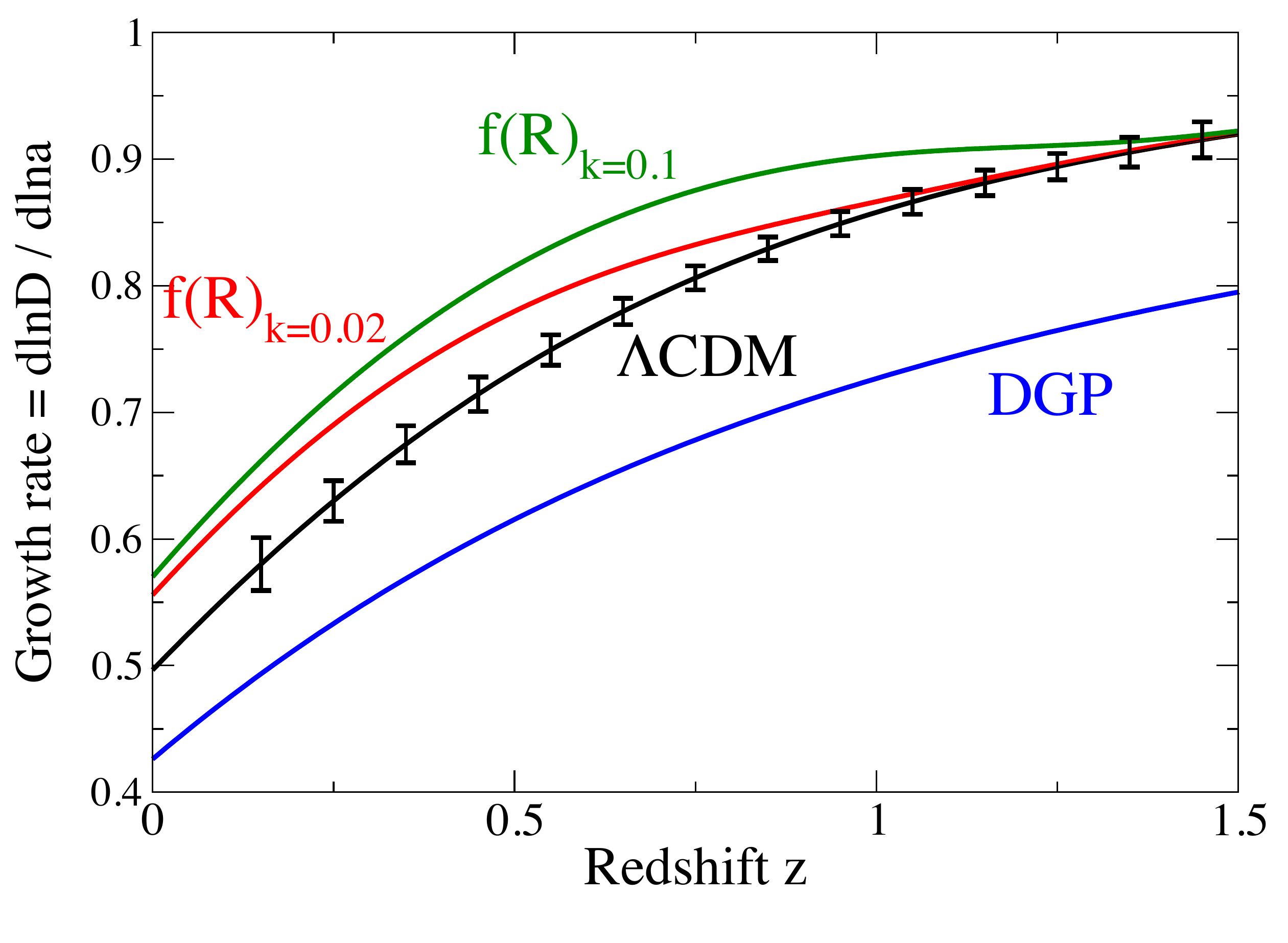}
%	\vspace{-0.2in}
	\caption{Growth rate of matter density fluctuations, $f(z)$. Theory prediction 
curves are shown for: the $\Lambda$CDM model; the Dvali--Gabadadze--Porrati 
braneworld model (the self-accelerating branch, see Sect.~\ref{sec:DGP}) \citep{DGP}; and the $f(R)$ (see Sect.~\ref{sec:f(R)}) modified gravity model \citep{HSFR2007} (model with $c=3$ from \cite{Linder2009}). Note that the growth in $f(R)$  models is scale-dependent so the authors show predictions at two wavenumbers, $k=0.02\hmpcinv$ and $k=0.1\hmpcinv$. 
Also shown are the error bars projected from a future galaxy spectroscopic 
redshift survey designed with DESI survey specifications \citep{DESI2016}. 
Image reproduced with permission from \cite{HutererEtAl2015}, copyright by Elsevier.}
	\label{fig:growth}
\end{center}
\end{figure*}

This time evolution equation for $\delta$ has a solution with decaying and growing modes. We are interested in the growing modes (denoted with a + subscript) that gave the structures that we observe today in the universe. One thus defines $D_{+}(t)$ as the linear growth factor of perturbations relating the overdensity $\delta(t)$ at some given time $t$ to its value at some initial time $t_i$.  That is 
\be
\delta(t)=\frac{D_{+}(t)}{D_{+}(t_i)} \delta(t_i),
\ee
where  ${D_{+}(t_i)}$ and $\delta(t_i)$ are constants set by initial conditions. The growth factor is often properly normalized as $G(z)\equiv D(a)/a$. 

A paramount quantity in probing the growth of large scale structure is the growth rate,  defined as the derivative of the logarithm of the growth factor with respect to the logarithm of the scale factor, i.e., 
\be
f(a)\equiv \frac{d \ln D}{d \ln a}.
\label{eq:growthrate}
\ee
As we will discuss further, some observations, such as Redshift Space Distortions (RSD), are directly sensitive to this function (or its product with the amplitude of matter fluctuation, $\sigma_8(a)$). The growth differential equation \eqref{eq:growthCosmic} above can be rewritten in terms of the growth rate \eqref{eq:growthrate} where the effect of modification to gravity can be encapsulated in an effective gravitational constant $G_{\rm eff}$ or a Modified Gravity parameter $\mu(k,a)$ (see Sect.~\ref{sec:MGparameters} further) and thus re-written as: 
\be
\label{eq:grwthfeq1}
\frac{df}{d \ln a}+f^2+\left(\frac{\dot{H}}{H^2}+2\right)f=\frac{3}{2}\frac{G_{\rm eff}^{\psi}}{G}\Omega_m \equiv \frac{3}{2}\,\mu\, \Omega_m
\ee
(for GR, $G_{\rm eff}=G$ and $\mu=1$ , recovering the standard expression). 

For illustration, we reproduce Fig.~2 from \cite{HutererEtAl2015} (Fig.~\ref{fig:growth} here) where it is shown how the function $f(z)$ can be a discriminator for various gravity theories. 

%%%%%%%%%%%%%%%%%%%%%%%%%%%%%%%%%%%%%%%%%%%%%%%%%%%%%%%%%%%%%%%%%%%%%%%%%%%%%%%%%%%
%%%%%%%%%%%%%%%%%%%%%%%%%%%%%%%%%%%%%%%%%%%%%%%%%%%%%%%%%%%%%%%%%%%%%%%%%%%%%%%%%%%
%%%%%%%%%%%%%%%%%%%%%%%%%%%%%%%%%%%%%%%%%%%%%%%%%%%%%%%%%%%%%%%%%%%%%%%%%%%%%%%%%%%
\subsubsection{Correlation function and matter power spectrum}
\label{sec:matter_spectrum}

The galaxy correlation function is a measure of the degree of clustering in a spatial or angular distribution of galaxies. If $\delta_g(\mathbf{r})$ represents the galaxy overdensity with respect to an expected mean density then the correlation function is given by the 2-point function 
\begin{equation}
  \xi(\mathbf{r_1},\mathbf{r_2})
    \equiv \langle
    \delta_g(\mathbf{r_1})\delta_g(\mathbf{r_2})
    \rangle\,,
\end{equation}
where $\langle \dots \rangle\,$ denotes the ensemble average. The galaxy correlation function can be further understood as follows \citep{Baugh2000}:
Let's consider two volume elements, $dV_1$ and $dV_2$ separated in space by $r_{12}$. The 2-point correlation can be defined as the excess probability, in comparison with a random distribution, of finding a galaxy in $dV_1$ and another in $dV_2$. That is:

\begin{equation}
  d P = \bar{n}^2\left[1+\xi(r_{12})\right]
    dV_1 dV_2\,,
  \label{eq:numberpair}
\end{equation}
where $\bar{n}$ is the mean galaxy number density. Due to the assumption of isotropy and homogeneity, the vector notation is dropped and only the distance $r_{12}$ has been kept. 

A closely related quantity is the galaxy power spectrum which is defined as the Fourier transform of the correlation function as
\begin{eqnarray}
  P_g(k) &= & 
    \int\xi(r)e^{i{\bf k\cdot r}}d^3r\,,\\  
  \xi(r) & = & 
    \int P(k)e^{-i{\bf k\cdot r}}
    \frac{d^3k}{(2\pi)^3}\,.
\end{eqnarray}

Note that we have again dropped the vector notation in the argument of $P_g(k)$ and $\xi(r)$ due to the statistical isotropy and homogeneity. In other words, they are only functions of the magnitudes of ${\bf k}$ and ${\bf r}$.
In this case, it is assumed that one of the two galaxies is at the origin and the other one is at a distance $r$.
It is worth noting that for a Gaussian random field, the power spectrum contains all the statistical information of the field which explains its wide use in cosmological studies. 

The correlation function can be measured from a galaxy survey using estimators taking into account observational subtleties \citep{LandySzalay1993}. Its theoretical counterpart is calculated from using the model predicted matter power spectrum that we discuss next. However, we use now the term matter because we refer to the dark matter field and its fluctuation, $\delta(\mathbf{k},z)$, which is traced by the galaxy fluctuation modulo some bias factor. The matter power spectrum, $P(k,z)$,  is defined by  
\be
  \langle \delta({\bf k},z)\delta({\bf k'},z)\rangle = (2\pi)^3P(k,z)\
\delta_D^3({\bf k}-{\bf k}'),
\label{eq:MPK}
\ee  
where $\delta_D^3$ is the delta function of Dirac.  $P(k,z)$ is determined from theoretical grounds as we discuss next. 

The standard picture of structure formation in the universe is that structures have grown by gravitational infall and clustering from primordial small fluctuations  in the matter density field. These seed fluctuations would have originated from microscopic quantum fluctuations that have been blown up to macroscopic scales by cosmic inflation \citep{Guth1981,Bardeen1983,AlbrechtSteinhardt1982}.
These primordial fluctuations would be scale invariant and described by the power spectrum \citep{Harrison1970,Peebles1970,Zeldovich1972}
\be 
P(k) \propto k^{n_s}.
\ee
with $n_s\approx 1$. This is consistent with current observations finding that $n_s=0.9652 \pm 0.0062$, see e.g., \cite{Planck2015,WMAP2003}.

The matter power spectrum today has evolved from this primordial spectrum while subject to a number of physical processes. During the radiation-dominated epoch, perturbations outside the horizon grow as the square of the expansion scale factor while those inside the horizon do not grow. This is due to the radiation pressure in the primordial plasma acting against gravity and preventing gravitational infall. Furthermore, as the universe expands, modes entering the horizon are also frozen.  This happens until the time of matter-radiation dominance equality where modes inside the horizon can then grow. Accordingly, the scale of the horizon at this matter-radiation equality is marked in the distribution of density fluctuations and appears as a turn-over in the shape of the matter power spectrum, see e.g., \cite{Peacock1999,dodelson2003modern}.  This and other processes about mode behaviors are formulated in the so-called transfer function, $T(k)$,  \citep{Bardeen1986,Sugiyama1995,EisensteinHu1998}. The primordial power spectrum is also enhanced by the growth factor of structure, $G(z)$ as described in Sect.~\ref{sec:growth}. In sum, the matter power spectrum today can be written as a product of the components discussed above plus a primordial amplitude determined by observations:

\be
P(k,z)=A_s\,k^{n_s}\,T^2(k)\,G^2(z).
\label{eq:matter_power_spectrum0}
\ee

In a last step, we need to connect the galaxy and matter power spectra. For that, we recall that galaxies trace the distribution of dark matter in the universe so the galaxy overdensity also traces the matter overdensity. However, this tracing is subject to some subtle galaxy bias that can be non-local and nonlinear encoding various processes and physics of structure formation, see for example discussion in  \cite{Percival2013} and references therein. On large scales, it is often assumed that one has a linear bias defined via $\delta_g(z,k)= b (z,k) \,\delta_m(z,k)$. Additionally, as we discuss in some detail in Sect.~\ref{sec:ClusteringRSD}, peculiar motion of galaxies adds distortions that can be accounted for via the factor $f(z)\mu^2$ where $\mu$ is the cosine of the angle to the line of sight. Consequently, the galaxy power spectrum can be written as 

\be
P^s_{gg}(k,\mu,z)=A_s\,k^{n_s}\,T^2(k)\,G^2(z)\left[b(z,k)+f(z)\mu^2\right]^2. 
\label{eq:galaxy_spectrum}
\ee

Finally, the linear matter power spectrum above under-predicts power on small scales, and must be modified to the nonlinear matter power spectrum $P_{nl}$ to include nonlinear effects on small scales using simulations or fitting formulas for specific class of models, e.g., \cite{PD1996,Smith2003HF} for $\Lambda$CDM and \cite{Zhao2014Halot,MGCAMB2,ZhaoEtAl2009}  for $f(R)$ MG models (see Sect.~\ref{sec:f(R)}). The presence of screening mechanisms also complicates the picture for nonlinear modes in MG. 
There have been some recent interesting developments on simulation codes for MG models.  \cite{COMPARENL} (and references therein) presents a comparative analysis of MG N-Body codes. See also \cite{Cola2017MG,Cola2017MG2} where a Comoving Lagrangian Acceleration (COLA) approach was used. This last method uses fewer time-steps and resources 
and trades some accuracy at small scales to obtain more efficiency. A parameterization for modified gravity on nonlinear cosmological scales was also proposed in \cite{Lombriser2016NL}.  

Relevant to our review, deviations from general relativity can affect the transfer function $T(k)$, the growth factor $G^2(z)$, and the growth rate $f(z)$. These can be reflected on the shape and amplitude of the galaxy power spectrum as a function of redshift and scale with some degeneracies. We discuss in the next section various observational probes, surveys and techniques that constrain and connect to the galaxy power spectrum.

%%%%%%%%%%%%%%%%%%%%%%%%%%%%%%%%%%%%%%%%%%%%%%%%%%%%%%%%%%%%%%%%%%%%%%%%%%%%%%%%%%%
%%%%%%%%%%%%%%%%%%%%%%%%%%%%%%%%%%%%%%%%%%%%%%%%%%%%%%%%%%%%%%%%%%%%%%%%%%%%%%%%%%%
%%%%%%%%%%%%%%%%%%%%%%%%%%%%%%%%%%%%%%%%%%%%%%%%%%%%%%%%%%%%%%%%%%%%%%%%%%%%%%%%%%%
%%%%%%%%%%%%%%%%%%%%%%%%                                  %%%%%%%%%%%%%%%%%%%%%%%%%
%%%%%%%%%%%%%%%%%%%%%%%%  COSMOLOGICAL PROBES OF GRAVITY  %%%%%%%%%%%%%%%%%%%%%%%%%
%%%%%%%%%%%%%%%%%%%%%%%%                                  %%%%%%%%%%%%%%%%%%%%%%%%%
%%%%%%%%%%%%%%%%%%%%%%%%%%%%%%%%%%%%%%%%%%%%%%%%%%%%%%%%%%%%%%%%%%%%%%%%%%%%%%%%%%%
%%%%%%%%%%%%%%%%%%%%%%%%%%%%%%%%%%%%%%%%%%%%%%%%%%%%%%%%%%%%%%%%%%%%%%%%%%%%%%%%%%%

\section{Cosmological probes of gravity theory}
\label{sec: Probes}

A well-appreciated ``break'' that nature has given us in cosmology is that we have two  categories of measurements and probes that we can use. One category of probes constrains the expansion history and geometry of the universe via, for example, distance measurements and expansion rate. The second category constrains the growth  and history of structure formation and clustering over space and time in the universe. Not only can we combine them, we can also contrast them for consistency. Indeed, combining probes from the two categories allows one to break further degeneracies between cosmological parameters and to tighten significantly the constraints, while contrasting their constraints can reveal systematics in some data sets or the need of some extensions to the underlying model. It is worth noting that some probes are sensitive to both the expansion and the growth such as CMB and weak lensing, however, for probing modifications to GR, it is rather the growth constraints that are the most useful.  

Modifications to gravity change the Friedmann equations and the functions derived from them for distance and expansion observables. We give in Sect.~\ref{sec:MGtheories} examples for some MG models. However, as we show there as well, the modified terms in the Friedmann equations can be cast into effective dark energy density and pressure leading to an effective equation of state. A number of MG models can then have an expansion history that is indistinguishable from that of $\Lambda$CDM (or a quintessence model closed to it), thus fitting cosmological distance and expansion observations equally well with the $\Lambda$CDM. However, such models can still exhibit a growth of structure that is different from that of $\Lambda$CDM so growth data can then be used as a discriminator between the theories.  For this reason, studies testing GR at cosmological scales then focused on deviations from GR (or MG models) that can mimic well the expansion history of $\Lambda$CDM but can still be distinguished from it using the growth rate of structure. For that, most studies assume a $\Lambda$CDM (or a quintessence $w$CDM) background model and then use the growth probes to constrain any deviation from GR. It has been argued though that one should implement and use both expansion and growth explicitly modified functions for consistency. Also, the background can be used to test GR based on spatial curvature consistency, see e.g. \cite{Zolnierowski2015}.

We briefly overview various probes of gravity below and refer the reader to corresponding review articles  in each sub-section. We start with probes of cosmic geometry and expansion and then follow with various probes of the growth of large-scale structure in the universe. 

%%%%%%%%%%%%%%%%%%%%%%%%%%%%%%%%%%%%%%%%%%%%%%%%%%%%%%%%%%%%%%%%%%%%%%%%%%%%%%%%%%%
%%%%%%%%%%%%%%%%%%%%%%%                                         %%%%%%%%%%%%%%%%%%%
%%%%%%%%%%%%%%%%%%%%%%%          SUB-SECTION                    %%%%%%%%%%%%%%%%%%%
%%%%%     Probes of cosmic geometry and expansion                              %%%%
%%%%%                                                                          %%%%
%%%%%%%%%%%%%%%%%%%%%%%%%%%%%%%%%%%%%%%%%%%%%%%%%%%%%%%%%%%%%%%%%%%%%%%%%%%%%%%%%%%
%%%%%%%%%%%%%%%%%%%%%%%%%%%%%%%%%%%%%%%%%%%%%%%%%%%%%%%%%%%%%%%%%%%%%%%%%%%%%%%%%%%

\subsection{Probes of cosmic geometry and expansion}
\label{sec:geometry}

Bearing in mind the strategy described above, probes of expansion and geometry have been very useful in constraining tightly background cosmological parameters such as the density parameters, the Hubble constant, the true or effective equation of state of dark energy,  and then setting the stage for growth probes to constrain any deviation from GR at cosmological scales.

%%%%%%%%%%%%%%%%%%%%%%%%%%%%%%%%%%%%%%%%%%%%%%%%%%%%%%%%%%%%%%%%%%%%%%%%%%%%%%%%%%%
%%%%%%%%%%%%%%%%%%%%%%%%%%%%%%%%%%%%%%%%%%%%%%%%%%%%%%%%%%%%%%%%%%%%%%%%%%%%%%%%%%%
%%%%%%%%%%%%%%%%%%%%%%%%%%%%%%%%%%%%%%%%%%%%%%%%%%%%%%%%%%%%%%%%%%%%%%%%%%%%%%%%%%%
\subsubsection{Standard candles: Type Ia Supernova}
One of the first compelling evidences for cosmic acceleration came from Supernovae type Ia (SN Ia) observations \citep{RiessEtAl1998,PerlmutterEtAl1999}. After some corrections, SN Ia can be considered as good standard candles with an average absolute bolometric magnitude of $M_{B}\approx-19.3$; see for example \cite{Phillips1993}. The ratio of their apparent brightness to their intrinsic one can provide a measure of their luminosity  distance while their redshift can be measured independently from spectroscopy. The theoretical model's function $d_L(z)$ (or $m(z)$) are then fit to the data points after further corrections on the data, see for example \cite{Hamuy1996,RiessEtAl1998,PerlmutterEtAl1999} and references therein. These and other similar plots are known as the popular Hubble plots. SN Ia Hubble plots provide relative measurements of distances that can be calibrated using low redshift distance measurements such as Cepheid variable stars in the host galaxies  building a distance ladder. A more practical function to use for distance estimation in cosmological analyses is the distance modulus 

\be
\mu(z)=m(z)-M=5 \log D_L + 25,
\ee
where $M$ is an effective absolute magnitude degenerate with the Hubble constant, $H_0$ and $D_L$ is the luminosity distance in units of Mpc given, for example, for a spatially flat $\Lambda$CDM universe by  
\be 
D_L(z)=\frac{(1+z)}{H_0}\int^{z}_0\frac{dz'}{\sqrt{\Omega_m^0 (1+z')^{3}+ \Omega_{\Lambda}^0 }}.  
\ee
$D_L(z)$ for spatially curved universes follows straightforwardly from equations \eqref{eq:d_L}, \eqref{f_K}, \eqref{eq:chi} and \eqref{eq:FriedmannOmegas}.  
Supernova data combined with other distance probe data sets can put tight constraints on background cosmological parameters. For example, supernova constraints on present time density parameters $\Omega_m^0$ and $\Omega_{\Lambda}^0$  have a degeneracy direction that is orthogonal to that from CMB constraints so when combined together they provide tight constraints on these parameters, see e.g., \cite{WMAP2003}. We list here a number of projects and popular compilations of supernova data that we will refer to in this review including:  Supernova Legacy Survey (SNLS) compilation \citep{ConleyEtAl2011}; Union2.1 compilation \citep{SuzukiEtAl2012}; Joint Light Curve Analysis (JLA) constructed from SNLS, SDSS and several low-redshift SN samples, e.g., \cite{BetouleEtAl2014}; Pan-STARRS sample, e.g., \cite{RestEtAl2014}; and most recently the Pantheon Sample compiled from a number of the above and other surveys which was provided in \cite{Pan-STARRS-2017}.
 
%%%%%%%%%%%%%%%%%%%%%%%%%%%%%%%%%%%%%%%%%%%%%%%%%%%%%%%%%%%%%%%%%%%%%%%%%%%%%%%%%%%
%%%%%%%%%%%%%%%%%%%%%%%%%%%%%%%%%%%%%%%%%%%%%%%%%%%%%%%%%%%%%%%%%%%%%%%%%%%%%%%%%%%
%%%%%%%%%%%%%%%%%%%%%%%%%%%%%%%%%%%%%%%%%%%%%%%%%%%%%%%%%%%%%%%%%%%%%%%%%%%%%%%%%%%
\subsubsection{Standard rulers: Angular Distance to CMB Last Scattering Surface and Baryon Acoustic Oscillations}
\label{sec:rulers}
The very early universe was made of a hot and dense plasma of electrons, baryons, mixed with a pressure-less dark matter component. Photons were trapped with this plasma via Thompson scattering. This is sometimes referred to as the baryon-photon fluid. As the universe expanded and cooled down, electrons and protons formed neutral hydrogen atoms. This is called recombination and happened at approximately 380,000 years after the Big Bang corresponding to a redshift of about 1090 \citep{Planck2015,WMAP2003}. Shortly after that, photons decoupled from the matter and traveled freely in the universe constituting the relic background radiation that we observe today as the CMB. 

Before decoupling, the baryon-photon fluid was subject to gravitational infall toward the center of overdense regions (dominated by dark matter)  but then pushed back outward by the building pressure of the photons. This process created spherical sound oscillations in the plasma fluid traveling at a sound speed $c_s$ that depends on the baryons and photon density parameters. The largest comoving distance that such sound waves could have traveled from the Big Bang time to decoupling time is denoted here as $r_{s,com,dec}$ and can be calculated as follows
\bea 
r_{s,com,dec}&=&\int^{t_{dec}}_{0}\frac{c_sdt}{a} \nonumber \\
             &=&\frac{c}{\sqrt{3}}\int^{t_{dec}}_{0}\frac{dt}{a\sqrt{1+(3\Omega_b)/4(\Omega_{\gamma})a}}  \nonumber \\
             &=& \frac{c}{\sqrt{3}H_0}\int^{a_{dec}}_{0}\frac{da}{\sqrt{\Omega_r+a\Omega_m}\sqrt{1+(3\Omega_b)/4(\Omega_{\gamma})a}}  \nonumber 
\label{eq:r_s_com_dec}
\eea
For example, if we use the values from \cite{Planck2015} as follows: $\Omega_{b} = 0.0492$, $\Omega_m=0.3156$, $\Omega_{\gamma} = 5.45\times 10^{-5}$, $\Omega_{r} = 9.16\times 10^{-5}$ for baryon, matter, photon, and radiation (photons+neutrinos) density parameters, respectively; $H_0=67.3$ \Hu and  $z_{dec}=1090$; then Eq.~\eqref{eq:r_s_com_dec} above gives $r_{s,com,dec}=144.7$ Mpc. 

The corresponding physical scale is given by $r_{s,dec}=a_{dec}\times r_{s,com,dec}=0.133$ Mpc and is called the crossing sound horizon at time of recombination. It corresponds to the largest scale at which an acoustic oscillation can be present in the baryon-photon fluid. After decoupling, these standing acoustic waves remained imprinted in the CMB temperature maps as well as in the distribution of matter structure in the universe. It constitutes a ``standard ruler''  that can be measured in the universe while taking into account the expansion scale factor (or redshift). 

For the CMB, this standard ruler and the angular diameter distance from the observer to the CMB last scattering surface can be combined to give the angular size of the sound horizon on such a surface as 
\be 
\theta_s\approx\frac{r_s}{d_A^{sls}}. 
\ee

This angle is particularly sensitive to the density and spatial curvature parameters, thus providing a good constraints on the geometry of the universe. 
This is related to the position of the CMB acoustic peaks (e.g., $\ell \approx \pi/\theta_s$ for the first peak). Planck has put a remarkably tight constraints on this angle as $\theta_s=(1.04106 \pm 0.00031)\times 10^{-2}$ \citep{Planck2015}. 
A concise description of how the distance to last scattering using the crossing sound horizon can be found in for example \cite{Wijenayake2015} and more detail in \cite{Bond1997}.  

On the side of Baryons, part of the pattern is the presence of shells of overdense regions with comoving radius equal to the sound crossing horizon. This pattern is called the Baryon Acoustic Oscillations (BAO) and was indeed detected in various galaxy surveys as we cite further below. In BAO geometry, one is  dealing with a spherical shell of matter so one can use the standard ruler along the line of sight (longitudinal) as well as in the transverse direction.   

For the line-of-sight part, one can write from the line element of spacetime
\be 
H(z) = \frac{\delta z}{\delta\chi_{\parallel}}.
\ee
One can measure $\delta z$ from spectroscopy in the survey while $\delta \chi_{\parallel}$ is the standard ruler, so one can constrain the Hubble function $H(z)$ at some effective redshift. 

For the transverse part, one can use the small angle approximation for the angle subtended by the standard ruler  $\delta \chi_{\bot}$ as 
\be 
d_A(z)=\frac{\delta \chi_{\bot}}{\delta \theta},
\ee
where $\delta \theta$ is measured from the survey while $\delta \chi_{\bot}$  is the known standard ruler so one can derive the angular diameter distance $d_A(z)$ at the effective redshift used.  

Some analyses like \cite{Casta2009,Chuang2012} have used this approach and made very low-signal-to-noise detection because extremely large volumes are necessary for a 2D correlation function \citep{BeutlerEtAl2011}. But a number of other analyses, e.g., \cite{ColeEtAl2005,BeutlerEtAl2011,BlackeEtAl2011,AndersonEtAl2012} made much stronger detections using rather a 1D correlation function and an effective projected distance defined as  
\be 
D_V(z) \equiv \left[ (1+z)^2 d_A^2(z)\frac{cz}{H(z)} \right].
\label{eq:DV}
\ee
In such analyses, what is fit to the data is then the ratio
\be 
d_z=\frac{r_s(z_{\rm drag})}{D_V(z)},
\ee
where $r_s(z_{drag})$ is specifically the comoving crossing sound horizon when baryons became dynamically decoupled from photons. This can be understood as after photons last scattering, the baryons encountered a baryon drag epoch until redshift of about 1060 \citep{Planck2015}. Other variations or definitions of useful effective distances like \eqref{eq:DV} have been defined and used in literature \citep{Bassett2010,Aubourg2015}.

A number of measurements of BAO have been made and have become very useful in constraining the background geometry providing important complementary data to that of CMB and SN measurements. These include measurements of the BAO effective projected distance (or other measures) by for example the 
SDSS at $z_{\rm eff}=0.15$ \citep{Eisenstein2005DO,RossEtAl2014}, 
the 2-degree-Field Galaxy Survey  (2dFGRS) at $z_{\rm eff}=0.32$  \citep{ColeEtAl2005}, 
BOSS LOWZ at $z_{\rm eff}=0.32$ and CMASS at $z_{\rm eff}=0.57$ \citep{AndersonEtAl2014},
the 6dFGS measured at $z_{\rm eff}=0.106$ \citep{BeutlerEtAl2011}, 
and WiggleZ survey at $z_{\rm eff}=0.6$  \citep{BlackeEtAl2011}. 

%%%%%%%%%%%%%%%%%%%%%%%%%%%%%%%%%%%%%%%%%%%%%%%%%%%%%%%%%%%%%%%%%%%%%%%%%%%%%%%%%%%
%%%%%%%%%%%%%%%%%%%%%%%%%%%%%%%%%%%%%%%%%%%%%%%%%%%%%%%%%%%%%%%%%%%%%%%%%%%%%%%%%%%
%%%%%%%%%%%%%%%%%%%%%%%%%%%%%%%%%%%%%%%%%%%%%%%%%%%%%%%%%%%%%%%%%%%%%%%%%%%%%%%%%%%
\subsubsection{Local measurements of the Hubble constant or measurements of H(z)}

The Hubble constant, $H_0$, is one of the oldest cosmological parameters describing the rate of expansion of the Universe and entering all distance and geometry measurements of the universe.

A direct measurement of the local Hubble constant is possible using the cosmic distance ladder (e.g., \citealt{Freedman2010}). Once this local measurement is accomplished, it can serve as a prior to further cosmological analyses.  This is in particular useful if one wants to fix the background cosmology to that of a fiducial $\Lambda$CDM while allowing for the growth parameter to vary. This is useful in the case of models that can mimic a $\Lambda$CDM expansion but can still have a distinct growth rate of structure, like for example some $f(R)$ models (see Sect.~\ref{sec:f(R)}). 

Furthermore, other cosmological probes such as the CMB infer the value of the Hubble constant by assuming and using a cosmological model. Therefore the comparison of the local measurement with that of the CMB provides an important consistency test for the underlying model. This highlights the importance of such a local measurement and we report here some of the values of the local measurements of $H_0$. 

We list here some measurements of $H_0$. First, using the Hubble Space Telescope (HST) Key Project and Cepheid calibration of distances to 31 galaxies and other calibrated secondary distance indicators (Type Ia and Type II Supernovae),  \cite{FreedmanEtAl2001} reported $H_0 = 72 \pm 8$ \Hu. 
A decade later, \cite{RiessEtAl2011} used HST new camera observations of over 600 Cepheids in host galaxies of 8 Type Ia SN. This allowed the authors to calibrate the SN magnitude-redshift relation and to obtain a much more precise value of $H_0 = 73.8 \pm 2.4$ \Hu.  \cite{Efstathiou2014} used different 
 outlier rejection criteria for the Cepheids and obtains 
 $H_0 = 70.6 \pm 3.3$ \Hu. He also obtained $H_ = 72.5 \pm 2.5$ \Hu when the H-band period-luminosity relation is assumed to be independent of metallicity using other combined distance anchors.  \cite{2012Freedman-et-al-Hubble} used HST with further calibrations from the Spitzer Space Telescope to measure $H_0 = 74.3 \pm 1.5 \,\rm{(statistical)}\, \pm 2.1\, \rm{(systematic)}$ \Hu. 
Most recently,   \cite{RiessEtAl2016}, used four geometric calibration methods of Cepheids to obtain $73.24\pm1.74$ \Hu.  

It is worth noting here that a tension seems to persist between the local measurement values and the lower value obtained from Planck, i.e., $H_0=66.93 \pm 0.62$ \Hu. This tension has been the subject of numerous discussions in recent literature offering different perspectives \citep{2016-Bernal-Verde-Riess-H0,LinIshakIOI-2,Lin2017A,2018-Lukovic-etal-H0-lowz,2017-Yuting-etal-BAO-H0,2017-Haridasu-etal-BAO-H0,2018-Zhang-Huang-Li-H0,Gomez2018,2017-DES-H0}. As we discuss further below in some of the sub-sections (see e.g., Sect.~\ref{sec:constraints_on_galileons} and Sect.~\ref{sec:constraintsonNL}), some authors find that some modified gravity models reduce or alleviate the tension in the Hubble parameter, (see e.g.,  \cite{BarreiraEtAl2014,Belgacem2017NL}) 

However, other approaches have been used to determine local measurement of $H_0$. Some time ago, \citep{Gott2001} developed and used a median statistics method that provides an alternative of $\chi^2$ likelihood methods and requires fewer assumptions about the data. They found at that time a median value of $H_0=67$\Hu with $\pm 2$\Hu statistical errors (95\% CL) and $\pm 5$\Hu statistical errors (95\% CL) from  
using 331 measurements of $H_0$ from by Huchra's compilation. 
Some time later \citep{Chen2011} used the same method and the final compilation of Huchra with 553 measurements finding a median of 
$H_0 = 68 \pm 5.5$\Hu (at 95\% CL) including statistical and systematics uncertainties.
Most recently, \citep{Chen2017} used rather the Hubble function $H(z)$ with 28 measurements at intermediate redshifts $0.07\le z \le 2.3$ in order to determine the local Hubble constant, $H_0$. They find for the spatially flat and non-flat $\lcdm$  model, $H_0=68.3^{+2.7}_{-2.6}$. The authors stress that this value is consistent with the low value obtained with the previous work using the median statistics. They also note that this value is consistent with the low value measured by Planck while it includes the  high value from local measurement in the previous paragraph within the 2$\sigma$ bound. 
Further work using, $H(z)$, was carried \citep{Moresco2016,Farooq2017,Yu2018} where the authors put constraints on a cosmological deceleration-acceleration transition with various levels of confidence.  \cite{Capozziello2014} made some first developments to constrain $f(R)$ models using the cosmological deceleration-acceleration transition redshift.  They required that the model reduces to $\lcdm$ at $z=0$ but they parametrize possible departures from it at higher redshifts in terms of a two-parameter logarithmic correction. They found that the transition in this model happens at a redshift consistent with  using type Ia supernova apparent magnitude data and Hubble parameter measurements. Finally, \cite{Gomez2018} followed on the $H(z)$ approach using 
cosmic chronometers, Type Ia supernovae, Gaussian processes and a novel Weighted Polynomial Regression method to find $H_0=67.06\pm 1.68$\Hu which is in agreement with low values and in 2.71-$\sigma$ tension with the local measurement of Riess et al. They also determine a more conservative value of $H_0=68.45\pm 2.00$ which is still about 2-$\sigma$ tension with the value from Riess et al. further above. With future precise data from for example, GAIA, and other experiments, one will hopefully get to the bottom of these tensions.  
 
 %%%%%%%%%%%%%%%%%%%%%%%%%%%%%%%%%%%%%%%%%%%%%%%%%%%%%%%%%%%%%%%%%%%%%%%%%%%%%%%%%%%
%%%%%%%%%%%%%%%%%%%%%%%                                         %%%%%%%%%%%%%%%%%%%
%%%%%%%%%%%%%%%%%%%%%%%          SUB-SECTION                    %%%%%%%%%%%%%%%%%%%
%%%%%                  Weak gravitational lensing                              %%%%
%%%%%                                                                          %%%%
%%%%%%%%%%%%%%%%%%%%%%%%%%%%%%%%%%%%%%%%%%%%%%%%%%%%%%%%%%%%%%%%%%%%%%%%%%%%%%%%%%%
%%%%%%%%%%%%%%%%%%%%%%%%%%%%%%%%%%%%%%%%%%%%%%%%%%%%%%%%%%%%%%%%%%%%%%%%%%%%%%%%%%%

\subsection{Weak gravitational lensing}
\label{sec:WL}

Trajectories of photons traveling to us from remote galaxies get deflected along the line of sight by matter overdensities in the intervening medium. This is called gravitational lensing. Depending on the positions of the sources and lenses relative to the observer, these deflections can result in strong, intermediate, or weak lensing. Strong and intermediate lensing provides spectacular multiple images such as Einstein rings and crosses \citep{Cabanac2005,Belokurov2009}, giant arcs, and arclets \citep{Hennawi2008A}. Less impressive but so abundant, weak lensing consists of tiny distortions to the shapes of millions and millions of galaxies that can be accounted for using statistical techniques and turned into a powerful cumulative signal which probes the cosmology of the intervening deflector medium including any modification to gravity theory at cosmological scales. 

Weak lensing at cosmological scales, also called cosmic shear, is quantified by the shear of images that tend to transform circular shapes into elliptical ones and is represented by the complex-quantity $\gamma$, and the convergence, $\kappa$, that represents the magnification of these images. In this weak regime, the two effects are very small, of the order of a few percent at most and equal, thus used interchangeably. To linear order, the shear is a good approximation to the reduced shear that is determined from the measured shapes (ellipticies) of galaxy images and on scales typically used in weak lensing analyses to date, see e.g., reviews \cite{Bartelmann2001WL,Kilbinger2015}.

Cosmic shear surveys measure ellipticities and positions of galaxies in the sky and then build from them pairs and triplets called 2- and 3-point correlation functions that can be compared to theoretical models using the lensing power spectrum and bispectrum that are derived from the formalism as follows (we use a mixture of steps from \citealt{Kilbinger2015,2015-Troxel-Ishak-lensing}). 

The mean convergence can be written as a weighted projection of the overdensities along the line of sight 
\begin{equation}
  \kappa(\bm{\theta}) = \frac{3 H_0^2 \Omega_{\rm m}}{2 c^2} \int\limits_0^{\chi_{_{\rm H}}}
  {{\rm d} \chi}
  \frac{g(\chi)}{a(\chi)} f_K(\chi) \, \delta(f_K(\chi)  \bm{\theta}, \chi),
  \label{eq:kappa_final}
\end{equation}
where, $\chi_{_{\rm H}}$ is is the comoving coordinate at the horizon, $f_K(\chi)$ is given by Eq.~\eqref{f_K}, and $g(\chi)$ is defined as 
\begin{equation}
  g(\chi) = \int\limits_\chi^{\chi_{_{\rm H}}} {\rm d} \chi^\prime \, n(\chi^\prime)
  \frac{f_K(\chi^\prime - \chi)}{f_K(\chi^\prime)},
  \label{eq:lens_eff}
\end{equation}
and represent the lensing efficiency at a distance $\chi$. The convergence 2-point correlation functions is constructed as  
\begin{equation}
\langle \kappa(\bm{\theta_1}) \kappa(\bm{\theta_2}) \rangle,
\end{equation}
where again $\langle \,\,\rangle$ denotes the ensemble average. Now, the convergence scalar field can be decomposed into multipole moments of the spherical harmonics as
\begin{equation}
\kappa(\bm{\theta})=\sum_{lm}\kappa_{lm}Y^m_l(\bm{\theta}),
\end{equation}
where
\begin{equation}
\kappa_{lm}=\int d\hat{\theta} \kappa(\bm{\theta},\chi) Y^{m*}_l(\bm{\theta}).
\end{equation}

The convergence power spectrum $P_\kappa(\ell)$ is then defined by
\begin{equation}
\left < \kappa_{lm} \kappa_{l'm'} \right >=\delta_{ll'} \delta_{mm'} P_\kappa(\ell).
\end{equation}

In the Limber approximation \citep{1953ApJ...117..134L}, it is given by \cite{Kaiser1992,Jain1997,Kaiser1998}:
\begin{equation}
  P_\kappa(\ell) = \frac 9 4 \, \Omega_{\rm m}^2 \left( \frac{H_0}{c} \right)^4
  \int_0^{\chi_{_{\rm H}}} {\rm d} \chi \,
  \frac{g^2(\chi)}{a^2(\chi)} P_\delta\left(k = \frac{\ell}{f_K(\chi)}, \chi \right),
  \label{eq:pskappa_in_limber}
\end{equation}
where $ P_\delta\left(k = \frac{\ell}{f_K(\chi)}, \chi \right)$ is the 3D nonlinear matter power spectrum (Sect.~\ref{sec:matter_spectrum}). 
  
As we discuss further below, modifications to gravity will alter the growth factor function and the matter power spectrum \eqref{eq:matter_power_spectrum0} as well as Weyl potential Eq.~\eqref{eq:PoissonModSumLR}.
A generalization of the above steps to the convergence 3-point correlation, $\langle \kappa(\bm{\theta_1}) \kappa(\bm{\theta_2}) \kappa(\bm{\theta_3}) \rangle$, provides the convergence bispectrum
\begin{align}
B_{\kappa}(\ell_1,\ell_2,\ell_3)&=\int_0^{\chi_{_{\rm H}}}d\chi\frac{W^3(\chi)}{f_K(\chi)^4(\chi)}B_{\delta}(k_1=\frac{\ell_1}{f_K(\chi)},k_2=\frac{\ell_2}{f_K(\chi)},k_3=\frac{\ell_3}{f_K(\chi)};\chi),\label{eq:bispec}
\end{align}
where we encapsulated the other factors into the  $W(\chi)$ as follows,
\begin{align}
W(\chi)&=\frac{3}{2}H_0^2\frac{\Omega_m}{a(\chi)}\int_{\chi}^{\chi_{_{\rm H}}}d\chi' n(\chi')f_K(\chi)\frac{f_K(\chi'-\chi)}{f_K(\chi')},\label{eq:weighting}
\end{align}
and $B_{\delta}(k_1=\frac{\ell_1}{f_K(\chi)},k_2=\frac{\ell_2}{f_K(\chi)},k_3=\frac{\ell_3}{f_K(\chi)};\chi)$ is the 3D matter bispectrum.

Next, we describe a few more steps on how comparison to observed ellipticities of galaxies is performed. We note that the ellipticity is also represented as a complex number field just like the shear. For a galaxy with intrinsic ellipticity $\epsilon^{\rm int}$, cosmic shear modifies this ellipticity [via combination with the reduced shear \cite{Kilbinger2015}]  such that the observed ellipticity in the weak-lensing regime is given by
\begin{equation}
 \varepsilon \approx \varepsilon^{\rm int} + \gamma .
  \label{eq:eps_gamma}
\end{equation}
If we average over a large number of galaxies, we expect the averaged first term to drop due to the assumed random intrinsic ellipticity of galaxies (any residual is usually put into a noise term) so the observed ellipticity components can be used as an estimator of the complex shear, i.e., $\gamma = \left\langle \varepsilon \right\rangle$.

Additionally, galaxies also have intrinsic alignments that provide signals contaminating the lensing signal. These intrinsic alignments are due to processes of galaxy formation in the gravitational field. They need to be isolated and mitigated for weak lensing to reach its full potential. See the following reviews for this topic \citep{2015-Troxel-Ishak-lensing,2015-KirK-etal-Galaxy-aligments}

\begin{figure*}[!th]
	\begin{center}
	\includegraphics[width=\textwidth]{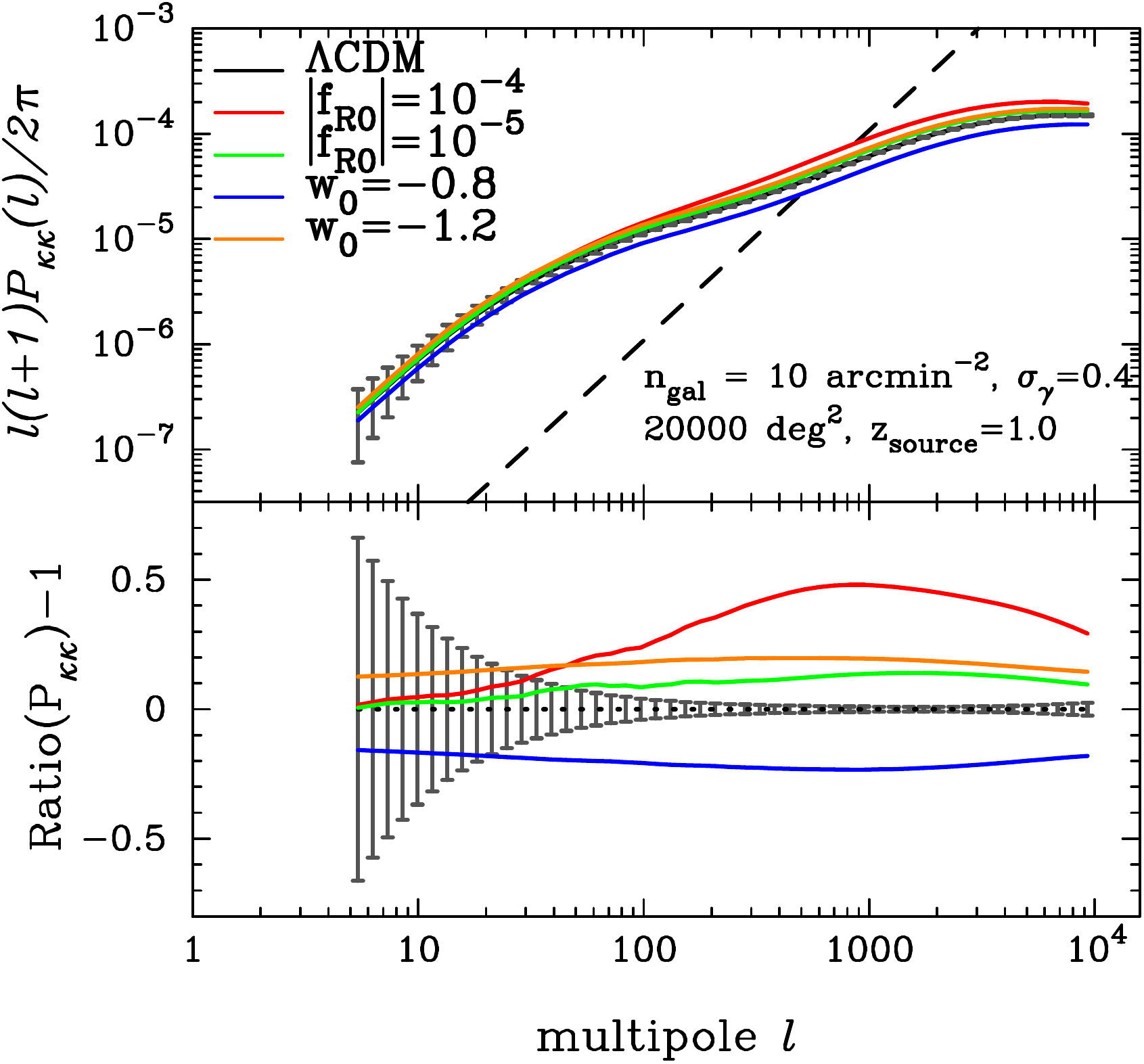}
	\caption{
	Top panel: Convergence power spectra for $f(R)$ models (see Sect.~\ref{sec:f(R)}), dynamical dark energy models and the $\Lambda$CDM standard model. Error bars are for the survey indicated on the figure -- sky coverage of 20,000 square degrees with a galaxy density number of 10 per arcminutes squared. The dashed line corresponds to the shot noise term of auto power spectrum. 
	Bottom panel: Ratio between $\Lambda$CDM model and  $f(R)$ models or $w$CDM models. Image reproduced with permission from \cite{ShirasakiEtAl2016}, copyright by the authors. 
	\label{fig:stat_comp_model}
	}
	\end{center}
\end{figure*}
	 
In practice, the two components of the shear can be identified as a tangential component with respect to the 1-axis, i.e., $\gamma_t=-\gamma_1$, and a cross-component, i.e., $\gamma_\times=-\gamma_2$, obtained by a rotation of an angle $+\pi/4$ from the tangential component. These components are used to build 2-point correlators that can be combined to construct two practical and often-used 2-point correlations from observations as follows   \citep{Miralda1991_1}, 
\begin{eqnarray}
  \xi_+(\theta) 
  & = \langle \gamma_{\rm t} \gamma_{\rm t} \rangle(\theta) + \langle \gamma_\times \gamma_\times \rangle(\theta); \quad
  \nonumber \\
  \xi_-(\theta)
  & = \langle \gamma_{\rm t} \gamma_{\rm t} \rangle(\theta) - \langle \gamma_\times \gamma_\times \rangle(\theta) . 
  \label{eq:xi_pm}
\end{eqnarray}
The explicit corresponding weighted estimators from ellipticities can be found in for example \cite{Kilbinger2015}.

Finally, in order to compare the correlation functions above to their theoretical counterparts, the shear 2-point correlations are related to the convergence power spectrum as follows

 \begin{eqnarray}
  \xi_+(\theta) & = \frac {1} {2\pi} {\displaystyle \int} {\rm d} \ell \, \ell \,{\rm J}_0(\ell
   \theta) P_\kappa(\ell), 
  \nonumber \\
   \xi_-(\theta)
  &   = \frac {1} {2\pi}  {\displaystyle \int}
   {\rm d} \ell \, \ell \,{\rm J}_4(\ell \theta)
   P_\kappa(\ell), 
   \label{eq:xi_lr_pm_pkappa}
\end{eqnarray}
where $J_n(x)$ are the n-th order Bessel function of the first kind.
 
Finally, it is worth mentioning that cosmic shear analyses perform a powerful technique called tomography where the data is split into redshift bins. This strongly probes the growth rate of large scale structure. With tomography, the 2-point correlation functions between two bins i and j is specialized as 
 \begin{eqnarray}
  \xi_{\pm}^{ij}(\theta) & = \frac {1} {2\pi} {\displaystyle \int} {\rm d} \ell \, \ell \,{\rm J}_{0/4}(\ell
   \theta) P_\kappa^{ij}(\ell),
   \label{eq:xi_lr_pm_pkappa_tomo}
\end{eqnarray}
where the corresponding power spectrum is given by 
\begin{equation}
  P_\kappa^{ij}(\ell) = \frac 9 4 \, \Omega_{\rm m}^2 \left( \frac{H_0}{c} \right)^4
  \int_0^{\chi_{\rm lim}} {\rm d} \chi \,
  \frac{g^i(\chi)g^j(\chi)}{a^2(\chi)}  P_\delta\left(k = \frac{\ell}{f_K(\chi)}, \chi \right).
  \label{eq:pskappa_in_limber_tomo}
\end{equation}

Modifications to gravity are constrained by weak lensing via the growth factor function and any other changes in the matter power spectrum \eqref{eq:matter_power_spectrum0} as well as the modifications to the Weyl potential equation \eqref{eq:PoissonModSumLR}. 
The latter change is usually captured phenomenologically by the addition of the MG parameter factor, $\Sigma(k,\chi)^2$, in the integrand of equation  \eqref{eq:pskappa_in_limber_tomo}.   
This highlights the sensitivity and importance of WL surveys in testing deviations from GR.  We reproduce here the right-top panel of Fig.~1 from \cite{ShirasakiEtAl2016} (Fig.~\ref{fig:stat_comp_model} here) comparing convergence power spectra of two $f(R)$ models, two dynamical dark energy models and the standard $\Lambda$CDM model.   

Recent cosmic shear surveys have already provided us with several analyses to constrain modification to GR or some classes of MG models that we discuss further below. These include, CFHTLenS \citep{2013CFHTlens,SimpsonEtAl2013}, KIDS \citep{JoudakiEtAl2017,HildebrandtEtAl2017}, and KIDS+2dFLenS \citep{KIDS2017EG,Joudaki2018}. It is expected that LSST \url{https://www.lsst.org/}  \citep{LSSTDESC} and WFIRST \url{https://wfirst.gsfc.nasa.gov/}  \citep{WFIRST}, and Euclid \url{http://sci.esa.int/euclid/} \citep{Euclid} will be particularly effective in constraining beyond  $\Lambda$CDM model including deviations from GR and a number of classes of MG theories \citep{JenningsEtAl2012,2015-Xu-RSD-fR,2012-Kwan-etal-Gravity-RSD,2015-Tsujikawa-RSD-Horndeski,Bellini-et-al-2016,2016-Okumura-etal-RSD-MG}

%%%%%%%%%%%%%%%%%%%%%%%%%%%%%%%%%%%%%%%%%%%%%%%%%%%%%%%%%%%%%%%%%%%%%%%%%%%%%%%%%%%
%%%%%%%%%%%%%%%%%%%%%%%                                         %%%%%%%%%%%%%%%%%%%
%%%%%%%%%%%%%%%%%%%%%%%          SUB-SECTION                    %%%%%%%%%%%%%%%%%%%
%%%%%   Galaxy surveys: Clustering and Redshift Space Distortions (RSD)        %%%%
%%%%%                                                                          %%%%
%%%%%%%%%%%%%%%%%%%%%%%%%%%%%%%%%%%%%%%%%%%%%%%%%%%%%%%%%%%%%%%%%%%%%%%%%%%%%%%%%%%
%%%%%%%%%%%%%%%%%%%%%%%%%%%%%%%%%%%%%%%%%%%%%%%%%%%%%%%%%%%%%%%%%%%%%%%%%%%%%%%%%%%

\subsection{Galaxy surveys: Clustering and Redshift Space Distortions (RSD)} 

\label{sec:ClusteringRSD}

In the recent years, a wealth of cosmological information has been provided to us from  spectroscopic redshift surveys such as SDSS, BOSS, 2dF, 6dF and WiggleZ.
From galaxy redshift surveys one can measure the isotropically averaged galaxy power spectrum or the galaxy correlation function and thus put constraints on cosmological parameters as well as MG parameters and models. This can be done via constraints on various factors in the galaxy power spectrum  \eqref{eq:galaxy_spectrum} discussed in Sect.~\ref{sec:matter_spectrum}. For example, we reproduce  Fig.~2 from \cite{BarreiraEtAl2014} (see Fig.~\ref{fig:bfs_lr} here) showing in the bottom panel the data points from the SDSS-DR7 Luminous Red Galaxy host halo power spectrum of  \cite{Reid2010SDSSLRG} against Galilean MG models and $\Lambda$CDM with massive neutrinos \citep{BarreiraEtAl2014}.

Additionally, there are Lyman-$\alpha$ surveys (sub-surveys) that can determine the frequency, density and temperature of matter clouds containing neutral hydrogen between the observer and remote quasars. Each spectrum gives information about multiple structures along the line of sight and that traces the distribution and growth of matter along the line of sight, see for example \cite{Weinberg2003,McDonald2006,Font-Ribera2013}. 
 
In regards to testing deviations from GR using galaxy redshift surveys, it seems that  ``the good comes from the bad''. Indeed, observations along the line of sight are also subject to distortions due to the fact that we make measurements in the redshift space and then convert them to the real space. It turns out that these distortions are a rich source of cosmological information which has at its forefront the redshift space distortions (RSD) that are very sensitive to the growth rate of structure and the gravity theory governing such a growth. We briefly describe below some aspects of the RSD formalism and refer the reader to specialized reviews on the topic  \citep{SamushiaEtAl2014,Blake2011RSD,Hamilton1998,PW2009,Percival2013} and references therein. 

Redshifts to remote cosmic objects such as galaxies are distorted by peculiar velocities of these objects with respect to the Hubble flow. These peculiar velocities follow large-scale infall of matter toward over dense regions in the cosmic web and by that they can trace the growth rate of large-scale structure. The distortions can be observed in the redshift space as two main effects. The first one is due to random peculiar velocity distribution of galaxies in clusters that produce a Doppler effect stretching out a cluster of galaxies in the radial direction on redshift maps. This radial stretching points to the observer and was dubbed by the ``fingers-of-god'' (FoG) effect, see e.g., the seminal papers by \cite{Kaiser1987,Hamilton1998}. See also earlier work by
\cite{Jackson1972}. The FoG effect happens at relatively smaller nonlinear scales. The second effect happens on larger scales where the peculiar velocities are not random but directed coherently toward the center of overdense regions (center of mass of clusters). It is a subtle blend of effects that combine to produce a flattening of the distribution on larger scales on redshift survey maps, sometimes dubbed as the ``pancakes-of-god'', see e.g., \cite{Hamilton1998,PW2009,Percival2013}. The related equations are as follows. 

A point in the redshift space can be related to the real space by 
\begin{equation}
\label{eq:stor_lr}
\mathbf{s}({\bf{r}} ) = {\bf{r}} + v_r ({\bf{r}} ) \hat {\bf{r}},
\end{equation}
where $v_r$ is the peculiar velocity projected in the radial direction. 
Next, we recall the linearized continuity equation 
\begin{equation}
\beta\delta_m + \bar{\nabla} \cdot \bar{v} = 0 \, 
\end{equation}
where $v$ is the matter velocity field, $\beta(z) \equiv f(z)/b(z)$ and $b(z)$ is the galaxy bias . 

Using the Jacobian between the redshift and real spaces, conservation of galaxy number in the two spaces,  the continuity equation and a few steps, it is straightforward to derive \citep{Kaiser1987,Hamilton1998}
\begin{equation}
\label{eq:kaiserRSD}
\delta^s_g(k) = \left( 1+\beta \mu^2 \right) \delta^r_g(k),
\end{equation}
where the $\mu$ is the cosine of the angle with the line of sight. 

Using \eqref{eq:kaiserRSD} and a linear galaxy bias, the corresponding power spectra are related as follows 
\begin{eqnarray}
\label{eq:kaiserRSD2}
P^s_g(k,\mu,z)&=& b(z)^2\left[ 1 + \beta(z) \mu^2 \right]^2 P_m^r(k,z)   \\
&=& \left[ b (z) + f(z) \mu^2 \right]^2 P_m^r(k,z) 
\, ,
\end{eqnarray}
where in the last line, we split $b(z)$ and $f(z)$ on purpose and note that from the matter power spectrum on the right comes its amplitude, e.g., $\sigma_8$ that is then degenerate with $f(z)$ in such a measurement.  This illustrates why RSD surveys probe $b(z)\sigma_8$ and $f(z)\sigma_8$, unless the degeneracies are broken by other means.

Equation \eqref{eq:kaiserRSD2} gives the linear RSD at large-scales\footnote{Expanding Eq.~\eqref{eq:kaiserRSD2} shows how each term relates to the respective power spectra under the assumptions of linearity in the density, velocity and galaxy bias. That is 
\be
P^{s}_{gg}(k,\mu)=P^r_{gg}(k)+2\mu^2 P^r_{g \theta}(k)+\mu^4 P^r_{\theta \theta}(k)
\label{eq:fullRSD}
\ee
where $\theta=\nabla .\textbf{v}$ is the divergence of the peculiar velocity field and where $P_{gg}(k)$, $P_{g\theta}(k)$, $P_{\theta\theta}(k)$, are the
galaxy--galaxy, galaxy--$\theta$ and $\theta$--$\theta$ power spectra
respectively.}, while the nonlinear FoG effect can be modeled by a damping factor multiplying the power spectrum and often chosen to be an exponential (Lorentzian) or Gaussian form \cite{PW2009}
\begin{equation}
F_{Lorentzian}(k,\mu^2)=[1+(k\sigma_p\mu)^2]^{-1},  
\label{eq:damping1}
\end{equation} 
\begin{equation}
F_{Gaussian}(k,\mu^2)=\exp[-(k\sigma_p\mu)^2].
\label{eq:damping2}
\end{equation} 

It is then customary to multiply Eq.~\eqref{eq:kaiserRSD2} and \eqref{eq:damping1} to combine the effect with caution though about some limitations and the need for some accurate simulations as discussed in for example \cite{PW2009}. Indeed, other combined models including contributions from nonlinear effects and numerical simulations are used to fully explore RSD modeling and observations and we refer the reader to the following RSD reviews in the literature \citep{Hamilton1998,PW2009,Percival2013} and references therein. 
 
%%%%%%%%%%%%%%%%%%%%%%%%%%%%%%%%%%%%%%%%%%%%%%%%%%%%%%%
% Figure 17 from 2016-Okumura-etal-RSD-MG
%%%%%%%%%%%%%%%%%%%%%%%%%%%%%%%%%%%%%%%%%%%%%%%%%%%%%%%
\begin{figure}[t!]
\begin{center}
\includegraphics[width=1.0\textwidth,height=3in]{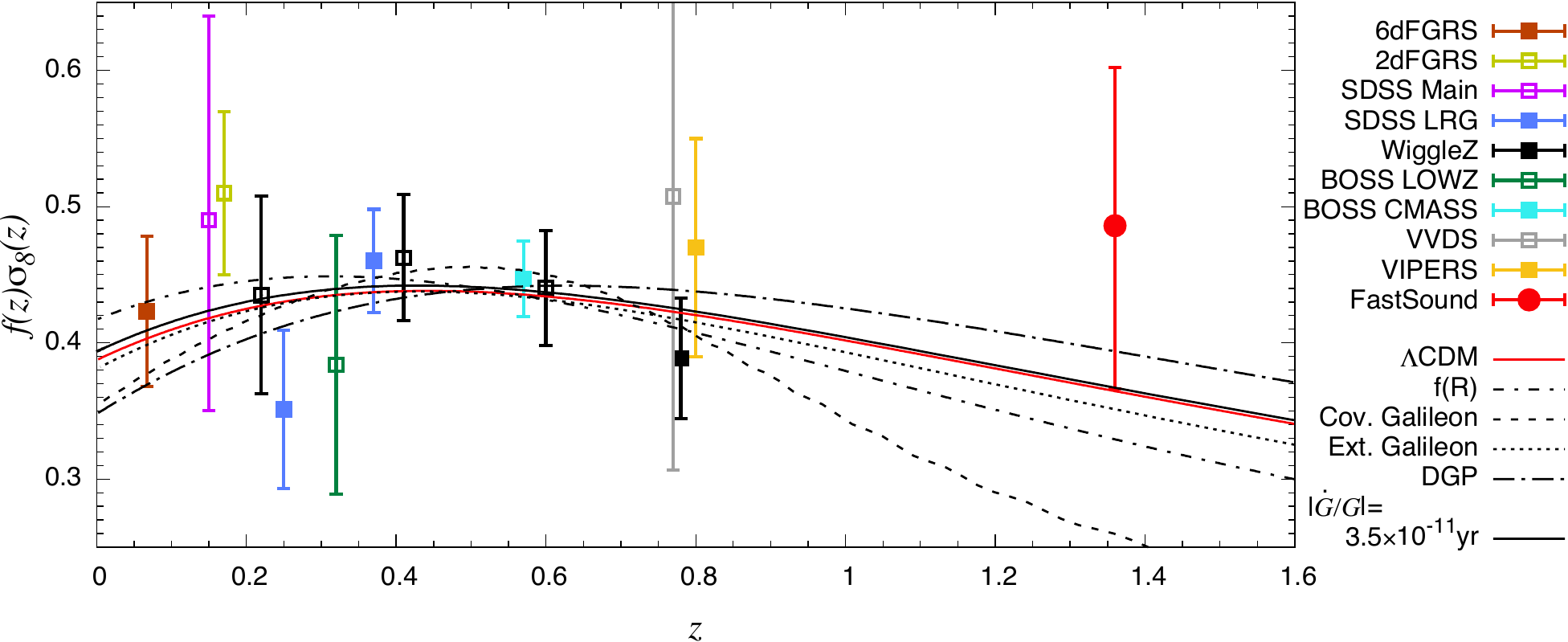}
\end{center}
 \caption{Growth rate $f(z)\sigma_8(z)$ measurements for redshift range $0<z<1.55$ and theoretical predictions from standard GR-$\Lambda$CDM model and MG models $f(R)$ (see Sect.\ref{sec:f(R)}), covariant Galileons (see Sect.~\ref{sec:Galileons}), extended Galileons, DGP (see Sect.~\ref{sec:DGP}), and models with varying gravitational constant. 
 The constraint obtained from Subaru FastSound sample at $1.19<z<1.55$  \citep{2016-Okumura-etal-RSD-MG} is plotted as the big red point.
The other results include the 6dFGS, 2dFGRS, SDSS main galaxies, SDSS LRG, BOSS LOWZ , WiggleZ, BOSS CMASS, VVDS, and VIPERS surveys at $z<1$. 
 Predicted $f\sigma_8$ from GR-$\Lambda$CDM with the amplitude determined by minimizing their $\chi^2$ is shown as the red solid line. The data points used for the $\chi^2$ minimization are denoted as the filled-symbol points. The other curves are predictions from MG models as indicated on the right. 
Image reproduced with permission from \cite{2016-Okumura-etal-RSD-MG}, copyright by the authors.
}
\label{fig:RSDMG}
\end{figure}
%%%%%%%%%%%%%%%%%%%%%%%%%%%%%%%%%%%%%%%%%%%%%%%%%%%%%%% 
 
Finally, it is worth noting that measurement of RSD are degenerate with another effect called the Alcock--Paczynski effect \citep{AP1979} which is caused by the conversion 
of angles and redshifts measured in redshift space to physical distances and Hubble function in the real space. If the theoretical cosmological model used is significantly different from the true model then further distortions are introduced in this process. 
These can be confused with the RSD effects and need to be accounted for. This results in a further multiplicative expression to Eq.~\eqref{eq:kaiserRSD2} with one or two more  parameters. See for example, treatments and discussions in \cite{Ballinger1996,SimpsonPeacock2010,SamushiaEtAl2012IL,MontanariDurrer2012}.
This is well summarized in the following equation from \cite{SKA2015}: 
\begin{eqnarray}
  P_g^{\rm s}(k',\mu',\alpha_\bot,\alpha_{||},{\bf p})=
  \frac{(b+\mu'^{2}f)^2} {\alpha_{\bot}^2\alpha_{||}} 
  P_m^{\rm r}\left[\frac{k'}{\alpha_\bot}\sqrt{1+\mu^{'2}\left(\frac{1}{F^2}-1\right)}\right],
  \label{eq:rsdap_lr}
\end{eqnarray}
\noindent
where ${\bf p}$ are the cosmological parameters of the real-space power-spectrum and the primed quantities are the observed quantities that have been introduced here to distinguish them from the real quantities as follows: $k'$ and $\mu'$ are the observed wavevector and angle; their relation to the real quantities is given by 
  $k'_{||}=\alpha_{||}k_{||}$,
  $k'_\bot=\alpha_\bot k_\bot$,
  $\mu'=\frac{k_{||}'}{\sqrt{k_{||}'+k_\bot'}}$; $F=\alpha_{||}/\alpha_\bot$, with
  $\alpha_{||}=\frac{H^{\rm fid}}{H^{\rm real}}$ and 
  $\alpha_\bot=\frac{D^{\rm real}}{D^{\rm fid}}$  the ratios of angular and radial
distances between the fiducial and real cosmological models, see \cite{SKA2015}. 

An important aspect of RSD analyses is to use measurements of the correlation function from galaxy redshift surveys and then compare them to galaxy theoretical power spectrum or its Legendre decomposition in order to estimate $f\sigma_8$ and $b\sigma_8$ at different effective redshifts. 

For our review, we stress that modifications to gravity enter into the $f(z)\sigma^8$ term in Eq. \eqref{eq:rsdap_lr} and also into the $G^2(z)$ contained in the matter power spectrum. RSD measurements are thus very important in constraining deviations from GR  affecting Poisson equation \eqref{eq:poisson_1}. While current error bars on measurements are still too large to exclude a number of contenders to GR, RSD is considered one of the most promising probes of gravity theories and has been used in a number of analysis as we discuss further below. For example, it has been shown in \cite{2013-Okada-etal-RSD-MG} that RSD can already exclude some covariant Galileon MG models (see Sect.~\ref{sec:Galileons}) to high level of confidence \citep{2013-Okada-etal-RSD-MG}. We reproduce Fig.~17 from  \cite{2016-Okumura-etal-RSD-MG} (see Fig.~\ref{fig:RSDMG}) for a number of $f\sigma_8$ measurements to date along with GR-$\Lambda$CDM and five MG models (see discussion in Sect.~\ref{sec:fsigma8constraints}.

Current RSD data include for example measurements from 6dFGS  \citep{BeutlerEtAl2012T6}, 
2dFGRS, \citep{ColeEtAl2005}, SDSS LRG \citep{SamushiaEtAl2012IL}, BOSS LOWZ \citep{Tojeiro2012TC}, BOSS CMASS \citep{AndersonEtAl2014}, VVDS \citep{GuzzoEtAl2008}, 
VIPERS \citep{delaTorreEtAl2013TV}, WiggleZ Dark Energy Survey \citep{BlakeEtAl2012,ParkinsonEtAl2012}, and Subaru FMOS galaxy redshift survey (FastSound) \citep{2016-Okumura-etal-RSD-MG}. A compilation of 34 points with corrections from model dependencies can be found in \cite{Nesseris2017RSD}. 
It is worth noting that when using $f \sigma_8$ data to constrain modified gravity models, one has to make sure no assumptions of the $\lcdm$ model are kept in the data points due to calibration using $\lcdm$ mocks. See for example the following papers that performed validation analyses of $f \sigma_8$ constraints in MG models and pointed out to possible biases \citep{Taruy2014,Barreira2016c,Bose2017}.

In addition to linear scales, RSD and velocity power spectra were shown to be a promising probe of deviations from gravity. \cite{JenningsEtAl2012} used large volume N-Body simulations to study dark matter clustering in redshift space in $f(R)$ modified gravity models (see Sect.~\ref{sec:f(R)}). The nonlinear matter and velocity fields were resolved to a high level of accuracy over a broad range of scales for f(R) models. The analysis found significant deviations from the clustering signal in GR, with an enhanced boost in power on large scales and stronger damping on small scales in the $f(R)$ models at redshifts z below 1. In particular, they found that the velocity power spectrum is a strong discriminator between $f(R)$ and GR suggesting that the extraction of the velocity power spectrum from future galaxy surveys is a promising method to constrain deviations from GR. See also \cite{Hellwing2014} on the galaxy velocity field and a signature of MG.

It is worth mentioning here that almost a decade ago RSD already attracted a lot of attention after a study in \cite{GuzzoEtAl2008} using the VIMOS-VLT Deep Survey (VVDS) measured the anisotropy parameter $\beta(z=0.77) = 0.70 \pm 0.26$, which corresponds to a growth rate of structure $f(z=0.77) = 0.91 \pm 0.36$ consistent with GR and $\Lambda$CDM, but with too large errors leaving room for other possibilities.
We present recent constraints from RSD on gravity in Sect.~\ref{sec:fsigma8constraints}. 

%%%%%%%%%%%%%%%%%%%%%%%%%%%%%%%%%%%%%%%%%%%%%%%%%%%%%%%%%%%%%%%%%%%%%%%%%%%%%%%%%%%
%%%%%%%%%%%%%%%%%%%%%%%                                         %%%%%%%%%%%%%%%%%%%
%%%%%%%%%%%%%%%%%%%%%%%          SUB-SECTION                    %%%%%%%%%%%%%%%%%%%
%%%%%           Cosmic Microwave Background Radiation                          %%%%
%%%%%                                                                          %%%%
%%%%%%%%%%%%%%%%%%%%%%%%%%%%%%%%%%%%%%%%%%%%%%%%%%%%%%%%%%%%%%%%%%%%%%%%%%%%%%%%%%%
%%%%%%%%%%%%%%%%%%%%%%%%%%%%%%%%%%%%%%%%%%%%%%%%%%%%%%%%%%%%%%%%%%%%%%%%%%%%%%%%%%%

\subsection{Cosmic Microwave Background Radiation} 
\label{sec: CMB}

This relic radiation that we call the CMB is among the most powerful cosmological probes. Not only does it constrain the background geometry (as discussed in Sect.~\ref{sec:rulers}) but it also constrains the growth of structure in the universe. The information in the CMB is expressed into temperature and polarization power spectra. These spectra have primary anisotropies that were imprinted at the surface of last scattering and also secondary anisotropies that happen later while the CMB photons are traveling in the intervening medium.  

CMB spectra provide via their primary anisotropies a powerful probe of the early universe to constrain cosmological parameters. It is complementary to other geometry probes such as supernova and BAO that probe the later times. CMB by itself can already tightly constrain background parameters such as the Hubble constant, the matter density and the effective dark energy density parameters. In combination with other probes, it can also tightly constrain an effective dark energy equation of state. 

Most relevant to dark energy and modification to GR at cosmological scales, are the secondary anisotropies that constrain scalar mode perturbations and the growth of large-scale structure. These are the Integrated Sachs--Wolfe--Effect (ISW) that affect the spectrum at small multipoles (large angular scales) \citep{SWeffect,Kofman1985}, Lensing of the CMB \citep{BlanchardCMBL1987,ColeCMBL1989,LinderCMBL1990,SeljakCMBL1996} that affects the spectrum progressively at high multipoles (small angular scales), and the Sunyaev--Zel'dovich (SZ) effect at even higher multipoles (smaller angular scales). We review the former two effects in the next sub-sections.  

Finally, it is worth mentioning that a general practice in using CMB in analysis where geometry constraints are compared to growth constraints, the spectra are split into low and high multipoles as follows. Low multipoles ($\ell < 30$) are used to constrain the growth while the higher multipoles ($30 \le \ell \le 2508$) are more sensitive to the background geometry via the position of the acoustic peaks and are used for that.  

%%%%%%%%%%%%%%%%%%%%%%%%%%%%%%%%%%%%%%%%%%%%%%%%%%%%%%%%%%%%%%%%%%%%%%%%%%%%%%%%%%%
%%%%%%%%%%%%%%%%%%%%%%%%%%%%%%%%%%%%%%%%%%%%%%%%%%%%%%%%%%%%%%%%%%%%%%%%%%%%%%%%%%%
%%%%%%%%%%%%%%%%%%%%%%%%%%%%%%%%%%%%%%%%%%%%%%%%%%%%%%%%%%%%%%%%%%%%%%%%%%%%%%%%%%%
\subsubsection{Integrated Sachs--Wolfe (ISW) effect}
\label{sec: ISW}
The Integrated Sachs--Wolfe (ISW) effect is a secondary anisotropy in the CMB temperature fluctuations that is caused by time variations in the gravitational potentials \citep{SWeffect,Kofman1985,RS1968}. In this review, we focus on the late-time ISW that can be caused by a Dark Energy component or a modification to gravity that can effect the evolution of the potentials associated with large-scale structures and voids. Namely, CMB photons traveling to us encounter potential wells due to large structures. They gain energy while falling down the potential wells but then lose it back while climbing out of them except for a small difference left due to a stretching in the potential well caused by repulsive Dark Energy or Modified gravity that happened during the photons' journey through the potential. This results in a net gain in energy for the photons coming out of the potential's well. The opposite scenario happens to photons when they travel across large voids (potential hills) causing a net loss in their energy.  The effect is given by 
\be
\frac{\delta T}{T}(\hat{n})= -\int ^{\eta_{*}}_{\eta_{0}}d\eta\frac{\partial (\Psi+\Phi)}{\partial \eta}
\label{eq:ISW}
\ee
where $T$ is the CMB temperature, $\eta_{*}$ is the conformal time at CMB surface and  $\eta_{0}$ at the observer. We note that spatial curvature can also cause such a variation \citep{Kamion1996} but we assume here spatial flatness in accordance with current observational constraints. 

\begin{figure}
	\centering
	\includegraphics[width=0.55\textwidth]{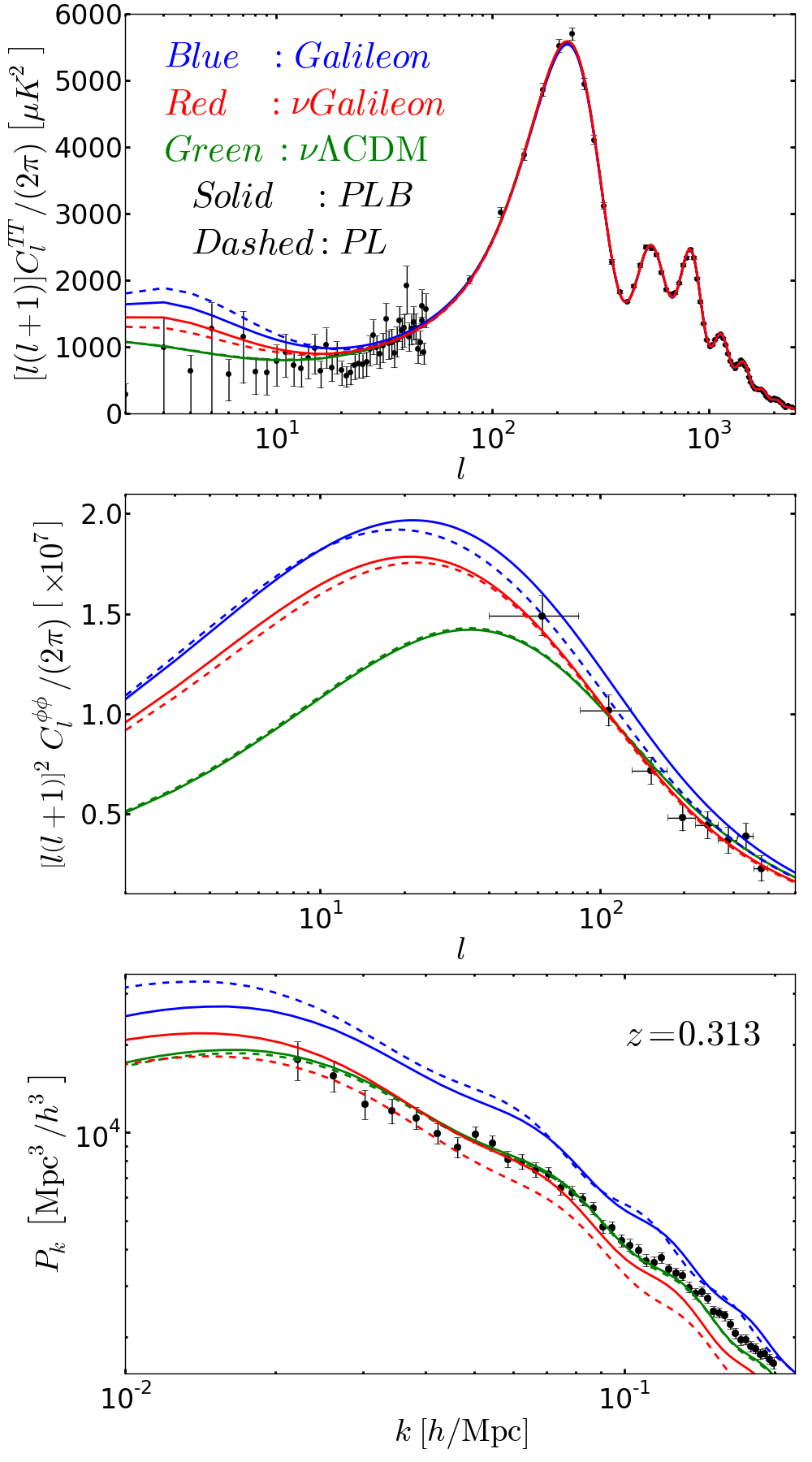}
	\caption{These plots illustrate the differences between $\lcdm$ and Galileon models (see Sect.~\ref{sec:Galileons}), with and without massive neutrinos. The Galileon models have background Friedmann equations that contain a scalar-field energy density contribution that generates late time cosmic acceleration and has an evolution consistent with observations and thus similar to that of a $\lcdm$  model. The Galileon scalar field here also affects linear perturbations and is not coupled to matter. The effect of the Galileon field considered here is focused on large-scale structure. The TOP: CMB temperature power spectra showing the ISW effect at low multipoles. MIDDLE: CMB lensing potential spectra. BOTTOM: linear matter power spectra. 
	The models plotted in dashed lines indicate their best fit models to \cite{Plankc2013Cosmo} temperature data, WMAP9 polarization data \citep{WMAP9}, and Planck-2013 CMB Lensing \citep{Planck2013WL}. They note these as PL models. The solid lines indicate their best fits to CMB data (i.e., PL) plus BAO measurements from 6dF, SDSS DR7 and BOSS DR9. They note these as PLB models. 
	The models correspond to best-fitting base Galileon modified gravity model (in blue), $\nucubic$ (in red) and $\nulcdm$ (in green). For the last two models, the authors added massive neutrino. In the upper and middle panels, the data points show the power spectrum measured by the Planck satellite \citep{Plankc2013Cosmo}. In the lower panel, the data points show the SDSS-DR7 Luminous Red Galaxy power spectrum of \cite{Reid2010SDSSLRG}, but scaled down to match the amplitude of the best-fitting $\nucubic$ (PLB) model \citep{BarreiraEtAl2014}. We refer to this figure from various parts of the text. Image reproduced with permission from \cite{BarreiraEtAl2014}, copyright by APS.} 
\label{fig:bfs_lr}\end{figure}

The ISW effect modifies the CMB temperature power spectrum at the largest angular scales with multipoles $\ell \le 10$ affecting the height of the left tail of the
 spectrum. The first detections of the ISW effect were done by cross-correlating the WMAP CMB temperature data with galaxy density surveys, see for example   \cite{BoughnEtAl2004,FosalbaEtAl2003,NoltaEtAl2004,CorasanitiEtAl2005, PadmanabhanEtAl2005,Vielva2006,Giannantonio2012} and later on by cross-correlating  Planck with large scale structure data \citep{Planck2013ISW,Planck2015ISW}. Other methods using stacking of CMB fields at coordinates coinciding with known superstructures have also led to detection, see for example \cite{GranettEtAl2008,PP2011,Planck2013ISW}. The ISW was also detected through the ISW-lensing bispectrum using Planck data only \citep{Planck2015ISW}.

By changing the gravitational potentials (as in \eqref{eq:PoissonModSumLR}) and their time evolution (growth), MG models affect the ISW and change the very-left end of the CMB power spectrum. We reproduce Fig.~2 from \cite{BarreiraEtAl2014} (see Fig.~\ref{fig:bfs_lr} here) where the top panel shows the ISW effect for various Galileon MG models (see Sect.~\ref{sec:Galileons}) and the $\Lambda$CDM model augmented by massive neutrinos. As we discuss further in Sect.~\ref{sec:constraints_on_galileons}, such an effect played a major role in ruling out the cubic Galileon models and putting very stringent constraints on the quartic and quintic ones. It is worth noting though that since the ISW effect enters only on the largest angular scales, its constraining power is limited by cosmic variance. However, cross-correlating CMB with large-scale structure tracers such as galaxies enhances its measurement significance and usefulness as we listed above.

As we describe further below, the ISW effect has been used extensively to constrain deviations from GR in conjunction with other data sets and plays a central role in obtaining such constraints. 

%%%%%%%%%%%%%%%%%%%%%%%%%%%%%%%%%%%%%%%%%%%%%%%%%%%%%%%%%%%%%%%%%%%%%%%%%%%%%%%%%%%
%%%%%%%%%%%%%%%%%%%%%%%%%%%%%%%%%%%%%%%%%%%%%%%%%%%%%%%%%%%%%%%%%%%%%%%%%%%%%%%%%%%
%%%%%%%%%%%%%%%%%%%%%%%%%%%%%%%%%%%%%%%%%%%%%%%%%%%%%%%%%%%%%%%%%%%%%%%%%%%%%%%%%%%
\subsubsection{CMB Lensing}
\label{sec: CMBLensing}
Just as in cosmic shear, CMB photons traveling to us from the surface of last scattering are subject to deflections by large-scale structure and mass concentrations along the intervening medium. These deflections change the trajectories of photons and affect the CMB temperature and polarization maps observed in the form of very small distortions  that can be statistically collected and analyzed from high-precision CMB experiments  \citep{BlanchardCMBL1987,ColeCMBL1989,LinderCMBL1990,SeljakCMBL1996}. This lensing smears out the CMB temperature power spectrum and produces non-guaussianities in the temperature and polarization maps, generating 3- and 4-point correlations \citep{Bernardeau1998,ZaldarriagaAndSeljak1999,OkamotoAndHu2003}, and converting E-mode polarization of the CMB photons into lensing B-mode \citep{ZaldarriagaAndSeljak1998}. CMB lensing and its effects have been measured by various experiments \citep{SPTCMBLensing2012,SPTCMBLensing2012b,SPTCMBLensing2012c,BOLARBEAR2014, BOLARBEAR2014b,BOLARBEAR2014c,ACTLensing2015,ACTLensing2015b,Planck2013WL,Planck2015XV}. For example, Planck-2015 measured the CMB lensing potential to an overwhelming 40-$\sigma$ confidence level \citep{Planck2015XV}. 

These deflections and the resulting observed lensed CMB are sensitive to the distribution and growth rate of large-scale structures and their associated gravitational potential. Modification to the gravitational potential due to deviations from general relativity are thus reflected on the CMB Lensing and can be used to constrain MG parameters and models.   

CMB lensing can be understood as a remapping of CMB temperature (or polarization) as follows. The lensed CMB temperature, noted as $\tT(\vnhat)$ in a direction $\vnhat$, is given by the unlensed temperature, $T(\vnhat') = T(\vnhat + \valpha)$ 
in the deflected direction $\vnhat'=\vnhat + \valpha$.  $\valpha$ is the deflection angle that is expressed at lowest order as $\valpha = \grad\psi_{_{\rm L}}$ where $\psi_{_{\rm L}}$ is the lensing potential, see e.g., \cite{LewisChallinor2006}. The latter is the result of an integration along the line of sight of the gravitational potential from the surface of last scattering all the way to us as an observer, that is
\begin{equation}
\psi_{_{\rm L}}(\vnhat) \equiv -2 \int_0^{\chi_*} \ud \chi\,
\frac{f_K(\chi_*-\chi)}{f_K(\chi_*)f_K(\chi)} \Psi_{\rm w}(\chi \vnhat; \tau_0 -\chi),  
\label{eq:psin_lr}
\end{equation}
where $\chi_*$ is the conformal distance to the surface of last scattering; $\tau_0 -\chi$ is the conformal time at which the photon was
at position $\chi \vnhat$; 
$\Psi_{\rm w}(\chi \vnhat; \tau)\equiv (\Psi+\Phi)/2$ is the Weyl gravitational potential at conformal distance  $\chi$, in direction $\vnhat$, and at conformal time $\tau$; 

Following a similar procedure as in Sect.~\ref{sec:WL}, the power spectrum of the CMB lensing potential, for a spatially flat cosmology and in the Limber approximation \citep{1953ApJ...117..134L} is given as (see, e.g., \citealt{LewisChallinor2006})
\begin{equation}
C_l^{\psi_{_{\rm L}}\psi_{_{\rm L}}} = \frac{8\pi^2}{l^3} \int_0^{\chi_*} \chi \ud \chi\,
\clp_\Psi(l/\chi;\tau_0-\chi) \left(\frac{\chi_*-\chi}{\chi_*\chi}\right)^2.
\end{equation}
The lensing potential power spectrum probes the matter power spectrum and its evolution and is thus sensitive to its amplitude, growth and how modification to GR affects these quantities. For example, it is very sensitive to modification to the second perturbed  equation \eqref{eq:poisson_2}. For example, we reproduce Fig.~2 from \cite{BarreiraEtAl2014} (Fig.~\ref{fig:bfs_lr} here) where the middle panel shows how CMB lensing power spectra for Galileon MG models (see Sect.~\ref{sec:Galileons}) versus $\Lambda$CDM model plus massive neutrinos.  

It is worth pointing out \cite{HojjatiAndLinder2016} where the authors showed that CMB Lensing will be particularly useful in constraining modified gravity models, massive neutrino models, or other new physical models that are scale dependent. Such signatures will show up in the CMB lensing power spectrum and provide an additional means to constrain MG models and other models beyond $w$CDM. 
They show that the shapes of the deviations of the CMB lensing power spectra from that of a $\Lambda$CDM model are fairly distinct between the various scale-dependent
physical origins. They  highlight the role of arcminute resolution polarization experiments such such as ACTpol, POLARBEAR/Simons Array, and SPT-3G, as well as the
next generation CMB-S4 will be able to distinguish between these models.

A number of analyses of CMB Lensing have provided already useful constraints on various cosmological parameters, see for example \cite{SPTCMBLensing2012,SPTCMBLensing2012b,SPTCMBLensing2012c,BOLARBEAR2014, BOLARBEAR2014b,BOLARBEAR2014c,ACTLensing2015,ACTLensing2015b,Planck2013WL,Planck2015XV}. 
 We will provide in Sect.~\ref{sec:constraints} further below, various constraints on deviations from GR and MG models based on CMB Lensing.

%%%%%%%%%%%%%%%%%%%%%%%%%%%%%%%%%%%%%%%%%%%%%%%%%%%%%%%%%%%%%%%%%%%%%%%%%%%%%%%%%%%
%%%%%%%%%%%%%%%%%%%%%%%%%%%%%%%%%%%%%%%%%%%%%%%%%%%%%%%%%%%%%%%%%%%%%%%%%%%%%%%%%%%
%%%%%%%%%%%%%%%%%%%%%%%%%%%%%%%%%%%%%%%%%%%%%%%%%%%%%%%%%%%%%%%%%%%%%%%%%%%%%%%%%%%
%%%%%%%%%%%%%%%%%%%%%%%%                                             %%%%%%%%%%%%%%
%%%%%%%%%%%%%%%%%%%%%%%%  FORMALISMS AND APPROACHES  TO MG PARAMS    %%%%%%%%%%%%%%
%%%%%%%%%%%%%%%%%%%%%%%%                                             %%%%%%%%%%%%%%
%%%%%%%%%%%%%%%%%%%%%%%%%%%%%%%%%%%%%%%%%%%%%%%%%%%%%%%%%%%%%%%%%%%%%%%%%%%%%%%%%%%
%%%%%%%%%%%%%%%%%%%%%%%%%%%%%%%%%%%%%%%%%%%%%%%%%%%%%%%%%%%%%%%%%%%%%%%%%%%%%%%%%%%

\section{Formalisms and approaches to testing GR at cosmological scales}

Modifications to GR at cosmological scales have been often proposed at the level of the action and its Lagrangian or at the level of the perturbed Einstein's equations. Accordingly, formalisms for deviations from GR in this context have been developed at these two levels as we discuss in the following sub-sections.  

%%%%%%%%%%%%%%%%%%%%%%%%%%%%%%%%%%%%%%%%%%%%%%%%%%%%%%%%%%%%%%%%%%%%%%%%%%%%%%%%%%%
%%%%%%%%%%%%%%%%%%%%%%%                                         %%%%%%%%%%%%%%%%%%%
%%%%%%%%%%%%%%%%%%%%%%%          SUB-SECTION                    %%%%%%%%%%%%%%%%%%%
%%%%%  Effective Field Theory (EFT) approach to dark energy and modified gravity %%
%%%%%                                                                          %%%%
%%%%%%%%%%%%%%%%%%%%%%%%%%%%%%%%%%%%%%%%%%%%%%%%%%%%%%%%%%%%%%%%%%%%%%%%%%%%%%%%%%%
%%%%%%%%%%%%%%%%%%%%%%%%%%%%%%%%%%%%%%%%%%%%%%%%%%%%%%%%%%%%%%%%%%%%%%%%%%%%%%%%%%%

\subsection{Effective Field Theory (EFT) approach to dark energy and modified gravity}
\label{sec:EFT}

The EFT approach to dark energy and modified gravity is often referred to as a ``Unified'' approach to dark energy since it includes in its action a broad spectrum of single field scalar-tensor dark energy and modified gravity models. It was applied first to inflation models using a Lagrangian derived from an EFT expansion \citep{CheungEtAl2008} and then to dark energy by for example \cite{GubitosiEtAl2013,BloomfieldEtAl2013,GleyzesEtAl2013,CreminelliEtAl2009}. 

The approach is based on constructing a Lagrangian that includes the scalar terms for a perturbed FLRW metric assuming a single field dark energy models with operators up to a given dimension and those that are invariant under spatial diffeomorphisms. This EFT formulation is also done in the unitary gauge where the foliations of constant time coincide with the hypersurfaces of uniform scalar field. This gauge allows one to write the action only in terms of the metric and its derivatives with no scalar field perturbations appearing there, however it brings limitation of a background-dependent EFT approach compared with the covariant EFT approach of, e.g.,  
\cite{Weinberg2008b,Bloomfield2012}. The action satisfying the above restrictions, that is up to quadratic order in the perturbations, and contains only operators that lead to at most second-order equations of motion, takes the following form in the Jordan frame \citep{GubitosiEtAl2013,BloomfieldEtAl2013,GleyzesEtAl2013,CreminelliEtAl2009}:
\begin{eqnarray}
\nonumber
 S &=& \int d^4x \sqrt{-g} \bigg\{ \frac{m_0^2}{2} \Omega(t) R+ \Lambda(t) - c(t) \delta g^{00} \\ \nonumber
 &+& \frac{M_2^4 (t)}{2} (\delta g^{00})^2 - \frac{\bar{M}_1^3 (t)}{2} \delta g^{00} \delta K^\mu_\mu
 - \frac{\bar{M}_2^2 (t)}{2} (\delta K^\mu_\mu)^2 \\ \nonumber
  &-& \frac{\bar{M}_3^2 (t)}{2} \delta K^i_j \delta K^j_i
    + \frac{\hat{M}^2 (t)}{2} \delta g^{00} \delta R^{(3)} \\ \nonumber
  &+& m_2^2(t)\left(g^{\mu\nu}+n^{\mu} n^{\nu}\right)\partial_{\mu}(g^{00})\partial_{\nu}(g^{00})   \bigg\} \\
 &+& S_{m} [g_{\mu \nu}, \chi_i] \ ,
\label{eq:EFT_action}
\end{eqnarray}
where $m_0^{-2} = 8\pi G$ is the reduced Planck mass;
 $\delta g^{00}$ is the perturbation of the time-time component of the inverse metric;
 $\delta {K}{^\mu_\nu}$, $\delta K$ are the perturbation of the extrinsic curvature and its trace; 
 $\delta R^{(3)}$ is the perturbation of the three dimensional spatial Ricci scalar of constant-time hypersurfaces; 
 $n^{\mu}$ is the 4-vector normal to the constant-time hypersurfaces; 
 and $S_m$ is the action for all matter fields $\chi_i$ minimally coupled to the metric $g_{\mu \nu}$.

The coefficients  $M^i_j(t)$ are functions of time and have dimensions of mass. 
The functions $c(t)$ and $\Lambda(t)$ (not to be confused with the cosmological constant) can be re-expressed in terms of the function $\Omega(t)$ and background functions such as the Hubble and density parameters by using the FLRW background evolution equations. Thus, the theories covered by action \eqref{eq:EFT_action} can be specified by the following 7 functions of time: 
\be
\{\Omega,\bar{M}_1^3,\bar{M}_2^4,\bar{M}_3^2,M_2^4,\hat{M}^2,
 m_2^2\}
 \label{eq:ListCoefEFT}
\ee
plus one function describing the background evolution such as the Hubble function. 

It is worth mentioning that the EFT approach covers both the background evolution and the linear perturbations of the metric so it provides equations and parameterization that can be compared to the background evolution as well as the growth of large-scale structure observations. However, in order to compare effectively the whole set to observations, one needs to do further useful parameterizations of the functions \eqref{eq:ListCoefEFT}. For example, for Horndeski models \citep{Horndeski1974}, these functions are mapped to the so-called $\alpha_x$ parameterization \citep{BelliniSawicki2014} which is then connected to the physical aspect of the theory as we discuss in Sect.~\ref{sec:horndeski} further below. See also another informative reconstruction of Horndeski from EFT of dark energy in \cite{Kennedy2017}. 

The EFT action \eqref{eq:EFT_action} is general enough to include broad classes of dark energy and modified gravity such as the Horndeski \citep{Horndeski1974} or generalized Galileons \citep{DeffayetEtAl2009b}, beyond Horndeski models \citep{Zuma2014,BeyondHorndeski1,BeyondHorndeski2}, Ho\v rava-Lifshitz gravity in its low energy limit \citep{Horava2009QG,KaseEFT2014}, ghost condensate models \citep{GCTheory2004}, and DGP braneworld models \citep{DGP}. We reproduce Table~I from \cite{LinderEtAl2015} (Table~\ref{tab:EFTTable} here) that shows the list of the function parameters \eqref{eq:ListCoefEFT}, the corresponding terms in the Lagrangian operators of the action  \eqref{eq:EFT_action}, and some gravity theories with the terms they involve from the EFT Lagrangian.  

While the EFT approach can be praised for its clear theoretical motivation and systematic nature, it has the disadvantage of requiring the use of a large number of parameters and functions. This number overwhelms the limited constraining power of current cosmological data. Nevertheless, some of the coefficients can be set to zero or can be shown to be interrelated in the case of some known dark energy or modified gravity models so one can reduce the number of parameters to a practical one. This of course affects the primary  motivation of the EFT approach in providing a systematics method but the hope is that as more orthogonal and precise data sets become available in the future this method will reach its aimed goals. Also, the effectiveness of the EFT approach was questioned in \cite{LinderEtAl2015} stating that the EFT functions used do not have a simple time dependence that can be fit to observations for different cosmic eras, but as they state, one can  nevertheless gain some general characteristics of such dependencies for early and late time limits of cosmic evolution. 

Most recently, \cite{Lagos2016,Lagos2018} followed on a previous effort of the  Parameterized-Post-Friedmann formalism of \cite{BFS2013PPF} in order to extend the EFT formalism to cover beyond scalar-tensor theories. The general approach they proposed recovers the standard $\alpha$-parameterization of \cite{BelliniSawicki2014} for Horndeski models (see Sect.~\ref{sec:horndeski}) but also applies to beyond-Horndeski models, vector-tensor theories, and tensor-tensor theories. In each of the more complicated theories, the formalism considers a few additional $\alpha_x$-parameters for up to 12 parameters in the most general case. We refer the reader to their papers for more information. 

Due to its broad application, the EFT approach has been implemented in several Einstein--Boltzmann solvers and Markov-Chain--Monte-Carlo codes to analyze CMB and other datasets, see for example \cite{EFTCAMB1,BelliniEtAl2017AC} and references therein, as well as our discussion in Sect.~\ref{sec:codes}.

\begin{table}[ht]
  \caption{The EFT formalism covers a number of different theories of dark energy and modified gravity. This table provides the list of coefficient functions \eqref{eq:ListCoefEFT}, related Lagrangian operators entering the theory from the action \eqref{eq:EFT_action} and some selected known cases in the literature (this table was adapted by \citealt{LinderEtAl2015} from \citealt{BloomfieldEtAl2013}). Horndeski theories are also known as generalized Galileons. Reproduced with permission from Table~I of \cite{LinderEtAl2015}, copyright by IOP. }
\label{tab:EFTTable}
  \centering
  \scriptsize
  \begin{tabular}{|l||c|c|c|c|c|c|c|c|c|}
    \hline
    \textbf{Model parameter} &
    \newline
    $\Omega$ &
    $\Lambda$ &
    $c$ &
    $M_2^4$ &
    $\bar{M}_1^3$ &
    $\bar{M}_2^2$ &
    $\bar{M}_3^2$ &
    $\hat{M}^2$ &
    $m_2^2$
    \\ \hline
    \textbf{Corresponding Operator} &
    R &
    &
    $\delta g^{00}$ &
    $(\delta g^{00})^2$ &
    $\delta g^{00} \delta {K}{^\mu_\mu}$ &
    $(\delta {K}{^\mu_\mu})^2$ &
    $\delta {K}{^\mu_\nu} {K}{^\nu_\mu}$ &
    $\delta g^{00} \delta R^{(3)}$&
    $\frac{\tilde{g}^{ij}}{a^2} \partial_i g^{00} \partial_j g^{00}$
    \\ \hline
    \hline
    $\Lambda$CDM & 1 & \checkmark & 0 & - & - & - & - & - & - \\ \hline
    Quintessence & 1/\checkmark & \checkmark & \checkmark & - & - & - & - & 
- & - \\ \hline 
    $f(R)$ & \checkmark & \checkmark & 0 & - & - & - & - & - & - \\ \hline 
    $k$-essence & 1/\checkmark & \checkmark & \checkmark & \checkmark & - & 
- & - & - & - \\ \hline 
    Galileon {(Kinetic Braiding)} & 1/\checkmark & 
\checkmark & \checkmark & \checkmark & \checkmark & - & - & - & - \\ \hline 
    DGP & \checkmark & \checkmark $\dagger$ & \checkmark $\dagger$ & 
\checkmark $\dagger$ & \checkmark & - & - & - & - \\ \hline 
    Ghost Condensate & 1/\checkmark & \checkmark & 0 & - & - & \checkmark & 
\checkmark & - & - \\ \hline 
    Horndeski  & \checkmark & 
\checkmark & \checkmark & \checkmark & \checkmark & \checkmark $\dagger$ & 
\checkmark $\dagger$ & \checkmark $\dagger$ & - \\ \hline
    Ho\v{r}ava--Lifshitz & 1 & \checkmark & 0 & - & - & \checkmark & - & - & 
\checkmark \\ \hline 
  \end{tabular}
    \begin{tabular}{p{2cm}cp{0.7 \textwidth}}
      &\checkmark & Operator is necessary\\
      &- & Operator is not included\\
      &1, 0 & Coefficient is unity or vanishes exactly\\
      &1/\checkmark & Minimally and non-minimally coupled versions of this 
model exist in the literature\\
      &$\dagger$ & Coefficients marked with a dagger are linearly related 
to other coefficients in that model by numerical coefficients 
    \end{tabular}
\end{table}

%%%%%%%%%%%%%%%%%%%%%%%%%%%%%%%%%%%%%%%%%%%%%%%%%%%%%%%%%%%%%%%%%%%%%%%%%%%%%%%%%%%
%%%%%%%%%%%%%%%%%%%%%%%                                         %%%%%%%%%%%%%%%%%%%
%%%%%%%%%%%%%%%%%%%%%%%          SUB-SECTION                    %%%%%%%%%%%%%%%%%%%
%%%%%                   Modified growth parameters                             %%%%
%%%%%                                                                          %%%%
%%%%%%%%%%%%%%%%%%%%%%%%%%%%%%%%%%%%%%%%%%%%%%%%%%%%%%%%%%%%%%%%%%%%%%%%%%%%%%%%%%%
%%%%%%%%%%%%%%%%%%%%%%%%%%%%%%%%%%%%%%%%%%%%%%%%%%%%%%%%%%%%%%%%%%%%%%%%%%%%%%%%%%%

\subsection{Modified growth parameters} 
\label{sec:MGparameters} 
We discussed in Sect.~\ref{sec:perturbations} how the growth of large scale structure can be described by the two equations \eqref{eq:poisson_1} and \eqref{eq:poisson_2} derived from  linear perturbations of the Einstein's Field Equations. Now, the effect of deviations from GR on the growth of large structure can be encapsulated in two parameters added to these equations. These are then often called the modified growth or Modified gravity (MG) equations. Usually, one of the MG parameters modifies the coupling between the gravitational potential and the energy-density source while the other parameter quantifies the difference between the two gravitational potentials. There are various related parameterizations notations and we review some of the most commonly used ones in the literature.  

One pair of such parameters is given by $Q(k,a)$ and $R(k,a)$ as follows, see e.g.,  \cite{CaldwellEtAl2007,AmendolaEtAl2008,BeanTangmatitham2010}: 
\bea
k^2{\Phi}  &=& -4\pi G a^2\sum_i \bar{\rho}_i \delta_i \,  Q(k,a)
\label{eq:PoissonModLR}\\
k^2(\Psi-R(k,a)\,\Phi) &=& -12 \pi G  a^2\sum_i \bar{\rho}_i(1+w_i)\sigma_i \, Q(k,a),
\label{eq:Mod2ndEinLR}
\eea
where each matter specie is denoted by the index $i$, $\bar{\rho}_i$ is the corresponding mass-energy density, $\delta_i$ is the rest-frame overdensity, and $\sigma_i$ is the shear stress. $Q(k,a)$ and $R(k,a)$ are scale and time dependent and both take the value of unity in GR.  

The parameter $Q(k,a)$ represents a modification to the ``Poisson equation'' \eqref{eq:poisson_1} (see comments in \citealt{ISITGR}), while the parameter $R(k,a)$ quantifies the inequality between the two potentials referred to as the gravitational slip \citep{CaldwellEtAl2007} (at late times, when anisotropic stress is negligible, Eq.\eqref{eq:Mod2ndEinLR} gives $R=\Psi/\Phi$).  \cite{CaldwellEtAl2007} noted the slip parameter as $\Psi = (1 + \varpi)\Phi$  based on a cosmological extension to the PPN formalism, see e.g., \cite{WillReview2014}.  

In order to avoid a strong degeneracy between the parameters $Q(k,a)$ and $R(k,a)$, Eqs. \eqref{eq:PoissonModLR} and \eqref{eq:Mod2ndEinLR} can be combined to introduce another MG parameter as follows (see, e.g., \citealt{AmendolaEtAl2008}):
\be
k^2(\Psi+\Phi) = -8\pi G a^2\sum_i \bar{\rho}_i \delta_i \,\Sigma(k,a)\, -12 \pi G  a^2\sum_i \bar{\rho}_i (1+w_i)\sigma_i \, Q(k,a),  
\label{eq:PoissonModSumLR}
\ee
where 
\be
\Sigma(k,a) \equiv \frac{Q(k,a)[1+R(k,a)]}{2}.
\label{eq:SigmaMG}
\ee
The parameter $\Sigma(k,a)$ enters the equation for the Weyl potential defined earlier (i.e., $\Psi_{\rm w}\equiv (\Psi+\Phi)/2$) which affects the propagation of light. The parameter is thus directly constrained by some observations such as weak gravitational lensing. Just like the parameters $Q$ and $R$, $\Sigma$ takes unity in General Relativity. 

A second pair of MG parameters often used in the literature is where a modification to Eq. \eqref{eq:poisson_1} is done indirectly by defining a modified field equation containing the parameter $\mu(k,a)$ plus a gravitational slip parameter, $\eta(k,a)$ \citep{ZhaoEtAl2010,ZhaoEtAl2009,MGCAMB2,CaldwellEtAl2007,AmendolaEtAl2008}. The modified growth equations then read:
\bea
k^2\Psi &=& -4\pi G a^2\sum_i \bar{\rho}_i \delta_i \, \mu(k,a).
\label{eq:MGmu}\\
\frac{\Phi}{\Psi} & = & \eta(k,a).
\label{eq:MGeta}
\eea
The generalization of these two equations for non-zero shear can be found in, for example, equations (13) and (14) of \cite{MGCAMB2}. Again, $\Sigma(k,a)$ is defined from their combination as 
\be
\Sigma (k,a)\equiv \frac{\mu(k,a)[1+\eta(k,a)]}{2}
\label{eq:Sigma2}
\ee
Similarly, these parameters have a scale and time dependencies and take the value of unity for GR. 

A third notation is one that associates MG parameters with effective gravitational constants in the growth equations (see, e.g., \citealt{Tsujikawa2007,SongKoyama2009,Linder2017}) so that the modified Poisson equations take the form  
\bea
k^2\Psi  &=& -4\pi G_{\rm eff}^{\Psi} a^2\sum_i \bar{\rho}_i \delta_i 
\label{eq:PoissonModG}\\
k^2(\Psi+\Phi) &=& -8\pi G_{\rm eff}^{\Psi+\Phi} a^2\sum_i \bar{\rho}_i \delta_i. 
\label{eq:PoissonModSumG}
\eea
Eq. \eqref{eq:PoissonModG} governs the coupling between the gravitational potential for non-relativistic particles to the source density fluctuation while Eq.  \eqref{eq:PoissonModSumG} governs the coupling of the gravitational potential for relativistic particles to the source density fluctuation and affects geodesics of relativistic particles such as light propagation and gravitational lensing. Often $G_{\rm eff}^{\Psi}$ is dubbed as $G_{matter}$ and $G_{\rm eff}^{\Psi+\Phi}$ as $G_{light}$.

It is worth concluding this sub-section by providing the relationships between the different parametrizations above during matter domination and  assuming zero anisotropic stress
\bea
\mu =Q R = \frac{G_{\rm eff}^{\Psi}}{G}=\frac{G_{matter}}{G} \qquad &\mbox{, }& \qquad \eta = \frac{1}{R}
\label{eq:comp1_LR}\\
\Sigma = \frac{Q(1+R)}{2} =\frac{G_{\rm eff}^{\Psi+\Phi}}{G}= \frac{G_{light}}{G} \qquad &\mbox{, }&\qquad \mu \eta = Q .
\label{eq:comp2_LR}
\eea
A more extended discussion of the relationship between MG parameters can be found in \cite{Daniel2010MG}. 

Finally, on super-horizon scales $k\ll a H$ and for adiabatic perturbations, there are further useful constraints from coordinates invariance that apply to GR and also MG theories \citep{Bertschinger2006P}. These provide a consistency relation between the two gravitational potential which reduces the two independent functions (MG parameters) above to only one parameter. The consistency relation plus the MG parameter $\eta(a)$ can be used to characterize deviation from GR at super-horizon scales. In other words, at these long wavelength, $\eta(a)$ is the only important degree of freedom for MG gravity \citep{Bertschinger2006P,BZ2008,HS2007PPF}. 

%%%%%%%%%%%%%%%%%%%%%%%%%%%%%%%%%%%%%%%%%%%%%%%%%%%%%%%%%%%%%%%%%%%%%%%%%%%%%%%%%%%
%%%%%%%%%%%%%%%%%%%%%%%                                         %%%%%%%%%%%%%%%%%%%
%%%%%%%%%%%%%%%%%%%%%%%          SUB-SECTION                    %%%%%%%%%%%%%%%%%%%
%%%%%             Evolution of MG parameters in time and scale                 %%%%
%%%%%                                                                          %%%%
%%%%%%%%%%%%%%%%%%%%%%%%%%%%%%%%%%%%%%%%%%%%%%%%%%%%%%%%%%%%%%%%%%%%%%%%%%%%%%%%%%%
%%%%%%%%%%%%%%%%%%%%%%%%%%%%%%%%%%%%%%%%%%%%%%%%%%%%%%%%%%%%%%%%%%%%%%%%%%%%%%%%%%%

\subsection{Evolution of MG parameters in time and scale}

Departures from General Relativity can evolve in time and/or scale and this has been included in parametrizations and studies. Mainly two approaches have been used in doing so. The first method employs generic functional forms while the second uses binning in redshift and scale. A third method combines the two previous ones into a hybrid method. 

\begin{itemize}
\item{\textbf{Functional forms for time and scale evolution:} For example, \cite{BeanTangmatitham2010} used:
\be
X(k,a) = \left[X_0 e^{-k/k_c}+X_\infty(1-e^{-k/k_c})-1\right]a^s +1,
\label{eq:BeanEvo}
\ee
where $X$ denotes, for example, $Q$ or $R$.  $Q_0$ and $R_0$ are the present-day asymptotic superhorizon values while $Q_\infty$ and $R_\infty$ are the present-day asymptotic subhorizon values of $Q(k,a)$ and $R(k,a)$. $k_c$ is a comoving transition scale. The time evolution is given by $a^s$. It was noted though in, for example \cite{ZhaoEtAl2010,Song2011s,DossettEtAl2011}, that such a functional exponential form causes a too strong dependence of MG parameters on the exponent $s$ and can exacerbate tensions between GR and data \citep{DossettEtAl2011}. 
It was found in these papers that a binning method in redshift avoids this problems. 
The model parameters that can be used to detect deviations from GR are now: $Q_0$, $R_0$, $Q_\infty$, $R_\infty$, $k_c$, and $s$. The parameters $s$ and $k_c$ take the values $s=0$ and $k_c=\infty$ in GR and the other parameters reduce to unity.  The constraints on $\Sigma(k,a)$ can then be derived using Eq. \eqref{eq:SigmaMG}.

In a similar way, the parameters, $\mu$ and $\eta$ have also been allowed to evolve, for example, in redshift. In \cite{DossettEtAl2011}, the two parameters have a redshift dependence transitioning to constant values below some redshift, $z_s$, and then take the GR value of unity following a hyperbolic tangent function with a transition width, $\delta z$:
\bea 
\mu(z) &=& \frac{1-\mu_{0}}{2}\Big(1 + \tanh{\frac{z-z_s}{\delta z}}\Big) + \mu_{0},
\label{eq:GBmuEvo1_LR}\\
\eta(z) &=& \frac{1-\eta_0}{2}\Big(1 + \tanh{\frac{z-z_s}{\delta z}}\Big) + \eta_{0}.
\label{eq:GBetaEvo_LR}
\eea 
The parameter $\Sigma(z)$ then follows from Eq. \eqref{eq:comp2_LR} above.  

Functional forms for MG parameters have been discussed to be less flexible than binning or hybrid methods in \cite{DossettEtAl2011,Daniel2010MG}. 
}

\item{\textbf{Time and scale binning method of MG parameters:}
An example of binning MG parameters in time (redshift) and scale is provided in   \cite{Dossett2015}. 
Two scale bins are defined as $k\le0.01\,h$ Mpc$^{-1}$ and $k>0.01\,h$ Mpc$^{-1}$. These are crossed with two other bins in redshift defined by $0<z\le1$ and $1<z\le 2$. In order to assure for the transition between the bins to be continuous and for numerical implementation stability, the following transition functions have been been used:  
\be
X(k,a) =\frac{1}{2}\big(1 + X_{z_1}(k)\big)+\frac{1}{2}\big(X_{z_2}(k) - X_{z_1}(k)\big)\tanh{\frac{z-1}{0.05}}+\frac{1}{2}\big(1 - X_{z_2}(k)\big)\tanh{\frac{z-2}{0.05}},\label{eq:ZBinEvo}
\ee
with
\bea
X_{z_1}(k) &=& \frac{1}{2}\big(X_2+X_1\big)+\frac{1}{2}\big(X_2-X_1\big)\tanh{\frac{k-0.01}{0.001}},
\label{eq:kBin} \\ \nonumber
X_{z_2}(k) &=& \frac{1}{2}\big(X_4+X_3\big)+\frac{1}{2}\big(X_4-X_3\big)\tanh{\frac{k-0.01}{0.001}},
\eea
where $X$ takes the values $Q$ or $\Sigma$ so in this parameterization a total of eight MG parameters are varied, $\Sigma_i$ and $Q_i$, $i=1,2,3,4$. Again, all these parameters take a value of unity in GR. 
}
 
\item{\textbf{Hybrid methods for MG parameters}:  Finally, the implementation of MG parameters can be optimized to take advantage of each of the two methods above. For that, hybrid methods have been employed in order to keep a functional form for the scale dependence while using bins of redshift for the time evolution as follows \citep{Dossett2015}. The redshift bins are similarly given by Eq.~\eqref{eq:ZBinEvo} above while the scale dependence is given the form:
\bea
X_{z_1}(k) &=& X_1 e^{-\frac{k}{0.01}}+X_2(1-e^{-\frac{k}{0.01}}), 
\label{eq:kHybridQ} \\ \nonumber
X_{z_2}(k) &=& X_3 e^{-\frac{k}{0.01}}+X_4(1-e^{-\frac{k}{0.01}}).
\eea
This gives again eight MG parameters, $\Sigma_i$ and $Q_i$, $i=1,2,3,4$ to be constrained by observations.}

\item{\textbf{f(R) guided time and scale parametrization}:

Guided by $f(R)$ formalism (see Sect.~\ref{sec:f(R)}), \cite{BZ2008} suggested a phenomenological time and scale parametrization as follows: 
\begin{eqnarray}
  \mu(a,k)&=&\frac{1+\alpha_1k^2a^s}{1+\alpha_2k^2a^s} \\ 
  \eta (a,k)&=&\frac{1+\beta_1k^2a^s}{1+\beta_2k^2a^s},  
  \label{eq:BZparamLR}
\end{eqnarray}
To construct such a parameterization, the authors required GR to hold at early times, so that $s>0$. They also noted that this parametrization describes $f(R)$ theories with $|f_R|\ll1$ for $\alpha_1=\frac{4}{3}\alpha_2=2\beta_1=\beta_2=4f_{RR}/a^{2+s}$. 
$(\alpha_1,\alpha_2,\beta_1,\beta_2)$ are arbitrary constants with 
$\alpha_2$ and $\beta_2$ positive so $\mu$ and $\gamma$ remains finite for all k. 
$\alpha_1$ must be positive as well to assure that $\mu$ is positive and gravity is attractive. 
}

\item{\textbf{Using rational functions of $k^2$ and five functions of time:}

\cite{Silvestri2013} showed that for local theories of gravity with one scalar degree of freedom with up to second order equation of motion and in the quasi-static approximation, the two MG parameter $\mu(k,a)$ and $\eta(k,a)$ can be written as rational functions of $k^2$ with at most 5 functions of time in all generality as follows: 
\be
\eta(a,k)=\frac{p_1(a)+p_2(a)k^2}{1+p_3(a)k^2},
\ee
\be
\mu(a,k)=\frac{1+p_3(a)k^2}{p_4(a)+p_5(a)k^2}.
\ee
They note that even if this parametrization has been derived for the quasi-linear limit, it is expected to work fine at the near- and super-horizon scales  since $\eta(a,k \rightarrow 0)=p_1(a)\ne 1$. They also note that $\mu(a,k\rightarrow 0)=1/p_4(a)\ne 1$ should be of no-consequences on observables and that super-horizon perturbations will have an evolution consistent with the background expansion \citep{Silvestri2013}. See also  discussions for this type of rational functions in \cite{deFelice2011} 
and for higher order in the wavenumber 
in \cite{Vardanyan2015}.

}
\end{itemize}

\begin{table}[t]
\caption{The layout of the binned parametrizations.  Specifically, for the first two binned methods this involves using $\{\mu_{1},\,\Sigma_{1}\}$ for the $0<z\leq1$ and  $0.0<k\leq k_x$ bin, $\{\mu_{2},\,\Sigma_{2}\}$ for the $1<z\leq 2$ and  $0.0<k\leq k_x$ bin, $\{\mu_{3},\,\Sigma_{3}\}$ for the $0<z\leq 1$ and  $k_x<k< \infty$ bin, and $\{\mu_{4},\,\Sigma_{4}\}$ for the $1<z\leq 2$ and  $k_x<k< \infty$ bin, and the third binned method uses $\{\mu_{1},\,\Sigma_{1}\}$ for the $0<z\leq1.5$ and  $0.0<k\leq k_x$ bin, $\{\mu_{2},\,\Sigma_{2}\}$ for the $1.5<z\leq 3$ and  $0.0<k\leq k_x$ bin, $\{\mu_{3},\,\Sigma_{3}\}$ for the $0<z\leq 1.5$ and  $k_x<k< \infty$ bin, and $\{\mu_{4},\,\Sigma_{4}\}$ for the $1.5<z\leq 3$ and  $k_x<k< \infty$ bin. Table reproduced with permission from \cite{Dossett2015}, copyright by APS.}
\label{tab:Grid}
\centering
\begin{tabular}{|c|c|c|}\hline 
&\multicolumn{2}{|c|}{Redshift bins}\\\hline
Scale bins & $0.0<z\leq 1,\,1.5$ & $1,\,1.5 <z \leq 2,\,3$\\\hline
$0.0 < k \leq k_x$& $\mu_{1},\,\Sigma_{1}$& $\mu_{2},\,\Sigma_{2}$\\\hline
$k_x < k< \infty$& $\mu_{3},\,\Sigma_{3}$&$\mu_{4},\,\Sigma_{4}$ \\\hline
\end{tabular}
\end{table}
%

%%%%%%%%%%%%%%%%%%%%%%%%%%%%%%%%%%%%%%%%%%%%%%%%%%%%%%%%%%%%%%%%%%%%%%%%%%%%%%%%%%%
%%%%%%%%%%%%%%%%%%%%%%%                                         %%%%%%%%%%%%%%%%%%%
%%%%%%%%%%%%%%%%%%%%%%%          SUB-SECTION                    %%%%%%%%%%%%%%%%%%%
%%%%%              The growth index parameter $\gamma$              %%%%
%%%%%                                                                          %%%%
%%%%%%%%%%%%%%%%%%%%%%%%%%%%%%%%%%%%%%%%%%%%%%%%%%%%%%%%%%%%%%%%%%%%%%%%%%%%%%%%%%%
%%%%%%%%%%%%%%%%%%%%%%%%%%%%%%%%%%%%%%%%%%%%%%%%%%%%%%%%%%%%%%%%%%%%%%%%%%%%%%%%%%%

\subsection{The growth index parameter $\gamma$} 

Another approach to use the linear growth of structure to constrain deviations from General Relativity is by defining the growth index parameter as follows. 
In some pioneering early work for a matter-dominated universe, the growth function $f$ was shown to be well-approximated by the following ansatz \citep{Peebles1980,Fry1985,Lightman1990}:
\be
f \equiv \Omega_m^\gamma
\ee
where $\gamma$ is the growth index parameter.   \cite{Peebles1980} introduced the approximation $f(z=0)\approx\Omega_0^{0.6}$ for matter dominated models. After that, \cite{Fry1985,Lightman1990} proposed more accurate approximations for such a model, i.e., $f(z=0) \approx \Omega_0^{4/7}$. 

Later on, the work was extended to dark energy models (GR-wCDM) with a slowly varying equation of state by \cite{Wang1998} deriving the following expression:
\begin{equation}
\label{eq:wcdm0LR}
\gamma(\Omega_m,w)=\frac{3(1-w)}{5-6w}+\frac{3}{125}\frac{(1-w)(1-3w/3}{(1-6w/5)^2(1-12w/5)}(1-\Omega_m)
\end{equation}
with an asymptotic early value of $\gamma_\infty^{wCDM}=3(1-w)/(5-6w)$ reducing to the well known $\Lambda$CDM model value of $\gamma^{LCDM}=\frac{6}{11}=0.545$.

\cite{Linder2005} extended this growth index approach to modified gravity theories and pointed out that it can be used as a discriminator between quintessence dark energy models and modified gravity models. For example, for the DGP model (see Sect.~\ref{sec:DGP}) has a growth index parameter of $\gamma^{DGP}=\frac{11}{16}=0.68$ \citep{Lue2004,Linder2005} and thus is clearly distinct from the value of the $\Lambda$CDM model. Indeed, despite some dispersion of $\gamma^{wCDM}$ for various values of $w$ and also some dispersion of $\gamma^{DGP}$ for various values of $\Omega_{m}(a)$, such fluctuations do not overlap and $\gamma$ remains a good discriminator for gravity theories, see e.g., \cite{LinderCahn2007,Gong2008,Polarski2008,IshakDossett2009} for spatially flat models and \cite{GongCurved2009,Mortonson2009} for curved models.  

Moreover, the growth index can be allowed to vary in redshift and provides more stringent constraints on gravity theories \citep{Polarski2008,IshakDossett2009}. For example, \cite{Polarski2008} proposed a redshift dependent parameterization of the form 
\be
\gamma(z)=\gamma_0 +  \gamma'\,z \,,
\ee
where $\gamma'\equiv \frac{d\gamma}{dz}(z=0)$. The study showed the usefulness of a variable growth index to distinguish between dark energy models and modified gravity models \citep{Polarski2008}. \cite{IshakDossett2009,WuEtAl2009} proposed a redshift dependent parameterization that covers a wide range of redshift highlighting that the sign of the slope $\gamma(z)$ can provide further discrimination between gravity theories.   

%%%%%%%%%%%%%%%%%%%%%%%%%%%%%%%%%%%%%%%%%%%%%%%%%%%%%%%%%%%%%%%%%%%%%%%%%%%%%%%%%%%
%%%%%%%%%%%%%%%%%%%%%%%                                         %%%%%%%%%%%%%%%%%%%
%%%%%%%%%%%%%%%%%%%%%%%          SUB-SECTION                    %%%%%%%%%%%%%%%%%%%
%%%%%                        The $E_G$-parameter test                          %%%%
%%%%%                                                                          %%%%
%%%%%%%%%%%%%%%%%%%%%%%%%%%%%%%%%%%%%%%%%%%%%%%%%%%%%%%%%%%%%%%%%%%%%%%%%%%%%%%%%%%
%%%%%%%%%%%%%%%%%%%%%%%%%%%%%%%%%%%%%%%%%%%%%%%%%%%%%%%%%%%%%%%%%%%%%%%%%%%%%%%%%%%

\subsection{The $E_G$-parameter test}
\label{sec:E_G}

  \cite{ZhangEtAl2007} proposed a measure they called $E_G$ to test deviations from GR's gravitational potentials in a way that is insensitive to the galaxy bias. The idea is to use a ratio of the galaxy--galaxy lensing angular cross power spectrum over the velocity--galaxy cross power spectrum. We use here a mixture of notation from  \cite{ZhangEtAl2007} and  \cite{Leonard2015} to describe this quantity. The corresponding estimator was defined in the original paper \citep{ZhangEtAl2007} as 
\begin{equation}
\hat{E}_{G}(\ell, \delta \ell)=\frac{C_{\kappa g}(\ell, \delta \ell)}{3H_{0}^2a^{-1}\sum\limits_{\alpha} j_{\alpha}(\ell, \delta \ell)P^{\alpha}_{vg}},
\label{eq:E_Zhang}
\end{equation}
where $C_{\kappa g}(\ell, \delta \ell)$ is the galaxy--galaxy lensing cross-power spectrum  in bins of $\delta \ell$; $P^{\alpha}_{vg}$ is the galaxy--velocity cross-power spectrum between $k_{\alpha}$ and $k_{\alpha+1}$; and $f_{\alpha}(\ell, \delta \ell)$ is a weighting function defined accordingly.  
The corresponding expectation value is then given by:
\begin{equation}
E_{G}(\ell)=\left[\frac{\nabla^2(\Psi+\Phi)}{3H_0^2a^{-1}f\delta_M}\right]_{k=\ell/\bar{\chi},\bar{z}}
\label{Expected_E}
\end{equation}
where $f$ is the linear growth rate of structure, $\delta_M$ is the matter overdensity field, $\bar{\chi}$ is the comoving distance corresponding to redshift $\bar{z}$. 
For GR $\Lambda$CDM, $E_G$ is independent of length scale and is given by  \cite{ZhangEtAl2007}
\be
E_{G}=\frac{\Omega_M(z=0)}{f(z)}. 
\ee
The scale independence holds for $w$CDM models with large-sound speed and negligible anisotropic stress like Quintessence. It also holds for some modified gravity models like DGP (see Sect.~\ref{sec:DGP}) but not for other MG models. The scale dependence of $E_G$ can be used as a further discriminator between MG models \citep{ZhangEtAl2007}. 

It is also worth providing a second definition of $E_G$ motivated by observations as given by \cite{ReyesEtAl2010}
\begin{equation}
E_G(R)=\frac{\Upsilon_{gm}(R)}{\beta \Upsilon_{gg}(R)} \,,
\label{eq:Alt_E_G}
\end{equation}
where $R$ is the transverse separation from the lens-galaxy; $\Upsilon_{gm}(R)$ and $\Upsilon_{gg}(R)$ are the galaxy-matter and galaxy-galaxy annular differential surface densities respectively, see e.g., \cite{Baladauf2010}. By construction, these are correlation functions that do not include any contribution from length scales smaller than some cut-off $R = R_0$.  
This second definition in Eq.~\eqref{eq:Alt_E_G} provides a ratio that is practically similar to the information content of Eq.~\eqref{eq:E_Zhang} and also factors out the galaxy bias. Most recently, \cite{Leonard2015} provided further insights on how theoretical uncertainties such as scale dependence of the bias, projection effects, and cut-off scale can affect measurements of $E_G$ using future high precision probes and the conclusions that can be drawn from them. We present further below in Sect.~\ref{sec:E_G_constraints} some constraints on the $E_G$ measure from recent data. 

We conclude this sub-section with some recent findings about the $E_G$ measure from \cite{KIDS2017EG} using the deep imaging data of KiDS with overlapping spectroscopic regions from 2dFLenS, BOSS DR12 and GAMA. The authors find that changing the metric potentials by as much as 10\% produces smaller differences in the $E_G$ predictions than changing the value of $\Omega_m^0$ between the values prefered by Planck and KiDS. They conclude that for this statistic to achieve its aim, the current tensions in cosmological parameters between Planck and large scale structure must be resolved first.

%%%%%%%%%%%%%%%%%%%%%%%%%%%%%%%%%%%%%%%%%%%%%%%%%%%%%%%%%%%%%%%%%%%%%%%%%%%%%%%%%%%
%%%%%%%%%%%%%%%%%%%%%%%                                         %%%%%%%%%%%%%%%%%%%
%%%%%%%%%%%%%%%%%%%%%%%          SUB-SECTION                    %%%%%%%%%%%%%%%%%%%
%%%%%                 Parameterized Post-Friedmann Formalism                   %%%%
%%%%%                                                                          %%%%
%%%%%%%%%%%%%%%%%%%%%%%%%%%%%%%%%%%%%%%%%%%%%%%%%%%%%%%%%%%%%%%%%%%%%%%%%%%%%%%%%%%
%%%%%%%%%%%%%%%%%%%%%%%%%%%%%%%%%%%%%%%%%%%%%%%%%%%%%%%%%%%%%%%%%%%%%%%%%%%%%%%%%%%

\subsection{Parameterized Post-Friedmann Formalism}

It appears that the parametrized post-Friedmann (PPF) formalisms at cosmological scales \citep{HS2007PPF,BFS2013PPF} has not yet reached the same popularity that its homologous, the  parameterized post-Newtonian (PPN), has received when testing GR at solar system levels or binary systems \citep{WillReview2014}.
This could be attributed perhaps to the context and the level of maturity of other methods developped to deal with the specific problems for which each formalism has been introduced. There are at least two major developments in PPF formalisms \citep{HS2007PPF,BFS2013PPF} but also a number of previous developments such as in \cite{Bertschinger2006P,CaldwellEtAl2007,Amin2008MG,2010-optimally-param-Pogosian,Baker2011}. It is also worth noting that the PPF work of \cite{BFS2013PPF} was followed by some of the same authors and others in \cite{Lagos2016,Lagos2018} where the approach was changed to an EFT one as we comment at the end of this subsection. 

While inspired by PPN, PPF needs to be formulated to account for cosmological Hubble scales where the exact form of the linearized metric is unknown and the redshift dependence must be taken into account. Therefore, PPF uses rather functions of the redshift and scale and is based on the parameterization of the perturbed field equations instead of the spacetime metric \citep{BFS2013PPF,Amendola2013}. We provide a very brief overview below and refer the reader to the original papers \citep{HS2007PPF,BFS2013PPF}. 

The first one was proposed in \cite{HS2007PPF} where the authors discuss super-horizon, quasi-static and nonlinear regimes of modified gravity with a particular attention to the transitions between them. They construct a PPF formalism for linear perturbations in MG models that joins the super-horizon regime and the sub-horizon quasi-static regime. They propose PPF functions that make the bridge between these two regimes at a scale parameterized by the Hubble length. They defined three functions and one parameter as follows: 
\begin{itemize}

\item{
The metric ratio 
\begin{equation}
g(\ln a,\kh) \equiv {\Phi -\Psi \over {\Phi} +\Psi}\,
\label{eqn:gdefinition}
\end{equation}
where $k_H \equiv k/aH$ is the wavenumber in units of the Hubble parameter.
Note that in terms of the post-Newtonian parameter $\eta = {\Phi}/\Psi$, 
$g= (\eta-1)/(\eta+1)$. 

The expansion history H and the metric ratio g define completely 
super-horizon scalar metric fluctuations for adiabatic
perturbations. 
}

\item{
The function $f_\zeta(\ln a)$ expressing the super-horizon relationship between the metric and density, see Eqs (16)-(19) in \cite{HS2007PPF}.  
As noted there, the exact form of $f_\zeta(\ln a)$
is rarely important for observable quantities. That is the case, for example, for the galaxy redshift surveys and gravitational lensing. Only observable quantities  that
depend on the comoving density scales beyond the quasi-static regime are affected by $f_\zeta(\ln a)$.  
}

\item{
The function $f_G(\ln a)$ that parameterizes a possible time-dependent modification of the Newton constant in the quasi-static regime. It is defined from the Poisson equation 
\begin{equation}
k^2 \Psi_{\rm w} = {4\pi G \over 1+f_G} a^2 \bar{\rho}_m \delta_{m} \, 
\label{eq:modpoissonLR}
\end{equation}
where $\Psi_{\rm w}$ is the Weyl potential defined earlier. 
}

\item{The parameter $c_\Gamma$ that characterizes the relationship between the transition scale and the Hubble scale. As shown in \cite{HS2007PPF}, the interpolation between the super-horizon regime and the quasi-static regime is given by  
 \begin{equation}
( 1+ c_\Gamma^2 \kh^{2})\left[ \Gamma' +\Gamma + c_\Gamma^2 \kh^{2 }\left(\Gamma-f_G\Psi_{\rm w}\right)
\right]= S \,.
\label{eq:gammaeomLR}
\end{equation}
where $\Gamma$ is added to the modified Poisson equation~\eqref{eq:modpoissonLR} in order to match the super-horizon scale behavior 
 \begin{equation}
k^2 [\Psi_{\rm w} + \Gamma] = 4\pi G a^2 \rho_m\Delta_m \,,
\label{eq:gammapoissonLR}
\end{equation}
and where $S$ is the source for the equation of motion of $\Gamma$ \citep{HS2007PPF}.
} 
\end{itemize}

For MG models affecting cosmic evolution after matter radiation equality, 
these 3 functions governing the relations for the metric, the density and the velocity, plus the usual transfer functions specify fully the linear observables of the model. 

They provided two examples, one for a $f(R)$ theory model (see Sect.~\ref{sec:f(R)}) and another for a DGP theory model (see Sect.~\ref{sec:DGP}). 
We reproduce their example for the former here. The square of the Compton length (inverse mass) in units of the Hubble length for $f(R)$ is proportional to
\begin{eqnarray}
B&\equiv& {f_{RR} \over 1+f_R} {R'}{H \over H'} \,,
\label{eq:Compton_FR}
\end{eqnarray}
where $^{\prime}=d/d\ln a$ and  $f_{RR}= d^2 f/dR^2$. The metric ratio parameter $g \rightarrow -1/3$ below the Compton length scale. They determine that 
the PPF metric ratio as $k_H\rightarrow0$ is given by 
\begin{equation} 
g(\ln a, \kh=0)= g_{\rm SH}(\ln a) = {\Phi-\Psi \over {\Phi} + \Psi}  \,.
\end{equation}
and 
\begin{equation}
f_\zeta = c_\zeta g
\label{eq:f_zeta}
\end{equation}
with $c_\zeta\approx-1/3$. 
They take for the transition to the quasi-static regime the interpolating function 
\begin{align}
 g(\ln a,k)  &= { g_{\rm SH} +  g_{\rm QS}(c_{g}\kh)^{n_{g}} \over 1+ (c_{g}\kh)^{n_{g}}}\,,
 \label{eq:g_interpolation}
\end{align}
where $g_{\rm QS}=-1/3$.   They find that $c_g=0.71 B^{1/2}$ and $n_g=2$ where they used $\Omega_m=0.24$ and $w_{\rm eff}=-1$. 

Last, they find that $f_R$ is the function that rescales the effective Newton constant and the quasi-static transition happens near the horizon scale. The two statements correspond to
\begin{equation}
f_G = f_R \,, \qquad c_\Gamma=1 \,.
\end{equation}

The second PPF formalism was proposed in \cite{BFS2013PPF} taking into account the recent exploding development in the area of dark energy and modified gravity models. 
A concise summary of the formalism was also given in \cite{Amendola2013} and we follow that presentation here. \cite{BFS2013PPF} start with scalar perturbations of the Einstein field equations of the form 
\begin{equation}
\label{abc}
\delta G_{\mu\nu} \;=\; 8\pi G\,\delta T_{\mu\nu}+\delta
U_{\mu\nu}^{\mathrm{metric}}+\delta
U_{\mu\nu}^{\mathrm{d.o.f}}+\mathrm{\ gauge\ invariance\ fixing\ terms}\,,
\end{equation}
where $\delta T_{\mu\nu}$ is the perturbed stress-energy tensor of cosmic fluids. $\delta U_{\mu\nu}^{\mathrm{metric}}$ contains new terms from metric perturbations due to modified gravity that constitute terms beyond those coming from $\delta G_{\mu\nu}$ in GR. $\delta U_{\mu\nu}^{\mathrm{d.o.f.}}$ contains terms from scalar perturbations of new degrees of freedom due to modified gravity. For example, such terms can come from perturbations of the scalar field from scalar-tensor theories or scalar modes from vector or tensor fields in MG models. 

\cite{BFS2013PPF}  then considered the expansion of $\delta U_{\mu\nu}^{\mathrm{metric}}$ in terms of two gauge-invariant perturbation variables. The first is simply the standard gauge-invariant Bardeen potentials, $\hat{\Phi}$. The second is a combination of the two Bardeen potentials as follows: $\hat\Gamma=1/k (\dot{\hat{\Phi}}+{\cal H}\hat\Psi)$. 
They provided then the equations further below where $\delta U_{\mu\nu}^{\mathrm{metric}}$ is expressed as a linear combination of $\hat{\Phi}$, $\hat\Gamma$ and their derivatives keeping the gauge-invariance of the field equations. The coefficient of such terms are then part of the PPF function set. They also expressed $\delta U_{\mu\nu}^{\mathrm{d.o.f.}}$ for the new degrees of freedom in terms of gauge-invariant potentials $\{\hat\chi_i\}$ with also coefficients providing other PPF functions. They write then the expanded four components of the perturbed field equations Eq.\,\eqref{abc}, 
%
%\begin{align}
%-a^2\delta G^0_0&=8\pi a^2 G\,%\bar{\rho}_M\delta_M+\pmb{A_0} k^2\hat{\Phi}+\pmb{F_0}k^2\hat\Gamma+\pmb{\alpha_0}k^2\hat\chi+\pmb{\alpha_1}k\dot{\hat\chi}+k^3 M_{\delta}(\dot\nu+2\epsilon)\label{eq:FE1LR}\\ 
%
%-a^2\delta G^0_i&=\nabla_i\left[8\pi a^2 G\,\bar{\rho}_M (1+\omega_M)\theta_M+\pmb{B_0} k\hat{\Phi}+\pmb{I_0}k\hat\Gamma+\pmb{\beta_0} k\hat\chi+\pmb{\beta_1}\dot{\hat\chi}+k^2M_{\Theta}(\dot\nu+2\epsilon)\right]\label{eq:FE2LR}\\
%
%a^2\delta G^i_i&=3\,8\pi a^2 G\,\bar{\rho}_M\Pi_M+\pmb{C_0} k^2\hat{\Phi}+\pmb{C_1} k\dot{\hat{\Phi}}+\pmb{J_0}k^2\hat\Gamma+\pmb{J_1} k\dot{\hat\Gamma}+\pmb{\gamma_0} k^2\hat\chi+\pmb{\gamma_1} k \dot{\hat\chi}+\pmb{\gamma_2} \ddot{\hat\chi}\\
%&\quad +k^3M_P (\dot\nu+2\epsilon)\label{eq:FE3LR}\\
%
%a^2\delta \hat{G}^i_j&=8\pi a^2 G\,\bar{\rho}_M (1+\omega_M)\Sigma_M+ \pmb{D_0}\hat{\Phi}+\frac{\pmb{D_1}}{k} \dot{\hat{\Phi}}+\pmb{K_0}\hat\Gamma+\frac{\pmb{K_1}}{k}\dot{\hat\Gamma}+\pmb{\epsilon_0}\hat\chi+\frac{\pmb{\epsilon_1}}{k}\dot{\hat\chi}+\frac{\pmb{\epsilon_2}}{k^2} \ddot{\hat\chi}  \label{eq:FE4LR}
%\end{align} 
%
where 22 PPF parameters 
%$\pmb{A_0}-\pmb{\epsilon_2}$
where used as functions of time (redshift).  

%$\delta \hat{G}^i_j=\delta G^i_j-\frac{\delta^i_j}{3}\delta G^k_k$; $\epsilon$ and $\nu$ are off-diagonal metric perturbations; $k$ is the Fourier wavenumber. Like $\pmb{A_0}-\pmb{\epsilon_2}$, the $M_\delta$, $M_\Theta$ and $M_P$ parameters are also functions of the background variables. However, the $M_i$ functions are fixed by the zeroth-order field equations, while $\pmb{A_0}-\pmb{\epsilon_2}$ are determined by the linearly perturbed field equations. The very last terms in Eqs.\,\eqref{eq:FE1LR}\,--\,\eqref{eq:FE3LR} are to ensure gauge invariance of the equations. Examples of specific functions for various MG theories can be found in \cite{BFS2013PPF}. See detailed discussions and forms for some of the functions in the extended paper of \cite{BFS2013PPF}.

The set of PPF parameters covers super-horizon and sub-horizon scales but the set simplifies significantly in the quasi-static regime reducing to what could be encapsulated in one of the pairs of parameters discussed in Sect.~\ref{sec:MGparameters}. It was argued in, for example \cite{Amendola2013}, that in such a regime, which is relevant to weak lensing surveys and galaxy surveys, such a minimal subset is more practical to compare with observation but \cite{BFS2013PPF} explains that such a PPF formalism can extend to horizon scales and can serve for comparisons to large-scale CMB modes contributions to the ISW effect and lensing-ISW cross-correlations, well beyond the quasi-static approximation \citep{Hu2013ISW,Hu2008MGCMB}. 

Most recently, some of the authors of \cite{BFS2013PPF} and others commented in \cite{Lagos2016,Lagos2018} that the expanded four components of the perturbed field equations with PPF parameters in \cite{BFS2013PPF} contain a lot of free functions because the parameterization is built directly at the level of the field equations. In other words, the coefficients PPF parameters are not all independent. To remove some of the redundancies, \cite{Lagos2016,Lagos2018} built a corresponding parametrization at the level of the action which they call the EFT of cosmological perturbations. As a result, the maximum needed number of parameters drops to 12 in this EFT parameterization compared to 22 in the EFT formalism above.  
This provides an extension to the scalar-tensor EFT approach that we discussed in Sect.~\ref{sec:EFT}.  

Finally, we conclude this section by a most recent work of \cite{Clifton2018} where the authors proposed a set of 4 parameters to model minimal deviations from GR (metric theories) that can be used to cover scales at solar systems, galactic, and cosmological scales all the way to super-horizon. Two of the parameters are the well-known effective gravitational constant ($\mu$) and the slip parameter (that they note $\zeta$). They apply consistency relations in order to connects the behavior of these parameters between small and large scales.  They show that using these conditions, $\mu$ and $\zeta$ can be expressed on small and large scales using 4 parameters $\{\alpha,\gamma,\alpha_c,\gamma_c\}$. The first two parameters are the same as the PPN parameters but allowed to vary at cosmological scales while the two other are specific to cosmological evolution and enters the two Friedmann equations. They refer to the set as PPNC.
It will be interesting to see applications of this set to currently available data.   

\subsection{Remarks on transition to nonlinear scales}
A legitimate question is to ask if the various parametrizations and approaches discussed above could deal (or be extended to deal) in some way with nonlinear scales. A related question is if any parametrizations can deal with the nonlinear  scales then can they reflect accurately any screening mechanism (see Sect.~\ref{screening}) at work in models.  

First, the phenomenological MG parameterization using $\mu$, $\eta$, $\Sigma$ and other related parameters have been proposed based on the linearly perturbed Einstein equations so they are constrained to only linear scales by construction. Most recently, \cite{Clifton2018} proposed a scheme (or parametrization) that is argued to link between MG parametrization at small scales and large scales. The idea is based on two parameters they put between quote marks as the ``slip'' and the ``effective Newton's constant'' that can be written in terms four functions of time. Two of these four functions are a direct generalization of the usual $\alpha$ and $\gamma$ parameters from PPN formalism at small scales, see e.g., \cite{WillReview2014}. This development uses concepts of averaging small scales to larger scales. This very recent proposal came in a short paper and is at a very early stage at the moment of writing this review. It will be interesting to follow further development of this work and any clarifications on how it could deal with any screening mechanisms and other relevant questions.

Second, when considering the measure $E_G$ at nonlinear scales, it was observed in \cite{Leonard2015} that there was a difference between $E_G(\ell)$ as given by Eq.~\eqref{Expected_E} and $E_G(R)$ as given by Eq.~\eqref{eq:Alt_E_G}. They state that while $E_G(\ell)$ is defined in Fourier-space and includes only linear scales, that is not necessarily the case for $E_G(R)$  which is defined in real space and scales are not separated in an easy way.
They found that the inclusion of of non-linearities in the correlation function
used into $E_G(R)$ do not cause the measure to deviate from the expected GR value at small scales. They attribute this to fact that nonlinearities enter
into $\Upsilon_{gm}(R)$ and $\Upsilon_{gg}(R)$ (i.e., the galaxy-matter and galaxy-galaxy annular differential surface densities) via the same combination of correlation function terms, so they effectively cancel out from the ratio. 
It remains an open question whether such a behavior is also expected for modified gravity models.

Third, the PPF formalism of \cite{HS2007PPF} was proposed with a prescription on how to derive the nonlinear matter power spectrum in modified gravity theories that should in principle capture the screening mechanism as well. The prescription is based on the assumption that such a nonlinear power spectrum should reduce to that of GR on small scales. The fitting formula they proposed is as follows 
\be\label{eq:ppf}
P(k,z)=\frac{P_{\rm non-GR}(k,z)+c_{\rm nl}\Sigma^2(k,z)P_{\rm GR}(k,z)}{1+c_{\rm nl}\Sigma^2(k,z)},
\ee
where $P_{\rm GR}$ is for the nonlinear power spectrum in a GR-$\Lambda$CDM model that has the same expansion history as that of the modified gravity model under consideration. $P_{\rm non-GR}$ is for the nonlinear power spectrum in this modified gravity but without the screening mechanism necessary to recover GR on small scales. In other words, the fitting formula corrects the MG power spectrum to fit GR at small scales. The weighting function, 
\be
\label{eq:SigmaW}
\Sigma^2(k,z)\equiv\frac{k^3}{2\pi^2}P_{\rm lin}(k,z),
\ee
represents the degree of nonlinearity and governs the degree of screening efficiency. $P_{\rm lin}$ is the linear power spectrum in the modified gravity model. The $c_{nl}$ are coefficient (but can also be time-dependent) to control the scale of the effect. See, e.g., \cite{HS2007PPF}. 

\cite{Koyama2009N} did further fitting using the PPF formalism with prescription   above and added an exponent $n$ on the right of Eq.~\eqref{eq:SigmaW}.  They found that $n=1$ for DGP and $n=1/3$ for $f(R)$ provide good fits to N-Body simulations of the models up to $k\sim0.5$ h/Mpc. They also determined values for $c_{nl}$ in their fitting work. \cite{Zhao2011Z} used an exponent $n$ as a function of $k$ and 3 parameters. They extended the good fit to N-body simulations up to $k=10$ h/Mpc for $f(R)$ models. These two studies and others found that the Chameleon mechanism at work was accurately reproduced by the implementation of this prescription.  

\cite{Lombriser2014c} and \cite{Lombriser2014} combined the spherical collapse model, the halo model, linear perturbation theory, quasi-nonlinear interpolation motivated by the $c_{nl} \Sigma^2(k,z)$ above and one-loop perturbations in order to derive a description of nonlinear the nonlinear matter power spectrum of f(R) gravity with chameleon screening on scales of up to $k\sim10$ h/Mpc. 
This encouraged \cite{Lombriser2016NL} to push further the method above of 
combining the perturbative approach with one-halo contributions obtained from a generalized modified spherical collapse model. The author proposed a parametrization based on the spherical collapse that enters into effect as one transitions into the deep nonlinear regime.  The formalism he proposed allows one to encode different screening mechanisms at work in scalar-tensor theories. 
This sophisticated parametrization is then combined with generalized perturbative approaches to give a formalism that constitutes a nonlinear extension to the linear PPF formalism discussed above. For a detailed description, see  \cite{Lombriser2016NL}. 

Finally, there have been some recent proposals of extending the EFT formulation of the  dark energy to nonlinear scales such as in, e.g., \cite{Cusin2018} for the Vainshtein mechanism, or to develop Post-Newtonian--Vainstein formalism that can be connected to it, see e.g., \cite{McManus2017,Bolis2018}. It was highlited in \cite{Lombriser2018a} that the EFT formulation of dark energy they explore in their paper can be connected to the nonlinear parameterization developped in \cite{Lombriser2016NL}. The topic of expanding the EFT formulation of dark energy to nonlinear regime is a subject of interest in the most recent literature and is to be followed very closely. 

%% ---> Done. A 2-page section has been added with 12 references for this.

%%%%%%%%%%%%%%%%%%%%%%%%%%%%%%%%%%%%%%%%%%%%%%%%%%%%%%%%%%%%%%%%%%%%%%%%%%%%%%%%%%%
%%%%%%%%%%%%%%%%%%%%%%%%%%%%%%%%%%%%%%%%%%%%%%%%%%%%%%%%%%%%%%%%%%%%%%%%%%%%%%%%%%%
%%%%%%%%%%%%%%%%%%%%%%%%%%%%%%%%%%%%%%%%%%%%%%%%%%%%%%%%%%%%%%%%%%%%%%%%%%%%%%%%%%%
%%%%%%%%%%%%%       CONSTRAINTS ON MG PARAMETERS                        %%%%%%%%%%
%%%%%%%%%%%%%%%%%%%%%%%%%%%%%%%%%%%%%%%%%%%%%%%%%%%%%%%%%%%%%%%%%%%%%%%%%%%%%%%%%%%
%%%%%%%%%%%%%%%%%%%%%%%%%%%%%%%%%%%%%%%%%%%%%%%%%%%%%%%%%%%%%%%%%%%%%%%%%%%%%%%%%%%
%%%%%%%%%%%%%%%%%%%%%%%%%%%%%%%%%%%%%%%%%%%%%%%%%%%%%%%%%%%%%%%%%%%%%%%%%%%%%%%%%%%

\section{Constraints and results on MG parameters (i.e., deviations from GR) from current cosmological data sets}
\label{sec:constraints}

In this section we describe current results on testing MG phenomenological parameters from cosmology. These are only a subset of selected available papers and results in the literature. We aimed here to focus on some of the recent results, or in some cases, on less recent constraints but those that helped exclude substantial regions of MG parameter spaces. We organize this section by the parameterizations described above and then by probes and surveys.

%%%%%%%%%%%%%%%%%%%%%%%%%%%%%%%%%%%%%%%%%%%%%%%%%%%%%%%%%%%%%%%%%%%%%%%%%%%%%%%%%%%
%%%%%%%%%%%%%%%%%%%%%%%                                         %%%%%%%%%%%%%%%%%%%
%%%%%%%%%%%%%%%%%%%%%%%          SUB-SECTION                    %%%%%%%%%%%%%%%%%%%
%%%%%                 Constraints on modified growth parameters                %%%%
%%%%%                                                                          %%%%
%%%%%%%%%%%%%%%%%%%%%%%%%%%%%%%%%%%%%%%%%%%%%%%%%%%%%%%%%%%%%%%%%%%%%%%%%%%%%%%%%%%
%%%%%%%%%%%%%%%%%%%%%%%%%%%%%%%%%%%%%%%%%%%%%%%%%%%%%%%%%%%%%%%%%%%%%%%%%%%%%%%%%%%

\subsection{Constraints on modified growth parameters}

%%%%%%%%%%%%%%%%%%%%%%%%%%%%%%%%%%%%%%%%%%%%%%%%%%%%%%%%%%%%%%%%%%%%%%%%%%%%%%%%%%%
%%%%%%%%%%%%%%%%%%%%%%%%%%%%%%%%%%%%%%%%%%%%%%%%%%%%%%%%%%%%%%%%%%%%%%%%%%%%%%%%%%%
%%%%%%%%%%%%%%%%%%%%%%%%%%%%%%%%%%%%%%%%%%%%%%%%%%%%%%%%%%%%%%%%%%%%%%%%%%%%%%%%%%%
\subsubsection{Constraints from Planck CMB, ISW, CMB Lensing, and other data sets}
\label{sec:PlanckMGparam}

\begin{figure*}[t!]
\begin{center}
\hspace*{-1cm}
\begin{tabular}{cc}
\includegraphics[width=.45\textwidth]{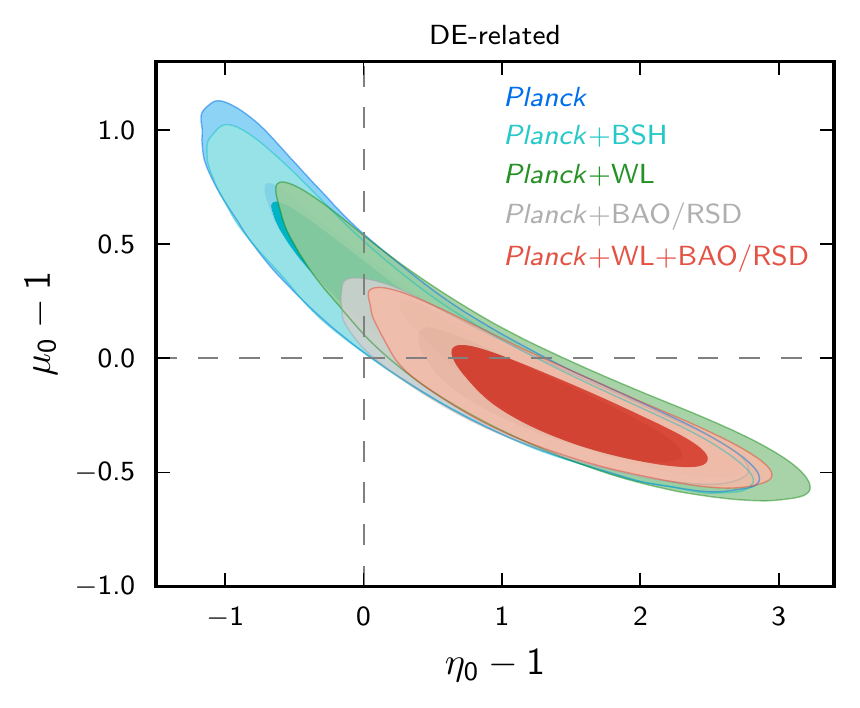}&
\includegraphics[width=.45\textwidth]{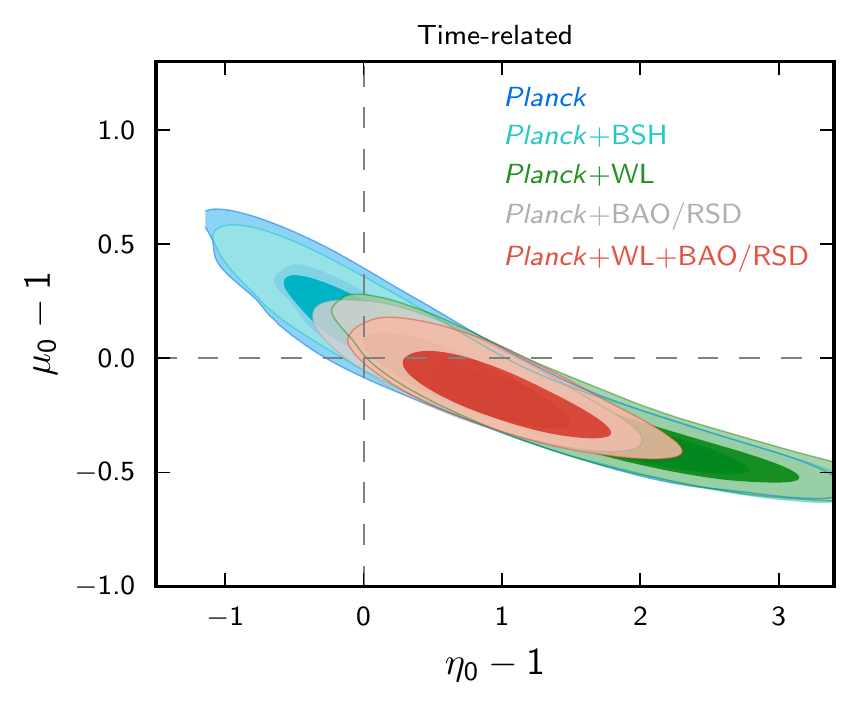}
\end{tabular}
\caption{Contour plots for marginalized posterior distributions for 68\% and 95\% C.L
for the two parameters
$\{\mu_0-1,\eta_0-1\}$ at the present time with no scale dependence. On the left, time dependence is considered via the effective dark energy density parameter. On the right panel, time evolution is considered by direct inclusion of the scale factor. Results discussed in text of Sect.~\ref{sec:PlanckMGparam}.
The label \textit{Planck} stands for \textit{Planck}\mksym{TT}+\mksym{TEB}.
Figure reproduced with permission from \cite{Planck2015MG}.
}
\label{fig:mu_vs_eta}
\end{center}
\end{figure*}

We start with the XIVth paper of the Planck 2015 data release \citep{Planck2015MG} that was dedicated to dark energy and modified gravity models beyond $\Lambda$CDM (we hereafter refer to the paper as Planck2015MG). The authors used Planck CMB temperature, polarization and CMB lensing data sets combined with several other data sets as follows. They defined Planck low-$\ell$ data their temperature and polarization multipoles with $\ell \le 29$ (noted therein as ``lowP''), and also the  high-$\ell$ temperature-only data (noted Planck-TT) with $30 \le \ell \le 2500$. They also used their CMB lensing data which is sensitive to dynamical dark energy and late-time modification to gravity \citep{Planck2015XV}. Planck2015MG considered BAO as the primary data set to be combined with CMB in order to break degeneracies among cosmological parameters constrained by the background evolution and used data from \cite{RossEtAl2014,AndersonEtAl2014,BeutlerEtAl2011}. They used supernova data from the (JLA) compilation \citep{BetouleEtAl2013,BetouleEtAl2014}. They also used a local measurement of the Hubble constant, $H_0=70.6\pm3.3$ \Hu, from  \cite{Efstathiou2014} who reanalyzed the results of \cite{RiessEtAl2011}. For constraints on the growth-rate of large scale structure, Planck2015MG used constraints on $f\sigma_8$ from the RSD data compilation of \cite{SamushiaEtAl2014} (see references therein) as well as weak lensing data from the CFHTLenS survey using the 2D data of \cite{KilbingerEtAl2013} and the tomography data from only blue-galaxies in order to avoid any intrinsic alignment contamination present in the red-galaxies \citep{2013CFHTlens}.

For MG parameters, Planck2015MG constrained $\mu(k,a)$, $\eta(k,a)$, and $\Sigma(k,a)$ as defined earlier in Eqs. \eqref{eq:MGmu}, \eqref{eq:MGeta}, and \eqref{eq:Sigma2} but added to them specific time and scale dependencies. They defined a parametrization that is similar to that described in \eqref{eq:BZparamLR} \citep{BZ2008} for the quasi-static regime but which is more general and covers a wider range of scales \citep{Planck2015MG}.    
For the time evolution they considered two cases, one where the dependence is expressed via the effective dark energy density $\Omega_{\rm{DE}}(a)$, and a second case where the scale factor appears directly in the parametrization. They also split the time evolution using $E_{ij}$ constants, $i,j-1,2$ to represent early and late time evolution.  The $E_{ij}$ parameters are constrained from the data and the parameters $\mu$, $\nu$ and $\Sigma$ are reconstructed from them.  

\begin{figure}[t!]
\begin{center}
\hspace*{-1cm}
\begin{tabular}{cc}
\includegraphics[width=.55\textwidth]{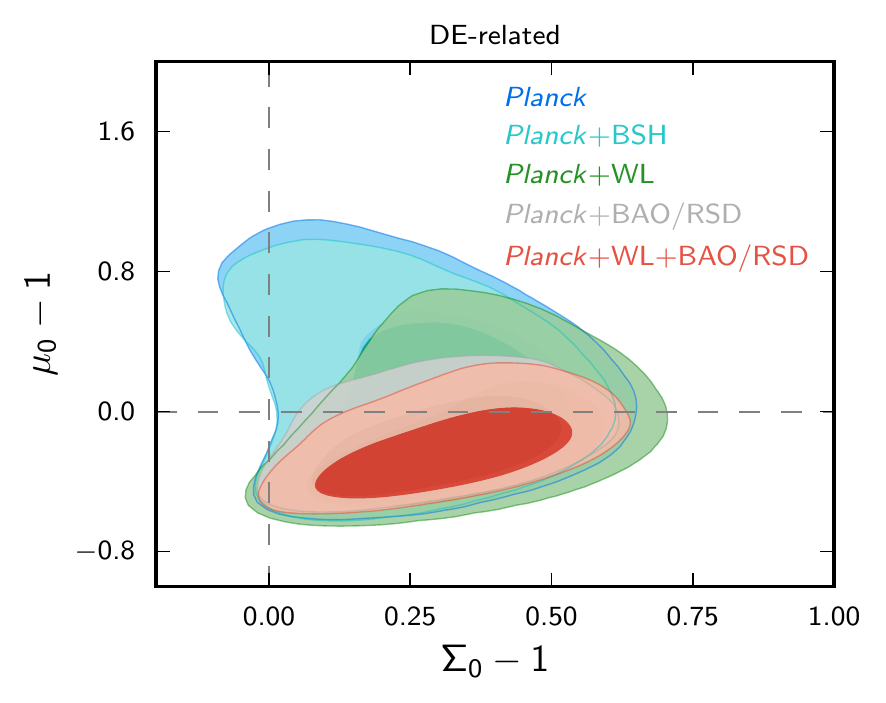}&
\end{tabular}
\caption{Contour plots for marginalized posterior distributions for 68\% and 95\% C.L for the two parameters $\{\mu_0-1,\eta_0-1\}$ at the present time with no scale dependence. The time dependence is considered via the effective dark energy density parameter. $\Sigma$ is obtained from Eq.~\eqref{eq:Sigma2}. Results discussed in text of Sect.~\ref{sec:PlanckMGparam}. In the labels, \textit{Planck} stands for \textit{Planck}\mksym{TT}+\mksym{TEB}.
Figure reproduced with permission from \cite{Planck2015MG}.
}
\label{fig:mu_vs_sigma} 
\end{center}
\end{figure}

However, Planck2015MG found that the current data can not meaningfully constrain the scale dependent MG parameters and that the inclusion scale dependence have very little effect on the $\chi^2$ value of the best fit. Therefore their main MG parameter analysis was carried out without scale dependence except for a small illustrative example. 

We reproduce here their Fig.~14 (see Fig.~\ref{fig:mu_vs_eta} here), their Fig.~15 (see Fig.~\ref{fig:mu_vs_sigma}) and their Table 6 (see Table~\ref{tab:MGDEres} here) showing constraints on $\mu(k,a)$, $\eta(k,a)$, and $\Sigma(k,a)$ from various combinations of Planck and other data sets. Note that Planck2015MG use on their figures or tables \textit{Planck} to refer to the combination Planck TT + lowP data. We expanded that in the header of Table~\ref{tab:MGDEres} for clarity.

Their reproduced figures \ref{fig:mu_vs_eta} and \ref{fig:mu_vs_sigma} show that while, $\mu(k,a)$, $\eta(k,a)$, and $\Sigma(k,a)$ are close to their GR value of $1$, some tension with GR is present and they provide some explanations for the source of such tension. This is indicated by the dashed horizontal and vertical lines in Fig.~\ref{fig:mu_vs_eta}. In case (1) above, with time evolution based on effective $\Omega_{\rm DE}(a)$, the tension is at the 2$\sigma$ level for Planck TT + lowP data and rises above 2$\sigma$ when the constraints are tightened by adding the BAO+RSD data. The tension reaches 3$\sigma$ level for Planck TT+lowP+WL+BAO+RSD combination. For case (2), with time evolution depending directly on $a$, there is less tension. It goes from 1-$\sigma$ for Planck TT + lowP data to 2-$\sigma$. They commented that the latter increase from 2 to 3-$\sigma$ in the tension is mainly driven by the additional external data sets and so is the goodness of the fit of the models with the two additional MG parameters that show an improvement that ranges from $\delta\chi^2=-6.3$ when using Planck+lowP to $\delta\chi^2=-10.8$ when combining Planck TT+lowP+WL+BAO+RSD, compared to the $\Lambda$CDM. 

\begin{table*}[t!]
\caption{\label{tab:MGDEres}Marginalized mean values and 68\,\% C.L. errors on cosmological parameters and the MG parameters 
$\{\mu_0-1,\eta_0-1\,\Sigma_0-1\}$ at the present time with no scale dependence. The time dependence is considered via the effective dark energy density parameter.
$\Sigma$ is obtained from Eq.~\eqref{eq:Sigma2}. Results discussed in text of Sect.~\ref{sec:PlanckMGparam}.
Table reproduced with permission from \cite{Planck2015MG}.}

\footnotesize 
\begin{tabular}{ccccccc}\hline 
Parameter&Planck TT&Planck TT+&Planck TT+&Planck TT+lowP&Planck TT+lowP&Planck TT,TE,\\
\omit& +lowP& lowP+BSH& lowP+WL& +BAO/RSD& +WL+BAO/RSD&EE+lowP+BSH\\\hline
$\mu_0-1$   & $0.07^{+0.24}_{-0.51}$ & $0.04^{+0.22}_{-0.48}$  & $-0.14^{+0.13}_{-0.34}$   & $-0.17^{+0.14}_{-0.23}$& $-0.21^{+0.12}_{-0.21}$ & $0.06^{+0.23}_{-0.48}$\\
$\eta_0-1$  & $0.70\pm 0.94$         & $0.72\pm 0.90$          & $1.36^{+1.0}_{-0.69}$     & $1.23\pm 0.62$         & $1.45\pm 0.60$          & $0.60\pm 0.86$\\
$\Sigma_0-1$& $0.28\pm 0.15$         & $0.27\pm 0.14$          & $0.34^{+0.17}_{-0.14}$    & $0.29\pm 0.13$         & $0.31\pm 0.13$          & $0.23\pm 0.13$\\
$\tau$      & $0.065\pm 0.021$       & $0.063\pm 0.020$        & $0.061^{+0.020}_{-0.022}$ & $0.062\pm 0.019$       & $0.057\pm 0.019$        & $0.060\pm 0.019$\\
$H_0$ $(\frac {\rm{km}}{s\,\rm{Mpc}})$& $68.5\pm 1.1$          & $68.17\pm 0.58$         & $69.2\pm 1.1$             & $68.26\pm 0.69$        & $68.55\pm 0.66$         & $67.90\pm 0.48$\\
$\sigma_8$  & $0.817^{+0.034}_{-0.055}$ & $0.816^{+0.031}_{-0.051}$ & $0.786^{+0.021}_{-0.037}$ & $0.792^{+0.021}_{-0.025}$ & $0.781^{+0.019}_{-0.023}$ & $0.816^{+0.031}_{-0.051}$ \\
\hline
\end{tabular}
\end{table*} 

Planck2015MG comment that the tension above can be understood from their Fig.~1 showing that the best fit power spectrum Planck TT+ lowP prefers models with slightly less power in the CMB at large scales (i.e., ISW effect) and models with a higher CMB lensing potential when compared to the $\Lambda$CDM model. 
They state that this point corroborates with the fact that MG parameters departing from GR values are found to be degenerate with the lensing amplitude parameter $A_L$.  This is simply a non-physical scaling parameter to check how the CMB power spectrum is affected by lensing. It should be equal to 1 for consistency. \cite{2008-lensing-anomaly} found that $A_L$ is not equal to 1 when using the $\Lambda$CDM model, but Planck2015MG find that if MG parameters  are allowed to vary then $A_L$ becomes consistent with unity again but then MG parameters move away from their $\lcdm$ value. However, Planck2015MG points out that CMB lensing analysis from the 4-point function of \cite{Planck2015XV} is consistent with $A_L=1$ and in agreement with $\lcdm$ with no requirement of a higher lensing potential. Therefore, when Planck2015MG use this CMB Lensing data, the MG parameter confidence contours are shifted to regions where the 
the tensions above are removed (fall to 1-$\sigma$ for CMB data only and below 2-$\sigma$ for all data combined). GR and $\Lambda$CDM provides a good fit then.  
It is worth noting though that recent work confirms some tension between Planck temperature and polarization data versus Planck CMB Lensing data \citep{Motloch2018}.

Their Fig.~16 and Table~7 provide a summary of the tensions with and without CMB Lensing where they present the tension using departure from the line of maximum degeneracy between the two MG parameters. 

Their Table~6 (Table~\ref{tab:MGDEres}) shows the corresponding  marginalized mean values and the 65\% CL errors  on the MG parameters for each combination of data sets. This shows the explicit  constraints on MG parameters and the tensions reported above. As commented in Planck2015MG, the addition of the BAO+SN+H does not improve significantly the MG constraints while the RSD data does provide a noticeable improvement, as expected. 
Finally, as shown in their Fig.~18, the current available data is not able to provide useful constraints when the scale dependence of the MG parameters is included in the analysis. 

%%%%%%%%%%%%%%%%%%%%%%%%%%%%%%%%%%%%%%%%%%%%%%%%%%%%%%%%%%%%%%%%%%%%%%%%%%%%%%%%%%%
%%%%%%%%%%%%%%%%%%%%%%%%%%%%%%%%%%%%%%%%%%%%%%%%%%%%%%%%%%%%%%%%%%%%%%%%%%%%%%%%%%%
%%%%%%%%%%%%%%%%%%%%%%%%%%%%%%%%%%%%%%%%%%%%%%%%%%%%%%%%%%%%%%%%%%%%%%%%%%%%%%%%%%%
\subsubsection{Constraints on MG parameters from mainly weak lensing data}

\textbf{KIDS-450 + other data sets:}

\cite{JoudakiEtAl2017} conducted a detailed analysis to test extensions to the standard $\Lambda$CDM cosmological model including constraints on deviations from GR using weak lensing tomography using 450 deg$^2$ of imaging data from the Kilo Degree Survey (KiDS) \citep{HildebrandtEtAl2017}. The authors also used the Planck temperature and polarization measurements on large angular scales ($\ell\le 29$) using low-$\ell$ (TEB likelihood)  and temperature only (TT) at smaller scales (PLIK TT likelihood) \citep{Planck2015}. They explored if any of the extensions to the standard model could alleviate the tension reported in  \cite{HildebrandtEtAl2017} between KiDS and Planck constraints. The extent and sources of these tensions has been put into question though by \cite{Efstathiou2017}.  

They used the parameterization $Q(k,z)$ and $\Sigma(k,z)$ as in \eqref{eq:PoissonModLR} and \eqref{eq:PoissonModSumLR}, and binned in scale and redshift similar to Table~\eqref{tab:Grid}, with transitions at $\rm k= 0.05 \rm h\, \rm Mpc^{-1}$ and $\rm z = 1$. They used as lensing statistics, the correlations functions in equations \eqref{eq:xi_lr_pm_pkappa_tomo}. They included in their analysis all of the key lensing systematics such as intrinsic alignments of galaxies and baryonic effects by modeling them and adding the corresponding parameters to be also constrained by the data. They used for the MG part of their analysis the ISiTGR software \citep{ISITGR} which is a modified version of CosmoMC and CAMB \citep{COSMOMC,CAMB} (see Sect.~\ref{sec:ISITGR}). 

\begin{figure*}[t!]
\begin{center}
{\includegraphics[width=0.49\textwidth]{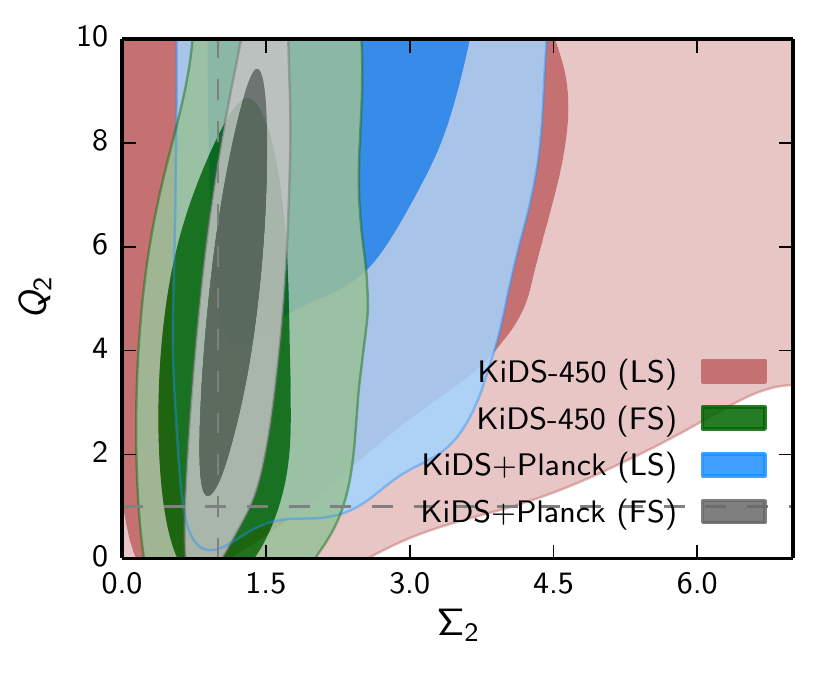}}
{\includegraphics[width=0.49\textwidth]{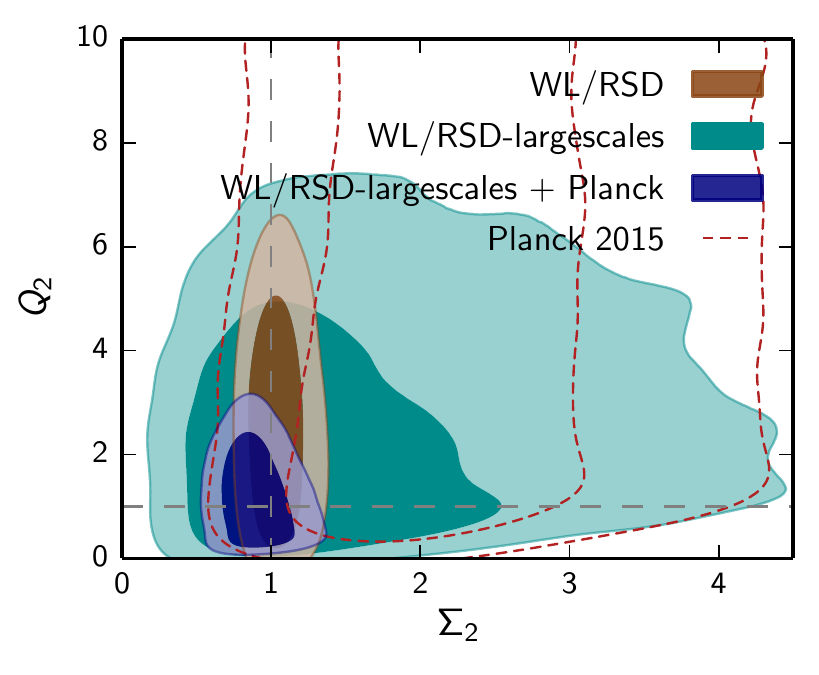}}
\end{center}
\vspace{-2.0em}
\caption{LEFT: Marginalized posterior contours (inner 68\%~CL, outer 95\%~CL) in the $Q_2 - \Sigma_2$ space for KiDS with fiducial angular scales shown in green (labeled by `FS'), KiDS keeping only the largest angular scales shown in pink (labeled by `LS'), and combined with Planck in grey and blue, respectively. The indices represent the combination of MG bins, such that $z < 1$ and $k > 0.05~h~{\rm{Mpc}}^{-1}$. The intersection of the dashed lines give the GR prediction (i.e., $Q = \Sigma = 1$). RIGHT: In addition to the cases described on th left, the constraints  include galaxy--galaxy lensing correlation with cosmic shear in WL and RSD data as described in the text.  'Large-scale cuts' mean that small scales have been excluded because of no adequate modeling for generic MG deviations in the nonlinear regime that can be utilized here. Again, the intersection of the horizontal and vertical lines is the GR prediction (i.e., $Q = \Sigma = 1$). Images reproduced with permission from  \cite{JoudakiEtAl2017}
}
\label{joudakiQ2S2}
\end{figure*}

We reproduce the right panel of their Fig.~13 (see left panel of Fig.~\ref{joudakiQ2S2} here) showing the constrains on $Q_2$ and $\Sigma_2$. As shown on the figure, KiDS constraints are consistent with GR and are mainly sensitive to $\Sigma_2$ as expected for lensing constraints. The authors report that this is also the case for the other 6  $Q_i$ and $\Sigma_i$ parameters. Furthermore, using $\chi^2$ and other Bayesian tests, they  find that the data has no significant preference for the model with additional MG parameters compared to $\Lambda$CDM. The tension between Planck and KiDS goes away but they attribute that to the weakening in the constraints due to the additions of 8 MG parameters. They conclude that their data (combined with Planck)  has no preference for a deviation from GR. They found instead that a model  with a dynamical dark energy and a time-evolving equation of state is moderately preferred by the data and alleviates the tension between their data and Planck.

In a subsequent study \citep{Joudaki2018}, the authors combined KiDS lensing tomography data and the overlapping areas from two spectroscopic redshift galaxy clustering surveys: 2dFLenS \citep{BlakeEtAl2016a} and BOSS  \citep{DawsonEtAl2013,AndersonEtAl2014}. The same Planck data as above was used again.  
They performed cosmological parameter constraints including MG parameters using three large-scale structure measurements: cosmic shear tomography, galaxy-galaxy lensing tomography, and redshift-space distortions (RSD) in the form of redshift-space multipole power spectra \citep{Taylor1996RSD}.  
This provided the analysis with significantly more constraining power and tightening of constraints on all parameters. However, this tightening of constraints also made the tension between large-scale constraints and Planck at the 2.6$\sigma$ level. 
They found that models with MG parameters could resolve the discordance in the linear/large-scale case, but are not favored by model selection. The same result stands for extended models with massive neutrinos, curvature or evolving dark energy. The big plus for constraints on MG parameters in their analysis comes from the complementarity between cosmic shear that is sensitive to the sum of the two potentials via light deflection, i.e., $\Psi+\Phi$, and the redshift space distortions that are sensitive to the potential $\Psi$ via the matter growth of large scale structure. 
They use the same bins in redshift and scale for MG parameters as above and keep the background cosmology as a $\Lambda$CDM one. 

We reproduce the right panel of their Fig.~11 (in the right panel of our Fig.~\ref{joudakiQ2S2}) showing the new constraints in $Q_2-\Sigma_2$ plane. These two parameters are in the second bin in redshift (i.e., $z<1$) and second bin in length-scale (i.e., $k>0.05\,h \rm {Mpc}^{-1}$). One can see a significant improvement in the constraints in the right panel compared to the left which highlights the importance of adding the RSD data and the galaxy-galaxy lensing correlation to cosmic shear data, as the authors stress in their conclusion. 

For this WL+RSD combined analysis, they find $Q_2 = 2.8^{+1.1}_{-2.0}$ and $\Sigma_2 = 1.04^{+0.11}_{-0.14}$, while for KiDS only in \cite{JoudakiEtAl2017} $\Sigma_2 = 1.23^{+0.34}_{-0.70}$ and unconstrained $Q_2$ within its prior range. 
These and all other constraints on the six other modified gravity parameters are all consistent with the GR values of unity. 
The tightest constraints in this analysis come from combining cosmic shear, galaxy -- galaxy lensing correlation, RSD and Planck  $Q_2 = 1.28^{+0.41}_{-1.00}$ and $\Sigma_2 = 0.90^{+0.14}_{-0.18}$. As they comment, these are conservative results since only large-scale cuts are used which are found consistent with Planck. 
This is a good improvement from the previous analysis above with large-scale `KiDS cosmic shear +Planck' constraints where $Q_2 > 2.2$ (restricted by the upper bound prior) and $\Sigma_2 = 2.13^{+0.58}_{-1.10}$. 
The authors conclude that as we will have more overlap between KiDS and 2dFLenS/BOSS, we will be able to obtain more stringent constraints using the data combination used here. 

\mbox{}\\
{\flushleft \textbf{CFHTLenS + other data sets:}}
  
Some years earlier, \cite{SimpsonEtAl2013} used combined structure growth data from the CFHTLenS tomographic cosmic shear survey \citep{2013CFHTlens,BenjaminEtAl2012}, the WiggleZ Dark Energy Survey \citep{BlakeEtAl2012}, and redshift space distortions from the 6dFGS \citep{BeutlerEtAl2012T6} to constrain MG parameters and deviations from the Newtonian potentials. They also used background data for $H_0$ from \cite{RiessEtAl2011}, BAO data from \cite{AndersonEtAl2012}, and \cite{PadmanabhanEtAl2012}, as well as CMB  temperature (TT) and polarization (TE) with data from WMAP7 \citep{WMAP7}.

They used a slightly modified parametrization  so that our $\mu (k, a)$ and $\Sigma (k, a)$ in \eqref{eq:MGeta} and \eqref{eq:Sigma2} are replaced by $[1 + \mu (k, a)]$ and $[1 + \Sigma (k, a)]$ respectively and now taking 0 value in the GR case instead of 1. 
They modeled the time-evolution of the MG parameter to scale with the background effective dark energy density as:
\[ \label{eq:mgparams}
\Sigma(a) = \Sigma_0 \frac{\Omega_\Lambda(a)}{\Omega_\Lambda}   \, , \, \, \, \, \mu(a) = \mu_0 \frac{\Omega_\Lambda(a)}{\Omega_\Lambda}  \, ,
\]
\noindent where $\Omega_\Lambda \equiv \Omega_\Lambda (a=1)$ is today's value so that $\mu_0$ and $\Sigma_0$ represent today's values of $\mu(a)$ and $\Sigma(a)$ as well, respectively.

They used measurements constraints on $(f\sigma_8,F)$ from the WiggleZ and 6dFGS  surveys where $F(z)$ represents the amplitude of the Alcock--Paczynski effect degenerate with the RSDs as we discussed in Sect.~\ref{sec:ClusteringRSD}. These measurements are from three effective redshift slices from the WiggleZ $z = 0.44$, $0.60$, and $0.73$, with $\sigma_8(z) = (0.41 \pm 0.08, 0.39 \pm 0.06, 0.44 \pm 0.07)$ and $F = (0.48 \pm 0.05, 0.65 \pm 0.05, 0.86 \pm 0.07)$ plus a fourth data point of $f \sigma_8 = 0.423 \pm 0.055$  at a lower redshift  $z=0.067$ from the 6dFGS with negligible sensitivity to the Alcock-Paczynski distortion. 

In their analysis they considered the $\Lambda$CDM, the flat and non-flat wCDM models all augmented with the MG parameters $\mu_0$  and $\Sigma_0$. In all cases, they found no indication of departure from general relativity on cosmological scales. They put the following limits on MG parameters:  $\mu_0 = 0.05 \pm 0.25$ and $\Sigma_0 = 0.00 \pm 0.14$ for a flat $\Lambda$CDM background model.  They note that these correspond to deviations in the present-day Newtonian potential and spatial curvature potential of $\delta \Psi/\Psi_{GR} = 0.05 \pm 0.25$ and $\delta {\Phi} / \Phi_{GR} = -0.05 \pm 0.3$ respectively, with significant correlations between the errors. When they allow for $w$ to vary for the background, these constraints change to  $\mu_0 = -0.59 \pm 0.34$ and $\Sigma_0 = -0.19 \pm 0.11$. They also constrained the growth index parameter to $\gamma=0.52 \pm 0.09$ for a $\Lambda$CDM background model, thus in agreement with the GR value of $6/11=0.545$.  

%%%%%%%%%%%%%%%%%%%%%%%%%%%%%%%%%%%%%%%%%%%%%%%%%%%%%%%%%%%%%%%%%%%%%%%%%%%%%%%%%%%
%%%%%%%%%%%%%%%%%%%%%%%%%%%%%%%%%%%%%%%%%%%%%%%%%%%%%%%%%%%%%%%%%%%%%%%%%%%%%%%%%%%
%%%%%%%%%%%%%%%%%%%%%%%%%%%%%%%%%%%%%%%%%%%%%%%%%%%%%%%%%%%%%%%%%%%%%%%%%%%%%%%%%%%
\subsubsection{Constraints on MG parameters from various probes and analyses}

\cite{Simone2017} Perform an extensive analytical and numerical analysis of the MG parameters $\Sigma$ and $\mu$ or equivalently $G_{light}/G$ and $G_{matter}/G$. They consider Horndeski models that are broadly consistent with background and perturbation tests of gravity and the cosmic expansion history with late time acceleration. They also take into account the recent result from GW170817 and its counterpart GRB170817A, setting $c_T=c$. They confirm a conjecture they made in their earlier work \citep{PogosianEtAl2016} about MG parameters in Horndeski models, that is $(\Sigma-1)(\mu-1)\ge0$ (that is the two factors must be of the same sign) must hold in viable Horndeski models in the quasi-static approximation. They also discussed in their previous work \citep{PogosianEtAl2016} consistency relations between the two MG parameters that, if broken would exclude some sub-classes of Horndeski models (e.g., $\Sigma \ne 1$ would rule out all models with a canonical form of kinetic energy). 
They remark that while the results of \cite{Planck2015MG} indicate $\mu<1$ and $\Sigma>0$   are not statistically significant, however, if such values will hold in more precise experiments in the future that would rule out all Horndeski models. 
In the latter paper, they show that requiring no ghosts and no gradient instabilities prevents from having values within the $\Sigma-1>0$ and $\mu-1<0$ range. They also examined the conjectured condition versus the Compton wavelengths considered. They also found that observations from background expansion also put constraints on gravitational coupling  which in turn re-enforces the conjecture limits. They also test the validity of the quasi-static approximation in Horndeski models finding that it holds well at small and intermediate scales but fails at $k\le 0.001h$/Mpc. They conclude in their analysis that despite the stringent result from GW, there remain Horndeski models with non-trivial modifications to gravity at the level of linear perturbations and large scale structure. They stress the complementarity of different approaches used to constrain modification to GR and the practicality of using the phenomenological $\Sigma$ and $\mu$ parameterization and their consistency relations, see also \cite{PogosianEtAl2016}.

Another analysis of these self-consistency relations between MG parameters and growth rate in Horndeski models was performed  by  \cite{Perenon2017}. 
They considered accelerating Horndeski models with $-1.1 \le w_{\rm eff} \le -0.9$ 
and classified them according to their early or late time effects as follows.  
Late-time dark energy where both dark energy energy momentum tensor and non-minimal gravitational couplings are negligible at early times. Early-time dark energy where the dark energy momentum tensor is at work even at early times but non-minimal coupling happens at late time only. Finally, they call early modified gravity where both dark energy momentum and non-minimal gravitational couplings are also present at early time during matter domination. They proposed a convenient way to represent the viability of the models using two diagnostic planes: the  $\mu(z)-\Sigma(z)$ and the $f(z)\sigma_8(z)-\Sigma(z)$ planes. They derived the following conclusions from their detailed analysis in the first plane. If model-independent measurements find either (i) $\Sigma-1<0$ at redshift zero or (ii) $\mu-1<0$ with $\Sigma-1>0$ at high redshifts ($z>1.5$) or (iii) $\mu-1>0$ with $\Sigma-1<0$ at high redshifts, Horndeski theories are ruled out.  
In the second plane, they found that: (i) If $f\sigma_8$ is found to be larger than that of $\Lambda$CDM model at $z>1.5$ then early dark energy models are ruled out. On the opposite case (for $f\sigma_8$), (ii) measuring $\Sigma<1$ will rule out late dark energy models, while, (iii) $\Sigma>1$, it is the early modified gravity case as described earlier in this paragraph that is allowed.

\begin{figure}[t]
\begin{center}
\includegraphics[scale=0.45]{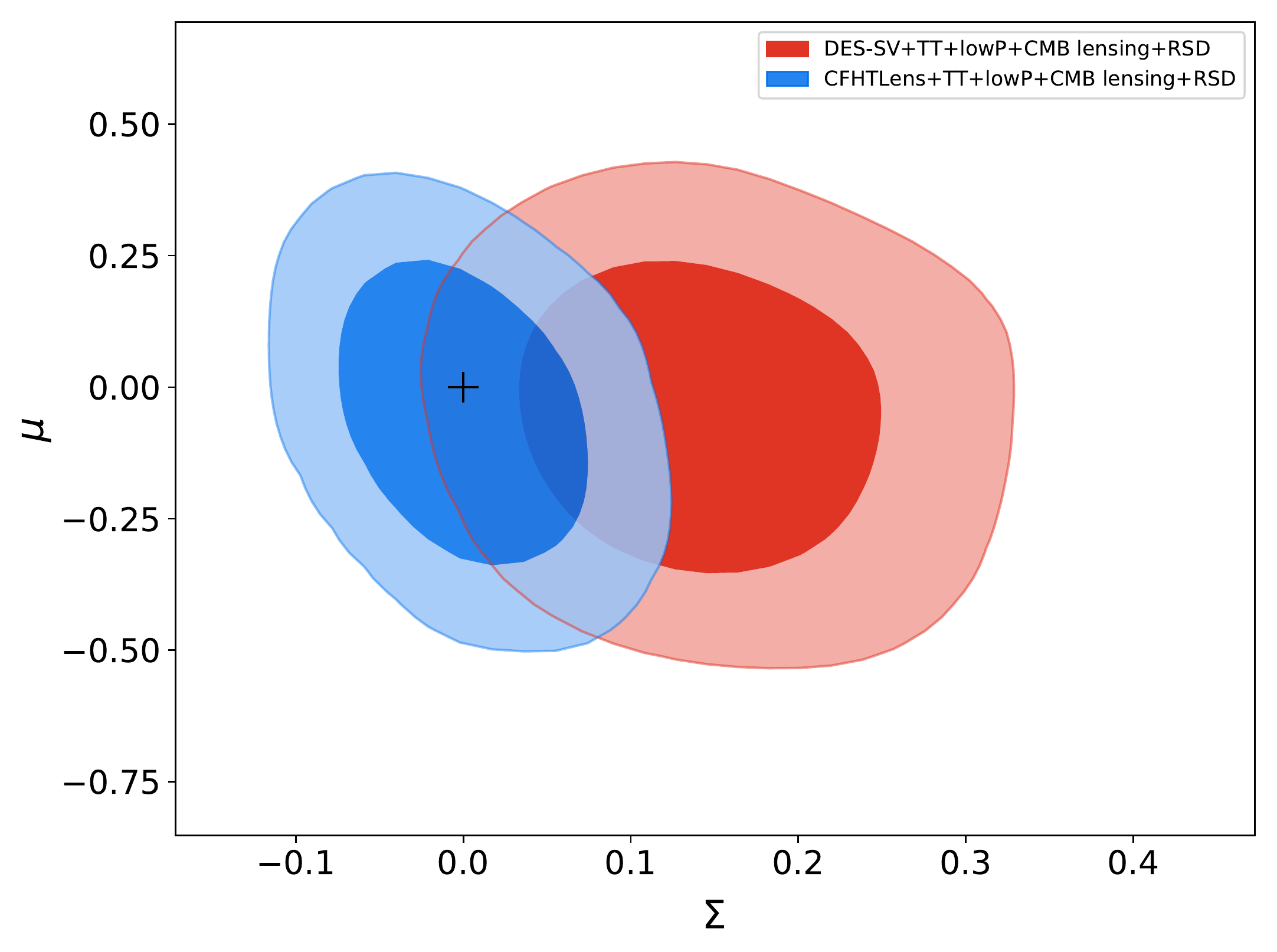}
\caption{Contour plots for 68\% and 95\% CL on MG parameters $\Sigma$ and $\mu$ combining \textit{Planck} CMB data (TT+lowP+CMB lensing), RSD data from BOSS DR12 and 6dFGS,  and cosmic shear data from CFHTLenS in blue and DES-SV in red. The cross point represent the GR values (0,0) according to the authors' definitions and show that GR is consistent with the data sets used. The combination for the contours in blue gives among the tightest current constraints on MG parameters as:  $\Sigma = -0.01_{-0.04}^{+0.05}$ and $\mu =  -0.06 \pm 0.18$ (68 \% confidence level). Figure reproduced with permission from \cite{Ferte2017}.}
\label{fig:combprobes}
\end{center}
\end{figure}

\cite{Ferte2017} performed an analysis to constrain the two MG parameters but using the definitions $[1 + \mu (a)]$ and $[1 + \Sigma (a)]$ to enter in the Poisson and lensing equations instead of of $\mu (a)$ and $\Sigma (a)$ so taking 0 values in the GR case instead of 1. They also used equations \eqref{eq:mgparams} to model their time evolution scaling with the effective dark energy density parameter with no scale dependence since current data cannot constrain their scale dependence.
They use CMB measurements from Planck, cosmic shear from CFHTLenS and DES science verification data, and RSD from BOSS DR12 and the 6dF galaxy survey. 
They derived constraints using this combination of probes but including either CFHTLenS or DES-SV separately for weak lensing finding the results shown in their Fig.~10 (reproduced here as Fig.~\ref{fig:combprobes}). The constraining power of CFHTLens is larger leading to tighter constraints. In the DES-SV data, they also marginalized over the amplitude intrinsic alignment of galaxies finding a positive value leading to a higher value of $\Sigma$ as shown on the figure. 
The constraints found using CMB Planck (TT+lowP+CMB Lensing), RSD data (BOSS DR12+), and CFHTLenS cosmic shear are: $\Sigma = -0.01_{-0.04}^{+0.05}$ and $\mu =  -0.06 \pm 0.18$ (68 \% confidence level) which are among tightest current constraints on MG parameters. GR is consistent with these tightened bounds, although there is still room for deviations from it, in particular, for the $\mu$ parameter. The authors then perform some forecast analysis for improvement using five years of DES and LSST data showing substantial improvement on the parameters and in particular $\Sigma$ that we present in Sect.~\ref{sec:forecasts}. 

%%%%%%%%%%%%%%%%%%%%%%%%%%%%%%%%%%%%%%%%%%%%%%%%%%%%%%%%%%%%%%%%%%%%%%%%%%%%%%%%%%%
%%%%%%%%%%%%%%%%%%%%%%%                                         %%%%%%%%%%%%%%%%%%%
%%%%%%%%%%%%%%%%%%%%%%%          SUB-SECTION                    %%%%%%%%%%%%%%%%%%%
% Constraints on $f \sigma_8$ from galaxy surveys and RSD measurements %
%%%%%                                                                          %%%%
%%%%%%%%%%%%%%%%%%%%%%%%%%%%%%%%%%%%%%%%%%%%%%%%%%%%%%%%%%%%%%%%%%%%%%%%%%%%%%%%%%%
%%%%%%%%%%%%%%%%%%%%%%%%%%%%%%%%%%%%%%%%%%%%%%%%%%%%%%%%%%%%%%%%%%%%%%%%%%%%%%%%%%%

\subsection{Constraints on $f \sigma_8$ from galaxy surveys and RSD measurements }
\label{sec:fsigma8constraints}
\cite{AlamEtAl2016} presented cosmological constraints from galaxy clustering data of the completed SDSS-III BOSS survey. The study used combined galaxy samples with 1.2 million galaxies. The spectroscopic survey used BAO methods to measure the angular diameter distance and the Hubble parameter. Most relevant to testing gravity, the survey constrained the growth of structure using the combination $f\sigma_8$ from RSD measurements. 
In this concluding analysis of SDSS-III BOSS, they combined individual measurements from seven previous companion SDSS papers into a set of consensus values for the angular diameter distance, the Hubble parameter and  $f\sigma_8$ at 3 redshifts: $z=0.38,0.51, \,\rm {and \,0.61}$. 

The method they employed to test deviations from GR was not based on using directly any MG parameters. They instead defined two parameters that rescale $f\sigma_8$ as follows:
\be
f \sigma_8 \rightarrow f \sigma_8 [A_{f \sigma_8}+B_{f \sigma_8}(z-z_p)]
\label{RescaledGrowth}
\ee
with a redshift pivot $z_p=0.51$ \citep{AlamEtAl2016}. GR will have $A_{f \sigma_8}=1$ and $B_{f \sigma_8}=0$.

\begin{figure*}[ht!]
\begin{center}
{\includegraphics[width=0.55\textwidth]{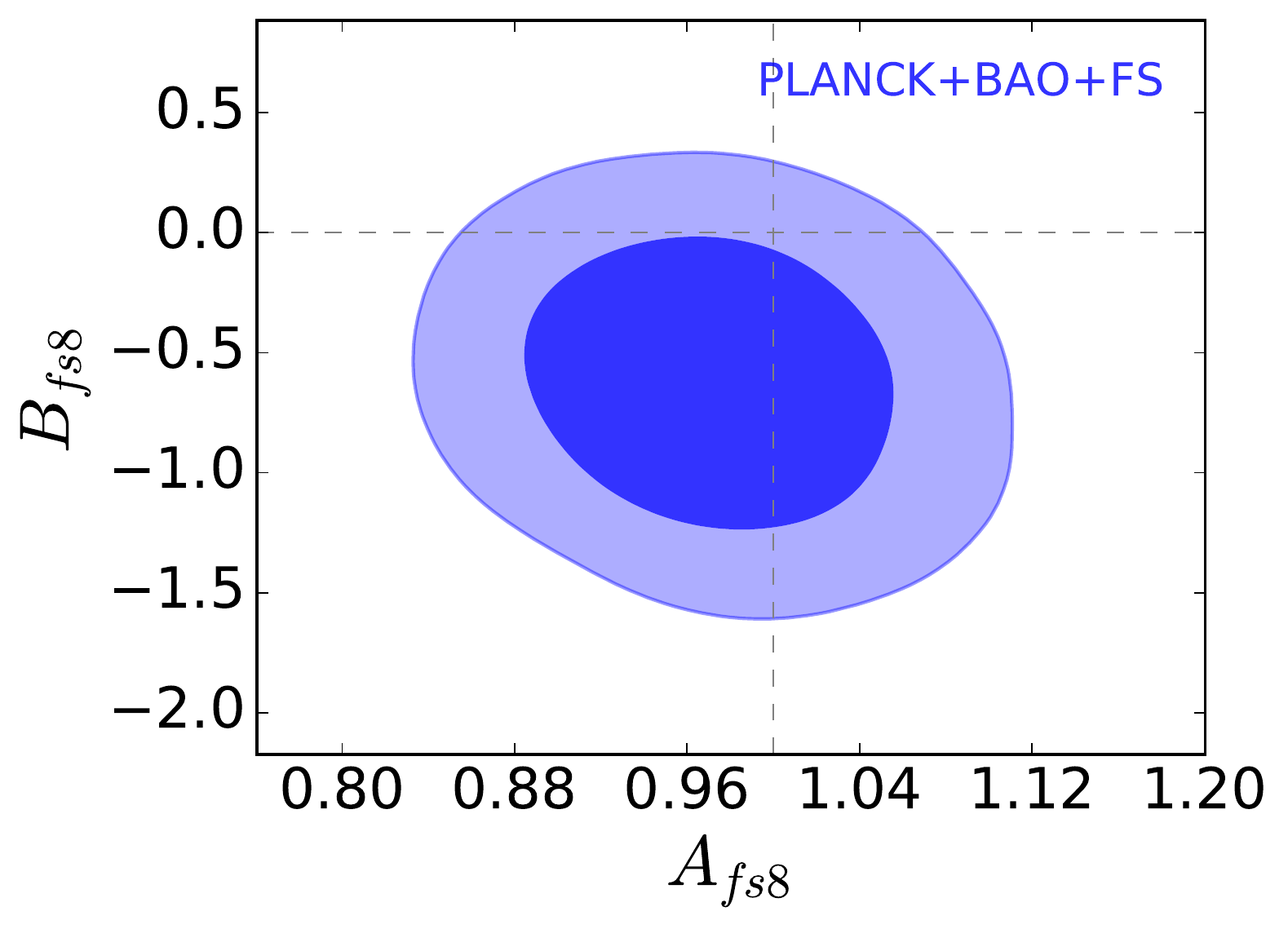}}
\end{center}
\vspace{-2.0em}
\caption{68\% and 95\% confidence contours on modification of the growth function in the 
    $\Lambda$CDM cosmological model using the form   
$f \sigma_8 \rightarrow f \sigma_8 [A_{f \sigma_8}+B_{f \sigma_8}(z-z_p)]$
with a redshift pivot $z_p=0.51$. The results are consistent with the predictions of General Relativity: $A_{f\sigma_8}=1$, $B_{f\sigma_8}=0$ (dashed grey lines). 
As explained in the text, the analysis used temperature and polarization data from Planck 2015, and a set of consensus values for BAO and RSD using full shape measurements (FS) from SDSS DR12. Figure reproduced with permission from \cite{AlamEtAl2016}. 
}
\label{fig:RSDBOSSLR}
\end{figure*}

They combined their RSD and BAO measurements along with temperature and polarization data from Planck-2015 \citep{Planck2015}. For a $\Lambda$CDM background model and a redshift independent rescaling, they find  $A_{f \sigma_8} = 0.96\pm 0.06$, so a growth amplitude value that is consistent with GR. When they allow for a redshift-dependent variation, they find $A_{f \sigma_8} = 0.97\pm 0.06$ and $B_{f \sigma_8} = -0.62\pm 0.40$. This is a 1.5-$\sigma$ deviation from a zero GR-value so they considered this as not statistically significant and concluded that their results are consistent with GR. They also found very little changes in these values when they allow for the equation of state $w$ and the spatial curvature parameter to vary. We reproduce their Fig. 20 (as Fig.~\ref{fig:RSDBOSSLR} here) showing consistency with GR of the two rescaling parameters. They also compile there 11 measurements of $f\sigma_8$ from their work and other surveys.  We note that they used only BOSS RSD data in the results for $f \sigma_8$ above as they state other data come from a variety of analysis and modeling approaches but are nevertheless consistent with those of BOSS within the error bars shown. The authors note that the current growth measurements of $f \sigma_8$ reaffirm the validity of GR. 
It is worth noting though that some other MG models such as nDGP (see Sect.~\ref{sec:DGP}) or ${\mathit{RR}}$ non-local gravity  are still also consistent with RSD data due to the large error bars. 

Another recent analysis in \cite{TorreEtAl2017} used RSD and galaxy-galaxy lensing from the final data set of VIMOS Public Extragalactic Redshift Survey (VIPERS) \citep{delaTorreEtAl2013TV} combined with CFHTLenS data \citep{2013CFHTlens} at a redshift range of at $0.5<z<1.2$. 
The joint analysis obtained measurements of $f\sigma_8(0.6)=0.48\pm0.12$ and
  $f\sigma_8(0.86)=0.48\pm0.10$. The galaxy-galaxy lensing does not add 
any improvement in constraining these values but alleviates the degeneracies with galaxy bias and $\sigma_8$. This allows the constraints to be separated as $\left[f(0.6),\sigma_8(0.6)\right]=[0.93\pm0.22,0.52\pm0.06]$ and  $\left[f(0.86),\sigma_8(0.86)\right]=[0.99\pm0.19,0.48\pm0.04]$ in consistency with GR but again with errors bars large enough to allow for other MG models. 

Most recently, \cite{2016-Okumura-etal-RSD-MG} made a high redshift ($z\sim 1.4$) measurement of $f\sigma_8$ using the FastSound survey using the Subaru Telescope. 
They obtained  $f(z)\sigma_8(z)=0.482\pm 0.116 $ at $z\sim 1.4$ after marginalizing over the galaxy bias parameter $b(z)\sigma_8(z)$. The background expansion was fixed to that of 
a $\Lambda$CDM model and using the RSD measurements on scales above $8\hinvmpc$.
This is a first measurement above redshift 1 and corresponds to $4.2\sigma$ detection of RSD. As shown in their Fig.~17 (Fig.~\ref{fig:RSDMG} here), this high redshift measurement is consistent with GR but models such as covariant or extended Galileons (see Sect.~\ref{sec:Galileons}), $f(R)$ (see Sect.~\ref{sec:f(R)}) and other MG models with varying gravitational constant were all found outside the 1-$\sigma$ bound. The figure shows the importance of high redshift RSD measurement in strongly constraining these models in the future. They note the combination of low-$z$ and high-$z$ RSD measurements will be useful in constraining gravity models without relying on CMB data. 

\cite{Nesseris2017RSD} gathered a compilation of 34 data points where they made corrections for model dependence. In order to avoid overlap and maximize independence of the data-points, they  also constructed a sub-sample from this compilation that they call the `Gold' growth data set with 18 data-points. They determine the best fit $w$CDM from the growth evolution equation using the gold data set and find it in 3-$\sigma$ tension with the best fit Planck-15/$\Lambda$CDM model parameters $w$, $\Omega_m^0$ and $\sigma_8$. They found that the tension disappears if they allow for the evolution of the effective gravitational constant.  

Finally, \cite{Kazantzidis2018} constructed an extended compilation of 63 data points of $f\sigma_8$ published between 2006 and 2013, They correct the data for the fiducial model and find that using the whole set gives a best fit $\Omega^0_m-\sigma_8$ that is in a 5-$\sigma$ tension with the Planck-2015 $\lcdm$ parameter values. However, they show that the tension drops to below 1-$\sigma$ when they use the 20 most recent values while using the 20 earliest data gives a 4.5-$\sigma$ tension. They find that the drop in the tension using the recent data is due to the fact that these are at high redshift with large enough errorbars that accommodate GR and many other theories. They argue that it is more effective to obtain more data at redshift below 1 and with higher precision to be able to distinguish more effectively between gravity theories.

\begin{figure*}[t!]
\begin{center}
{\includegraphics[width=0.50\textwidth]{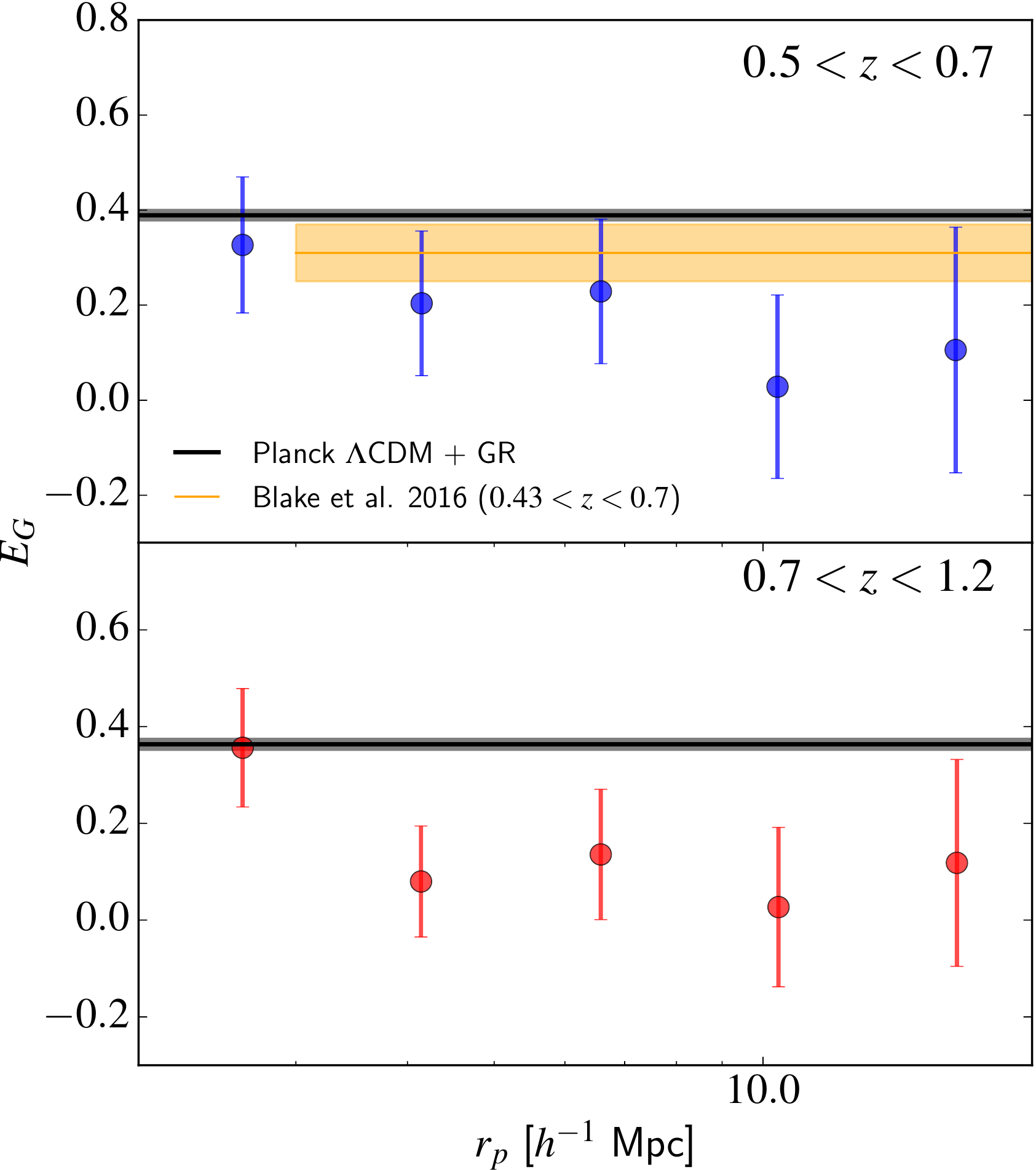}}
\end{center}
\vspace{-2.0em}
\caption{$E_G$ parameter as function of scale in redshift range $0.5<z<0.7$ (top panel) and $0.7<z<1.2$ (bottom
  panel) as measured in \cite{TorreEtAl2017}. In the two panels, the solid curves and shaded areas around them correspond to the prediction and the $68\%$ uncertainty (thin grey) band for GR with a $\lcdm$ background set to the best fit model of TT+lowP+lensing Planck 2015 \citep{Planck2015}. In the top panel, the horizontal (light brown) stripe shows the averaged
  $\smash{\overline{E}_G}$ over the range $3h^{-1} \rm {Mpc}<r_p<50h^{-1} \rm {Mpc}$
  from \cite{BlakeEtAl2016b} at redshift $0.43<z<0.7$. Figure reproduced with permission from \cite{TorreEtAl2017}. As they comment there, these measurements of $E_G$ are
slightly lower (1--2$\sigma$) than expected for the standard $\lcdm$ model of GR and one way to remedy to this is to lower the matter-density parameter.
  }
\label{fig:E_G_VIPERS}
\end{figure*}
%
%
%%%%%%%%%%%%%%%%%%%%%%%%%%%%%%%%%%%%%%%%%%%%%%%%%%%%%%%%%%%%%%%%%%%%%%%%%%%%%%%%%%%
%%%%%%%%%%%%%%%%%%%%%%%                                         %%%%%%%%%%%%%%%%%%%
%%%%%%%%%%%%%%%%%%%%%%%          SUB-SECTION                    %%%%%%%%%%%%%%%%%%%
%%%%%                         Constraints on $E_G$                             %%%%
%%%%%                                                                          %%%%
%%%%%%%%%%%%%%%%%%%%%%%%%%%%%%%%%%%%%%%%%%%%%%%%%%%%%%%%%%%%%%%%%%%%%%%%%%%%%%%%%%%
%%%%%%%%%%%%%%%%%%%%%%%%%%%%%%%%%%%%%%%%%%%%%%%%%%%%%%%%%%%%%%%%%%%%%%%%%%%%%%%%%%%

\subsection{Constraints on $E_G$}
\label{sec:E_G_constraints}

\cite{ReyesEtAl2010} provided one of the first measurements of $E_G$ at redshift $z=0.3$ finding $E_G=0.39\pm0.06$ in agreement with GR-$\Lambda$CDM value of $0.408 \pm 0.029$ although the uncertainties did not exclude some other possible alternative gravity theories such as $f(R)$ (see Sect.~\ref{sec:f(R)}) that predict a range of $E_G$ between 0.328 and 0.365. Nevertheless, the measured value was in a 2.5-$\sigma$ tension with the TeVeS models. The authors used 70,205 luminous red galaxies (LRGs) from the SDSS survey \citep{Eisenstein2005DO} and the RSD measurement from this sample from \cite{Tegmark2006} of $\beta = 0.309\pm 0.035$ on large scales and at $z = 0.32$.  The authors used galaxy-galaxy lensing and galaxy clustering of the LRG sample on Mpc scales and at this effective redshift. They used the slightly different version of $E_G$ as present in Eq.~\eqref{eq:Alt_E_G}.

A more recent measurement of $E_G$ was achieved in \cite{BlakeEtAl2016b}  using 
deep and overlapping imaging and spectroscopic datasets by combining the Red Cluster Sequence Lensing Survey (RCSLenS) \citep{RCSLenS2016}, the Canada-France-Hawaii Telescope Lensing Survey (CFHTLenS) \citep{2013CFHTlens}, the WiggleZ Dark Energy Survey \citep{2011WiggleZ-BAO} and the Baryon Oscillation Spectroscopic Survey (BOSS) \citep{Font-Ribera-etal-2014}.  They converted their measurements of galaxy-galaxy lensing, galaxy clustering and redshift space distortions into galaxy-matter annular differential surface densities ready to be used for the second definition of $E_G$ as given by \eqref{eq:Alt_E_G}. They found $E_G=0.48 \pm 0.10$ at $z=0.32$ and $E_G=0.30 \pm 0.07$ at $z=0.57$ when averaging over scales $10<R<50h^{-1}\rm {Mpc}$. These are both consistent with the perturbed GR-$\Lambda$CDM values of $E_G = 0.41$ and $0.36$ at these respective redshifts. This confirms again GR but the uncertainties are still wide enough to allow for other MG theories.
Next, a high-redshift measurement of $E_G$ came from \cite{TorreEtAl2017} who combined redshift space distortions from VIPERS and galaxy--galaxy lensing using the same portion of the sky from CFHTLenS. They found 
$E_G(z = 0.6) = 0.16 \pm 0.09$ and $E_G(z = 0.86) = 0.09 \pm 0.07$, when $E_G$  is averaged over scales above 3 Mpc/h. We reproduce their figure 17 as Fig.~\ref{fig:E_G_VIPERS}  here. As they comment, these measurements of $E_G$ gives values that are slightly lower than expected for the standard $\lcdm$ of GR, but the results are consistent with GR within 1--2$\sigma$.

Another interesting value of $E_G$ comes from \cite{PullenEtAl2016} where the authors combined measurements of CMB lensing and galaxy velocity field. Unlike previous measurements of $E_G$, this one used CMB lensing instead of galaxy--galaxy lensing. The authors state that this will be less sensitive to contamination by intrinsic alignments of galaxies and will allow for the largest scale measurement of $E_G$ averaging over scales up to 150 $h^{-1}$ Mpc. They used cross-correlations of the Planck CMB lensing map with the SDSS III CMASS galaxy sample along with the CMASS galaxy auto-power spectrum and RSD. They used a definition of $E_G$ adapted to these probes (see their Eqs.~(3) and (15)). They find $E_G(z = 0.57) = 0.243 \pm 0.060$ (stat) $\pm 0.013$ (sys) 
The authors note that this measurement is in tension with GR at a level of 2.6-$\sigma$. Taking cosmological values from Planck-2015 and BOSS BAO, the GR value at $z=0.57$ is $0.402 \pm 0.012$. The authors noted that small tensions with GR start only when considering scales above 80 Mpc/h. They also comment that some deficit at very large scale in the CMB-Lensing galaxy cross power spectrum is present so they do not consider this as an indication of significant deviation from GR. 

\cite{Alam2016} combined data from BOSS CMASS sample DR11 galaxy clustering, CFHTLenS lensing and RSD of $\beta$ measurement from BOSS.  They found 
$E_G(z=0.57)=0.42\pm0.056$ which is in agreement (at 13\% level) with the prediction of GR, $E_G(z=0.57)=0.396\pm0.011$, using the Planck 2015 cosmological parameters. They corrected their results for systematics effects including scale dependence bias affecting its complete cancellation, difference in lensing and clustering windows and redshift weighting, intrinsic alignment of galaxies on lensing, cosmic variance, calibration bias in lensing, and limitations due to the choice of cutoff scale $R_0$.  They run simulations and found that these theoretical observational systematic errors are smaller than the statistical errors in the measurement.

\cite{KIDS2017EG} used the deep imaging data of the KiDS survey combined with overlapping spectroscopic areas from 2dFLenS, BOSS DR12 and GAMA surveys. They found $E_G(z = 0.267) = 0.43 \pm 0.13$ from using GAMA, $E_G(z = 0.305) = 0.27 \pm 0.08$ from using (BOSS LOWZ+2dF Low Z) and $E_G(z = 0.554) = 0.26 \pm 0.07$ from using (CMASS+2dF High Z). The results are consistent with GR with a $\Lambda$CDM background and linear perturbations. However, they found that their result and other measurements of $E_G$ favor a lower value of the matter density $\Omega_m^0$ than the one preferred by Planck. They caution that the statistic $E_G$ is very sensitive to such a tension in the cosmological parameters which can have more effect than a deviation in GR and a change of as much as 10\% in the gravitational potentials. 

Most recently, \cite{Singh2018} used  galaxy clustering from BOSS LOWZ sample with galaxy lensing from SDSS finding $\left<E_G\right>=0.37^{+0.036}_{-0.032}$ (statistical) $\pm 0.026$ (systematic) which is consistent with the GR predicted value ($0.46$) using Planck $\lcdm$ parameters and when both statistical and systematic errors are considered. Then they used BOSS LOWZ and Planck CMB lensing finding $\left<E_G\right>=0.43^{+0.068}_{-0.073}$ (stat). This is statistically consistent with SDSS galaxy lensing result and also with GR predictions. They found $\left<E_G\right>=0.39^{+0.05}_{-0.05}$ (stat) when using the CMASS sample and CMB lensing. The result is consistent with the GR prediction of $0.40$ at the higher redshift of the CMASS sample. They also split the LOWZ sample into two redshift samples and found results on $E_G$ that are consistent with GR predictions at 
 $2.5\sigma$ level (stat) or better.  They found that nonlinear corrections and systematic effects can introduce errors $\sim 1$--$2$\% so below the statistical errors while shear calibration and photometric uncertainties add another  $\sim5\%$ error for the SDSS galaxy lensing. 

%%%%%%%%%%%%%%%%%%%%%%%%%%%%%%%%%%%%%%%%%%%%%%%%%%%%%%%%%%%%%%%%%%%%%%%%%%%%%%%%%%%
%%%%%%%%%%%%%%%%%%%%%%%%%%%%%%%%%%%%%%%%%%%%%%%%%%%%%%%%%%%%%%%%%%%%%%%%%%%%%%%%%%%
%%%%%%%%%%%%%%%%%%%%%%%%%%%%%%%%%%%%%%%%%%%%%%%%%%%%%%%%%%%%%%%%%%%%%%%%%%%%%%%%%%%
%%%%%%%%%%%%%%%%%%%%%%%%                                         %%%%%%%%%%%%%%%%%%
%%%%%%%%%%%%%%%%%%%%%%%%  TYPES OF MODIF TO GR AND MG MODELS     %%%%%%%%%%%%%%%%%%
%%%%%%%%%%%%%%%%%%%%%%%%                                         %%%%%%%%%%%%%%%%%%
%%%%%%%%%%%%%%%%%%%%%%%%%%%%%%%%%%%%%%%%%%%%%%%%%%%%%%%%%%%%%%%%%%%%%%%%%%%%%%%%%%%
%%%%%%%%%%%%%%%%%%%%%%%%%%%%%%%%%%%%%%%%%%%%%%%%%%%%%%%%%%%%%%%%%%%%%%%%%%%%%%%%%%%

\section{Types of modifications to GR at cosmological scales and corresponding MG models}
\label{sec:MGtheories}

\begin{figure*}[h]
\includegraphics[width=\textwidth,height=2in,angle=0,fbox]{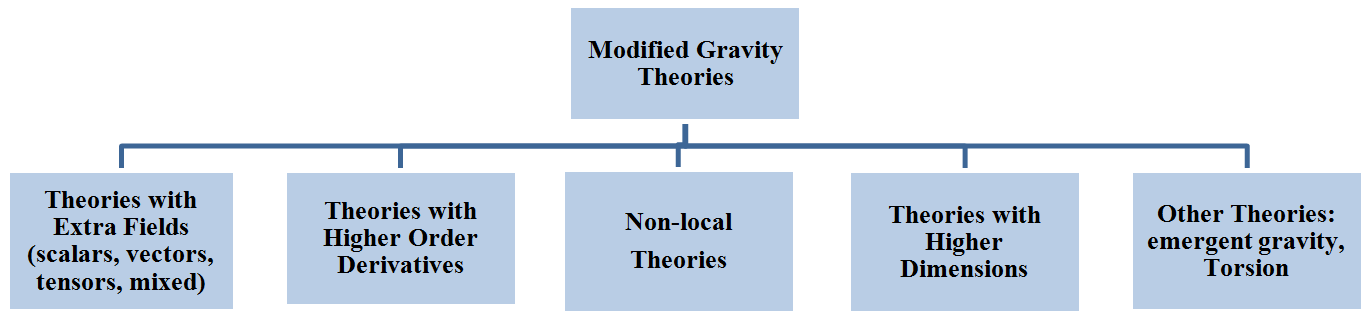}
\caption{Various categories of modified gravity (MG) theories according to the principle or requirement they violate. It is worth noting that some models can belong to more than one category here like for example some higher dimensional models that have extra fields. 
\label{fig:MGTypes}}
\end{figure*}

%%%%%%%%%%%%%%%%%%%%%%%%%%%%%%%%%%%%%%%%%%%%%%%%%%%%%%%%%%%%%%%%%%%%%%%%%%%%%%%%%%%
%%%%%%%%%%%%%%%%%%%%%%%                                         %%%%%%%%%%%%%%%%%%%
%%%%%%%%%%%%%%%%%%%%%%%          SUB-SECTION                    %%%%%%%%%%%%%%%%%%%
%%%%%                   Cartan-Weyl-Lovelock theorem                           %%%%
%%%%%                                                                          %%%%
%%%%%%%%%%%%%%%%%%%%%%%%%%%%%%%%%%%%%%%%%%%%%%%%%%%%%%%%%%%%%%%%%%%%%%%%%%%%%%%%%%%
%%%%%%%%%%%%%%%%%%%%%%%%%%%%%%%%%%%%%%%%%%%%%%%%%%%%%%%%%%%%%%%%%%%%%%%%%%%%%%%%%%%

\subsection{Cartan-Weyl-Lovelock theorem}

General relativity is based on well-defined principles and physical requirements discussed in Sect.~\ref{sec:General_Relativity}, most of which, are encapsulated in the structure of the Einstein tensor and field equations. Einstein used guidance from such principles and requirements to shape and propose his theory \citep{Einstein1915}. After that, Cartan \citep{Cartan1922}, Weyl \citep{Weyl1922}, Vermeil \citep{Vermeil1917}, and Lovelock \citep{Lovelock1971TE,Lovelock1972TF} worked on a succession of assertions and theorems about the uniqueness of Einstein's tensor and the field equations. These led ultimately to the Lovelock theorem \citep{Lovelock1971TE,Lovelock1972TF} that can be summarized as follows, e.g., \cite{2007-Ishak-Remarks-DE,BertiElAlReview2015}: 

\textit{In a spacetime of four dimensions, the only divergence free tensor of valence two that is constructed only from the metric tensor and its derivatives up to second order, and preserves diffeomorphism invariance, is the Einstein tensor plus a cosmological constant term.}

Theories that deviate from GR can, in most cases, be delineated into categories according to what principle or requirement they violate. A modification to GR can thus happen by allowing (i) extra fields, (ii)  higher-order metric derivatives, (iii) extra dimensions, (iv) non-locality or violation of Lorentz-invariance, see Fig.~\ref{fig:MGTypes}. Accordingly, MG models can be classified into the sub-categories described in the sub-sections that follow. 

However, as mentioned already in the introduction, the scope of this review is to give an overview of the current status of testing GR at cosmological scales rather than providing a review of MG models. Therefore, we only provide some outlines of models of interest or  models making a good illustrative case for a given category, while we refer the reader in each case to other specialized reviews in the literature. We refer the reader to some of the thorough reviews of MG models including \citep{2012-Clifton-MG,2015-rev-Joyce-et-al,BertiElAlReview2015} and references therein.  

%%%%%%%%%%%%%%%%%%%%%%%%%%%%%%%%%%%%%%%%%%%%%%%%%%%%%%%%%%%%%%%%%%%%%%%%%%%%%%%%%%%
%%%%%%%%%%%%%%%%%%%%%%%                                         %%%%%%%%%%%%%%%%%%%
%%%%%%%%%%%%%%%%%%%%%%%          SUB-SECTION                    %%%%%%%%%%%%%%%%%%%
%%%%%                  Modified gravity versus dark energy                     %%%%
%%%%%                                                                          %%%%
%%%%%%%%%%%%%%%%%%%%%%%%%%%%%%%%%%%%%%%%%%%%%%%%%%%%%%%%%%%%%%%%%%%%%%%%%%%%%%%%%%%
%%%%%%%%%%%%%%%%%%%%%%%%%%%%%%%%%%%%%%%%%%%%%%%%%%%%%%%%%%%%%%%%%%%%%%%%%%%%%%%%%%%

\subsection{Modified gravity versus dark energy}

A question that keeps coming back in the community is what is the distinction between dark energy and modified gravity models. How to distinguish between the two as a cause of cosmic acceleration. There is more than one answer to this question but with some possible clear guidelines and prescriptions that can be set. 

\cite{2016-Joyce-Lombriser-Schmidt-DEvsMG} and possibly others, propose to use the strong equivalence principle (SEP) (see Sect.~\ref{sec:Basic_principles}) to draw a distinction between GR + dark energy models versus MG models. They suggest to call any model that satisfies the SEP as a dark energy model and any model that violates SEP to be an MG model. They state that, heuristically, the SEP forbids the presence of a fifth force which motivates the use of such a discriminant. They state that using the SEP to make this distinction can be motivated further by the conjecture that GR is the only metric theory that satisfy the SEP, see \cite{WillReview2014}. They then state that a more pragmatic distinction is to rather use directly the presence (or not) of a fifth force to identify a model as being an MG model (or not) but with some grey zone as observed in for example \cite{Kunz2007}.

Next, \cite{Amendola2013} provided a phenomenological prescription to this question. 
First, they point to the simple case of quintessence that is straightforwardly referred to as dark energy model. In such a model, the scalar field is minimally coupled to curvature (see Sect.~\ref{sec:GBD}). In quintessence models, the scalar field also has a standard kinetic energy and the scalar potential represents the only functional degree of freedom. However, things get more ambiguous when moving beyond quintessence. 
The difficulty is that different models can have the same observables \citep{Kunz2007}. Also, some modified field equations can be recast into GR with extra source terms. 
Additionally, some scalar field dark energy models such as k-essence can have perturbations and clustering that can change the Poisson equation and induce a modified gravity parameter $Q(k,a)$ signaling a deviation from GR. Therefore, they suggested and used the following prescription:  

\begin{itemize}
\item{
Standard dark energy models: the scalar field here is non-minimally coupled to curvature in the Einstein's equations and has standard kinetic energy. The dark energy has no clustering on sub-horizon with a sound speed equal to the speed of light. Quintessence is a well-known example or perhaps definition. 
}
\item{
Clustering dark energy: In this case dark energy has fluctuations and can cluster on sub-horizon scales. These perturbations in the dark energy modify the Poisson equation  \eqref{eq:PoissonModLR} by inducing an MG parameter $Q(k,z)\ne 0$. But in this case, no gravitational slip is allowed. That is $\eta(k,a)=0$ and the clustering dark energy does not cause any anisotropic shear.  A good example is k-essence \citep{kessence1,kessence2}. This is also the case for the no-slip gravity \citep{Linder2018NSG}.
}
\item{Modified gravity models: These are models where the Field equations are changed leading to changes in the Poisson equations with non vanishing slip parameter $\eta(k,a)$. 
These are characterized by the presence of fifth force and violate the SEP.
Particles and bodies do not follow geodesics of the physical metric in the Einstein frame. This includes for example $f(R)$ (see Sect.\ref{sec:f(R)}), DGP (see Sect.~\ref{sec:DGP}), non-minimal coupled scalar-tensor theories and ``dark energy'' models with anisotropic clustering. 
}
\end{itemize}

\cite{Amendola2013} chose to follow the common practice of calling modified gravity models where GR is modified or where dark energy clusters. In other words, the last two items above. So models with $Q=\eta=1$ are dark energy models while if any of them departs from unity then it is an MG model. Of course, as they stress, this is not meant to be a fundamental classification but rather a convenient and useful phenomenological prescription.

%%%%%%%%%%%%%%%%%%%%%%%%%%%%%%%%%%%%%%%%%%%%%%%%%%%%%%%%%%%%%%%%%%%%%%%%%%%%%%%%%%%
%%%%%%%%%%%%%%%%%%%%%%%                                         %%%%%%%%%%%%%%%%%%%
%%%%%%%%%%%%%%%%%%%%%%%          SUB-SECTION                    %%%%%%%%%%%%%%%%%%%
%%%%%            Modified gravity theories with extra fields                   %%%%
%%%%%                                                                          %%%%
%%%%%%%%%%%%%%%%%%%%%%%%%%%%%%%%%%%%%%%%%%%%%%%%%%%%%%%%%%%%%%%%%%%%%%%%%%%%%%%%%%%
%%%%%%%%%%%%%%%%%%%%%%%%%%%%%%%%%%%%%%%%%%%%%%%%%%%%%%%%%%%%%%%%%%%%%%%%%%%%%%%%%%%

\subsection{Modified gravity theories with extra fields}
\label{sec: ExtraFieldMG}

\begin{figure*}
\includegraphics[width=\textwidth,height=8in,angle=0,fbox]{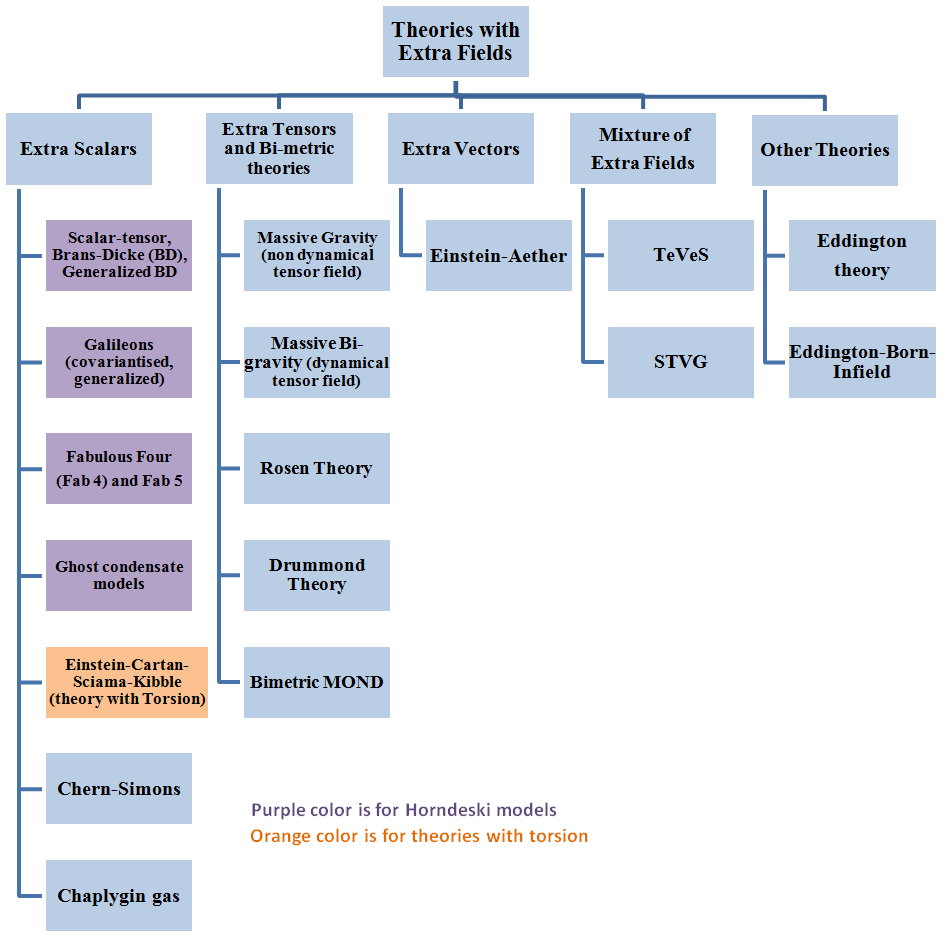}
\caption{Extra-Fields modified gravity models. Purple color is for Horndeski models and orange is for models with torsion. This table follows the models as listed in \cite{2012-Clifton-MG}
\label{fig:ExtraFieldsTab}}
\end{figure*}

In this category, the modification comes from adding scalar, vector or tensor field(s) to the metric. Fig.~\ref{fig:ExtraFieldsTab} provides examples of models for each sub-category and we provide below some illustrative examples for each sub-category.

%%%%%%%%%%%%%%%%%%%%%%%%%%%%%%%%%%%%%%%%%%%%%%%%%%%%%%%%%%%%%%%%%%%%%%%%%%%%%%%%%%%
%%%%%%%%%%%%%%%%%%%%%%%%%%%%%%%%%%%%%%%%%%%%%%%%%%%%%%%%%%%%%%%%%%%%%%%%%%%%%%%%%%%
%%%%%%%%%%%%%%%%%%%%%%%%%%%%%%%%%%%%%%%%%%%%%%%%%%%%%%%%%%%%%%%%%%%%%%%%%%%%%%%%%%%
\subsubsection{Theories with extra scalar field}  
\label{sec:scalar-tensor}
Scalar-tensor theories have been extensively studied in the literature from a theoretical point of view as well as comparison to observations, see for example \cite{Fujii2007} and references therein. Here a dynamical scalar field is added to the metric tensor hence the popular name. Let's survey the following examples. 

%%%%%%%%%%%%%%%%%%%%%%%%%%%%%%%%%%%%%%%%%%%%%%%%%%%%%%%%%%%%%%%%%%%%%%%%%%%%%%%%%
\paragraph{Illustrative example 1: Generalized Jordan--Fierz--Brans--Dicke (GJFBD)}\label{sec:GBD}\mbox{}\\ 

The GJFBD models have been very popular as scalar-tensor theories of gravity physics at various regimes, see e.g., the reviews \cite{WillReview2014,2012-Clifton-MG,KOYAMA2016TestGR}. In cosmology, the interest recently shifted to Galileon (see Sect.~\ref{sec:Galileons}) and Horndeski models because they can provide self-accelerating models. The Lagrangian for the GJFBD models can be written as, 
\begin{equation}
\label{eq:ScalarTensorL}
\mathcal{L} = \frac{1}{16 \pi} \sqrt{-g} \left[\phi R-\frac{\omega(\phi)}{\phi} \nabla_\mu \phi \nabla^{\mu}\phi -2 \Lambda(\phi)\right] + \mathcal{L}_m(\psi_{\rm m}, g_{\mu\nu}),
\end{equation}
where $\omega(\phi)$ is a coupling function, $\Lambda(\phi)$ is a potential or a function generalizing the cosmological constant, and  
$\mathcal{L}_m(\psi_{\rm m}, g_{\mu\nu})$ is the Lagrangian of the matter field $\psi_{\rm m}$.  

Variation of \eqref{eq:ScalarTensorL} with respect to the metric gives the first set of field equations, 
\begin{equation}
\label{ScalarTensorFE1}
\phi G_{\mu\nu} +\left[ \square \phi +  \frac{1}{2}\frac{\omega}{\phi} (\nabla \phi)^2+ \Lambda\right]   g_{\mu\nu} 
-    \nabla_\mu \nabla_\nu \phi -\frac{\omega}{\phi} \nabla_\mu \phi \nabla_\nu \phi
 = 8\pi T_{\mu\nu}.
\end{equation}
while variations with respect to the scalar field provides, after some steps, the remaining equations, 
\begin{equation}
\label{ScalarTensorFE2}
(2 \omega +3) \square \phi + \omega' (\nabla \phi)^2 
+4 \Lambda -2 \phi \frac{d\Lambda}{d\phi} = 8 \pi T.
\end{equation}

The action \eqref{eq:ScalarTensorL} is written in the Jordan frame where the scalar field is non-minimally coupled to the Ricci curvature scalar. It is assumed that there exist in this frame a metric $g_{\mu\nu}$ to which all matter species are universally coupled and the particles follow geodesics of this metric. The scalar field does not couple directly to the matter fields. 

One can transform \eqref{eq:ScalarTensorL} to the Einstein frame using a conformal transformation $g_{\mu \nu} =A(\phi)^2 \bar{g}_{\mu \nu} $ and by redefining the scalar field. In such an Einstein frame the scalar field is now minimally coupled to the Ricci scalar of $\bar{g}_{\mu \nu}$. However, the scalar field is directly coupled to the matter fields and test particles do not follow geodesics of $\bar{g}_{\mu \nu}$. The scalar field acts as an effective potential and isolated test particles feel a universal 4-acceleration. 

A popular sub-case of the theory is the Jordan--Fierz--Brans--Dicke (JFBD) theory \citep{BD1961,Will1993} obtained by setting $\omega$ as a constant noted as the Brans-Dicke coupling parameter $\omega_{_{_{\rm BD}}}$ and setting $\Lambda=0$, so \eqref{eq:ScalarTensorL} reduces to 
\begin{equation}
\label{BDAction}
\mathcal{L} = \frac{1}{16 \pi} \sqrt{-g} \left[\phi R-\frac{\omega_{_{\rm BD}}}{\phi} \nabla_\mu \phi \nabla^{\mu}\phi \right] + \mathcal{L}_m(\psi_{\rm m}, g_{\mu\nu}) 
\end{equation}
where the Brans-Dicke field gives an effective gravitational constant. The theory approaches General Relativity when $\omega \rightarrow \infty$. 

Exact solutions for spherically symmetric vacuum in Brans--Dicke theory have been derived and compared to solar system observations, see for example \cite{WillReview2014}. The Cassini-Huygens mission \citep{Cassini2003} sets the constraints $\omega_{\rm BD}>40,000$ so Brans--Dicke must be very close to GR. Unless there is a successful screening mechanism at work at small scales, this bound makes it difficult for Brans--Dicke theories to depart from GR at cosmological scales. For example, \cite{Bisabr2012} discuss Chameleon screened Generalized Brans--Dicke cosmology. However, as we discuss in Sect.~\ref{sec:Large-mass}, \cite{WangEtALNG} showed that such Chameleon screened models cannot explain cosmic acceleration unless we add a cosmological constant to them. 

As for the cosmology of JFBD, the field equations for an FLRW metric and a perfect fluid source give, the following Friedmann equations: 
\bea
\label{eq:BDFRW1}
H^2 &=& \frac{8 \pi \bar{\rho}}{3 \phi} - \frac{k}{a^2} - H
\frac{\dot{\phi}}{\phi} + \frac{\omega}{6} \frac{\dot{\phi}^2}{\phi^2}\\
\frac{\ddot{\phi}}{\phi} &=& \frac{8 \pi}{\phi} \frac{(\bar{\rho}-3\bar{P})}{(2
\omega +3)} -3 H \frac{\dot{\phi}}{\phi},
\label{eq:BDFRW2}
\eea
where over-dots are for derivatives with respect to proper time. The general solutions to the Brans-Dicke equations above have been fully explored in e.g., \cite{Gurevich1993,Barrow1993BD}.

In addition to the background equations, linear perturbations have been worked out in \cite{Nariai1969,Wu2010BD1,Nagata2002,Chen1999} so the theory can be compared to large scale structure and CMB data. For the perturbed FLRW metric \eqref{eq:PFLRW} in the Newtonian conformal gauge, a dust source, scalar field perturbation $\phi=\phi_0+\delta\phi$, and assuming the quasi-static approximation, the following scalar perturbation equations are obtained, e.g., \cite{KOYAMA2016TestGR}:
\begin{eqnarray}
\nabla^2 \Psi &=& 4 \pi G a^2 \delta\rho - \frac{1}{2} \nabla^2 \delta\phi, \label{BDPert1}\\
(3 + 2 \omega_{\rm BD}) \nabla^2 \delta\phi &=&- 8 \pi G a^2 \delta\rho, \label{BDPert2}\\ 
{\Phi} - \Psi &=&  \delta\phi \label{BDPert3}. 
\end{eqnarray}
The perturbations of the scalar field act as an effective anisotropic stress producing a slip between the two potentials. Inserting \eqref{BDPert2} into \eqref{BDPert1} shows that the presence of  the second term in Eq.~\eqref{BDPert1} is equivalent to a modification to the Newton gravitational constant.

%%%%%%%%%%%%%%%%%%%%%%%%%%%%%%%%%%%%%%%%%%%%%%%%%%%%%%%%%%%%%%
\paragraph{Illustrative example 2: Galileon and Covariant Galileon models}\mbox{}\\
\label{sec:Galileons}

Galileon models were introduced by \cite{NicolisEtAL2009Galileons} with some inspiration from DGP models (see Sect.~\ref{sec:DGP}) and aiming to go beyond their limitations. The models are constructed such that their action in flat spacetime is invariant under the following Galilean shift symmetry for the scalar field \citep{NicolisEtAL2009Galileons}
\begin{equation}
\partial_{\mu} \phi \to \partial_{\mu} \phi + c_{\mu},
\label{eq:gal-sym-lr}
\end{equation}
where $c_{\mu}$ is a constant vector. It turns out that with such conditions, the most general Lagrangian, that gives second order derivative equations of motion, has only 3 terms, in addition to the scalar field term and its canonical kinetic term. 
The terms are \citep{NicolisEtAL2009Galileons,DeffayetEtAl2009}:
\begin{eqnarray}
{\cal L}_1^{\rm gal}&=& \phi, \\
{\cal L}_2^{\rm gal}&=&-\frac{1}{2}(\partial \phi)^2, \\
{\cal L}_3^{\rm gal}&=&-\frac{1}{2}(\partial \phi)^2 \Box \phi, \\
{\cal L}_4^{\rm gal} &=&-\frac{1}{2} (\partial \phi)^2 \left[(\Box\phi)^2-(\partial_\mu\partial_\nu\phi)^2\right],  \\
{\cal L}_5^{\rm gal} &=& -\frac{1}{4}(\partial \phi)^2 
\bigl[(\Box\phi)^3 -3\Box\phi(\partial_\mu\partial_\nu\phi)^2
+2(\partial_\mu\partial_\nu\phi)^3 \bigr].
\label{eq:Galileon_LR}
\end{eqnarray}

The corresponding equations of motion can be found in \cite{NicolisEtAL2009Galileons}. A concise discussion on how this Lagrangian or other equivalent forms give only second order derivative equations of motion plus other properties of the models can be found in \cite{KOYAMA2016TestGR,2012-Clifton-MG,NicolisEtAL2009Galileons}. Galileon models can also result from conformal invariance  \citep{NicolisEtAL2009Galileons,CreminelliEtAl2013Galileons} or from the action of a brane in a higher dimensional spacetime \citep{deRham2010DBI:GAL}. Galileon models can also be constructed with more than one scalar field \citep{Trodden2011GG,Trodden2015CG,deRham2012GS}, or in terms of vector fields \citep{Tasinato2014CA,Heisenberg2014PROCA,Hull2014VG}.   

The next logical thing that was considered in literature was to convert Galileon models to curved spacetime. It was shown though that simply turning partial derivatives into covariant derivatives (covariantization) causes the appearance of third order derivatives in the equation of motion with the associated ghost fields \citep{DeffayetEtAl2009b}. To eliminate such higher order terms, \cite{DeffayetEtAl2009} introduced counter terms to write the covariant Galileon action as:
\begin{eqnarray}
{\cal L}_3 &=& -\frac{1}{2}(\nabla \phi)^2 \Box \phi, 
\label{eq:covGalileon}
\\
{\cal L}_4 &=& \frac{1}{8} (\nabla \phi)^4 R - \frac{1}{2}(\nabla \phi)^2
\left[(\Box\phi)^2-(\nabla_\mu\nabla_\nu\phi)^2\right], \\
{\cal L}_5 &=& -\frac{3}{8} (\nabla \phi)^4 G^{\mu\nu} \nabla_\mu\nabla_\nu\phi
\nonumber\\
&&- \frac{1}{4}  (\nabla \phi)^2
\bigl[(\Box\phi)^3 -3\Box\phi(\nabla_\mu\nabla_\nu\phi)^2
+2(\nabla_\mu\nabla_\nu\phi)^3 \bigr].
\end{eqnarray}  

Models from Lagrangian up to ${\cal L}_3$ are referred to as Cubic Galileons, up to  to ${\cal L}_4$ as Quartic, and up to ${\cal L}_5$ as Quintic. The models are self-accelerating with no need for a cosmological constant. An example of a concise practical formulation of the models to compare to cosmological data can be found in for example \cite{Barreira2014GAL3}. 

The covariant Galileon equations of motion are second and only second order derivatives. They are considered to be a subclass of the Horndeski models. A further generalization of the covariant Galileons to include zeroth and first order derivative equations of motion was carried out in \cite{DeffayetEtAl2009b} leading to the Horndeski action \eqref{eq:Horndenski_L} \citep{Horndeski1974}.   

It is worth mentioning that \cite{BeyondHorndeski2,BeyondHorndeski1} have shown that the addition of the counter terms in the covariantization of Galileon action is not strictly necessary to obtain healthy models. The equations of motion can thus be still cast into second order due to further constraints. In fact, these led to the proposal of the so-called beyond Horndeski models \citep{Zuma2014,BeyondHorndeski2,BeyondHorndeski1}.  

%%%%%%%%%%%%%%%%%%%%%%%%%%%%%%%%%%%%%%%%%%%%%%%%%%%%%%%%%%%%%%%%%%%%%%%%%%%%%%%%%
\paragraph{Illustrative example 3: Horndeski models and beyond ($\alpha_x$ parameterization)}\label{sec:horndeski}\mbox{}\\

This is the most general single-field scalar-tensor theory with second-order derivative equations of motion in (3+1) dimensions. A while ago, Horndeski \citep{Horndeski1974} derived the corresponding general Lagrangian and field equations but such work went quiet for sometime until their re-discovery within studies of generalized Galileon models, e.g. \cite{DeffayetEtAl2009,DeffayetEtAl2009b,DeffayetEtAl2011,Koba2011}, finding that the generalized covariant Galileon models are equivalent to Horndeski models. 
Most recently, Horndeski models have been extensively studied analytically and phenomenologically but fully constraining them observationally remains a challenge, e.g. \cite{Planck2015MG} due to their large number of parameters.

The Horndeski action is given by, e.g., \cite{Horndeski1974,DeffayetEtAl2011,Koba2011}
\begin{equation}
S = \int d^4x \sqrt{-g} \left[\sum_{i=2}^{5}{\cal L}_{i} + {\cal L}_M(g_{\mu \nu},\psi_{\rm m}) \right]
\label{Horndeski_Action}
\end{equation}
where
\begin{eqnarray}
\nonumber
{\cal L}_{2} & = & K(\phi,X),\\ \nonumber
{\cal L}_{3} & = & -G_{3}(\phi,X)\Box\phi,\\ \nonumber
{\cal L}_{4} & = & G_{4}(\phi,X)\, R+G_{4X}\,[(\Box\phi)^{2}-(\nabla_{\mu}\nabla_{\nu}\phi)\,(\nabla^{\mu}\nabla^{\nu}\phi)]\,,\\ \nonumber
{\cal L}_{5} & = & G_{5}(\phi,X)\, G_{\mu\nu}\,(\nabla^{\mu}\nabla^{\nu}\phi) \\ \nonumber
&-&\frac{1}{6}\, G_{5X}\,[(\Box\phi)^{3}-3(\Box\phi)\,(\nabla_{\mu}\nabla_{\nu}\phi)\,(\nabla^{\mu}\nabla^{\nu}\phi) \\
&+&2(\nabla^{\mu}\nabla_{\alpha}\phi)\,(\nabla^{\alpha}\nabla_{\beta}\phi)\,(\nabla^{\beta}\nabla_{\mu}\phi)] \ ,
\label{eq:Horndenski_L}
\end{eqnarray}
where $K$ and $G_{3}$-$G_{5}$ are functions of the scalar field $\phi$ and that of its kinetic energy, $X=-\partial^{\mu}\phi\partial_{\mu}\phi/2$, $R$ is the Ricci scalar, $G_{\mu\nu}$ is the Einstein tensor. $G_{iX}$ and $G_{i\phi}$ are the partial derivatives of $G_{i}$ with respect to $X$ and $\phi$, respectively.
The four functions, $K$ and $G_{3}$-$G_{5}$ characterize completely this class of theories. The corresponding equations of motion can be found in \cite{BelliniSawicki2014,DeffayetEtAl2011,Koba2011}. It is worth noting that there are no a-priori mass or energy scales that are associated with the functions $G_{3}$-$G_{5}$ that would put them in some hierarchical order. When a model is specified, these functions may feature a mass scale that will determine at what scale they contribute to the dynamics. This mass scale is usually chosen so the terms have an effect at cosmological scales. The appearance of such mass scales differ though from one model to another within the Hordeski models. A brief discussion for Galileon models can be found after Eq.~3 in \cite{Baker2018}. 

A physically meaningful parameterization for the Horndeski models was introduced by \cite{BelliniSawicki2014} from applying and specializing the EFT approach discussed in Sect.~\ref{sec:EFT} to this class of models. First, for Horndeski models, the following relations between the functions of the EFT action \eqref{eq:EFT_action} must hold:

\be
m_2^2=0; \ 2{\hat M}^2 = {\bar M}^2_2 =-{\bar M}^2_3.
\ee

Consequently, the nine EFT functions can be replaced by four functions of time only noted as $\alpha_M, \alpha_K, \alpha_B$ and $\alpha_T$ plus the effective Planck mass $M_\ast^2$ and an additional function of time for the background such as for example $H(a)$ \citep{BelliniSawicki2014}. These 5 functions of time and the effective Planck mass fully characterize the linear dynamics of the Horndeski models. 

The relationships between the $\alpha_x$ parameters and the set of EFT functions of \eqref{eq:ListCoefEFT} have been provided in \cite{BelliniSawicki2014} and can be summarized as follows:

\begin{eqnarray}
\textit{M}_\ast^2 &=& m_0^2\Omega + \bar{\textit{M}}_2^2 \label{eq:M_ast}; \\
\textit{M}_\ast^2 H \alpha_M &=& m_0^2 \dot{\Omega} + \dot{\bar{\textit{M}}}_2^2;
  \label{eq:alphaM} \\
\textit{M}_\ast^2 H^2 \alpha_K &=& 2c + 4 \textit{M}_2^4; \\ \label{eq:alpha_K}
\textit{M}_\ast^2 H \alpha_B &=& -m_0^2 \dot{\Omega} -\bar{\textit{M}}_1^3; \label{eq:alphaB} \\
\textit{M}_\ast^2 \alpha_T &=& -\bar{\textit{M}}_2^2 \label{eq:alpha_T}.
\end{eqnarray}

The authors also provided a connection between the physical properties of the theory and  the $\alpha_x$ parameters as follows:

\begin{itemize}

\item {$\alpha_K$: quantifies the \textit{kineticity} of the scalar field originating from the presence of its kinetic energy term in the Lagrangian. For example, minimally coupled models such as quintessence or k-essence all have a scalar field kinetic term and thus $\alpha_K \ne 0$. On the other hand $f(R)$ (see Sect.~\ref{sec:f(R)}) or $f(G)$ models have no such a term and thus $\alpha_K = 0$. In the general Horndeski models, $\alpha_K$ receives contributions from the Lagrangian functions $K$, $G_3$, $G_4$ and $G_5$, see Appendix~A in \cite{BelliniSawicki2014}.}

\item {$\alpha_T$: quantifies the excess of tensor (gravity waves) speed from the speed of light (i.e., $c^2_T-1$) and thus the deviation of gravitational waves speed from that of light. This also affects the coupling between the matter and the Newtonian potential resulting in anisotropic stress regardless of scalar perturbations. In the general Horndeski models, $\alpha_T$ receives contributions from the functions $G_4$ and $G_5$.}

\item {$\alpha_B$: quantifies the \textit{braiding} or mixing of the kinetic terms of the scalar field and the metric. Can cause dark energy clustering. $\alpha_B = 0$ for minimally coupled models of dark energy such as quintessence and k-essence but non-zero for all modified gravity models, i.e., all models where a fifth force is present \citep{PogosianEtAl2016}.  $\alpha_B$ receives contributions from the functions $G_3$, $G_4$ and $G_5$ in Horndeski models.}

\item {$\alpha_M$: quantifies the \textit{running rate} of the effective Planck mass. It is generated by a restricted non-minimal coupling. It creates anisotropic stress.  
$\alpha_M = -\alpha_B \ne 0$ for $f(R)$ models.  $\alpha_M = 0$ for minimally coupled models of dark energy models such a quintessence and k-essence.} 

\end{itemize}

It is worth noting that because the $\alpha_x$-parameterization can be connected very well to physical properties of the of the models, it can then serve well the task of assessing the stability criteria of the models, see for example a recent discussion in \cite{Kennedy2018} and references therein.  

Later on, \cite{BeyondHorndeski1,BeyondHorndeski2} added to Eqs.~\eqref{eq:M_ast}--\eqref{eq:alpha_T} and the $\alpha_x$ parameterization the following relation 
\be
M_\ast^2 \alpha_H = 2 \hat{M}^2 - \bar{M}_2^2,
\ee
where the authors introduced $\alpha_H\ne 0$ to parameterize a deviation from Horndeski models. Their formalism thus included viable models with a single scalar field but with higher-order equations of motion referred to as beyond-Horndeski models. However, the authors showed how internal constraints in the theory assures that it is free of Ostrogradski instabilities. Some of beyond Horndeski models are obtained by a disformal transformation. However, see some reservation and discussion in \cite{Crisostomi2016BH} about the beyond-Horndeski characterization. 

\begin{table}[h!]
\caption{Parameter functions $\alpha_{i}$ for various dark energy and modified gravity model sub-classes of Horndeski models. 
Simple dark energy models are described mostly by one or two functions while covariant Galileons and $f(G)$ require three. Reproduced with permission from Table~1 of \cite{BelliniSawicki2014}, copyright by IOP.}
\label{tab:alphas}
\centering
\begin{tabular}{lrrrr}
\hline
\textsf{\textbf{\small{}Model Class}}&\textsf{\textbf{\small{}$\boldsymbol{\alpha_{\textrm{K}}}$}}&{\small{}$\boldsymbol{\alpha_{\textrm{B}}}$}& \textsf{\textbf{\small{}$\boldsymbol{\alpha_{\textrm{M}}}$}}&\textsf{\textbf{\small{}$\boldsymbol{\alpha_{\textrm{T}}}$}}\\
\hline
\hline
\textbf{\emph{\small{}$\boldsymbol{\Lambda}$CDM }}&\textbf{\emph{\small{}0}}& \textbf{\emph{\small{}0}}&\textbf{\emph{\small{}0}} & \textbf{\emph{\small{}0}}\\
\hline
{\small{}cuscuton ($w_{X}\neq-1)$}&{\small{}0}&{\small{}0}&{\small{}0}&{\small{}0}\\
{\small{}\citep{Afshordi2007CF}}&&&&\\
\hline
{\small{}quintessence}&{\small{}$(1-\Omega_{\text{m}})(1+w_{X})$}&{\small{}0}&{\small{}0} &{\small{}0}\\
{\small{}\citep{Ratra1988CC,Wetterich1988}}&&&&\\
\hline
{\small{}k-}\emph{\small{}essence}{\small{}/perfect
fluid}&{\small{}$\frac{(1-\Omega_{\text{m}})(1+w_{X})}{c_{\text{s}}^{2}}$}&{\small{}0}& {\small{}0}&{\small{}0}\\
{\small{}\citep{kessence1,kessence2}}&&&&\\
\hline
{\small{}kinetic gravity braiding}&{\small{}$\nicefrac{m^{2}\left(n_{m}+\kappa_{\phi}\right)}{H^{2}M_{\text{Pl}}^{2}}$}&{\small{}$\nicefrac{m\kappa}{HM_{\text{Pl}}^{2}}$}&{\small{}0}&{\small{}0}\\
{\small{}\citep{Deffayet2010,Kobayashi2010}}&&&&\\
\hline
{\small{}covariant Galileon cosmology}&{\small{}$-\nicefrac{3}{2}\alpha_{\textrm{M}}^{3}H^{2}r_{\text{c}}^{2}e^{2\phi/M}$}&{\small{}$\nicefrac{\alpha_{\textrm{K}}}{6}-\alpha_{\textrm{M}}$}&{\small{}$\nicefrac{-2\dot{\phi}}{HM}$}&{\small{}0}\\
{\small{}\citep{Chow2009}}&&&&\\
\hline
{\small{}Imperfect fluid scalar-tensor}&{\small{}$\nicefrac{\dot{\phi}^{2}K_{,\dot{\phi}\dot{\phi}}e^{-\kappa}}{H^{2}M^{2}}$}&{\small{}$-\alpha_{\textrm{M}}$}&{\small{}$\nicefrac{\dot{\kappa}}{H}$}&{\small{}0}\\
{\small{}\citep{Sawicki2013}}&&&&\\
\hline
{\small{}metric $f(R)$}&{\small{}0}&{\small{}$-\alpha_{\textrm{M}}$}&{\small{}$\nicefrac{B\dot{H}}{H^{2}}$}&{\small{}0}\\
{\small{}\citep{Carroll2004,Song2007}}&&&&{\small{}}\\
\hline
{\small{}MSG/Palatini $f(R)$}&{\small{}$-\nicefrac{3}{2}\alpha_{\textrm{M}}^{2}$}& {\small{}$-\alpha_{\textrm{M}}$}&{\small{}$\nicefrac{2\dot{\phi}}{H}$}&{\small{}0}\\
{\small{}\citep{Carroll2006,Vollick2003}}&&&&\\
\hline
{\small{}$f(G)$}&{\small{}0}&{\small{}$\frac{-2H\dot{\xi}}{M^{2}+H\dot{\xi}}$}&{\small{}$\frac{\dot{H}\dot{\xi}+H\ddot{\xi}}{H\left(M^{2}+H\dot{\xi}\right)}$}&{\small{}$\frac{\ddot{\xi}-H\dot{\xi}}{M^{2}+H\dot{\xi}}$}\\
{\small{}\citep{Carroll2005,DeFelice2010}}&&&&\\
\hline
\end{tabular}
\end{table}

It is worth recalling here the definitions of conformal and disformal transformations of the metric given by  
\bea
\label{eq:conformal_disformal}
\bar{g}_{\alpha\beta} \; = \; A(\phi,{X}) \, {g}_{\alpha\beta} \, + \, B(\phi,{X})\,  \partial_\alpha\phi\, \partial_\beta\phi
\eea
where ${X} \equiv -\frac{1}{2}{g}^{\alpha\beta}\partial_\alpha\phi\,  \partial_\beta\phi$. 
The first term on the right of (\eqref{eq:conformal_disformal}) represents a conformal transformation rescaling the metric tensor. The second term is a pure disformal transformation stretching the metric in the direction given by  $\partial_\alpha\phi$.

Further efforts continued to explore models beyond Horndeski. Remarkably,  \cite{Langlois2016D,LangloisNoui2016Deg} identified the degeneracy conditions that assure that the theory is free from Ostrogradsky ghost even if their equations of motion have higher order derivatives. This allowed \cite{Langlois2016D,Crisostomi2016BH} to identify viable beyond-Horndeski theories and even new classes of ghost free degenerate higher order theories in\cite{Langlois2016D,Crisostomi206EH,Achour2016a,BenAchour2016DH,Crisostomi2017a}.
The models introduced in \cite{Langlois2016D} are now known as the degenerate higher derivative theories beyond Horndeski (DHOST), as dubbed in \cite{Langlois2017D3} and the concise review \citep{Langlois2017D2}. They generalize Horndeski and beyond-Horndeski models and are the most general class of ghost-free scalar-tensor theories. In these theories, it was shown in the vacuum in absence of matter coupling that the presence of a special degeneracy of the Lagrangian ensures the absence of ghosts even if the equations of motion are higher order. Also, if the matter coupling is disformal then it can not break this degeneracy but that is not the case for minimal coupling of matter. This is an interesting class of models that remain to be studied and compared to cosmological observations. We discuss some constraints on these models from neutron-star-merger event GW170817 and GRB170817A in Sect.~\ref{sec:GW170817}.

If the coupling of matter is disformal (it can be minimal of course), then it could not break the degeneracy, and the ghost is indeed absent.

We reproduce for illustration purposes, Table~1 from \cite{BelliniSawicki2014} where the $\alpha_x$ functions are given for known dark energy or modified gravity models. 

%%%%%%%%%%%%%%%%%%%%%%%%%%%%%%%%%%%%%%%%%%%%%%%%%%%%%%%%%%%%%
%\mbox{}\\
\paragraph{Other scalar-tensor theories}\mbox{}\\

An interesting scalar-tensor theory is that of Mimetic gravity that was originally proposed as mimetic dark matter in \cite{MimeticDM2013}, see specialized review \citep{MDMReview2016}. It was extended to produce inflation and late-time cosmic acceleration as well as to address cosmological or astrophysical singularities \citep{Chamseddine2014,Chamseddine2017,Chamseddine2017b,Achour2017MG}.
The theory and its extensions can be constructed from the action, e.g., \cite{Langlois2018Mimetic}  
\bea\label{mimetic}
S[\tilde{g}_{\alpha\beta},\phi] = \int d^4x \, \sqrt{ -g} \, 
{\cal L}(\phi,\partial_\alpha \phi, \nabla_{\alpha}\!\nabla_\beta \phi\, ; g_{\alpha\beta}) \, ,
\eea
where the variation must be taken with respect to scalar field $\phi$ and the auxiliary metric $\tilde{g}_{\alpha\beta}$ which is related to the physical metric  by a non-invertible disformal transformation,
\bea\label{disformal}
g_{\alpha\beta} \; = \; \tilde A(\phi,\tilde{X}) \, \tilde{g}_{\alpha\beta} \, + \, \tilde B(\phi,\tilde {X})\,  \partial_\alpha\phi\,  \partial_\beta\phi  \, ,\qquad \tilde{X} \equiv \tilde{g}^{\alpha\beta}\partial_\alpha\phi\,  \partial_\beta\phi\,.
\eea
The original mimetic dark matter theory had the Einstein--Hilbert term for $g_{\alpha \beta}$ as Lagrangian so in \eqref{mimetic} it would depend only on $g_{\alpha \beta}$ and not on $\phi$ explicitly. We refer the reader to the review \cite{MDMReview2016} for various formulations and discussions. 

\cite{MDMReview2018} performed a dynamical analysis of the theory showing that Mimetic Gravity can have successive radiation and matter dominated epochs followed by an accelerating phase. 
Interestingly, the dark matter and dark energy parameter have the same order of magnitude thus addressing the cosmic coincidence problem. These and other features were also stressed in \cite{Chamseddine2014} and references therein.
\cite{Mirzagholi2015} introduced a novel simple mechanism to
produce mimetic DM during radiation epoch.
 Perhaps the most interesting feature is that of a possible unified scenario for inflation, dark matter and dark energy. 

However, a very recent study \cite{Langlois2018Mimetic} showed that mimetic gravity theories 
can be viewed/formulated as degenerate higher-order scalar theories (DHOST) \citep{Langlois2016D} with an extra local gauge symmetry. They study linear perturbations about a homogeneous and isotropic background for all mimetic theories and find that they have either gradient instabilities or an Ostrogradsky type of instability in the scalar sector coupled to matter. The matter they included was in the form of k-essence scalar field.  It will be interesting to  see further development on this particular point and if there are ways around it in this unifying scenario of the dark sector.   

Another interesting scalar-tensor theory is the ghost condensation model as proposed by \cite{GCTheory2004}. A scalar ghost field is added but the theory is kept stable because the terms in the action push the kinetic terms to a fixed condensation value avoiding instability. The theory has spontaneous breaking of Lorentz invariance. In such a theory, the ghost condensate field plays a role in the gravitational sector that is similar to that of the Higgs field in particle physics. That is gravitational fields propagating through the ghost condensate scalar field acquire a mass just like particles acquire mass while propagating through the Higgs field. The ghost condensate field fills space in the universe and is equivalent to a fluid with the same equation of state, $w=-1$, as a cosmological constant, and thus can drive the observed cosmic acceleration. But, such a ghost condensate fluid has physical scalar excitations and can be described as an effective field theory. The theory has interesting features such as attractive or repulsive gravity and has been used for problems in inflation, dark matter and cosmic acceleration \citep{GCTheory2004}. 
The models lead to an interesting cosmological phenomenology including Friedmann equations that can be fit to observations and scalar perturbations that lead to a growth equation with additional terms that can be compared to large scale structure observations \citep{Mukohyama2006,Krause2006}. 

\cite{Charmousis2012GS} proposed what they called the $F_{ab}$ four scalar-tensor theory and its cosmology in \cite{Fab4Cosmo}. Interestingly, this theory proposed a self-tuning mechanism that screens the contribution of the cosmological constant to curvature through phase transition. The name is given because the theory is based on 4 specific terms from the Horndeski action they call Fab Four (in analogy with the Beatles, they named the terms as John, Paul, George, and Ringo). The fabulous aspect is the self-tuning screening of the cosmological constant with a way around Weinberg's no-go theorem \citep{Weinberg1988CC} by allowing the scalar field to break Poincar\'e invariance on the self-tuning vacua. However, it was argued in \cite{Fab5A,Fab5B} that such models fail to provide a viable cosmic evolution for the whole cosmic history. Furthermore, these models have been ruled out by the requirement of $c_T=c$ from the neutron star merger event GW170817/GRB170817A.

\cite{Afshordi2007,Afshordi2007CF} introduced a scalar-tensor model they call Cuscuton that is based on the infinite sound speed limit of k-essence model fluid \citep{ArmendarizPicon1999}. k-essence is a model where the late-time acceleration is caused by the kinetic energy of the scalar field and not its potential energy as is the case in quintessence dark energy models (this the special case of 
Eq.~\eqref{eq:Horndenski_L} where only the second is present and the function $K$ depends only on $X$ there). k-essence field has perturbations which cause a change in the GR Poisson Eq.~\eqref{eq:PoissonModLR} (i.e., $Q(k,z)\ne 0$) but no gravitational slip, i.e., $\eta(k,z)=1$. \cite{Afshordi2007} show that Cuscuton model is causal and perturbations do not introduce any additional dynamical degree of freedom but just obey only a constraint equation. They state that the model can be viewed as an effective modification of gravity on large scales. They also remark that this is the only modification of Einstein gravity that does not introduce any additional degrees of freedom and is not conformally equivalent to GR. They studied two models with specific potentials finding that one can mimic $\Lambda$CDM expansion history but has some early time departure from it. The second model has an expansion history similar to that of DGP (see Sect.~\ref{sec:DGP}) but is consistent with ISW effect from WMAP data. Cuscuton modes have a free potential and constraining them using observations will depend on the choice of such a potential.

Finally, it is worth including a different type of scalar-tensor theory known as the Einstein--Cartan--Sciama--Kibble theory. It constitutes an interesting development in gravity theories in which the torsion tensor is not vanishing and the affine connection is not symmetrical  \citep{Cartan1922b,Cartan1923,Cartan1924,Sciama1962,Sciama1964,Kibble1961}. The torsion is related to the angular momentum (spin) of matter and the theory differs from GR only when the spin effects are important.  \cite{Hehl1973} evaluated that for electron the density that will make the spin effect relevant is $\sim 13^{38}\,\rm{Kg/m}^{3}$ and $\sim 13^{45}\,\rm{Kg/m}^{3}$ for neutrons. These high densities can exist in the dense early universe but not any later during cosmic evolution. We refer the reader to the reviews \cite{Hehl1974,Trautman2006,Desai2016}.

%%%%%%%%%%%%%%%%%%%%%%%%%%%%%%%%%%%%%%%%%%%%%%%%%%%%%%%%%%%%%%%%%%%%%%%%%%%%%%%%%%%
%%%%%%%%%%%%%%%%%%%%%%%%%%%%%%%%%%%%%%%%%%%%%%%%%%%%%%%%%%%%%%%%%%%%%%%%%%%%%%%%%%%
%%%%%%%%%%%%%%%%%%%%%%%%%%%%%%%%%%%%%%%%%%%%%%%%%%%%%%%%%%%%%%%%%%%%%%%%%%%%%%%%%%%
\subsubsection{Extra Vector field(s)}

%%%%%%%%%%%%%%%%%%%%%%%%%%%%%%%%%%%%%%%%%%%%%%%%%%%%%%%%%%%%%%%%%%%%%%%%%%%%%%%%%%%
\paragraph{Illustrative example: Generalized Einstein-Aether Theories}\mbox{}\\
\label{sec:EAther}

In this theory, a unit timelike vector is added to the tensor metric. The vector field provides a preferred reference frame and constitutes an Aether-like field making the violation of Lorentz invariance.  \cite{Gasperini1987} first revived the idea of an Aether-like theory with the use of a scalar field and a preferred reference frame. 
\cite{Kostelecky1989} developed a framework for spontaneous Lorentz-symmetry breaking in higher dimensions that served for many related purposes later.
\cite{Jacobson2001EA,Jacobson2004EA,Eling2004} followed a decade later by proposing a theory where in addition to the metric, a unit timelike vector field is added to the theory providing a direction of time and a preferred frame breaking Lorentz invariance under boosts. This was then generalized further by \cite{Zlosnik1,Zlosnik2}. 
We outline some aspects of the field equations and cosmology in this theory following \cite{Zlosnik1,Zlosnik2,Meng2012EA}

The action for Generalized Einstein-Aether theory is given by  
\begin{equation}
S=\int d^{4}x \sqrt{-g} \left[\frac{R}{16 \pi G}+{\cal L}_{A}+{\cal L}_{M}\right],
\label{eq:EAactionLR}
\end{equation}
where the additional ${\cal L}_{A}$ term is the Lagrangian for the vector field, $A^\alpha$ given by \cite {Zlosnik1,Zlosnik2}:
\begin{eqnarray}
{\cal L}_{A}&=&\frac{M^{2}}{16 \pi G} {\cal F}({\cal K})+\frac{1}{16 \pi G} \lambda
(A^{\alpha}A_{\alpha}+1)\nonumber
\\
{\cal K}&=&M^{-2} {\cal K}^{\alpha \beta}_{\phantom{\alpha \beta} \gamma \sigma}
\nabla_{\alpha}A^{\gamma}\nabla_{\beta}A^{\sigma}\nonumber
\\
{\cal K}^{\alpha \beta}_{\phantom{\alpha \beta} \gamma \sigma}&=&c_{1}g^{\alpha
\beta}g_{\gamma \sigma}+c_{2}\delta^{\alpha}_{\gamma}\delta^{\beta}_{\sigma}+
c_{3}\delta^{\alpha}_{\sigma}\delta^{\beta}_{\gamma}-c_{4}A^{\alpha}A^{\beta}g_{\gamma\sigma},
\label{eq:EAlagrangianLR}
\end{eqnarray}
where $c_{i}$ are dimensionless constants and $M$ is a coupling constant
with mass dimension and typical scale value of the order of $H_0$ for cosmological purposes. $\lambda$ is a Lagrange multiplier to ensure the vector field is unit time-like, i.e., $A^{\alpha}A_{\alpha}=-1$.
The ${\cal F}({\cal K})$ is a free function. For the case of linear Einstein-Aether theory it is simply equal to ${\cal K}$. 

Variation of the action \eqref{eq:EAactionLR} with respect to $g^{\alpha\beta}$ and $A^{\beta}$ respectively gives 
\begin{eqnarray}
G_{\alpha\beta}&=&\tilde {T}_{\alpha\beta}+8 \pi G T^{matter}_{\alpha\beta}
\label{eq:eqn_g}
\\
\nabla_{\alpha}({\cal F}' J^{\alpha}_{\phantom{\alpha}\beta})&=& 2\lambda A_{\beta},
\label{eq:eqn_A}
\end{eqnarray}
where $\tilde {T}_{\alpha\beta}$ is the energy-momentum tensor for the vector field,
${\cal F}'=\frac{d{\cal F}}{d{\cal K}}$, and $J^{\alpha}_{\phantom{\alpha}\sigma}=2 {\cal K}^{\alpha \beta}_{\phantom{\alpha\beta} \sigma \gamma}\nabla_{\beta}A^{\gamma}$.
For ${\cal K}^{\alpha \beta}_{\phantom{\alpha \beta} \gamma \sigma}$ given by \eqref{eq:EAlagrangianLR}, $\tilde {T}_{\alpha\beta}$ is given by \cite{Zlosnik1,Zlosnik2}
\begin{equation}
\tilde {T}_{\alpha\beta}=\frac{1}{2}\nabla_{\sigma} \left[{\cal F}'
(J_{(\alpha}^{\phantom{\alpha}\sigma}A_{\beta)}-J^{\sigma}_{\phantom
{\sigma}(\alpha}A_{\beta)}-J_{(\alpha\beta)}A^{\sigma})\right]-{\cal F}'
Y_{(\alpha\beta)}+\frac{1}{2}g_{\alpha\beta}M^2{\cal F}+\lambda A_{\alpha}
A_{\beta},
\label{eq:EATab}
\end{equation}
where the $_{(\dots)}$ denotes symmetry with respect to the indices. $Y_{\alpha\beta}$ is given for the particular choice of \eqref{eq:EAlagrangianLR} (but setting $c_4=0)$ by 
\begin{equation}
Y_{\alpha\beta}=-c_{1}\left[(\nabla_{\nu}A_{\alpha})(\nabla^{\nu}A_{\beta})-
(\nabla_{\alpha}A_{\nu})(\nabla_{\beta}A^{\nu})\right].
\label{eq:EAYab}
\end{equation}

Next, we outline some aspects of the cosmological evolution in the theory. We consider the general FLRW metric \eqref{eq:flrw}, the unit time-like vector 
\be
A^{\alpha}=(1,0,0,0). 
\label{eq:EAA}
\ee
For the matter source, we consider a perfect fluid with velocity field $u^{\alpha}$ and energy momentum tensor given by 
\begin{equation}
T^{matter}_{\alpha\beta}=\rho u_{\alpha}u_{\beta}+p (u_{\alpha}u_{\beta}+g_{
\alpha\beta}).
\label{eq:EAMfluid}
\end{equation}
The results are as follows \citep{Zlosnik1}: 
\begin{eqnarray}
\nabla_\beta A^{\beta}&=&3 H
\nonumber
\\
{\cal K}&=&3\alpha\frac{H^2}{M^2},
\label{eq:EAKsim}
\end{eqnarray}
where $\alpha \equiv c_{1}+3c_{2}+c_{3}$ and $H\equiv\dot{a}/a$ is the usual Hubble parameter. 
As shown in \cite{Zlosnik1,Zlosnik2},  the energy momentum tensor, \eqref{eq:EATab}, also takes the form of a perfect fluid with effective state variables given by 
\begin{equation}
\bar{\rho}_{A}=3\alpha H^2({\cal F}'-\frac{\cal F}{2 \cal K})
\label{eq:EArho_A}
\end{equation}
and 
\begin{equation}
\bar{p}_{A}=3\alpha H^2(-\frac{2}{3}{\cal F}'+\frac{\cal F}{2\cal K})-\alpha\dot{
\cal F}'H-\alpha{\cal F}'\frac{\ddot{a}}{a},
\label{eq:EAp_A}
\end{equation}
satisfying the energy conservation equation $\dot{\rho}_A+3H(\rho_A+p_{A})=0$. 

Next, the field equations give the modified Friedmann equations \citep{Zlosnik1,Zlosnik2}
\begin{eqnarray}
(1-\alpha{\cal F}'+\frac{1}{2}\frac{\alpha{\cal F}}{\cal K})H^2+\frac{k}{a^2}&=&\frac{
8\pi G}{3}\bar{\rho}
\label{eq:Friedmann1}
\\
\frac{d}{dt}(-2H+\alpha{\cal F}'H)+\frac{2k}{a^2}&=&8\pi G(\bar{\rho}+\bar{p}).
\label{eq:Friedmann2}
\end{eqnarray}
We can see that additional terms are present from the function ${\cal F}({\cal K})$ and its derivatives that can be encapsulated to play the role of an effective cosmological constant due to the presence of the Aether field vector. The theory also contains a modified effective gravitational constant. Specific examples can be found in \cite{Zlosnik2,Zuntz2010EA,Lim2005} where specific choices of ${\cal F}({\cal K})=\gamma(-{\cal K})^n$ can lead to late time acceleration with $n=0$ corresponding to a $\lcdm$ model. \cite{Meng2012EA} proposed other models leading to other effective dark energy models. \cite{Battye2017EA} also developed a designer approach to generalized Einstein-Aether to mimic any $w$CDM background. Instead of specifying a specific
${\cal F}({\cal K})$  function, other parameters such as $w$ and $\Omega_{\rm de}$ can be specified to find a functional form for ${\cal F}({\cal K})$. This essentially amounts to solving the generalized Friedmann equations \eqref{eq:Friedmann1} and \eqref{eq:Friedmann2}.

Linear perturbations for generalized Einstein-Aether theory have been worked out in, e.g., \cite{Zuntz2010EA,Picon2010EA,Battye2017EA} taking into account perturbations of the metric and the vector field. This gives modified Poisson equations where the vector field leads to a different source for the Poisson equations and also induces a slip between the two gravitational potentials. This provides a means to test the models using large scale structure as well as CMB and to distinguish them from the $\lcdm$ model.

\paragraph{Other vector-tensor theories}\mbox{}\\

Some of the first vector-tensor theories were those of \cite{Will1972} where the authors derived and explored the models within an extended PPN formalism.  
Other vector-tensor theories include the recent generalized Proca theories where the vector field is promoted to a Proca massive vector field with ghost free models and accelerating cosmologies \citep{Heisenberg2014PROCA,DeFelice2016Proca1,Jimenez2017Proca,DeFelice2016Proca2,Heisenberg2016BG}. 
A number of other developments on vector-tensor theories can be found in \cite{Jimenez2016CF,Heisenberg2014PROCA,Tasinato2014CA,Allys2016GP,BeltranJimenez2016DS,
Heisenberg2016BG,Kimura2017EV,DeFelice2016Proca1,DeFelice2016EG,Jimenez2017Proca,
Emami2017SS,BeltranJimenez2013SO,BeltranJimenez2017II,Hull2014VG,Hull2016CV,Allys2016GS,
Nakamura2017Proca}. A concise review on generalized Proca theories can be found in \cite{Heisenberg2017Review}

%%%%%%%%%%%%%%%%%%%%%%%%%%%%%%%%%%%%%%%%%%%%%%%%%%%%%%%%%%%%%%%%%%%%%%%%%%%%%%%%%%%
%%%%%%%%%%%%%%%%%%%%%%%%%%%%%%%%%%%%%%%%%%%%%%%%%%%%%%%%%%%%%%%%%%%%%%%%%%%%%%%%%%%
%%%%%%%%%%%%%%%%%%%%%%%%%%%%%%%%%%%%%%%%%%%%%%%%%%%%%%%%%%%%%%%%%%%%%%%%%%%%%%%%%%%
\subsubsection{Extra Vector and Scalar fields}

%%%%%%%%%%%%%%%%%%%%%%%%%%%%%%%%%%%%%%%%%%%%%%%%%%%%%%%%%%%%
\paragraph{Illustrative example: TeVeS Theory}\mbox{}\\
\label{sec:TEVES}

A Tensor-Vector-Scalar theory known as TeVeS in the literature was introduced by Bekenstein in \cite{Bekenstein2004}  as a relativistic generalization  of Modified-Newtonian-Dynamics (MOND) theory \citep{MOND1,MOND333333}. MOND and TeVeS both aim at addressing some observations such as the flat rotation curves of galaxies without the need for Dark Matter. MOND has been criticized for not fitting other astrophysical observations but see discussion and debate in for example \cite{Scott2001MOND,Foreman2012,McGaugh2011}. 

TEVES provides a more complex theory where the additional vector field could for example cause a stronger gravitational infall of baryons during the early universe epoch and thus alleviates the need for dark matter to create strong gravitational potential wells, see e.g., \cite{DodelsonLigori2006}. 

The TeVeS action is commonly written in two frames and we follow that here (a single frame formulation can be found in \citealt{Zlosnik2}). The gravitational fields are written in the Einstein frame (sometime also referred to as the Bekenstein frame for this specific theory) while the matter fields are written in the frame of the physical metric, $g_{\mu\nu}$. The three gravitational fields of the theory are the Bekenstein metric tensor, $\metE_{\mu\nu}$, the  
the Sanders vector field, $A_\mu$, and the scalar field, $\phi$. The matter metric is related to the Bekenstein metric by ~\cite{Bekenstein1993} 
\begin{equation}
   \metM_{\mu\nu} = e^{-2\phi}\metE_{\mu\nu} - 2\sinh(2\phi)A_\mu A_\nu.
   \label{eq:metric_relation}
\end{equation}

The TeVeS theory is defined by the sum of the following four actions: 
\begin{enumerate}

\item{For the metric field,
\begin{equation}
   S_{\metE} = \frac{1}{16\pi G}\int d^4x \; \volE \; \RiemE,
\label{eq:S_EH}
\end{equation}
where $G$ is the bare gravitational constant related to Newton's constant, $G_N$, via the solution to the quasistatic spherically symmetric solution of the TeVeS field equations \citep{Bekenstein1993}. See also \cite{2012-Clifton-MG} for a concise discussion.
}

\item{For the vector field,
\begin{equation}
S_A = -\frac{1}{32\pi G} \int \mathrm{d}^4 x\sqrt{-\tilde{g}} [K_B F_{\mu\nu} F^{\mu\nu} -2\lambda(A_{\mu} A^{\mu} + 1)],
\end{equation}
where $F_{\mu\nu} \equiv 2\tilde{\nabla}_{[\mu} A_{\nu]}$, $F^{\mu\nu} =\tilde{g}^{\mu\alpha} \tilde{g}^{\nu\beta} F_{\alpha\beta}$, $A^{\mu} = \tilde{g}^{\mu\nu}A_{\nu}$,   $\lambda$ is a Lagrange multiplier to ensure $\tilde{g}^{\mu\nu} A_{\mu} A_{\nu} = -1$, and $K_B$ is a dimensionless constant related to the vector field. $K_B$ constitutes one of the additional parameters of the TeVeS models.
}

\item{
For the scalar field, 

\begin{equation}
S_{\phi} = -\frac{1}{16\pi G} \int \mathrm{d}^4 x
\sqrt{-\tilde{g}} [\mu (\tilde{g}^{\mu\nu} -
A^{\mu}A^{\nu}) \tilde{\nabla}_{\mu}\phi
\tilde{\nabla}_{\nu}\phi + V(\mu)],
\end{equation}

where $\mu$ is a  dimensionless non-dynamical scalar field. $V(\mu)$ is a free function which typically depends on a scale $l_B$ (this is a second parameter of the TeVeS model). The action for the scalar field is constructed such that TeVeS theory has a MOND non-relativistic limit, under some conditions and for some specific forms of the function $V(\mu)$. For example, the function in \cite{Bekenstein2004} is given by 
\begin{equation}
\frac{\mathrm{d}V}{\mathrm{d}\mu} = -\frac{3}{32\pi l_B^2 \mu_0^2} \frac{\mu^2(\mu-2\mu_0)^2}{\mu_0-\mu}, \label{eq.VBek}
\end{equation}
where $\mu_0$ is a dimensionless constant (the third parameter of the TeVeS model) and leads to a MOND limit. Similarly, other more general functions leading to MOND can found in \cite{Bourliot2007,Sanders2006,Angus2006}.  
}

\item{
For the matter fields, $\psi_{\rm m}$,

\begin{equation}
S_m = \int \mathrm{d}^4 x \sqrt{-g} \mathcal{L}[g, \psi_{\rm m}, \partial \psi_{\rm m}].
\end{equation}
where here the matter frame metric is used. We note that if arbitrary matter fields (including for instance fermions) are allowed and a  Lagrangian that can depend on the matter field derivative, then by covariance that Lagrangian would also need to involve the derivative of the metric as well.
}

\end{enumerate}

The corresponding field equations for the metric tensor, the vector field and the scalar field are given respectively by:  
\begin{eqnarray}
   \EinE_{\mu\nu} &=& 8\pi G\left[ T_{\mu\nu} + 2(1 -
  e^{-4\phi})A^{\alpha}T_{\alpha(\mu} A_{\nu)}\right] 
  \nonumber \\ 
&&        
+ \mu \left[ \connE_\mu \phi \connE_\nu\phi  - 2
  A^\alpha\connE_\alpha\phi \; A_{(\mu}\connE_{\nu)}\phi  \right]
        + \frac{1}{2}\left(\mu V' -  V\right) \metE_{\mu\nu}   
  \nonumber \\ 
&&        
+   K\left[F^\alpha_{\;\;\mu} F_{\alpha\nu} - \frac{1}{4} F^{\alpha
  \beta} F_{\alpha \beta} \metE_{\mu\nu}\right] - \lambda A_\mu A_\nu,
\end{eqnarray}

\begin{equation}
K \connE_\alpha F^\alpha_{\;\;\mu}
=  -\lambda A_\mu - \mu A^\nu\connE_\nu\phi \connE_\mu\phi  + 8\pi G
(1 - e^{-4\phi})A^{\nu}T_{\nu\mu} \label{eq:A_eq_LR},
\end{equation}
and 
\begin{eqnarray}
   \connE_\mu \left[   \mu \metS^{\mu\nu} \connE_\nu\phi \right] &=&
 8\pi G e^{-2\phi}\left[\metM^{\mu\nu} + 2e^{-2\phi}
 A^{\mu}A^\nu\right] T_{\mu\nu}.
 \label{eq:Phi_eq_LR}
\end{eqnarray}
In addition to the field equations, the theory has two constraints. The first is the usual timelike constraint on the vector field, i.e., $A^\alpha A_\alpha=-1$. This is obtained by varying the action with respect to the Lagrange multiplier, $\lambda$. The second constraint fixes the non-dynamical field ,$\mu$, in terms of the other fields in the theory. It derives from from varying the action with respect $\mu$. The above field equations and constraints are all used in what follows. 

Next, we proceed to describe some aspects of the cosmology of TeVeS. Some studies based on the homogeneous and isotropic FLRW metric can be found in \cite{Bekenstein2004, Diaz-Rivera2006, Bourliot2007, Ferreira2008, Zhao2007, Hao2009}. 
We follow here \citep{Xu2015TEVES,2012-Clifton-MG} and give some key cosmological equations for a spatially flat FLRW background. The metric in conformal time and the matter frame reads
\begin{equation}
\mathrm{d}s^2 = a^2(\tau) (-\mathrm{d}\tau ^2 +
\mathrm{d}r^2),
\end{equation}
while in the Einstein frame, 
\begin{equation}
\mathrm{d}\tilde{s}^2 = b^2(\tau)
(-e^{-4\phi}\mathrm{d}{\tau}^2 + \mathrm{d}{r}^2 ).
\end{equation}
The two scale factors $a$ and $b$ are related by the disformal relation $a=be^{-\phi}$. 

The Friedmann equation in the Einstein frame is given by \cite{Skordis2006TC}: 
\begin{equation}
3\left(\frac{b^\prime}{b}\right)^2=a^2 \bigg[ \frac{1}{2}e^{-2\phi}(\mu
\frac{dV}{d\mu}+V)+8\pi G e^{-4\phi}\bar{\rho} \bigg],
\end{equation}
where $\bar{\rho}$ is the matter energy density. 
It should be noted that the vector field does not contribute to the dynamic of an FLRW background which is then completely described by the scalar field evolution equation
\begin{equation}
\phi^{\prime\prime} = \phi^\prime \bigg( \frac{a^\prime}{a}-\phi^\prime \bigg) - \frac{1}{U} \bigg[ 3\mu\frac{b^\prime}{b}\phi^\prime + 4\pi Ga^2e^{-4\phi}(\bar{\rho}+3\bar{P}) \bigg],
\end{equation}
where $U \equiv \mu+2\frac{dV}{d\mu}/\frac{d^2 V}{d\mu^2}$ and $\bar{P}$ denotes the
pressure from the matter sources (but not the scalar field).

In the matter frame, the physical Hubble parameter is defined as usual as $H \equiv \frac{a^\prime}{a^2}$. The corresponding equivalent of the Friedmann equation is then given by, see e.g. \cite{Skordis2006TC} 
\begin{equation}
3H^2 = 8\pi G_{\mathrm{eff}} (\bar{\rho} +
\bar{\rho}_{\phi}), \label{eq.Friedmann}
\end{equation}
where the effective gravitational constant is given by 
\begin{equation}
G_{\mathrm{eff}} = G \frac{e^{-4\phi}}{(1 +
\frac{\mathrm{d}\phi}{\mathrm{d}\ln{a}})^2}, 
\end{equation}
the energy density of the scalar field is given by 
\begin{equation}
\bar{\rho}_{\phi} = \frac{1}{16\pi G} e^{2\phi} (\mu \frac{dV}{d\mu} + V)
\end{equation} 
and its pressure by 
\begin{equation}
\bar{P}_\phi = \frac{e^{2\phi}}{16\pi G}\left( \mu \frac{dV}{d\mu} - V \right). 
\end{equation}

An effective density fraction can be defined as $\Omega_{\phi} = \frac{\bar{\rho}_{\phi}}{\bar{\rho} + \bar{\rho}_{\phi}}$. When the function $V$ takes the form of Eq.~\eqref{eq.VBek}, the scalar field energy density is found to track the matter energy density \citep{DodelsonLigori2006, Skordis2006LS, Skordis2009} with 
\be
\Omega_{\phi} = \frac{(1+3w)^2}{6(1-w)^2 \mu_0}, 
\ee
where $w$ is the equation of state of the background matter field and the scalar field contribution is always subdominant since $\mu_0$ is of the order of $10^2$.  

Adding a constant to the free function $V$, is equivalent to adding a cosmological constant to the effective Friedmann equation \eqref{eq.Friedmann} and thus producing cosmic acceleration. 

Finally, a concise  description of the perturbation equations in TeVeS can be found in, e.g., \cite{Skordis2006TC,Skordis2008GT,Skordis2006LS} and we refer the reader to those. Mainly, the matter overdensity and velocity field keep the same evolution equations as in GR but are supplemented by perturbation equations for the scalar and vector fields \citep{Skordis2006TC,Skordis2008GT,Skordis2006LS}. 
However, an important difference exists in the processes of growth of structures between  $\Lambda$CDM and TeVeS. In $\Lambda$CDM, baryons fall after decoupling into deeper potential wells caused by dark matter. But in TeVeS, it is rather the rapidly growing perturbations of the vector field that drives the growth of perturbations.  
Such a difference in the processes leads to differences in the growth rate of baryon perturbations as well as the amplitude of their peculiar velocity power spectrum, see \cite{Skordis2006LS,DodelsonLigori2006,Xu2015TEVES}. Unlike the $\Lambda$CDM model, the  growth rate in TeVeS is scale dependent which provides a further test to constrain the models \citep{Skordis2006TC,Skordis2008GT,Skordis2006LS,Xu2015TEVES}. 

TeVeS shows how adding a scalar and vector field to the metric tensor can add further complexity and sophistication to gravity, However, the theory has been recently found to be in tension with latest large scale structure and CMB data sets, e.g., \cite{ReyesEtAl2010,Xu2015TEVES}, although often disputed by its proposers, e.g., \cite{Bekenstein2012,Milgrom2017}.

\paragraph{Other Scalar-Vector-Tensor theories}\mbox{}\\

\cite{Moffat2006} proposed a scalar-tensor-vector gravity (STVG) theory, also referred to as MOG that allows the gravitational constant $G$, a vector field with coupling $\omega$, and the vector field mass $\mu$ to vary in space and time. This theory has modified equations of motion for test particles that have a modified gravitational acceleration law that can fit rotation curves of galaxies and also data from clusters of galaxies without the need for dark matter.  The theory is consistent with solar system tests of gravity and is ghost free. Gravitational waves and electromagnetic waves both travel on null geodesics of the metric with equal speeds so the theory is not ruled out by the GW event GW170817 and its electromagnetic counterpart GRB170817A \citep{Moffat2017}. The theory is reported in \cite{Moffat2011} to fit gravitational lensing of observations and to be consistent with some cosmological observations with no need of dark matter, however oscillations of the matter power spectrum in MOG are not suppressed \citep{Moffat2011}.
It remains to be tested against full LSS data or CMB data.

%%%%%%%%%%%%%%%%%%%%%%%%%%%%%%%%%%%%%%%%%%%%%%%%%%%%%%%%%%%%%%%%%%%%%%%%%%%%%%%%%%%
%%%%%%%%%%%%%%%%%%%%%%%%%%%%%%%%%%%%%%%%%%%%%%%%%%%%%%%%%%%%%%%%%%%%%%%%%%%%%%%%%%%
%%%%%%%%%%%%%%%%%%%%%%%%%%%%%%%%%%%%%%%%%%%%%%%%%%%%%%%%%%%%%%%%%%%%%%%%%%%%%%%%%%%
\subsubsection{Extra Tensor fields:}

Last but not least, it turned out that adding an extra metric tensor to GR can be a  very lucrative extension. For example, a first accomplishment in doing so was to achieve a gravity theory where the graviton has an effective  mass or a resonance (massive gravity)  \citep{FierzEtAl1939OR,deRhamEtAl2010GO,deRhamEtAl2011RO,HassanEtAl2012BG}. Moreover, some of such theories can provide self-accelerating cosmological models with no need for a cosmological constant. These massive gravity theories change the coupling between curvature of spacetime and its source and the idea behind generating cosmic acceleration is that gravity is weakened at the graviton's mass Compton wavelength which is comparable to Hubble scales. 

On more point which is worth highlighting is that such massive gravity theories allow for proposals of degravitation mechanisms of the cosmological constant \citep{Arkani-Hamed2002NM,Dvali2003DT}. 
The idea is that the massive graviton acts as a high-pass filter with filter scale, $L$, set by the inverse of the mass of the graviton. Sources with wavelengths $\ll L$ pass the filter and gravitate normally. However, sources with with wavelengths $\gg L$, like the cosmological constant, are filtered out leading to their degravitation \citep{Dvali2007DEGRAV}. This and other related ideas are very interesting but unfortunately  so far there is no realistic realization of such a degravitation mechanism. 

We outline below some selected aspects of two illustrative examples of   these tensor-tensor or bimetric theories, list some other models, and refer the reader to the specialized reviews \citep{deRhamReviewMG1,Hinterbichler2017,2012-Clifton-MG}.

%%%%%%%%%%%%%%%%%%%%%%%%%%%%%%%%%%%%%%%%%%%%%%%%%%%%%%%%%%%%%%%%%%%%
\paragraph{Illustrative example 1: massive gravity}\mbox{}\\

The idea goes back at least to the early attempts of Fierz and Pauli \citep{FierzEtAl1939OR} in simply deriving a theory of gravity with a massive graviton. Fierz and Pauli considered a non-dynamical background flat metric $\eta_{\alpha\beta}$ (Minkowski) and a dynamical linear perturbation, $h_{\alpha\beta}$ resulting in the dynamical  metric 
\begin{equation}
g_{\alpha\beta}=\eta_{\alpha\beta}+h_{\alpha\beta}.
\end{equation}
They derived and added a (PF)-term at linear order to the Einstein--Hilbert action that generates the massive graviton as follows 
\citep{FierzEtAl1939OR}
\be
{\cal L}_{FP}=m^2 [h^{\mu\nu}h_{\mu\nu}-(\eta^{\mu\nu}h_{\mu\nu})^2],
\label{PF_Action}
\ee
where $m$ is the mass parameter. They showed that this term is the only linear-order term that leads to no-ghost mode at this order. Therefore, the Fierz--Pauli is the unique consistent linear theory of massive gravity.

However, at nonlinear order, the story is different. The action with an FP term can be generalized to nonlinear order as \citep{Boulware1973CG}
\be
S=\frac{1}{16\pi G} \int d^4x \sqrt{-g}R(g)+\frac{m^2}{4} \sqrt{-g}
\left[g^{\mu\nu}g^{\alpha\beta}-g^{\mu\alpha} g^{\nu\beta}\right]
h_{\mu\nu} h_{\alpha\beta}. \label{GPF_Action}
\ee
It was shown by \cite{Boulware1973CG} that the Fierz--Pauli theory at nonlinear order acquires a scalar ghost mode and is thus unstable. 
Another problem with the theory is known as the van Dam, Veltman, and Zakharov (vDVZ) discontinuity, see \cite{vanDam1970MA,Zakharov1970LG}. Namely, that solutions to the theory cannot be continuously connected to their analog GR solutions when the graviton mass is taken to the zero limit, as one would naively expected from the action. To explain, let's consider the spherically symmetric vacuum solution representing the gravitational field around a concentric mass such as the Sun. Then, taking the limit of the graviton mass going to zero does not give back a solution analog to the GR Schwarzschild solution and is thus inconsistent with local observations such as the deflection angle of light, precession of planets, or light travel time delays.    

To deal with these two problems, some possible solutions were proposed in \cite{Vainshtein1972TT,Arkani-Hamed2003EF} where one could solve two problems with one stratagem. First, in order to deal with ghost modes appearing at higher orders, one would introduce tuned higher order interaction terms that would remove the ghost terms order by order. Second, \cite{Vainshtein1972TT} suggested his mechanism (see Sect.~\ref{sec:Kinetic}) where such higher order interaction terms would serve at small scales to shield additional-field interactions and lead to observations indistinguishable from GR.

A \textit{tour de force} came from de~Rham, Gabadadze and Tolley (dRGT) \citep{deRhamEtAl2010GO,deRhamEtAl2011RO} who succeeded in generalizing Fierz--Pauli  theory and formulating a stable massive gravity. For that, they considered  
$g_{\mu \nu}$ and $f_{\mu \nu}$ as the dynamical and non-dynamical metrics, respectively, and wrote the action:  
\begin{eqnarray}
\label{eq:dRGTaction}
\mathcal{S} = \int d^4 x \frac{\sqrt{-g}}{16 \pi G} R(g)-  
m^2\int d^4 x \frac{\sqrt{-g}}{8 \pi G} 
\sum\limits_{n=0}^4 \beta_n e_n \left(\mathbb{X} \right) 
+ \mathcal{S}_M(g_{\mu\nu}, \psi_{\rm m}) \,,
\end{eqnarray}
where the first part is the usual Hilbert--Einstein term and the third part is the matter action term, while the middle part gives the dRGT terms with $\beta_n$ as arbitrary constants and $e_n$ are functions defined by,
\begin{eqnarray}
\label{eq:polynomial}
e_0\left(\mathbb{X}\right) &=& 1, \nonumber\\ 
e_1\left(\mathbb{X}\right) &=& \left[\mathbb{X}\right],\nonumber \\
e_2\left(\mathbb{X}\right) &=& \frac{1}{2} \left( \left[\mathbb{X}\right]^2 - \left[\mathbb{X}^2\right] \right), \nonumber \\
e_3\left(\mathbb{X}\right) &=& \frac{1}{6} \left( \left[\mathbb{X}\right]^3 -3 \left[\mathbb{X}\right] \left[\mathbb{X}^2\right] + 2 \left[\mathbb{X}^3\right] \right) , \nonumber \\
e_4\left(\mathbb{X}\right) &=& \det \mathbb{X},
\end{eqnarray}
where $\mathbb{X} \equiv \sqrt{g^{\alpha \beta} f_{\beta \gamma}}$ and $\left[ \mathbb{X} \right]$ is its trace (i.e., $(\mathbb{X}^2)^\alpha\,_\gamma = g^{\alpha \beta} f_{\beta \gamma}$)

The equations of motion can be found in \cite{deRhamEtAl2010GO,deRhamEtAl2011RO}. 
Interestingly, massive gravity can have cosmological solutions that can self-accelerate,
however, the cosmological solutions have to be Minkowski type open FLRW with strongly coupled perturbations making them not analyzable by standard methods. There are other cosmological solutions with well-behaved perturbations but they require non-isotropy or preferred directions making them cosmologically less attractive, see e.g., \cite{Hinterbichler2017,deRhamReviewMG1}. 
It was then realized soon after that it would be interesting to have the second metric to be a non-Minkowski and dynamical like an FLRW metric \citep{Hassan2011MG1,HassanEtAl2012BG}.   

From a cosmological point of view, it turned out that adding a dynamical metric provides a richer phenomenology and the possibility to have stable and viable self-accelerating solutions \citep{AkramiHassan2015}, although in this case there are also bounds and conditions that must hold to avoid further instabilities as we discuss in the next section \citep{Koennig2014b,Lagos2014,Konning2015H}. 

It is worth noting that most recently \cite{Heisenberg2017MG} performed a thorough analysis of perturbations in massive gravity with SO(3) rotation invariance. 
The models violate Lorentz invariance and it was argued there that this makes it possible to avoid some problems in massive gravity. The models and their cosmology have been studied and reviewed in \cite{Dubovsky1,Dubovsky2,Bebronne2007,Blas2009,Domenech2017,Comelli2014}. 
It was shown in \cite{Dubovsky2,Comelli2014,Heisenberg2017MG} healthy models can have late-time self-acceleration.  \cite{Heisenberg2017MG} worked out perturbations in and FRLW background and with a perfect fluid source. They found models that have no ghosts nor gradient instabilities for effective dark energy equation of state $w_{\rm DE} > -1$ and $w_{\rm DE} < -1$. They also derived expressions for the effective gravitational constant and the slip parameter. Implementation of this formalism into full CMB code and large scale structure will allow for the comparison of these models to current and future cosmological data.

%%%%%%%%%%%%%%%%%%%%%%%%%%%%%%%%%%%%%%%%%%%%%%%%%%%%%%%%%%%%%%%%%%%%
\paragraph{Illustrative example 2: Bimetric massive gravity or bigravity}\mbox{}\\
\label{sec:bigravity}
In addition to realizing a massive gravity theory, adding a second dynamical metric has been shown to provide stable self-accelerating cosmological solutions with no need for a dark energy component, see e.g.  \cite{Hassan2011MG1,HassanEtAl2012BG,HassanEtAl2012CO,Koennig2014a,Koennig2014b,AkramiHassan2015}. 
These theories have a branch of models that admit a limit in which the Planck mass associated to the second metric is small and any scalar instabilities can be pushed to very early times where they are not observable \citep{Koennig2014b,Lagos2014,Cusin2015,AkramiHassan2015,Cusin2015IP,Schmidt-MayEtAl2016RD,Cusin2016AG}. Even if in this limit the background evolution becomes indistinguishable from that of the $\Lambda$CDM, \cite{AkramiHassan2015} state that it provides a technically natural value for the effective cosmological constant.

The action for bimetric massive gravity reads \citep{Hassan2011MG1,HassanEtAl2012BG},
\begin{eqnarray}
\label{eq:BGaction}
\mathcal{S} =&&  \int d^4 x \left( \frac{\sqrt{-g}}{16 \pi G} R(g) +
              \frac{\sqrt{-f}}{16 \pi G_f} R(f) \right) \nonumber \\
             &-&  m^2\int d^4 x \frac{\sqrt{-g}}{8 \pi G} \sum\limits_{n=0}^4 \beta_n e_n \left(\mathbb{X} \right) \nonumber \\
             &+& \mathcal{S}_M(g_{\mu\nu}, \psi_{\rm m}),
\end{eqnarray}
where here we note the additional action term with the Ricci scalar, $R(f)$, built out of the second metric, $f$, compared to the action ~\eqref{eq:dRGTaction}. 
Variation of Eq.~\eqref{eq:BGaction} with respect to $g_{\mu \nu}$ and $f_{\mu \nu}$ gives the field equations, 
\begin{eqnarray}
\label{eq:fieldeq1LR}
&& G_{\mu \nu} + \sum\limits_{n=0}^3 (-1)^n\beta_n g_{\mu \lambda} (Y_n)^\lambda_\nu = \kappa^2 T^M_{\mu\nu} \,, \\
\label{eq:fieldeq2LR}
&& F_{\mu \nu} +\sum\limits_{n=0}^3 (-1)^n\beta_{4-n} f_{\mu \lambda} (Y_n)^\lambda_\nu = 0 \,,
\end{eqnarray}
where $m^2$ has been absorbed into $\beta_n$, and $8\pi G_f$ was set to 1, following the notation of ~\cite{KhosraviEtAl2012MC, Koennig2014a,GengEtAl2017CP}. 
$G_{\mu \nu}$ and $F_{\mu \nu}$ are the Einstein tensors built from the metrics $g_{\mu \nu}$ and $f_{\mu \nu}$, respectively. $T^M_{\mu\nu}$ is the matter energy-momentum tensor and $(Y_n)^{\lambda}_\nu$ are matrices defined by
\begin{eqnarray}
\label{eq:polynomial2LR}
Y_0 &=& \mathbb{I} \,, \nonumber \\
Y_1 &=& \mathbb{X} -\mathbb{I} \left[\mathbb{X}\right] \,,\nonumber \\
Y_2 &=& \mathbb{X}^2 - \mathbb{X} \left[\mathbb{X}\right] + \frac{1}{2} \mathbb{I}\left( \left[\mathbb{X}\right]^2 - \left[\mathbb{X}^2\right] \right) \,, \nonumber \\
Y_3 &=& \mathbb{X}^3 - \mathbb{X}^2 \left[\mathbb{X}\right] + \frac{1}{2} \mathbb{X} \left( \left[\mathbb{X}\right]^2 - \left[\mathbb{X}^2\right] \right) - \frac{1}{6}\mathbb{I} \left( \left[\mathbb{X}\right]^3 - 3 \left[\mathbb{X}\right] \left[\mathbb{X}^2\right] + 2 \left[\mathbb{X}^3\right] \right) \,.
\end{eqnarray}

Next, the field equations are applied to the FLRW metrics in, e.g. \cite{Koennig2014a,Koennig2014b,GengEtAl2017CP},
\begin{eqnarray}
\label{eq:g_bg_LR}
&& ds^2 = g_{\mu \nu} dx^{\mu} dx^{\nu} = -dt^2 + {a(t)}^2 dx^i dx_i \\
\label{eq:f_bg_LR}
&& ds_f^2 = f_{\mu \nu} dx^{\mu} dx^{\nu} = -\frac{\dot{b}^2}{\dot{a}^2} dt^2 + b(t)^2 dx^i dx_i
\end{eqnarray}
to obtain the Friedmann-like equations \citep{GengEtAl2017CP} 
\begin{eqnarray}
\label{eq:eom1_LR}
&& H^2 = \frac{1}{3} \left( \bar{\rho}_M + \beta_0 + 3 \beta_1 \frac{b}{a} + 3 \beta_2 \frac{b^2}{a^2} + \beta_3 \frac{b^3}{a^3} \right) \,, \\
\label{eq:eom2_LR}
&& \dot{H} = -\frac{1}{2} \left( \bar{\rho}_M + \bar{P}_M +  \beta_1 \frac{b}{a}+2\beta_2\frac{b^2}{a^2}+\beta_3{b^3}{a^3}-\beta_1\frac{\dot{b}}{\dot{a}}-2\beta_2\frac{b}{a}\frac{\dot{b}}{\dot{a}}-\beta_3\frac{b^2}{a^2}\frac{\dot{b}}{\dot{a}} \right) \,,
\end{eqnarray}
for $g_{\mu\nu}$ and
\begin{eqnarray}
\label{eq:eom3_LR}
&& H^2 = \frac{1}{3} \frac{a}{b} \left( \beta_1 + 3 \beta_2\frac{b}{a} + 3 \beta_3 \frac{b^2}{a^2} + \beta_4 \frac{b^3}{a^3} \right) \,, \\
\label{eq:eom4_LR}
&& H^2+2\frac{H}{H_f}\frac{\ddot{a}}{a}= \left(\beta_2+2\beta_3\frac{b}{a}+\beta_4\frac{b^2}{a^2}+\beta_1 \frac{\dot{a}}{\dot{b}}+2\beta_2\frac{b}{a}\frac{\dot{a}}{\dot{b}}+\beta_3\frac{b^2}{a^2}\frac{\dot{a}}{\dot{b}} \right)\,\,,
\end{eqnarray}
for $f_{\mu\nu}$, where  $H=(\dot{a}/a)$ while $H_{(f)}=(\dot{b}/b)$ is the Hubble constant of $f_{\mu\nu}$, $\rho_M = \rho_r + \rho_m$ is the energy density of the radiation and matter and ($P_M = P_r + P_m$) is the sum of their pressures. $\kappa^2$ was set to 1. Note that the presence of $\dot{a}$ in \eqref{eq:f_bg_LR} allows \eqref{eq:eom3_LR} and \eqref{eq:eom4_LR} to be written using $H$. See also ~\cite{Koennig2014a,Koennig2014b} where a compact encapsulation of these equations is given. As can be seen from Equations \eqref{eq:eom3_LR} and \eqref{eq:eom4_LR}, and stressed in \cite{Koennig2014a,Koennig2014b}, the background dynamics depend entirely on the Hubble parameter of the metric $g_{\mu\nu}$ and the ratio of the two scale factors. 

The $\beta_0$ term represents a cosmological constant term. \cite{Koennig2014b}  performed a stability analysis finding that the only single parameter models without instabilities at early times are models with $\beta_2$ or $\beta_4$. They found there are no self-accelerating models (i.e., $\beta_0=0$) with a viable background evolution and stable perturbations on the finite branch. For the infinite branch, they found only models with non-vanishing $\beta_1$ and $\beta_4$ are self-accelerating, viable and stable for all cosmic evolution and they focused their analysis on those models. 

\cite{GengEtAl2017CP}, presented a minimum nontrivial case by setting $\beta_2 = \beta_3 = \beta_4 =0$ that we reproduce here for mere illustration purposes. 
Consequently, Eqs.~\eqref{eq:eom3_LR} and \eqref{eq:eom4_LR} reduce to
\begin{eqnarray}
\label{eq:hboa_LR}
\frac{b}{a} = \frac{ \beta_1}{3 H^2} \,, \qquad \mathrm{and} \qquad H_b \equiv \frac{H_f}{H} = 1 - 2 \frac{\dot{H}}{H^2} \,.
\end{eqnarray}
The authors defined an effective energy density and pressure from \eqref{eq:eom1_LR} and \eqref{eq:eom2_LR}
\begin{eqnarray}
\label{eq:rhode_LR}
&& \rho_{\rm DE} = \beta_0 + 3 \beta_1 \frac{b}{a} = \rho_{\rm DE}^{(0)} \left( \bar{\beta}_0 + \bar{\beta}_1 \frac{H_0^2}{H^2} \right) \,, \\
\label{eq:Pde_LR}
&& P_{\rm DE} = - \beta_0 - \beta_1 \left( 2 \frac{b}{a} + \frac{\dot{b}}{\dot{a}} \right) = \rho_{\rm DE}^{(0)} \left[ -\bar{\beta}_0 + \bar{\beta}_1 \frac{H_0^2}{H^2} \left( \frac{2\dot{H}}{3H^2} - 1 \right) \right]  \,,
\end{eqnarray}
that satisfy the continuity equation, $\rho_{\rm DE} + 3 H \left( \rho_{\rm DE} + P_{\rm DE} \right) = 0$ and where they defined 
\begin{eqnarray}
\label{eq:modelpar_LR}
\bar{\beta}_0 = \frac{ \beta_0 }{ \rho_{\rm DE}^{(0)}} \quad \mathrm{and} \quad \bar{\beta}_1 = \frac{\beta_1^2 }{ H_0^2 \rho_{\rm DE}^{(0)}} \,,
\end{eqnarray}
with $\bar{\beta}_0 + \bar{\beta}_1 = 1$ and $\rho_{\rm DE}^{(0)}$ being the corresponding  effective dark energy density at present. They noted that from Eqs.~\eqref{eq:rhode_LR} and \eqref{eq:Pde_LR}, $e_0 \left(\mathbb{X}\right)$ with the free parameter $\beta_0$ in the action plays the role of an effective cosmological constant. 
They also analyzed the evolution of the effective dark energy density and found that the model has a phantom-type  equation of state, $w_{\rm DE} < -1$.

Bimetric massive gravity has branches that are not ruled out by current observations, see Sect.~\ref{sec:constraints_massive}. The structure of the theory gives extra terms in the evolution equation that can be encapsulated as effective dark energy density and negative pressure thus producing late-time cosmic acceleration without the need for a cosmological constant. The models fit well background observational data and some growth 
(see Sect.~\ref{sec:constraints_massive}) and constitute a competitor to the $\Lambda$CDM GR model.
Comparison of these models to full CMB and large scale structure data is needed. 

However, \cite{Konning2015H,Lagos2014,Koennig2014b,2012-Comelli-etal} pointed out to some further instabilities that must to be avoided by requiring some conditions to hold. For example, Higuchi instability can occur in theories with massive spin-2 particles (here the massive graviton) where the mass must satisfy specific bounds in order to avoid modes with negative norm and the appearance of a Higuchi ghost \citep{Higuchi1986FM,Higuchi1989MS}.  Higuchi provided mass bounds for the de Sitter space while \cite{Fasiello2012H} derived mass bounds for massive gravity in flat FLRW spacetimes, see also \cite{Woodard2007}. 
Higuchi instability and scalar gradient instabilities for cosmological solutions in massive gravity are discussed in  \cite{Lagos2014,Cusin2015,Konning2015H}.  
\cite{Konning2015H} analyzed general models in singly coupled bimetric gravity around a FLRW background and found that all models that are not equivalent to $\lcdm$ suffer from either gradient or Higuchi instabilities. 

Other self-accelerating solutions in massive gravity with inhomogeneous fiducial metric were discussed in \cite{Koyama2011MG,Gratia2012,Khosravi2013}. The physical metric in these solutions is an FLRW and can be flat, however, these solutions were shown to suffer from instabilities as recapitulated in \cite{Khosravi2013}.

Perturbation and growth of structure equations for bigravity can be found in for example \cite{Koennig2014b,Kobayashi2016PO,LagosEtAl2017AG}. It was shown in \cite{Koennig2014a,Koennig2014b} that bimetric gravity has several classes of models with unstable linear perturbations. However, they also found that a particular class of models, named the infinite-branch, has a viable background evolution and stable linear perturbations. The infinite-branch refers simply to a specific evolution of the ratio of the scale factors $b/a$ of the two metrics of the theory \citep{Solomon2014,Koennig2014a,Koennig2014b}. Further cosmological constraint studies have since focused on the infinite-branch models of bimetric massive gravity as we discuss those in Sect.~\ref{sec:constraints_massive}. 
However, \cite{Konning2015H} showed that the infinite branch suffers from Higuchi instability which compromises its viability. Detailed discussions about massive gravity and bigravity, their phenomenology and cosmology can be found in the following review papers \cite{deRhamReviewMG1,Schmidt-MayEtAl2016RD}.

%%%%%%%%%%%%%%%%%%%%%%%%%%%%%%%%%%%%%%%%%%%%%%%%%%%%%%%%%%%%%%%%%%%%%%%%%%%
\paragraph{Other tensor-tensor theories}\mbox{}\\

Models with an extra 2-rank tensor include Rosen's theory \citep{Rosen1940GR,Rosen1973AB} with an extra non-dynamical flat metric. The theory is known to pass solar system tests of GR where it is indistinguishable from it \citep{Lee1976TF}. However, the theory has problems when it comes to pulsar and binary pulsar observations \citep{Lee1976TF,WillEtAl1977DG}. Namely, the theory allows for states with energy unbound from below and the emission of gravitational waves with negative energy. This would cause an increase of the spin of pulsars that is not compatible with observations of millisecond pulsars as shown in \cite{Lee1976TF}. Similarly,  \cite{WillEtAl1977DG} found that such a theory predicts large emission of dipole gravitational radiation that will increase the orbital period of the binary pulsar system to a level again inconsistent with observations of such systems. 

Another bimetric gravity theory is that of Eddington-Born-Infield (EBI) \citep{Eddington1924TM,Banados2007TG,Banados2008EA,Banados2008EAb}.
It is based on extentions to  Eddington theory of affine connections. It combines the metric tensor plus a connection. It was shown in \cite{Banados2009NO} that the connection can be replaced by a corresponding metric and thus expressing the theory as a bimetric gravity with interesting cosmological features that can account for dark matter and dark energy and thus a possible unifying theory \citep{Banados2009NO,Banados2009EG,Hu1998SF}. It was shown in \cite{Banados2009EG} that if one wants to keep the unification of dark matter and dark energy in this model, the integrated Sachs-Wolf effect is thus too large and becomes inconsistent with observations. Also, such a theory would also predict an angular power spectrum and galaxy power spectrum that are not consistent with current observations. 

Next, \cite{Drummond2001BG} proposed a tensor-tensor theory formulated using two sets of dynamical tetrads (vierbeins). The theory has a length scale of galactic size. Below such a length scale, it passes the standard test of GR but beyond such a scale it acquires an effective gravitational constant larger than Newton's constant. The author argues that the transition galactic scale can explain the flat velocity rotation curves of galaxies and can account for an alternative to dark matter. It is not clear from current literature whether this theory suffer from the same constraints as the EBI theory above as very little work has been done on its cosmological constraints. 

\cite{Gabadadze2012} proposed models to implement galileons on curved spacetime by 
by coupling a scalar with the galilean symmetry to a massive graviton. The models can 
maintain second order equations of motion, maintain the galilean shift symmetries, and allow the background metric to be dynamical. The models can be viewed as an extension of the ghost-free massive gravity, or as a massive graviton-galileon scalar-tensor theory. They have higher order equations of motion and infinite powers of the field, but are ghost-free. We refer the reader to the original paper. 
Finally, \cite{Milgrom2009BM,Milgrom2010CF} proposed a bimetric extension to MOND that reduces to MOND on small scales and the low acceleration regime of the theory. Cosmological aspects of the theory were studied in \cite{Clifton2010FC,Milgrom2010CF} finding that it can reproduce an FLRW evolution in the high acceleration limit. Some solutions can have cosmic acceleration due to a cosmological constant term in the theory but with some problems. Namely, \cite{Clifton2010FC} found that the solutions that remain in such a high acceleration regime for the entire evolution require either non-baryonic dark matter or extra terms in the original action, or else they fail observational constraints of $\Omega_{\Lambda}$ and do not predict the  right position of the first peak of the CMB temperature spectrum.

%%%%%%%%%%%%%%%%%%%%%%%%%%%%%%%%%%%%%%%%%%%%%%%%%%%%%%%%%%%%%%%%%%%%%%%%%%%%%%%%%%%
%%%%%%%%%%%%%%%%%%%%%%%                                         %%%%%%%%%%%%%%%%%%%
%%%%%%%%%%%%%%%%%%%%%%%          SUB-SECTION                    %%%%%%%%%%%%%%%%%%%
%%%%%              Modified gravity theories with higher-order derivatives     %%%%
%%%%%                                                                          %%%%
%%%%%%%%%%%%%%%%%%%%%%%%%%%%%%%%%%%%%%%%%%%%%%%%%%%%%%%%%%%%%%%%%%%%%%%%%%%%%%%%%%%
%%%%%%%%%%%%%%%%%%%%%%%%%%%%%%%%%%%%%%%%%%%%%%%%%%%%%%%%%%%%%%%%%%%%%%%%%%%%%%%%%%%

\subsection{Modified gravity theories with higher-order derivatives}
\label{sec:HigherDerivatives}

\begin{figure*}
\includegraphics[width=\textwidth,height=2in,angle=0,fbox]{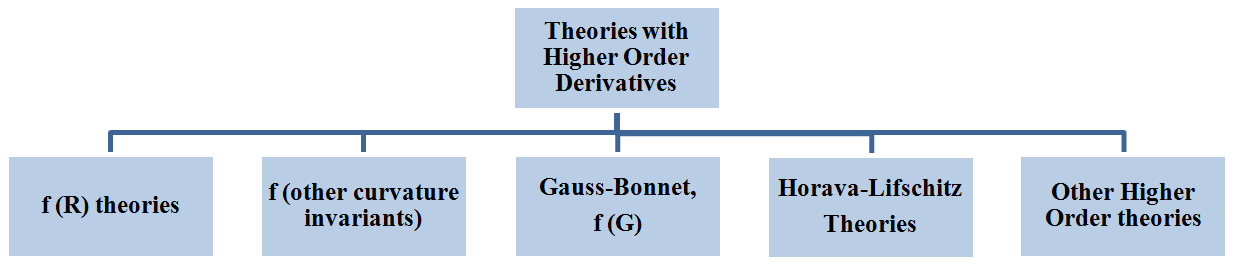}
\caption{Higher-Order-Derivatives modified gravity models.  
\label{fig:HigherDerivatives}}
\end{figure*}

Modification to GR can also be realized by allowing for higher order derivatives of the metric to be present in the equations of motion. 
Such theories can for example be derived from higher-order invariants built from the Riemann curvature tensor and the metric. Shortly after Einstein proposed GR, other theories of gravity using scalar invariants more general than the Ricci scalar were proposed \citep{Weyl1918}. In addition to an interesting phenomenology, it has been argued that the models have theoretical motivations within unification
theories of fundamental interactions and within field quantization on curved space-times \citep{Utiyama1962,Stelle1977,Birrell1982}. Figure \ref{fig:HigherDerivatives} shows some sub-categories of higher-order derivative theories.

However, the problem with this route is that it leads to theories that have problematic fields with states that admit negative unbound energy when quantized, known as ghost fields \citep{Stelle1978,Calcagni2005DE,Hindawi1996CS,Hindawi1996NV,Chiba2005GG,Navarro2006CL,
DeFelice2006,Barth1983QF,Nunez2005GC}. This is formulated as Ostrogradski's instability theorem stating that for a nondegenerate Lagrangian which depends on higher derivatives, the Hamiltonian is necessarily unbounded, see \cite{Woodard2007,Ostrogradski1850}. \cite{Dolgov2003CM,Faraoni2006MI,Seifert2007SO} Other instabilities for such theories have also been raised in \cite{Frolov2008}. 

Luckily, there is yet a limited number of higher-order derivative theories that by some particular construction avoid the presence of ghosts. These include, for example, the popular $f(R)$, Ho\v{r}ava--Lifschitz, and Gauss-Bonnet theories. Additionally, there has been most recently some developments in models beyond Horndeski with higher-order derivatives but some degeneracies making them ghost-free, see Sect.~\ref{sec:horndeski}.

We describe below some aspects of $f(R)$ and Ho\v{r}ava--Lifschitz theories as two illustrative examples for this category and list further below other selected models. 

%%%%%%%%%%%%%%%%%%%%%%%%%%%%%%%%%%%%%%%%%%%%%%%%%%%%%%%%%%%%%%%%%%%%%
%%%%%%%%%%%%%%%%%%%%%%%%%%%%%%%%%%%%%%%%%%%%%%%%%%%%%%%%%%%%%%%%%%%%%
%%%%%%%%%%%%%%%%%%%%%%%%%%%%%%%%%%%%%%%%%%%%%%%%%%%%%%%%%%%%%%%%%%%%%
\subsubsection{Illustrative example 1: $f(R)$ theories}
\label{sec:f(R)}

These theories derive from using a general function $f(R)$ in the action instead of simply $R$ as is the case in GR. The action reads 
\be
S=\int d^4 x \left( \sqrt{-g}f(R) +  {16 \pi G}\,\mathcal{L}_m(\psi_{\rm m}, g_{\mu\nu})\right)
\label{eq:fR_action_LR}
\ee
where $\mathcal{L}_m(\psi_{\rm m}, g_{\mu\nu})$ is the Lagrangian of the matter field, $\psi_{\rm m}$. Varying \eqref{eq:fR_action_LR} with respect to the metric gives the field equations

\begin{equation}
\label{eq:R16_LR}
f_R R_{\mu\nu} - \frac{1}{2} f g_{\mu\nu} - f_{R;\mu\nu}+g_{\mu\nu} \square f_R=
8 \pi G T_{\mu\nu} .
\end{equation}
where $f_R \equiv \partial f(R)/\partial R$ and $\square \equiv \nabla^\mu\nabla_\mu$ is the d'Alembertian operator. Obviously, when $f(R)=R$, the above reduces to Einstein's equations. It is worth mentioning that, unlike the case of GR, variation of the action \eqref{eq:fR_action_LR} with respect to the metric and the connection independently (known as the Palatini approach) leads to a different set of Field equations and thus different theories.  

We have put $f(R)$ in this section but it is fair to mention that $f(R)$ models can also be classified under scalar-tensor theories due to the equivalence between this formulation and that involving an additional scalar-field as we show further in this section. 

For an FLRW curved background metric, the field equations above generalized Friedmann equations read
\bea
\label{eq:frF1_LR}
H^2 &=& \frac{1}{3 F} \left[ 8 \pi G \bar{\rho} - \frac{1}{2} (f-RF)-3 H \dot{F}
  \right] - \frac{\kappa}{a^2}\\
\label{eq:frF2_LR}
\dot{H} &=& -\frac{1}{2F} ( 8 \pi G (\bar{\rho} +\bar{P}) +\ddot{F} - H \dot{F})
+ \frac{\kappa}{a^2}
\eea
where we further note $F\equiv f_R$, the Ricci scalar is given by $R = 6 ( 2 H^2 +\dot{H} + \kappa/a^2 )$, over-dots denote derivatives with respect to cosmic time $t$, and $\kappa$ is the curvature of spatial sections.

$f(R)$ theories have been shown to exhibit a very interesting cosmological phenomenology as they can have solutions that are self-accelerating and thus produce early time inflation or late-time observed cosmic acceleration with no requirement for a cosmological constant \citep{Nojiri2007UI,Nojiri2008MF,Nojiri2008MN,Nojiri2008FE,Bamba2008IA}.  
Furthermore, other proposals aim to provide a unifying scenario by producing early time inflationary acceleration as well as late-time cosmic acceleration \citep{Cognola2008CO,Elizalde2011NE}. Self-acceleration results from a more complex coupling between matter and curvature in such models that can be expressed, as we write below, as conditions on the functions $f(R)$ and its derivatives. 

In the context of cosmic acceleration, one can re-write the generalized Friedmann equations \eqref{eq:frF1_LR} and \eqref{eq:frF2_LR} as those of GR plus additional terms that can be recast into terms associated with state variables of an effective dark fluid as follows 
\bea
8 \pi G \rho_{\textrm{eff}} &=& \frac{RF-f-6 H \dot{F}}{2 F}\\
8 \pi G P_{\textrm{eff}} &=& \frac{2 \ddot{F} +4 H \dot{F} +f -RF}{2F}.
\eea
From the corresponding equation of state, 
\be
\label{eq:freqstate_LR}
w = \frac{2 \ddot{F} +4 H \dot{F} +f -RF}{RF-f-6 H \dot{F}},
\ee
one can then impose the condition for cosmic acceleration, $w < -1/3$, and find the conditions required on the function $f(R)$ and its derivatives to produce late-time self-accelerating models. 

Some limitations of the emergence of cosmic acceleration in $f(R)$ models were discussed in \cite{Clifton2015}. 

At least three self-accelerating $f(R)$ models have been popular and compared extensively to observations. These are the models of Starobinsky \citep{Starobinsky2007SJ} for early inflation 
\be
\label{eq:frstarobinsky_LR}
f(R)=R-\mu R_c \left[ 1- \left(1+\frac{R^2}{R_c^2} \right)^{-n}
  \right],
\ee
Hu and Sawicki model \citep{HSFR2007} with 
\be
f(R)=R- \frac{ \mu R_c }{1+(R/R_c)^{-2n}},
\ee
and Battye and Appleby \citep{Appleby2007DC} with 
\be
f(R)=R+R_c \log\left[e^{-\mu}+(1-e^{-\mu})e^{-R/R_c} \right]
\ee
for late-time acceleration, where $\mu$, $n$ and $R_c$ are positive constants.  

It is also worth mentioning the designer approach to models of $f(R)$ as in, for example, \cite{Song2007,Pogosian2008,Nojiri2006,Nojiri2007}. In this approach, the $f(R)$ model is reconstructed from a predefined background expansion history such as that given by the Hubble function of the $\lcdm$. Using this as a requirement for $f(R)$ Friedmann's equations gives an inhomogeneous second-order differential equation that can be solved numerically and using specific initial conditions, $f_{R0}$, or a Compton wavelength parameter. 

While $f(R)$ models brought some excitements in the community as being serious contenders to GR, their viable models require the chameleon screening mechanism in order to pass solar system constraints. However, as we discuss in Sect.~\ref{sec:Large-mass}, models requiring a chameleon-like screening mechanism were shown in \cite{WangEtALNG} to fail to produce the observed cosmic acceleration unless added with a cosmological constant. Thus, the models fail to be properly screened and simultaneously possess a self-acceleration feature.

It is worth noting that in order to assure well behaved initial conditions in $f(R)$ models at early epochs when curvature is high, the following condition is required 
\begin{equation}
\lim_{R\rightarrow \infty} f(R)/R \rightarrow 0 \,.
\end{equation}
In this way, any modification to gravity in $f(R)$ viable models happens well after 
radiation is negligible. As we will see further below, $f(R)$ theory can be expressed as a scalar-tensor theory with a scalaron field. As mentioned above, a second common practice used for $f(R)$ models is to parameterize them using the dimensionless Compton wavelength (of the scalaron) in Hubble units given by 
\cite{Song2007,HSFR2007}
\be\label{eq:ComptonfR}
B\equiv\frac{f_{RR}}{1+f_R}R'\frac{H}{H'} \,,
\ee
where $f_{RR}={\rm d}^2 f/{\rm d} R^2$ and ${~}'={\rm d} /{\rm d} \ln a$ here. 
This allows $f_{RR}$ to control the modification to gravity and solutions with a given expansion history can be characterized by $ B_0 \equiv B(\ln a=0)$ \citep{Song2007}. 
For GR, $B(a)=0$. It is customary for cosmological analyses to constrain the $B_0$ parameter. 

Cosmological perturbations for $f(R)$ have been fully worked out in a number of studies. See for example the reviews \citep{2012-Clifton-MG,Felice2010FR} for a summary and references. Using the flat perturbed FLRW metric in Newtonian gauge, the following informative relations can be obtained in the quasi-static approximation \citep{Felice2010FR}.
First, the gravitational potentials are given by 
\be
\Psi \simeq \frac{1}{2F} \left( \delta F-\frac{a^2}{k^2}  
\kappa^2 \delta \rho_m \right),
\ee
and 
\be
{\Phi} \simeq -\frac{1}{2F} \left( \delta F +\frac{a^2}{k^2}
\kappa^2 \delta \rho_m \right)\,.
\ee
where $\delta F$ satisfies $(k^2/a^2+M^2)\delta F \simeq \kappa^2\delta \rho_m/3$ and 
the mass parameter $M$ is given by 
\be
M^2 = \frac{F - R F_{R}}{3 F_{R}}.
\ee
The modified Poisson equations are given by 
\be
\frac{k^2}{a^2}\Psi \simeq -\frac{\kappa^2 \delta \rho_m}{2F} 
\frac{2+3M^2a^2/k^2}{3(1+M^2a^2/k^2)}\,,
\ee
and 
\be
\frac{k^2}{a^2}\Phi \simeq
-\frac{\kappa^2 \delta \rho_m}{2F}
\frac{4+3M^2a^2/k^2}{3(1+M^2a^2/k^2)}\,.
\ee
The Weyl potential $\Psi_{\rm w} \equiv (\Phi+\Psi)/2$ that enters observations of for example gravitational lensing and the ISW effect is given by  
\begin{equation}
\Psi_{\rm w} \simeq -\frac{\kappa^2}{2F}
\frac{a^2}{k^2} \delta \rho_m\,.
\label{eq:Phieff_LR}
\end{equation}
The difference of the two potentials is given by 
\be
\Psi-\Phi = - \frac{8 \pi a^2 (\bar{\rho}+\bar{P}) \sigma }{F} - \frac{\delta F}{F}
\label{eq:frperteq9_LR}
\ee
so that even in absence of shear, there is still a slip parameter between the two potentials due to modification to gravity. 

Before we end this section, it is worth showing how $f(R)$ theories can be formulated in terms of equivalent scalar field actions and what implications that has. 
First, we observe that the action \eqref{eq:fR_action_LR} is equivalent to that of a scalar field as 
\begin{equation}
S=\int  d^4 
x\sqrt{-g} \left(f(\phi) + (R- \phi) \frac{df(\phi)}{d\phi} \right) + {16 \pi G}\, \mathcal{L}_m(\psi_{\rm m}, g_{\mu\nu}).
\label{eq:fRaction2_LR}
\end{equation}
Varying with respect to the scalar field $\phi$ gives $(R - \phi) \frac{d^2f(\phi)}{d\phi^2}=0$
so $R=\phi$ for all $\frac{d^2f(\phi)}{d\phi^2}\ne 0$, showing that the action \label{eq:fRaction2_LR} is indeed equivalent to  \eqref{eq:fR_action_LR}. 

Next, one can introduce an auxiliary field, $\psi\equiv \frac{df(\phi)}{d\phi}$, and define a potential $V(\psi)$ as the Legendre transform of the function $f(\phi)$ given by 
\be
V(\psi)=f(\phi(\psi))-\phi(\psi) \psi
\ee
so the action \eqref{eq:fR_action_LR} can be written in the Jordan frame as 

\begin{equation}
S=\int  d^4 
x\sqrt{-g} \left( \psi R - V(\psi) \right) + {16 \pi G}\, \mathcal{L}_m(\psi_{\rm m}, g_{\mu\nu}).
\label{eq:fRaction3_LR}
\end{equation}

This action has now taken the well-known form of a non-minimally coupled scalar-tensor theory as discussed in the Generalized Brans--Dicke theory \ref{sec:GBD} with the parameter $\omega_{\rm BD}=0$. Importantly, the equations of motion of the theory are second order in the metric derivatives instead of fourth order and no ghost modes are present. 

Finally, one can show that the $f(R)$ action can be transformed to the Einstein frame using a conformal transformation $g_{\mu \nu} =A(\phi)^2 \bar{g}_{\mu \nu} $ and by redefining the scalar field. One can then write 
\begin{equation}
S = \int d^4 x \sqrt{-\bar{g}}\Big(\bar R - \frac{1}{2} (\partial \phi)^2- \bar{V}(\phi) \Big)
 +16 \pi G \int d^4 x \sqrt{-g} {\cal L}_m \Big( \psi_{\rm m}, A(\phi)^2 \bar{g}_{\mu \nu} \Big).
\label{eq:Eframe_FR}
\end{equation}
Now, the theory appears in this frame as that of a scalar minimally coupled to curvature, however, the scalar field couples now directly to the matter fields.

We refer the reader to the specialized reviews of $f(R)$ theories and their cosmology in  \cite{Felice2010FR,Sotiriou2010f(R)} and references therein.

%%%%%%%%%%%%%%%%%%%%%%%%%%%%%%%%%%%%%%%%%%%%%%%%%%%%%%%%%%%%%%%%%%%%%%%%%%%%%%%%%%%
%%%%%%%%%%%%%%%%%%%%%%%%%%%%%%%%%%%%%%%%%%%%%%%%%%%%%%%%%%%%%%%%%%%%%%%%%%%%%%%%%%%
%%%%%%%%%%%%%%%%%%%%%%%%%%%%%%%%%%%%%%%%%%%%%%%%%%%%%%%%%%%%%%%%%%%%%%%%%%%%%%%%%%%
\subsubsection{Illustrative example 2: Ho\v{r}ava--Lifshitz} 

Ho\v{r}ava--Lifshitz  gravity theory was proposed with the motivation to quantize gravity \citep{Horava2009QG,Horava2009JO,Horava2009SD}. The idea was to provide an ultraviolet (UV) completion of GR at the expense of breaking Lorentz invariance. 
For this, this theory also belongs to the category of Lorentz breaking theories. 
Such an invariance is however approximately recovered (i.e., staying below experimental constraints) in the infrared (IR) regime.  Following early studies of scalar fields by Lifshitz \citep{Lifshitz1941}, Ho\v{r}ava proposed to use an anisotropic scaling between space and time dimensions as 
\be 
\mathbf{x}\rightarrow l \mathbf{x}; \,\,\,\,t \rightarrow l^z t,
\ee 
where $z$ is called the dynamical critical exponent and the theory is often referred to as Ho\v{r}ava's gravity at a Lifshitz point $z$. This anisotropic treatment of space and time allowed the theory to avoid the Ostrogradski's ghost problem by allowing it to have higher order spatial derivatives but no time higher order derivatives.  

A convenient formalism to express a theory with such a split between time and space is 
the Arnowitt--Deser--Misner (ADM) decomposition of spacetime, see e.g., \cite{MTW1973},  given by 
\begin{equation}
\label{eq:admmetric_LR}
d s^2 = - N^2 c^2 \, dt^2 + g_{ij}(dx^i + N^i \, dt) (dx^j + N^j \, dt),
\end{equation}
where $N(t,x^i)$ is the lapse function and $g_{ij}$ is the 3-space metric. One is restricted to pick a preferred foliation of spacetime due to the anisotropy discussed above. Another important feature, but also source of problems, is that the GR invariance under diffeomorphisms is replaced by the more restrictive foliation preserving diffeomorphisms as  
\be
t\rightarrow \tilde{t}(t),\qquad x^i\rightarrow \tilde{x}^i(t,x^i).
\ee

The most general action for such a theory with second-order only time derivatives is given by  
\begin{equation}
S=\frac{M_{\rm pl}^2}{2}\int d^3x d t N \sqrt{g} \left\{ K^{ij} K_{ij} - \lambda K^2 -V(g_{ij},N)\right\}\, ,
\end{equation}
where $M_{\rm pl}$ is a constant that can be identified with the Planck mass, $\lambda$ is a dimensionless running coupling constant, and $V$ is a potential function depending on the spatial metric, the lapse function and their spatial derivatives. $K_{ij}$ is the extrinsic curvature given by 
\begin{equation}
K_{ij} = {1\over2N} \left\{  \dot g_{ij} - \nabla_i N_j - \nabla_j N_i \right\},
\end{equation}
where an overdot is for differentiation with respect to the time coordinate and $\nabla_i$ is the covariant derivative associated with the spatial metric. 

The Ho\v{r}ava--Lifshitz theory can have different versions. One version is said to have a \textit{detailed balance} property and is based on specific symmetry properties of the potential function $V$ \citep{Horava2009QG,Horava2009JO}. Ho\v{r}ava proposed \textit{detailed balance} to simplify the theory by reducing the number of curvature invariants needed to describe its formalism. Also, depending on whether the lapse function, $N$, is a function of time or a function of time and space coordinates, the theory is said to be projectable or non-projectable, respectively, with a number of implications including the cosmological evolution.   

The original Ho\v{r}ava--Lifshitz theory had a number of problems and various improvements have been proposed, see e.g. \cite{BPS2010,BPS2011,Horava2010imp1,Zhu2012HL,Zhu2011HL,Lin2014HL}. 
The theory and its improved versions have interesting phenomenological implications in the infra-red (low-energy) regime where, for example, it was shown that an integration constant can play the role of dark matter \citep{Mukohyama2009DM}. It was also argued in \cite{Appignani2010} that the presence of a bare geometrical cosmological constant in Ho\v{r}ava--Lifshitz with detailed balance can be used to address the cosmological constant problem, although still with some fine tuning. They do that by cancellation of the negative geometrical cosmological constant term from the theory against the vacuum energy term, leaving only a very small observed value \citep{Appignani2010}.

Cosmology of Ho\v{r}ava--Lifshitz theory has been discussed in a number of other papers including the reviews \citep{Sotiriou2011HL,Calcagni2009CO}; 
\citep{Gong2010,Misonoh2017} for the projectable case; 
\citep{Kobayashi2010HL} for non-projectable case; 
\citep{Huang2011,Huang2012,Huang2012b} for the projectable case with $U(1)$ local symmetry \citep{Horava2010imp1}; 
and \citep{Zhu2013,Zhu2013b} for the non-projectable case with the same $U(1)$ local symmetry. 

For cosmology, we provide a few general illustrative results following \cite{Sotiriou2011HL}. Under the foliation preserving diffeomorphisms, the theory is written in a prefered foliation. One can choose 
\be
N=1,\quad N^i=0,\quad g_{ij}={a(t)}^2\delta_{ij},
\ee
so that the ADM line element \eqref{eq:admmetric_LR} coincides with the FLRW metric
\be
ds^2=-dt^2+a(t)^2 \left[\frac{dr^2}{1-k r^2}+r^2\left(d\theta^2+\sin^2 \theta d\phi^2\right)\right].
\ee

For the cosmological solution, the difference here  between projectable and non-projectable theory manifests itself on the Hamiltonian constraint which is global in the projectable theory and local in the non-projectable one. This subtlety needs to be taken into account when studying the background cosmology dynamics as we delineate below. Indeed, the Ho\v{r}ava--Lifshitz field equations give two generalized Friedmann equations as follows. 

In the projectable case, the Hamiltonian constraint is global and gives the first equation as an integral
\begin{equation}
\label{eq:f1p_LR}
\int d^3 x a^3 \left\{\frac{3\lambda-1}{2}\; {\dot {a}^2\over a^2}  -  {V(a)\over 6}  - {8\pi G_N {\rho}\over 3}\right\}=0,
\end{equation}
where ${\rho}\equiv-g^{-1/2} \delta S_M/\delta N$ and  $S_M$ is the matter action, 
Of course, for the FLRW metric, the integrand in \eqref{eq:f1p_LR} is a function of time only and gets out of the space integral so it gives Eq.~\eqref{eq:f1np_LR} below. But  we assume that when \cite{Sotiriou2011HL} writes \eqref{eq:f1p_LR} in his review, he meant the case where the universe is not globally isotropic and homogeneous so the integrand in \eqref{eq:f1p_LR} does depend on spatial coordinates. 

The potential above is given by 
\begin{eqnarray}
V(a) = g_0\, M_{\rm pl}^2 +  {6 g_1 k  \over  a^2} + {12(3g_2+g_3)  k^2 \over M_{\rm pl}^2\,a^4} 
+ {24(9 g_4 + 3g_5+ g_6) k\over M_{\rm pl}^4 \, a^6}. 
\end{eqnarray}
where the $g_{i}$ are dimensionless couplings from the action \citep{Sotiriou2011HL}.

In the non-projectable case, the Hamiltonian constraint is local so we can get rid of the integral to write \citep{Sotiriou2011HL}
\begin{equation}
\label{eq:f1np_LR}
\frac{3\lambda-1}{2}\; {\dot {a}^2\over {a}^2}  -  {V(a)\over 6}  = {8\pi G_N \bar{\rho}\over 3}.
\end{equation}

The second Friedmann equation is the same for the two cases and is given by \cite{Sotiriou2011HL}
\begin{eqnarray}
\label{eq:f2_LR}
- \frac{3\lambda-1}{2} \; {\ddot {a}\over {a}}   &=&   {1\over2}\frac{3\lambda-1}{2} {\dot {a}^2\over {a}^2}   
-  {1\over12 {a}^2} {d[V(a)\, {a^3}]\over d {a}}
+ 4\pi G_N \bar{p},
\end{eqnarray}
where $\bar{p}\equiv -g^{ij}(2/N\sqrt{g})\delta S_m/\delta g^{ij}$.

In the non-projectable case, $\dot{a}$ can be eliminated from \eqref{eq:f2_LR} by use of \eqref{eq:f1np_LR} to write
\begin{eqnarray}
\label{eq:f3np_LR}
- \frac{3\lambda-1}{2} \; {\ddot {a}\over {a}}   &=&    -{1\over 12 {a}} {d[V({a}) {a}^2] \over d {a}}  +{4\pi G_N\over 3} (\bar{\rho}+3\bar{p}).
\end{eqnarray}
Next, differentiating \eqref{eq:f1np_LR} and subtracting it from \eqref{eq:f3np_LR} gives the usual conservation law as in GR and that will be used further below \citep{Sotiriou2011HL}
\be
\label{eq:mc_LR}
\dot{\bar{\rho}}+3\frac{\dot{a}}{a} (\bar{\rho}+\bar{p})=0.
\ee

\cite{Mukohyama2009DM} argued that the global nature of the Hamiltonian constraints in the projectable case has some specific implications for cosmic evolution. That is Eq.~\eqref{eq:f1p_LR} is irrelevant locally inside the Hubble horizon. Therefore, one has to work only with Eq.~\eqref{eq:f2_LR} and ignore \eqref{eq:f1p_LR}. Following the argument and integrating Eq.~\eqref{eq:f2_LR} gives
\begin{equation}
\label{eq:f1p2_LR}
\frac{3\lambda-1}{2}\; {\dot {a}^2\over {a}^2}  -  {V({a})\over 6}  = \frac{8\pi G_N}{3} \left(\bar{\rho}+\frac{C(t)}{a^3}\right),
\end{equation}
where the form of $C(t)$ depends on the conservation law satisfied by the matter source. Using \eqref{eq:f1p2_LR} in \eqref{eq:f2_LR} to eliminate $\dot{{a}}$, one writes  \citep{Sotiriou2011HL}
\begin{eqnarray}
\label{eq:f3p_LR}
- \frac{3\lambda-1}{2} \; {\ddot {a}\over {a}}   &=&    -{1\over 12 {a}} {d[V({a}) {a}^2] \over d {a}}  +{4\pi G_N\over 3} \left(\bar{\rho}+\frac{C(t)}{{a}^3}+3\bar{p}\right).
\end{eqnarray}
Interestingly, now the only difference between the two pairs of equations describing the background cosmological evolution, i.e., (\eqref{eq:f1p2_LR} and \eqref{eq:f3p_LR} for the projectable case versus (\eqref{eq:f1np_LR} and \eqref{eq:f3np_LR}) for the non-projectable case) is the presence of the function $C(t)$. 
 
In the projectable case, if we suppose that the state variables ($\rho$ and $p$) satisfy the conservation equation \eqref{eq:mc_LR} then $C(t)$ reduces to a constant and the corresponding term in the Friedmann equations \eqref{eq:f1p2_LR} and \eqref{eq:f3p_LR} above play the role of a pressureless dark matter component as shown in \cite{Mukohyama2009DM}. 
Last, it is worth noting that in the case of spatially curved geometry, two additional terms are present in the potential $V(a)$ and thus the Friedmann equations. The first of the last two terms in the potential is referred to as dark radiation and is proportional to $a^{-4}$ and the very last term is associated with a stiff matter  and is proportional to $a^{-6}$ \citep{Sotiriou2011HL}.

Finally, we close the discussion of the Ho\v{r}ava--Lifshitz theory by pointing out to a recent development of what is now called the \textit{healthy extension} theory proposed by \cite{BPS2010}. The theory avoids persistent instabilities  in the original theory and remains power-counting renormalizable \citep{BPS2010}. The theory reduces in the low-energy limit to a scalar-tensor theory with deviations from GR that can be made small by some choice of the parameter space. This healthy theory admits a solution around a static mass that has a gravitational potential of the same form as the GR Schwarzschild solution with \cite{BPS2010}
\be 
\Psi=\Phi=-\frac{m}{8\pi M_p^2(1-\alpha/2)r}
\ee
with an effective gravitational constant 
\be
G_N=[8\pi M_p^2(1-\alpha/2)]^{-1}.
\ee 

For a cosmological homogeneous background, the theory gives dynamical equations that differ from GR only by the presence of the $\lambda$ coupling. The first Friedmann equation is then given by
\be 
H^2=\frac{8 \pi G_{eff-cosmo}}{3}{\bar{\rho}}
\ee
with an effective gravitational constant at cosmological scales given by 
\be 
G_{eff-cosmo}=\frac{2}{2\pi M^2_p}(3\lambda-1),
\ee
so $\lambda=1$ restores GR Friedmannian cosmological evolution. 

The observational bound on the deviation from GR is provided from the measurement
of the primordial abundance of He$^4$ which gives \citep{Jacobson2008EG,Carroll2004LV}
\be
|G_{eff-cosm}/G_N  - 1| \le 0.13,
\ee
thus putting only mild constraints on the parameters $\alpha$ and $\lambda$ of the theory.

Perturbations for the Ho\v{r}ava--Lifshitz theory have been worked out in a number of papers including \citep{Mukohyama2009HLP,Gao2010CP,Wang2010HLP,Wang2010VTHL}. The growth equations are different from those of GR and offer the possibility to test these theories using large-scale structure. 

Recent reviews, papers and progress reports on Ho\v{r}ava--Lifshitz gravity and cosmology can be found in \cite{WangA2017HL,Sotiriou2011HL,Calcagni2009CO} and references therein. Discussions about the implications of Lorentz symmetry violations in Ho\v{r}ava--Lifshitz theory can be found in \cite{Sotiriou2009PV,Sotiriou2009QG,Visser2009LS,Nikolic2010HG,Cai2009DS,Charmousis2009SC,Li2009AT,Blas2009OT} while more about its cosmology  can be found in \cite{Kiritsis2009HC,Brandenberger2009MB,
Mukohyama2009PA,Saridakis2010HD,Mukohyama2009DM,Gao2010CP,Wang2009TA,
Wang2010CP,Mukohyama2010HL,Ferreira2012}.

%%%%%%%%%%%%%%%%%%%%%%%%%%%%%%%%%%%%%%%%%%%%%%%%%%%%%%%%%%%%%%%%%%%%%
%%%%%%%%%%%%%%%%%%%%%%%%%%%%%%%%%%%%%%%%%%%%%%%%%%%%%%%%%%%%%%%%%%%%%
%%%%%%%%%%%%%%%%%%%%%%%%%%%%%%%%%%%%%%%%%%%%%%%%%%%%%%%%%%%%%%%%%%%%%
\subsubsection{Other higher order derivative theories}

Conformal Weyl gravity has been actively pursued and developed by \cite{Mannheim1,Mannheim3} but the theory goes way back to the early work of Weyl \citep{Weyl1918}. The gravitational action is built solely from the Weyl tensor contracted with itself as  
\be
S_{CG}=\int d^4
x\sqrt{-g}C_{\alpha\beta\gamma\delta}C^{\alpha\beta\gamma\delta},
\ee
where the Weyl conformal tensor is given by
\begin{eqnarray}
C_{\alpha\beta\gamma\delta}&=&R_{\alpha\beta\gamma\delta}+\frac{1}{6}R 
[g_{\alpha\gamma}g_{\delta\beta}-g_{\alpha\delta}g_{\gamma\beta}] \nonumber
\\&&+\frac{1}{2}[g_{\alpha\delta}R_{\gamma\beta} 
+g_{\beta\gamma}R_{\delta\alpha}-g_{\alpha\gamma}R_{\delta\beta}-g_{\beta\delta}R_{\gamma\alpha}].
%\nonumber 
\end{eqnarray}

Even if the theory contains fourth order derivatives in the metric, it has been argued in \cite{Mannheim4NG,Mannheim2NG,Pavsic2013} that it is free of ghosts, although with some further open discussions in \cite{Pavsic2016}. 
  
The theory has an interesting phenomenology with a solution for a spherically symmetric field that has a metric component potential with two extra terms compared to that of the GR Schwarzschild's solution, e.g., \cite{Mannheim3}. It was argued that one of the additional terms can explain galaxy flat rotation curves as an alternative to dark matter while the second term can play the role of a cosmological constant \citep{MannheimDMDE}. The theory was often discussed in the context of rotation curves and as an alternative to dark matter \citep{Mannheim6,MannheimDMDE}, however, it was also argued in, for example,  \cite{Mannheim5,Mannheim3} that the theory could help to address the cosmological constant problem.

Some work has been done showing that the theory passes some solar system tests such as the bending of light, e.g. \cite{Cattani2013,Sultana2010}, although debated in, e.g. \cite{Campigotto2017}. It was also claimed in \cite{Yoon2013} that Mannheim's conformal gravity potential is problematic as it cannot reduce to a proper Newtonian limit at short distances without singularities in the mass density source. This was refuted in an extended response by \cite{Mannheim2016}. 
It was also found to fit some astrophysical distance tests in \cite{Yang2013CG,Diaferio2011} but was criticized as its Big Bang Nucleosynthesis predictions are not consistent with observations \citep{Knox1993CG}. 
\cite{Caprini2018} investigated very recently the gravitational radiation from Pulsar binary systems in conformal gravity using the system PSR~J1012+5307. They found that when fixing the graviton mass in conformal gravity so that the theory fits galaxy rotation curves without dark matter, the gravitational radiation from the system is much smaller than in GR and cannot explain the orbital decay of the binary system. At the cosmology level, more work remains to be done to compare conformal gravity to full data of CMB and large scale structure.

Another theory worth mentioning in this section is that built from the Gauss-Bonnet invariant,
\be
G=R^2-4R^{\alpha\beta}R_{\alpha\beta}+R^{\alpha\beta\gamma\delta}R_{\alpha\beta\gamma\delta},
\label{eq:GB}
\ee
constructed from this specific combination of the Ricci scalar squared, the Ricci 
tensor and Riemann tensor contracted with themselves. Albeit being quadratic in the Riemann and Ricci tensors, the Gauss--Bonnet  combination gives equations of motion that are ghost free, 
e.g., \cite{DeWitt1965,Li2007GB,Akbar2006FE}. Furtheremore, the graviton itself may still become a ghost in the FLRW background, so further no-ghost conditions must be imposed on the background, see e.g., \cite{DeFelice2006}.
Some models have been shown to be also free from other instabilities 
due to superluminal propagations and fit cosmological expansion constraints \cite{deFelice2009GB,MI2009GB1,MI2010GB}. The action is given by 
\be
S=\int{d^{4}x \sqrt{-g}\left[\frac{1}{2}R+f(G)\right]+\int{d^{4}x}\sqrt{-g}L_m}+\int{d^{4}x}\sqrt{-g}L_{\rm rad},
\label{eq:ActionGB}
\ee
where we has set in this sub-section $8 \pi G \equiv 1$, $L_m$ and $L_{rad}$ are the matter and radiation Lagrangians, respectively. Here, the Gauss--Bonnet term is effective at cosmological scales. 

Varying the action with respect to the metric gives the field equations 
\bea
&&{8[R_{\alpha\gamma\beta\delta}+R_{\gamma\beta}g_{\delta\alpha}-R_{\gamma\delta}g_{\beta\alpha}-R_{\alpha\beta}g_{\delta\gamma}+R_{\alpha\delta}g_{\beta\gamma} +\frac{1}{2}R(g_{\alpha\beta}g_{\delta\gamma}-g_{\alpha\delta}g_{\beta\gamma})]\nabla^{\gamma}\nabla^{\delta}}f_G\nonumber\\
&&+\,(Gf_G-f)g_{\alpha\beta} + R_{\alpha\beta}-\frac{1}{2}g_{\alpha\beta}R= T_{\alpha\beta},
\label{eq:FieldEquationsGB}
\eea
where we use the definition $f_G\equiv\frac{\partial{f}}{\partial{G}}$.

Now using the FLRW flat metric and a universe filled with matter and radiation, one derives the generalized Friedmann equation
\be
3H^2=Gf_G-f{G}-24H^3\dot{f}_{G}+\bar{\rho}_{m} +\bar{\rho}_{\rm rad}.
\label{eq:FriedmannGB1}
\ee
where $\bar{\rho}_m$ and $\bar{\rho}_{rad}$ are the matter and radiation energy densities, respectively, a dot represents $d/dt$. Also, in terms of $H$, 
\be
R=6(\dot{H}+2H^2)
\ee
and
\be
G=24H^2(\dot{H}+H^2).
\ee

Several models were proposed in \cite{deFelice2009GB} and shown to be consistent with observations of supernova magnitude-redshift data, distance to the CMB surface, Baryon Acoustic Oscillations (BAO), and Hubble Key project and age constraints in \cite{MI2009GB1,MI2010GB}. However, it was shown in \cite{InstabilityGB} that $f(G)$ models have generic divergent modes for matter linear perturbations which ruled them out. However, other theories based on using the Gauss--Bonnet invariant in higher dimensions known as Einstein--Gauss--Bonnet models are not ruled out and are still subject to discussions (see Sect.~\ref{sec:HigherDimensions}).  

A third theory worth listing here is the Chern--Simons theory which is also based on using specific combination of higher order curvature invariants. 
The theory can arise from particle physics, string theory or from geometrical considerations. An extensive review of the theory is given by \cite{Alexander2009}.

Finally, it is worth saying a few more words about models built from more general invariants than the Ricci scalar. A revived interest was raised into them in the early 2000s as some models were shown to exhibit late-time self-accelerating expansion without a dark energy component 
\citep{Carroll2005,EASSON2004MG,Easson2005CC}. It was also shown in \cite{Sotiriou2007TM,Meng2004PF,Nojiri2003MG} that they can have early-time inflation as well, thereby providing a possible unification scenario for the two accelerating phases. See for example reviews in \cite{Lobo2008TD,Nojiri2006IT} and references therein. However, previous studies stressed that the models considered were chosen somewhat randomly due to the large spectrum of possible curvature invariants (e.g., \citealt{Carroll2005,Dolgov2003CM}), and a systematic approach to these models was highly desirable \citep{Faraoni2006SS,Nojiri2008MF}. 
Accordingly, \cite{IshakMoldenhauer2009} proposed a systematic method to classify such models based on minimal sets of invariants. They explore an idea based on theorems from the theory of invariants in GR \citep{Debever1964,Carminati1991,Zakhary1997}. The idea was that curvature invariants are not independent from each other and, for a given algebraic type of the Ricci tensor (see, e.g., the Segre classification \citealt{Segre1884,StephaniEtAl2003}) and for a given Petrov type of the Weyl tensor (i.e., symmetry classification of space-times) -- e.g., \cite{Petrov2000,Pirani1957,Penrose1960,StephaniEtAl2003} -- there exists a complete minimal independent set (basis) of these invariants in terms of which all the other invariants may be expressed. As an immediate consequence of the connection made and the proposed approach, the number of independent invariants to consider is reduced from an infinite number to six in the worst case and to only two independent invariants in the case of primary interest of cosmology, i.e., all FLRW metrics.  

Although this was an interesting idea for classification of this class of models, the determining factor to limit the number of physically acceptable models came from considering their quantization. It was quickly recognized that only a small subset of such models are free of ghost instabilities as we discussed in Sect.~\ref{sec:HigherDerivatives}.

%%%%%%%%%%%%%%%%%%%%%%%%%%%%%%%%%%%%%%%%%%%%%%%%%%%%%%%%%%%%%%%%%%%%%%%%%%%%%%%%%%%
%%%%%%%%%%%%%%%%%%%%%%%                                         %%%%%%%%%%%%%%%%%%%
%%%%%%%%%%%%%%%%%%%%%%%          SUB-SECTION                    %%%%%%%%%%%%%%%%%%%
%%%%%           Modified gravity  theories with higher-dimensions              %%%%
%%%%%                                                                          %%%%
%%%%%%%%%%%%%%%%%%%%%%%%%%%%%%%%%%%%%%%%%%%%%%%%%%%%%%%%%%%%%%%%%%%%%%%%%%%%%%%%%%%
%%%%%%%%%%%%%%%%%%%%%%%%%%%%%%%%%%%%%%%%%%%%%%%%%%%%%%%%%%%%%%%%%%%%%%%%%%%%%%%%%%%

\subsection{Modified gravity  theories with higher-dimensions}
\label{sec:HigherDimensions}

\begin{figure*}
\includegraphics[width=\textwidth,height=5.5in,angle=0,fbox]{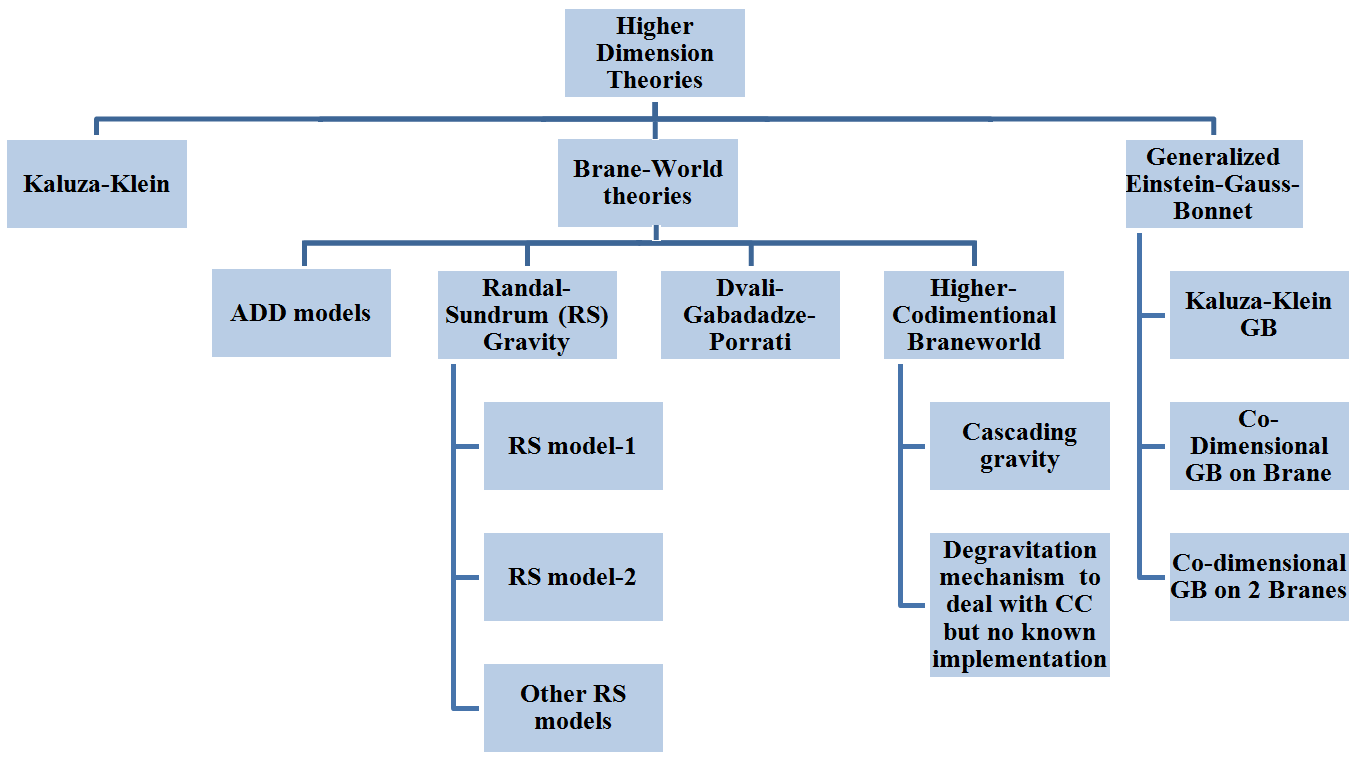}
\caption{Higher-Dimension modified gravity models.  
\label{fig:HigherDimensions}}
\end{figure*}

This class of models has been popular both in the scientific literature as well as in the media and public scene since extra dimensions beyond the 3+1 dimensions of GR has been the subject of much fantasy and fascination. Mathematically, studies of higher dimensional geometry have a long history going back to Riemannian geometry, a century and a half ago. Additionally, it is also worth mentioning that unification theories of physics such as superstring theory and supergravity require such higher dimensional spaces. Figure \ref{fig:HigherDimensions} shows some sub-categories of gravity theories with
higher dimensions.

Accordingly, a number of MG models have been proposed with higher dimensions along with their corresponding cosmologies. We provide here a brief overview following the presentation of \cite{2012-Clifton-MG} and refer the reader to this and other MG extensive reviews \citep{BertiElAlReview2015,2015-rev-Joyce-et-al}. After outlining some seminal or major developments on the topic, we present the Dvali--Gabadadze--Porrati (DGP) theory \citep{DGP} as an illustrative case.

%%%%%%%%%%%%%%%%%%%%%%%%%%%%%%%%%%%%%%%%%%%%%%%%%%%%%%%%%%%%%%%%%%%%%%%%%%%%%%%%%%%
%%%%%%%%%%%%%%%%%%%%%%%%%%%%%%%%%%%%%%%%%%%%%%%%%%%%%%%%%%%%%%%%%%%%%%%%%%%%%%%%%%%
%%%%%%%%%%%%%%%%%%%%%%%%%%%%%%%%%%%%%%%%%%%%%%%%%%%%%%%%%%%%%%%%%%%%%%%%%%%%%%%%%%%
\subsubsection{Theories with compact dimensions versus braneworld models}
\label{sec:compact_vs_brane}

One of the first developments of higher dimensional theories is that of Kaluza--Klein that aimed at unifying gravity and electromagnetism by the use of a fifth small (compact) dimension \citep{Nordstrom2007OT,Kaluza1921,Kaluza1926,Klein1926TA}. Such an additional dimension is compactified in a way so that the theory is effectively a four dimensional one and it does not alter observations in the solar system or at galactic scales but it may have effects at very tiny scales where it is probed using sub-milliliter gravitational experiments, or at cosmological scales where it may have some observational effects. This idea of compactified dimensions is adopted in string theory and supergravity that use 10 and 11 dimensions respectively but where the additional dimensions are compactified to leave only the usual 3+1 standard dimensions of space and time. Interestingly, Kaluza--Klein theory with one or more additional dimensions generically corresponds to an effective 4 dimensional theory with extra scalar fields such as the dilaton field. A dilaton field is scalar field that appears in theories with higher dimensions and compactification. Also, if the Newton constant or Planck mass are promoted  to a scalar field in a given theory then it is a dilaton. A concise discussion about compactification and stabilization of such additional dimensions can be found in \cite{2012-Clifton-MG} and references therein. 

A second distinct development came from the braneworld approach \citep{Akama2000AE,Rubakov1983DW,Arkani-Hamed199TH,Antoniadis1994SC,Arkani-Hamed1999CA} where the extra dimensions can be large or even infinite. In such a scenario the usual 3+1 dimensional hypersurface is called the brane and is embedded in a higher dimensional space called the bulk. The extra large dimensions are said to be non universal in the sense that particles and fields are now constrained only to the brane and only gravity is felt through the bulk. The Braneworld approach is well-motivated by string theory and M-theory and their use of D-branes embedded in higher dimensional spaces \citep{Horava1996ES,Lukas1999UA,Antoniadis1998ND,Polchinski1995DB}. This second approach has seen models with interesting cosmological applications that we discuss next, following \citep{2012-Clifton-MG}. 

A seminal proposal came from \cite{Arkani-Hamed199TH} who proposed an extra dimension to solve the so-called hierarchy problem between the electro-weak scale and the Planck scale being separated by 16 orders of magnitude. The idea is that such a hierarchy can be related to the hierarchy between the scale of the new dimension introduced and the electro-weak scale. The bulk here is a flat 5 dimensional Minkowski space. This is commonly referred to ad the ADD model for the initials of its proposers Arkani-Hamed, Dimopoulos, and Dvali. The ADD model and the alike have been subject to stringent constraints from solar system tests, see e.g., \cite{Overduin2000}.   

Following the work of \cite{Arkani-Hamed199TH}, Randall and Sundrum developed two braneworld models with an anti-de Sitter space as bulk. The first model has two branes \citep{Randall1999LM} and the second model with one brane \citep{Randall1999AA}. The first Randal Sundrum model also aimed at solving the hierarchy problem using two branes separated by a 5-dimensional anti-de Sitter space and is well motivated by string and supergravity \citep{Horava1996ES,Lukas1999UA,Duff2001AS}. However, it is considered incomplete because, on the two branes, the effective theory of gravity is not GR but rather Brans-Dicke gravity with a BD parameter related to the branes and is being not consistent with current observational constraints on the BD parameter \citep{Garriga2000GI}. The stabilization of moduli in the Randall--Sundrum model using the mechanism of \cite{GW1999} have been discussed in \cite{Lesgourgues2004,Brummer2006,Dey2007}. The second Randall Sundrum model has only one brane with a positive tension where GR is recovered as an effective theory and thus consistent with observations.   
Other models based on Braneworld constructions include the Karsh-Randal model \citep{Karch2001LL,Kaloper1999BD}, the Gregory, Rubakov and Sibiryakov model \citep{Gregory2000OU}, the asymmetric brane model \citep{Padilla2005CA,Padilla2005IM,Stoica2000CI}, and the Charmousis, Gregory and Padilla model \citep{Charmousis2007SA}.

The cosmology of braneworld scenarios has been discussed in  \cite{Binetruy2000NC,Bowcock2000GB,Padilla2005CA,Stoica2000CI,Shtanov2009IC,Maartens2001AD,Niz2008BI,Kiritsis2003CE,Kiritsis2005HA,Umezu2006OC,Ichiki2002OC,Liddle2003OC}  and cosmological perturbations have been worked out in \cite{Maartens2000CD,Langlois2001LC,Maartens2000CI,Copeland2001SI,Sahni2002RG,Nunes2002TQ,Liddle2003CR,Bridgman2002CP,Bridgman2001CV,Gordon2001DP,Langlois2000GW,Gorbunov2001GW,Mukohyama2000GG,Mukohyama2000PO,Mukohyama2002DF,Mukohyama2001IE,Hawking2000BN,Kodama2000BW,Langlois2000BC,Bruck2000CP,Koyama2000EO,Bruck2000CC,Kobayashi2001QF,Kobayashi2003PG,Kodama2001BO,Langlois2001EO,Hawking2001TA,Deruelle2001PO,Brax2001BC,Dorca2001CP,Chen2002RP,Chung2003LD,Deffayet2002OB,Riazuelo2002GI,Leong2002CD,Cardoso2007SP}. 

In braneworlds, the difference between the dimension of the bulk and the dimension of the brane is called the co-dimension of the brane. The discussions above are all for a co-dimension of one and that is what has been explored the most in the literature, however other models with higher co-dimensions have been discussed in \cite{Arkani-Hamed199TH,Cline2003CO,Carroll2003ST,Vinet2004CC,Nilles2004SA,Aghababaie2004TA,Dvali2001GO,Dvali2003SM,Dvali2003DT,Gabadadze2004SM,Dubovsky2003BG,Kaloper2007CT,Kaloper2008BI,Rham2008CGE,Rham2010CG,Rham2009CG,Rham2009F3,Minamitsuji2010SS,Agarwal2010CC,Corradini2008IG,Corradini2008IG2,Bostock2004EG,Charmousis2008SO,Charmousis2009PO,Charmousis2010TC,Papantonopoulos2008IC,Papantonopoulos2007RO,Cuadros-Melgar2009BH,Cuadros-Melgar2011PO}. Moreover, cascading gravity has been proposed and is based on scenarios where higher dimensional gravity goes through steps cascading from $(4+n)D$ down to $4D$ gravity, as for example one transitions from larger scales to smaller and smaller scales, see for example \cite{Rham2008CGE,Rham2008CGA,Rham2009CG,Corradini2008IG,Corradini2008IG2,Kaloper2007CT,Kaloper2008BI}. 

These higher co-dimensional or cascading gravity models have been useful in trying to address the cosmological constant problem \citep{Carroll2003ST,Cline2003CO,Vinet2004CC,Aghababaie2004TA,Dvali2003DT,Gabadadze2004SM}  by for example using a degravitation mechanism for the cosmological constant leading to a ``small'' dynamical effect as measured by current cosmological observations \citep{Arkani-Hamed2002NM}. 

Finally, we mention here briefly another class of higher dimensional theories called the Einstein--Gauss--Bonnet gravity. It is built in 5 or 6 dimensions and has as  an action the Ricci scalar, a higher-D cosmological constant term (not the same as the 4D one), and a Gauss--Bonnet invariant (see Eq.~\eqref{eq:GB}). A variation of such an action with respect to the metric gives the Einstein equations in a higher dimension plus a Lovelock tensor term \citep{Lovelock1971TE,Lanczos1938}. This has found applications in string theory \citep{Gross1987TQ,Mannheim3}. A concise discussion with examples can be found in \cite{2012-Clifton-MG}.

%%%%%%%%%%%%%%%%%%%%%%%%%%%%%%%%%%%%%%%%%%%%%%%%%%%%%%%%%%%%%%%%%%%%%
%%%%%%%%%%%%%%%%%%%%%%%%%%%%%%%%%%%%%%%%%%%%%%%%%%%%%%%%%%%%%%%%%%%%%
%%%%%%%%%%%%%%%%%%%%%%%%%%%%%%%%%%%%%%%%%%%%%%%%%%%%%%%%%%%%%%%%%%%%%
\subsubsection{Illustrative example : Dvali--Gabadadze--Porrati Gravity (DGP)}
\label{sec:DGP}

The DGP model is a popular braneworld theory of gravity and is based on a 3+1 brane embedded in a five dimensional Minkowski space \citep{DGP} and is known to have two branches. One is a self-accelerating branch that can produce cosmic acceleration without the need for a cosmological constant or brane tension and has been the subject of much interest earlier in the literature, see extended discussions in \cite{Deffayet2001CO,Deffayet2002AU,Lue2006DGP}. This branch is however plagued with instabilities and the presence of ghost degrees of freedom, e.g., \cite{Charmousis2006DGP,Luty2003DGP}. It also turned out at the end that this self-accelerating branch is not consistent with cosmological observations as we discuss further below. The second \textit{normal} branch (noted as nDGP) does not self-accelerate but has better stability properties than the former branch \citep{Charmousis2006DGP,Padilla2007AS,Gregory2007DGP,Gorbunov2006}. In order to exhibit acceleration this branch uses a brane tension that leads to a phantom type effective dark energy equation of state, $w_{_{EDE}}<-1$. At the perturbation level, the fifth force in these models is screened by the Vainshtein mechanism, see Sect.~\ref{sec:Kinetic}.  

We follow here the presentation of \cite{2012-Clifton-MG} to describe the model and its cosmology. The action is given by 
\be\label{dgp-action}
S = M_5^3 \int d^5 x \sqrt{-\gamma}  \R + \int d^4 x \sqrt{-g}\left[-2 M_5^3 K+\frac{M_4^2}{2} R-\sigma
  +{\cal L}_{\textrm{matter}}\right],
\ee
where $\R$ is the Ricci scalar built from the bulk metric $\g_{ab}$. 
$M_4$ and $M_5$ are the Planck scales in the brane and bulk, respectively.    
$g_{\mu\nu}$ is the metric on the brane, $R$ its Ricci scalar, and $K=g^{\mu\nu} K_{\mu\nu}$ is the trace of extrinsic curvature, $K_{\mu\nu}$. 
$\sigma$ is the tension or bare vacuum energy on the brane. 
${\cal L}_{matter}$ is the matter lagrangian. 

The two different mass scales give rise to a characteristic scale 
\be 
r_c\approx \frac{M_4^2 }{M_5^3 }. 
\ee
At scales shorter than $r_c$, gravity is 4 dimensional and reduces to GR but at scales larger than $r_c$, the 5 dimensional physics is involved and contributes to the dynamics. The field equations can be found in \cite{2012-Clifton-MG}.

With an FLRW metric for the brane, a Minkowski metric for bulk and zero tension, $\sigma$, one gets first a modified Friedmann equation from the 4D Einstein's equations \citep{Deffayet2001CO,Lue2003GS,Dick2001SC} as
\begin{equation}
H^2 + \frac{\kappa}{a^2} - \frac{\epsilon}{r_c}\sqrt{H^2 +
  \frac{\kappa}{a^2}} = \frac{8\pi G}{3}\bar{\rho} ,
\label{eq:DGP_friedman}
\end{equation}
where $\epsilon=-1$ is for the normal branch, and $\epsilon =1$ is for the
self-accelerating  branch.  

The other components of the Einstein equations combined with \eqref{eq:DGP_friedman} above, give the second evolution equations as
\begin{equation}
2\frac{dH}{dt} + 3 H^2 + \frac{\kappa}{a^2}  = -\frac{ 3 H^2 +
  \frac{3\kappa}{a^2} - 2 \epsilon r_c \sqrt{H^2 + \frac{\kappa}{a^2}}
  \;  8\pi G P}{1 - 2 \epsilon r_c \sqrt{H^2 + \frac{\kappa}{a^2}} } .
\label{eq:DGP_ray}
\end{equation}

Although clearly a modified gravity model, the DGP can be formulated using state variables of an effective dark energy 
\begin{equation}
8\pi G \rho_{E} =  \frac{3\epsilon}{r_c} \sqrt{H^2 + \frac{\kappa}{a^2}} ,
\end{equation}
\begin{equation}
8 \pi G P_{E} =  -\epsilon\frac{ \frac{dH}{dt} + 3 H^2 +
  \frac{2\kappa}{a^2} }{ r_c \sqrt{H^2 + \frac{\kappa}{a^2} }},
\end{equation}
and an effective equation of state, 
\begin{equation}
w_{DGP} = \frac{P_E}{\rho_E} = -\frac{ \frac{dH}{dt}  + 3 H^2 +
  \frac{2\kappa}{a^2}}{ 3H^2 + \frac{3\kappa}{a^2}  },
  \label{eq:DGPEOS}
\end{equation}
satisfying the conservation law. 
$$
\frac{d\rho_E}{dt} +3H (\rho_E + P_E) \equiv 0.
$$
During the late time self-accelerating epoch $w_E \rightarrow -1$ in \eqref{eq:DGPEOS}, mimicking a cosmological constant. 

Cosmological linear perturbations for DGP have been worked out in \cite{Deffayet2002OB}. \cite{Koyama2006SF} assumed the small-scale (quasi-static) approximation $k/a\gg r_c \cal H$ and obtained 
\begin{equation}
 - k^2 {\Phi} = 4 \pi G \left( 1 - \frac{1}{3\beta}\right)  \bar{\rho} a^2 \delta_M,
\label{eq:DGP_QS_Phi}
\end{equation}
and
\begin{equation}
 - k^2 \Psi = 4 \pi G \left( 1 + \frac{1}{3\beta}\right)  \bar{\rho} a^2 \delta_M,
\label{eq:DGP_QS_Psi}
\end{equation}
where $\beta = 1 + 2 \epsilon H r_c w_E$. 
It was shown in a number of studies that, unlike the normal branch, the self-accelerating branch of the DGP theory suffers from ghost instabilities \citep{Luty2003DGP,Charmousis2006DGP,Gregory2007DGP,Gorbunov2006,Koyama2007DGP}. 

It is also informative, for comparison with observations, to note a result from \cite{Wei2008GI,Ferreira2010LG} that the linear growth rate index parameter for the DGP model is given by 
\be
\gamma =
\frac{11}{16} - \frac{7}{5632} \Omega_{\rm DGP} +
\frac{93}{4096}\Omega_{\rm DGP}^2 +
O(\Omega_{\rm DGP}^3).
\ee

Nonlinear growth and simulations in DGP models include the works of \cite{Lue2004,Koyama2009N,Scoccimarro2009LS,Chan2009LS,Schmidt2009SC,
Schmidt2009CS,Khoury2009NS,Wyman2010EP,Schmidt2010SC,Seahra2010AD,
Chan2009LS,Cola2017MG2,Bose2018DGP}. 

While the DGP self-accelerating branch models have been now excluded by observation (see Sect.~\ref{sec:constraints_DGP}) and have been shown to be plagued by ghost instabilities, interest continues in the stable normal branch which is often referred to as the nDGP model. Models in this branch are not self-accelerating since the acceleration is due to the brane tension term playing the role of a cosmological constant. nDGP are used as benchmark to develop and test frameworks for MG studies as in for example the nonlinear regime, e.g., \cite{Hellwing2017,Bose2018DGP}.

%%%%%%%%%%%%%%%%%%%%%%%%%%%%%%%%%%%%%%%%%%%%%%%%%%%%%%%%%%%%%%%%%%%%%%%%%%%%%%%%%%%
%%%%%%%%%%%%%%%%%%%%%%%                                         %%%%%%%%%%%%%%%%%%%
%%%%%%%%%%%%%%%%%%%%%%%          SUB-SECTION                    %%%%%%%%%%%%%%%%%%%
%%%%%               Non-local modified gravity theories                        %%%%
%%%%%                                                                          %%%%
%%%%%%%%%%%%%%%%%%%%%%%%%%%%%%%%%%%%%%%%%%%%%%%%%%%%%%%%%%%%%%%%%%%%%%%%%%%%%%%%%%%
%%%%%%%%%%%%%%%%%%%%%%%%%%%%%%%%%%%%%%%%%%%%%%%%%%%%%%%%%%%%%%%%%%%%%%%%%%%%%%%%%%%

\subsection{Non-local modified gravity theories}

A different approach to gravity has been undertaken sometime ago with the idea to introduce some non-locality aspects. A recent class of such theories is where the fundamental action of gravity is local, but the corresponding quantum effective action is not. Non-local gravity in the recent context of cosmic acceleration has been introduced by Wetterich \citep{Wetterich1998EN} with the additional term  R$\Box^{-1}R$ to the Einstein--Hilbert action. Despite interesting features and being ghost-free, the model did not have a viable cosmological evolution. This was followed by a generalization proposed by Deser and Woodard in \cite{Deser2007NC} that made the additional term as $R f(\Box^{-1}R)$. It is possible to adjust their model so it can fit well the background accelerating expansion with no need for a dark energy component \citep{Woodard2014NM}, although in a non-predictive way state  \citep{Belgacem2017NL}. Cosmological perturbations and growth of structure equations have been worked out and the model compared to large scale structure data \citep{Park2013SF,Nersisyan2017SF,Park2017RO}. 
It was found in \cite{Park2013SF} that the $R f(\Box^{-1}R)$ is not in agreement with such large-scale structure data but \cite{Nersisyan2017SF} found the opposite and reported that the model are consistent with such data. This was confirmed in \cite{Park2017RO} so it is agreed now that the model is consistent with the growth of large-scale structure data. For the moment, the model has been compared only to structure formation data and need further comparison to full CMB and other data sets. As stated in \cite{Woodard2014NM}, their model paved the road for further developments in non-local gravity. 
A good discussion of non-local gravity and its cosmology was also given in \cite{Koivisto2008NL} where models similar to \cite{Deser2007NC} were carefully analyzed. The author found that even simple models can drive late-time cosmic acceleration without affecting early time cosmology. 
Furthermore, Barvinsky proposed a theory with an additional term of the form $\RMN\iBox\Gmn$ \citep{Barvinsky2003NA,Barvinsky2012DE,Barvinsky2012SD}. The author connects the theory he proposed to the paradigms of dark matter and dark energy. It is hoped to see more detailed developments of these models with observable functions and comparison to data. 
Most recently, \citep{Maggiore2014NG} followed the path of non-local gravity  but with a different approach where a new mass scale is generated in the IR limit and is associated with the non-local term in the theory. Their theory provides an interesting phenomenology for cosmic acceleration and is found to fit current observations \citep{Maggiore2014NG,Maggiore2014PD}. 

%%%%%%%%%%%%%%%%%%%%%%%%%%%%%%%%%%%%%%%%%%%%%%%%%%%%%%%%%%%%%%%%%%%%%%%%%%%%%%%%%%%
%%%%%%%%%%%%%%%%%%%%%%%%%%%%%%%%%%%%%%%%%%%%%%%%%%%%%%%%%%%%%%%%%%%%%%%%%%%%%%%%%%%
%%%%%%%%%%%%%%%%%%%%%%%%%%%%%%%%%%%%%%%%%%%%%%%%%%%%%%%%%%%%%%%%%%%%%%%%%%%%%%%%%%%
\subsubsection{Illustrative example: ${\mathit{RR}}$ model}

We use here for illustration, the specific model called ``${\mathit{RR}}$'' that was proposed in \cite{Maggiore2014NG} and was based on their earlier work of \cite{Maggiore2014PD}. The quantum effective action derived from the fundamental Einstein--Hilbert action is postulated to have the form
\be
\label{eq:RRaction}
\Gamma_{\rm {\mathit{RR}}}=\frac{\mplr^2}{ 2}\int d^4 x \sqrt{ -g}\,\,\left[R-\frac{1}{6} m^2 R\frac{ 1}{\Box^2 } R\right],
\ee
where the nonlocal term is assumed to capture non-perturbative  infrared effects due to quantum fluctuations, and corresponds, physically, to a dynamical mass generation for the conformal mode. Here $\mplr^2$ is the reduced Planck mass squared and $m$ is a mass parameter related to the generated fundamental mass scale $\Lambda_{\rm\scriptscriptstyle {\mathit{RR}}}$ by $\Lambda_{\rm\scriptscriptstyle {\mathit{RR}}}^4=(1/12)m^2\mplr^2$. The model has been reviewed in some detail in \cite{Maggiore2016NI,Belgacem2017NL} with comparison to available cosmological data in \cite{Dirian2014CP,Dirian2015NG,Dirian2016NG,Dirian2017CT} as we summarize further below. 

Recently, evidence for the emergence of the nonlocal term in \eqref{eq:RRaction} has been found by using nonperturbative results from lattice gravity \cite{Knorr2018}.
Also, as pointed out in \cite{Maggiore2014NG}, analogous nonlocal terms, proportional to $m^2 F^{\mu\nu} \Box^{-1} F_{\mu\nu}$,  have also been postulated to arise in the quantum effective action of QCD, where they reproduce results for the gluon propagator from lattice simulations \citep{Boucaud2001,Capri2005,Dudal2008}.

We present the model dynamics and cosmology following  \citep{Maggiore2016NI,Belgacem2017NL}. The  model can be written in a local form by the use of two auxiliary fields $U$ and $S$, \citep{Maggiore2014NG}
\be\label{defU}
U=-\iBox R\, ,\qquad 
S=-\iBox U.
\ee
The action can then be re-written by the use of the two Lagrange multipliers $\xi_1,\xi_2$ as follows \citep{Maggiore2014NG}
\be\label{S2}
\Gamma_{\rm {\mathit{RR}}}=\frac{\mplr^2}{2}\int d^4x \sqrt{-g}\, \left[ R\( 1-\frac{m^2}{6} S\)-\xi_1(\Box U+R)-\xi_2 (\Box S+U) \right]\nn.
\ee
Variation with respect to the metric gives the field equations 
\be\label{Gmn}
\Gmn=\frac{m^2}{6} K_{\mu\nu}+8\pi G\Tmn\, ,
\ee
where the tensor $K_{\mu\nu}$ is given in terms of the metric and the auxiliary fields as  \citep{Maggiore2014NG}
\be\label{defKmn}
K^\mu_\nu \equiv 2 S G^\mu_\nu - 2 \nabla^\mu \partial_\nu S + 2 \delta^\mu_\nu \Box_g S + \delta^\mu_\nu \partial_\rho S \partial^\rho U - \frac{1}{2} \delta^\mu_\nu U^2 - \big( \partial^\mu S \partial_\nu U + \partial_\nu S \partial^\mu U \big).
\ee
Variation with respect to the Lagrange multipliers $\xi_1,\xi_2$  gives the further equations that $U$ and $S$ must satisfy 
\be\label{BoxUS}
\Box U=-R\, ,\qquad
\Box S =-U\,.
\ee
This localization thus makes the theory appear as a scalar-tensor theory with two dynamical fields $U$ and $S$. However, as discussed in \cite{Belgacem2017NL} and references therein, upon quantization, the theory remains ghost-free because there are no quanta associated to these two fields. In fact, the classical instability develops rather on a cosmological timescale producing the late cosmic acceleration with an effective phantom-like dark energy component that is found to be consistent with cosmological background and growth of structure observations. 

Again following \cite{Belgacem2017NL}, we summarize some aspects of ${\mathit{RR}}$ model cosmology. The field equations \eqref{Gmn}--\eqref{BoxUS} are applied to the flat FLRW metric in Cartesian coordinates 
\be   
ds^2=-dt^2+a^2(t)d\textbf{x}^2
\ee
and the time evolution is parametrized using $x\equiv\ln a$. In addition to $U(x)$ and $S(x)$, the following dimensionless functions are defined 
\be
W(x)\equiv H^2(x)S(x),\,\,\,\,h(x)\equiv H(x)/H_0
\ee 
where $H(t)=\dot{a}/a$ is the Hubble function with $H_0$ its present value. The background evolution equations are then obtained as \citep{Maggiore2014NG}
\bea
&&h^2(x)=\Omega_M e^{-3x}+\Omega_R e^{-4x}+\g Y\label{syh}\\
&&U''+(3+\zeta) U'=6(2+\zeta)\, ,\label{syU}\\
&&W''+3(1-\zeta) W'-2(\zeta'+3\zeta-\zeta^2)W= U\, ,\label{syW}
\eea
where prime here denotes differentiation with respect to $x$, $\gamma\equiv m^2/(9H_0^2)$, $\zeta \equiv h'/h$ and
\be\label{defY}
Y\equiv \frac{1}{2}W'(6-U') +W (3-6\zeta+\zeta U')+\frac{1}{4}U^2\, .
\ee
In the modified Friedmann equation \eqref{syh}, an effective dark-energy density is identified as $\rho_{\rm DE}=\rho_0\gamma Y$ where $\rho_0=3H_0^2/(8\pi G)$ is the usual critical density given in \eqref{eq:criticalDensity}. 

As discussed in \cite{Maggiore2014NG,Belgacem2017NL}, a choice of boundary conditions can be made for the auxiliary fields $U$ and $W$ that can provide a minimal model with a background evolution that depends on the Hubble constant, $H_0$ the matter density parameter $\Omega_M$, and one additional parameter that is the mass $m$ that replaces the cosmological constant. So the model has the same background parameters as the $\Lambda$CDM model. Interestingly, just like for $\Omega_{\Lambda}$ in the  $\Lambda$CDM model, in the ${\mathit{RR}}$ model, the flatness condition allows one to derive the mass parameter $m$. The authors then proceeded to fit this minimal model to CMB, BAO, and SN data and found $\Omega_M\simeq 0.299$ and $h_0\simeq 0.695$ \citep{Dirian2017CT}. With these values fixed, they then integrated numerically \eqref{syh}-\eqref{syW} \citep{Maggiore2014NG} and studied the evolution of the corresponding effective dark energy equation of state, $w_{\rm DE}(x)$. Translating their results into the common CPL parameterization \citep{Chevallier2001IAU,Linder2003ET}, they found that the model has an effective phantom-like equation of state with $w_0\simeq -1.15$ and  $w_a\simeq 0.09$ \citep{Maggiore2014NG}.  

\cite{Belgacem2017NL} followed with an interesting discussion relating the evolution of the field $U$ to that of the effective dark energy of the model. While it is zero in the radiation dominated era, the field $U$ grows in the matter dominated era resulting in the growth of the effective dark energy density as indicated by \eqref{defY}. Moreover, this implies that both the effective dark energy density and its variation are positive resulting in a phantom effective equation of state caused by the ghost-like field $U$. The good news is that the ghost-like feature of the field in this case is only classical with no associated quanta nor instability when quantizing the theory. On the contrary, the classical instability is a plus and would be responsible for the observed late-time cosmic acceleration. 

\cite{Belgacem2017NL} compared the comoving distances between the minimal ${\mathit{RR}}$ model and the $\Lambda$CDM and found that for the same values of a fiducial cosmological model these can be different by up to 2.5\%, however, if each models is fed its best-fit values of cosmological parameters, the difference can be brought down to below 1\% up to redshift 6 (see Fig.~2 there). 

Discussions of ${\mathit{RR}}$ models with other initial conditions than the minimal models discussed above can be found in \cite{Maggiore2016NI,Belgacem2016SI,Maggiore2014PD}. It was found that some models can mimic the background with an equation of state that is very close to that a cosmological constant and different from it by less than a 1\%, on the phantom side again. This will make it challenging to distinguish such particular models from the $\Lambda$CDM model using observations, although the growth of structure remain to be explored.  

Cosmological perturbations for ${\mathit{RR}}$ models have been worked out in detail in \cite{Dirian2014CP}. They have been recapitulated in \cite{Maggiore2016NI} and \cite{Belgacem2017NL} and we refer the reader to those papers. We provide here a few remarks following \cite{Belgacem2017NL}. From using the flat FLRW perturbed metric in the Newtonian gauge plus linear perturbations of the auxiliary fields $U$ and $W$, with adiabatic initial conditions, the resulting perturbation evolutions  were found to be stable \citep{Dirian2014CP}. They also found in their analysis that, for the minimal ${\mathit{RR}}$ model, the perturbations are close to those of the $\Lambda$CDM model with a difference below 10\% \citep{Dirian2014CP}. This makes the comparison to data interesting in a sense that it is close to the $\Lambda$CDM and so it is expected to be found in a viable range, but it is also distinct from the $\Lambda$CDM so a comparison to find which model fits better the data will be possible and important. 

%%%%%%%%%%%%%%%%%%%%%%%%%%%%%%%%%%%%%%%%%%%%%%%%%%%%%%%%%%%%%%%%%%%%%%%%%%%%%%%%%%%
%%%%%%%%%%%%%%%%%%%%%%%%%%%%%%%%%%%%%%%%%%%%%%%%%%%%%%%%%%%%%%%%%%%%%%%%%%%%%%%%%%%
%%%%%%%%%%%%%%%%%%%%%%%%%%%%%%%%%%%%%%%%%%%%%%%%%%%%%%%%%%%%%%%%%%%%%%%%%%%%%%%%%%%
\subsubsection{Other Non-Local gravity theories}

Another interesting proposal of non-local gravity is that of Mashhoon, see e.g. \cite{Mashhoon1990,Mashhoon2008,Mashhoon2009,Mashhoon2009b}. Recently, the authors applied the theory to Newtonian cosmology with the aim to model structure formation without the need for dark matter \citep{Chicone2016}. It will be interesting  to see this theory tested using large scale structure and CMB data. A review can be found in \cite{Mashhoon2017}.

%%%%%%%%%%%%%%%%%%%%%%%%%%%%%%%%%%%%%%%%%%%%%%%%%%%%%%%%%%%%%%%%%%%%%%%%%%%%%%%%%%%
%%%%%%%%%%%%%%%%%%%%%%%%%%%%%%%%%%%%%%%%%%%%%%%%%%%%%%%%%%%%%%%%%%%%%%%%%%%%%%%%%%%
%%%%%%%%%%%%%%%%%%%%%%%%                                  %%%%%%%%%%%%%%%%%%%%%%%%%
%%%%%%%%%%%%%%%%%%%%%%%%  SCREENING MECHANISMS            %%%%%%%%%%%%%%%%%%%%%%%%%
%%%%%%%%%%%%%%%%%%%%%%%%                                  %%%%%%%%%%%%%%%%%%%%%%%%%
%%%%%%%%%%%%%%%%%%%%%%%%%%%%%%%%%%%%%%%%%%%%%%%%%%%%%%%%%%%%%%%%%%%%%%%%%%%%%%%%%%%
%%%%%%%%%%%%%%%%%%%%%%%%%%%%%%%%%%%%%%%%%%%%%%%%%%%%%%%%%%%%%%%%%%%%%%%%%%%%%%%%%%%

\section{Screening mechanisms}
\label{screening}

Most recent developments and proposals of MG models have been motivated by the problem of cosmic acceleration. Modifications to GR happen in a way to affect cosmological dynamics at large scales and to produce an accelerating expansion. 
However, any such modification at cosmological scales must survive well-established stringent solar system tests of gravity \citep{WillReview2014}. 

Therefore, MG models must either reduce to GR at small scales, by construction, or must have a mechanism that suppresses any deviation from GR at small scales. These are known as screening mechanisms. Some of them are  are based on ideas that relate the scalar field potential to the local matter density within planetary systems or galaxies since it is higher than the average cosmological density. 

Most MG models generate a fifth force acting at the level of perturbations due to the coupling of the scalar field to the matter in the Einstein frame. We use here the behavior of factors or components of such a fifth force potential to classify the corresponding screening mechanism following one of the classifications of, e.g., \cite{deRham2012GS,Jain2013,2015-rev-Joyce-et-al}. For that, we consider a simplified lagrangian for a scalar field conformally coupled to matter as  
\be
{\cal L} = -\frac{1}{2}Z^{\mu\nu}(\phi, \partial\phi,\ldots)\partial_\mu\phi\partial_\nu\phi-V(\phi)+\beta(\phi)T^\mu_{\;\mu}~,
\label{Lintro}
\ee
where the components of $Z^{\mu\nu}$ contain functions up to second order derivatives associated with self-interactions of the field, and $\beta(\phi)$ is a coupling function to the trace of the energy-momentum tensor, $T^\mu_{\;\mu}$. We can consider non-relativistic pressure-less sources and specifically a point source so that $T^\mu_{\;\mu}=-\rho=-{\cal M}\delta^3(\bm{x})$. 
The background value of the field, $\bar \phi$, is determined by this local density. 
We then consider a field perturbation, $\delta\phi$, around the background value, $\bar\phi$, with an equation of motion given by
\be
Z(\bar\phi)\Big(\ddot{\delta\phi}-c_s^2(\bar\phi)\nabla^2\delta\phi\Big)+m^2(\bar\phi)\delta\phi = \beta(\bar\phi){\cal M} \delta^3(\bm{x}) \,,
\label{generalscalareom}
\ee
where $c_s$ is an effective sound speed of perturbations, $m(\bar \phi)$ is the scalar effective mass. Assuming negligible spatial variations for the background  field over the scales of interest, the corresponding potential is given by \cite{2015-rev-Joyce-et-al}
\be
V(r) = -\frac{\beta^2(\bar\phi)}{Z(\bar\phi) c_s^2(\bar\phi)}\frac{e^{-\frac{m(\bar\phi)}{\sqrt{Z(\bar\phi)}c_s(\bar\phi)}r}}{4\pi r}{\cal M} ~.
\label{scalarpoteqn}
\ee
This Yukawa potential corresponds to a fifth force. This force is Yukawa-suppressed 
(via the exponential) at some large scales but needs to be suppressed at small 
scales such as the solar systems and inside galaxies. There are at least three 
mechanisms to produce such a screening in high density environments and to 
produce dynamics that complies with local tests of gravity and also do not 
perturb star motions and distributions in galaxies. We briefly overview them 
in the next subsections and refer the reader to specific papers and reviews 
on each mechanism. The topic of screening is extensively covered in the reviews 
\cite{2015-rev-Joyce-et-al,Khoury2010,JK2010SM,Burrage2018}. 
It was pointed out in \citep{Burrage2018} 
that Eq.~\eqref{scalarpoteqn} captures very well how 
screening may happen in interactions between fundamental particles,  
but is not particularly effective in explaining how screening happens around 
more macroscopic sources. For example they explain that the thin-shell mechanisms 
of chameleon and symmetron models that we will discuss further below are much 
better described in this equation by making curly M in Eq.~\eqref{scalarpoteqn} 
as a function of the true mass of the source like in equation 2.11 in 
\citep{Burrage2018}.

We use below a classification based on the MG model fifth force, however, other classifications have been proposed. See the review \cite{2015-rev-Joyce-et-al} for two other classifications of screening mechanisms. One they qualify as more phenomenological and more suited for astrophysical and cosmological observations. It is based on classifying the screening mechanisms in three types where the screening is set by the field itself, its first derivatives, or its second derivatives. The last classification they provide there is based on an effective field theory approach that they present as a unifying description for these mechanisms. The authors provide a large number of examples and organize their review around such screening mechanisms and we refer the reader to their review and references therein.  

%%%%%%%%%%%%%%%%%%%%%%%%%%%%%%%%%%%%%%%%%%%%%%%%%%%%%%%%%%%%%%%%%%%%%%%%%%%%%%%%%%%
%%%%%%%%%%%%%%%%%%%%%%%                                         %%%%%%%%%%%%%%%%%%%
%%%%%%%%%%%%%%%%%%%%%%%          SUB-SECTION                    %%%%%%%%%%%%%%%%%%%
%%%%%                   Large-mass based screening                             %%%%
%%%%%                                                                          %%%%
%%%%%%%%%%%%%%%%%%%%%%%%%%%%%%%%%%%%%%%%%%%%%%%%%%%%%%%%%%%%%%%%%%%%%%%%%%%%%%%%%%%
%%%%%%%%%%%%%%%%%%%%%%%%%%%%%%%%%%%%%%%%%%%%%%%%%%%%%%%%%%%%%%%%%%%%%%%%%%%%%%%%%%%

\subsection{Large-mass based screening}
\label{sec:Large-mass}

In this case, the scalar field mass, $m(\bar \phi)$ depends on the environment and causes the change. In a high density region, the scalar field acquires a large mass so the fifth force becomes very short range and highly suppressed as can be seen from the potential \eqref{scalarpoteqn}. At the opposite, in a low density region such as cosmological volume scales, the scalar field becomes light and the fifth force reaches the strength of the gravitational force manifesting itself in the growth perturbation equations (see Sect.~\ref{sec:perturbations}). An example of a field with such a behavior is the aptly named chameleon field \citep{KW2004CS,KW2004CS2}. See also the recent review \cite{Burrage2018} and references therein. We review some aspects of it following \cite{KW2004CS2,KOYAMA2016TestGR}. 

We consider a class of MG models using the chameleon mechanism and for which the action can be written in the following form in the Einstein frame 
\begin{equation}
S= \int d^4 x \sqrt{-g} \left[
\frac{1}{16 \pi G} R - \frac{1}{2} (\nabla \phi)^2 - V(\phi)
 \right]
 + S_{m} ( A^2(\phi) g_{\mu \nu},\psi_m)
\label{Einsteinframe}
\end{equation}
where the field is coupled to the matter via the metric $A^2(\phi) g_{\mu \nu}$. The matter particles do not follow geodesics in this frame and feel a fifth force generated by the scalar field as 
\be 
F_5=\nabla \ln A(\phi).
\ee
The scalar field dynamics are governed by an effective potential that depends on the local density $ T^{\mu}_{\mu}=-\rho$ as
\begin{equation}
V_{\rm eff} = V(\phi) +[A(\phi)-1]\rho .
\label{effectiveV}
\end{equation}
An example of a typical choice of the potential $V(\phi)$ and the coupling function $A(\phi)$ for the chameleon mechanisms is given by  
\begin{eqnarray}
A(\phi) &=& 1 + \xi \frac{\phi}{M_{\rm pl}}, 
\quad V(\phi) = \frac{M^{4 +n}}{\phi^n}.
\label{eq:typical_chameleon}
\end{eqnarray}
where $M$ is the mass scale parameter. 
The scalar field dynamics are characterized by the coupling function
\bea
\beta=\left. \mpl \frac{d \ln A}{d \phi} \right |_{\phi=\bar{\phi}}
\eea
and its mass around the minimum of the potential $\phi=\bar{\phi}$ given by 
\be
m^2 = \left. \frac{d^2V_{\rm eff}}{d\phi^2} \right |_{\phi=\bar{\phi}}
\ee
Thus the dependence of the scalar field mass on its environment will be determined by an appropriate choice of the potential $V(\phi)$. For example the potential\eqref{effectiveV} depends explicitly on the density. 

To understand a little bit better how this is implemented, let one recall the equation of motion of the scalar field given by
\be
\nabla^2 \phi= V_{{\rm eff,\phi}}( \phi)=  V_{,\phi}(\phi) + 8 \pi G \beta \rho.
\label{eq:phi_eom}
\ee

Now, following \cite{KW2004CS2}, one considers a spherically symmetric body with radius $R_c$ and homogeneous density $\rho_c$. The body is assumed to be embedded in a larger environment with homogeneous smaller density $\rho_{\infty}$. This is like the solar system in the galaxy, or a galaxy in the Hubble volume. For the spherically symmetric body, the Eq.~\eqref{eq:phi_eom} becomes
\be
\frac{d^2\phi}{dr^2} + \frac{2}{r} \frac{d\phi}{dr} = V_{,\phi}(\phi) + 8 \pi G \beta \rho. 
\label{eq:phi_eom_ss}
\ee
with $\rho(r)=\rho_c$ for $r<R_c$ and $\rho(r)=\rho_{\infty}$ for $r>R_c$.

A qualitative discussion followed by a quantitative derivation of the solution to \eqref{eq:phi_eom_ss} taking into account proper boundary conditions is carried out in some detail in \cite{KW2004CS2} giving
\begin{equation}
\phi(r) = -
\left( \frac{3 \delta R_c}{R_c}  \right) \frac{2 G M \beta }{r} e^{- m_{\infty} r} + \phi_{\infty},
\label{chameleon-force}
\end{equation}
where 
\begin{equation}
\frac{\delta R_c}{R_c}  = \frac{\phi_{\infty} - \phi_{s}}{6 \beta \mpl |\Psi_{N}| } \ll 1, 
\label{eq:thin_shell_sol}
\end{equation}
where $\phi_s$ is the field  value which minimizes $V_{\rm eff}$ inside the source, $\phi_{\infty}$ is another minimum outside the source, $r_{scr}$ delimits the screened area, $R_c$ is the radius outside $r_{scr}$, 
and $\Psi_{N}$ is the gravitational potential of the source with $|\Psi_{N}| = G M/R_c$. 

\begin{figure}[ht]
  \centering{
  \includegraphics[width=0.8\textwidth]{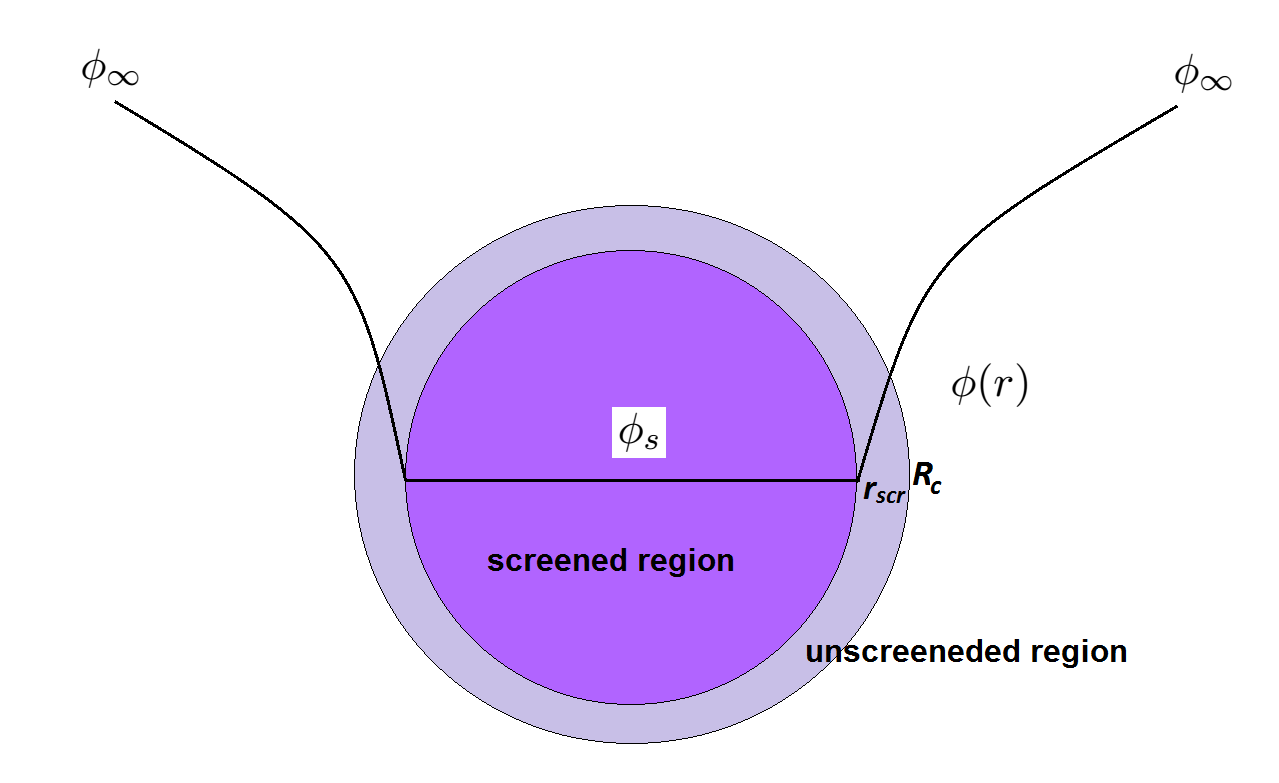}
  }
  \caption{Thin shell profile of the scalar field for the chameleon screening mechanism. Only the mass of the thin shell contributes to the fifth force outside the shell while the force is suppressed inside the shell. See text and \cite{KW2004CS2,KOYAMA2016TestGR} for discussion. 
   }
   \label{fig:chameleon}
\end{figure}

This solution is valid under the thin shell condition \citep{KW2004CS2}, 
\be
\delta R_c/R_c \ll 1. 
\ee
In such a case, only the mass within the thin-shell defined by $\delta R_c$ contributes to the fifth force outside the shell because in the interior of the source the scalar field mass is large and the fifth force is suppressed by the Yukawa exponential realizing the chamelon mechanism. This brings us to an important point stressed in \cite{KW2004CS2,KOYAMA2016TestGR} which is the gravitational potential profile from a dense region to a less dense region that matters rather than the dense region alone. We depict in Fig.~\ref{fig:chameleon}  the above picture for the chameleon thin-shell mechanism.  

%%%%%%%%%%%%%%%%%%%%%%%%%%%%%%%%%%%%%%%%%%%%%%%%%%%%%%%%%%%%
%%%%%%%%%%%%%%%%%%%%%%%%%%%%%%%%%%%%%%%%%%%%%%%%%%%%%%%%%%%%%

We conclude this sub-section by giving some informative constraints on the chameleon scalar field that can be obtained from applying the screening mechanism and the thin shell condition to the solar system, galactic scale and cosmic scale following \cite{KW2004CS2,KOYAMA2016TestGR}. In order to compare with observations though, we need to consider the Jordan frame where the MG parameters then relate to the coupling parameter $\beta$ in the thin shell condition as follows:  
\begin{equation}
\mu = 1 + 2 \beta_{\rm eff}^2 \quad \rm {and } \quad \eta = \frac{1 - 2 \beta_{\rm eff}^2}{1 + 2 \beta_{\rm eff}^2}. 
\label{eq:MGparamvsbeta}
\end{equation}
where
\be 
\beta_{\rm eff}^2 \equiv \beta^2 \frac{3 \delta R}{R}.
\label{eq:beta_eff}
\ee
For example, taking the sun density as $\rho_{sun} = 10$ g cm$^{-3}$ and the Milky way density as $\rho_{gal}=10^{-24}$ g cm$^{-3}$, one can conclude that the scalar field in the vicinity of the sun is suppressed compared to the galaxy. If one assumes the thin shell condition holds and uses equations \eqref{eq:MGparamvsbeta} and \eqref{eq:beta_eff}, one can write  $\eta -1 = - 4 \beta_{\rm eff} ^2$. Now, using Eq.~\eqref{eq:thin_shell_sol}, setting $\Psi_{N \rm gal} \approx 10^{-6}$ for the Milky Way's potential, and employing the constraints  $|\eta -1| =(2.1 \pm 2.3) \times 10^{-5}$ \citep{Cassini2003}, one can derive the constraint on the galaxy scalar field as 
\begin{equation}
\frac{\beta \phi_{\rm gal}}{\mpl} <  10^{-11}.
\label{solar}
\end{equation}
This is a model independent constraint but see \citep{KW2004CS2,KOYAMA2016TestGR} for model dependent stringent constraints on the scalar field in a cosmological environment and many other examples in \cite{KW2004CS2}. It was also pointed out in \cite{Sakstein2017} that \eqref{solar} only applies if one has a point particle and that PPN bound on chameleon screening is actually weaker than previously claimed due to the WEP. 

We end this sub-section with a significant result from \cite{WangEtALNG} who proved two no-go theorems limiting the cosmological impact on chameleon-like mechanisms such as one above but also the symmetron (a scalar field that couples to the ambient matter density) and dilaton (defined in Sect.~\ref{sec:compact_vs_brane}) mechanisms in the next sub-section. The first theorem states that the Compton wavelength (effective ``range'') of such a scalar field can be at most of Mpc scale so it limits its impact on large-scale structure reducing it to nonlinear scales only. So it will have no effect on the linear growth rate of large scale structure and its observables such as RSD, WL or clustering. The second theorem states that, in these theories, the conformal factor relating the Einstein and the Jordan frames and their scale factors is essentially constant in a Hubble time meaning that such theories cannot provide self-acceleration and rather require a form of dark energy for that. In other words, a broad class of chameleon, symmetron, and dilaton screened theories cannot have any significant effect on large scale density perturbations and cannot explain cosmic acceleration. This was quite a strong result leaving then only kinetic-terms based screening mechanisms such as Vainstein and k-mouflage, discussed further below, for consideration. 

%%%%%%%%%%%%%%%%%%%%%%%%%%%%%%%%%%%%%%%%%%%%%%%%%%%%%%%%%%%%%%%%%%%%%%%%%%%%%%%%%%%
%%%%%%%%%%%%%%%%%%%%%%%                                         %%%%%%%%%%%%%%%%%%%
%%%%%%%%%%%%%%%%%%%%%%%          SUB-SECTION                    %%%%%%%%%%%%%%%%%%%
%%%%%                 Weak-coupling based screening                            %%%%
%%%%%                                                                          %%%%
%%%%%%%%%%%%%%%%%%%%%%%%%%%%%%%%%%%%%%%%%%%%%%%%%%%%%%%%%%%%%%%%%%%%%%%%%%%%%%%%%%%
%%%%%%%%%%%%%%%%%%%%%%%%%%%%%%%%%%%%%%%%%%%%%%%%%%%%%%%%%%%%%%%%%%%%%%%%%%%%%%%%%%%

\subsection{Weak-coupling based screening} 
\label{Weak-coupling}

In this mechanism, it is the field's coupling $\beta(\phi)$ in Eq.~\eqref{scalarpoteqn} that depends on the environment. In a dense region such as the solar system, it becomes weak and causes the suppression of the fifth force. However, in low density environments such as at cosmological scales, the coupling strengthens and makes the fifth force of the order of the gravitational force affecting Poisson equations as in Sect.~\ref{sec:perturbations}. 

Examples of fields using this mechanism are the dilaton \citep{Damour1994TS,Brax2011NS} 
with a typical choice of the potential and coupling functions given by 
\begin{eqnarray}
A(\phi) &=& 1 + \frac{1}{2 M} 
(\phi - \bar{\phi})^2, 
\quad V(\phi) = V_0 e^{- \phi/M_{\rm pl}}
\end{eqnarray}
and the symmetron \citep{Hinterbichler2010SL,Olive2008ED}  with 
\begin{eqnarray}
A(\phi) &=& 1 + \frac{1}{2 M^2} 
\phi^2, 
\quad V(\phi) = -\frac{\mu^2}{2} \phi^2 + \frac{\lambda}{4} \phi^4 ,
\end{eqnarray}
where the action and effective potentials are given by Eqs.~\eqref{Einsteinframe} and \eqref{effectiveV}, respectively. We refer the reader to the original papers above for these mechanisms and the specialized review \cite{2015-rev-Joyce-et-al}. 

%%%%%%%%%%%%%%%%%%%%%%%%%%%%%%%%%%%%%%%%%%%%%%%%%%%%%%%%%%%%%%%%%%%%%%%%%%%%%%%%%%%
%%%%%%%%%%%%%%%%%%%%%%%                                         %%%%%%%%%%%%%%%%%%%
%%%%%%%%%%%%%%%%%%%%%%%          SUB-SECTION                    %%%%%%%%%%%%%%%%%%%
%%%%%                 Large kinetic terms based screening                      %%%%
%%%%%                                                                          %%%%
%%%%%%%%%%%%%%%%%%%%%%%%%%%%%%%%%%%%%%%%%%%%%%%%%%%%%%%%%%%%%%%%%%%%%%%%%%%%%%%%%%%
%%%%%%%%%%%%%%%%%%%%%%%%%%%%%%%%%%%%%%%%%%%%%%%%%%%%%%%%%%%%%%%%%%%%%%%%%%%%%%%%%%%

\subsection{Large kinetic terms based screening} 
\label{sec:Kinetic}

Another possibility is to make the kinetic function $Z(\bar\phi)$ in Eq.~\eqref{scalarpoteqn} dependent on the environment. These are in particular derivatives of the field corresponding to its nonlinear interactions. 
When such terms becomes large, they can effectively suppress the fifth force as can be seen in \eqref{scalarpoteqn}. Namely, this can happen when the first derivatives of the field become large as in the case of the k-mouflage mechanism \citep{kmouflage1}, see also reviews \cite{kmouflage2,kmouflage3}, or when the second derivatives become important realizing the Vainshtein mechanism \citep{Vainshtein1972TT}, see also review \cite{2015-rev-Joyce-et-al}.  

A typical choice of action that leads to the k-mouflage mechanism is of the form of the Horndeski class of models \eqref{Horndeski_Action} with only  
\be
{\cal L}_{2} = K(\phi,X)=X
+ \frac{\alpha}{4 \Lambda^4} X^2.
\label{eq:Kamouflage}
\ee
If one considers a solution to a spherically symmetric field around a source with a given gravitational potential then the k-mouflage screening occurs when the first derivative of the gravitational potential exceeds some critical value $\Lambda_c$.  
We chose the form of \eqref{eq:Kamouflage} just for illustrative purposes as it was shown in \cite{Barreira2015} that it does not pass some solar system and cosmological constraints. 

Vainshtein mechanisms can be realized by the typical choice of the Horndeski action with only
\bea
{\cal L}_{2} & = & K(\phi,X)=X,\\ \nonumber
{\cal L}_{3} & = & -G_{3}(\phi,X)\Box\phi=  \frac{1}{\Lambda^3} X \Box\phi. \nonumber
\eea
The Vainshtein screening occurs when the second derivatives of the gravitational potential exceed some critical value $\Lambda_c^3$.

We provide here a simple illustrative example of how the Vainshtein  mechanism works. 
Following \cite{2015-rev-Joyce-et-al}, we use the cubic Galileon with Lagrangian: 
\be
{\cal L} = -3(\partial\phi)^2 - \frac{1}{\Lambda^3}\Box\phi(\partial\phi)^2 +\frac{g}{M_{\rm Pl}}\phi T^\mu_{\;\mu}\ ,
\label{L3}
\ee
where gravitational strength coupling, $g$, is taken of the order unity and $\Lambda$ is the strong-coupling scale of the theory. The Vainshtein mechanism is realized by the $(\partial\phi)^2\Box\phi/\Lambda^3$ term becoming large compared to the term $(\partial\phi)^2$ near massive objects so that $\partial^2\phi \gg \Lambda^3$ is achieved. Varying~\eqref{L3} with respect to $\phi$ gives the equation of motion
\be
6\Box\phi + \frac{2}{\Lambda^3} \bigg( (\Box\phi)^2 - (\partial_\mu\partial_\nu\phi)^2\bigg) = - \frac{g}{M_{\rm Pl}} T^\mu_{\;\mu} \,.
\label{L3eom}
\ee
Next, the field is considered around a static point source with $T^\mu_{\;\mu} =-M\delta^{(3)}(\bm{x})$ and assumed to have a static spherically-symmetric profile, $\phi(r)$. Equation \eqref{L3eom} then becomes ~\citep{Nicolis2004CA}
\be
\bm{\nabla} \cdot \left(6\vec\nabla\phi + \hat{r} \frac{4}{\Lambda^3}\frac{(\vec\nabla\phi)^2}{r}\right) = \frac{gM}{M_{\rm Pl}} \delta^{(3)}(\bm{x})\,.
\ee
Upon integration, one obtains 
\be
6\phi' + \frac{4}{\Lambda^3}\frac{\phi'^2}{r} = \frac{gM}{4\pi r^2 M_{\rm Pl}}\, .
\ee
One can then solve this equation algebraically for $\phi'$ and use the stable solution for which $\phi'\rightarrow 0$ at $r \rightarrow \infty$. This reads
\be
\phi'(r) = \frac{3\Lambda^3  r}{4} \left( -1+ \sqrt{1 + \frac{1}{9\pi}\left(\frac{r_{\rm V}}{r}\right)^3 }\right)\, ,
\label{galfieldprofile}
\ee
where
\be
r_{\rm V} \equiv \frac{1}{\Lambda } \left(\frac{gM}{M_{\rm Pl}}\right)^{1/3}\ 
\ee
is the Vainshtein radius. 
%

%%%%%%%%%%%%%%%%%%%%%%%%%%%%%%%%%%%%%%%%%%%%%%%%%%%%%%%%%%%%%%%%%%%%%%%%%%%%%%%%%
\begin{figure}[ht]
  \centering{\includegraphics[width=0.8\textwidth]{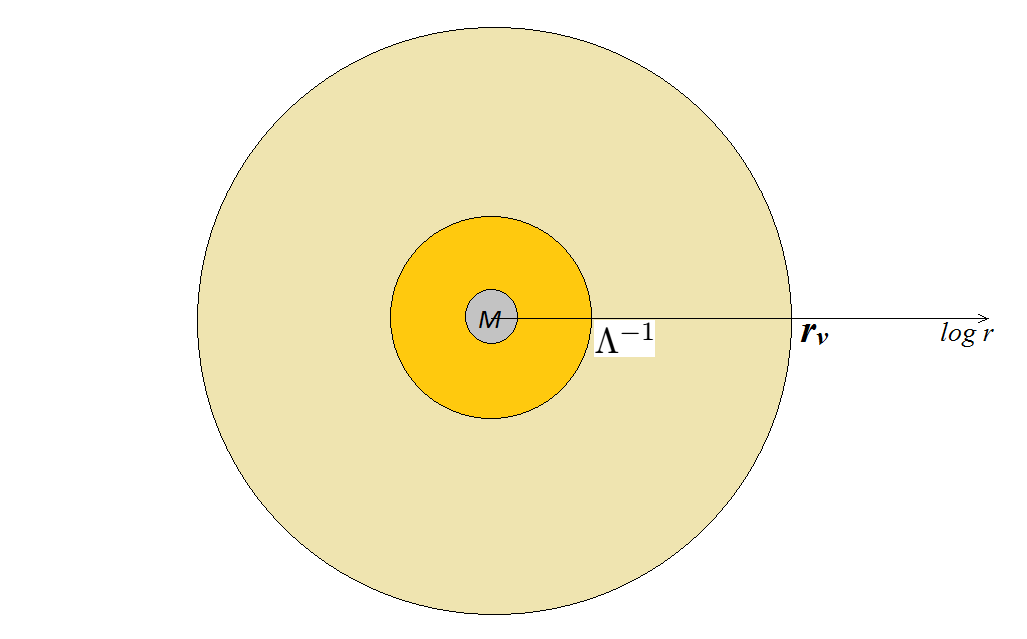}}
  \caption{Overall picture of the Vainshtein screening mechanism. 
There are three regions of interest: ($r\gg r_v$) far outside the Vainshtein radius and away from the source, the fifth force is not screened; ($r \ll r_v$) near the source and below $r_v$, the fifth force is suppressed (this includes $\Lambda^{-1}\ll r \ll r_v$). $r=\Lambda^{-1}$ represent the cut-off scale of the theory and is typically very large compared to the Schwarzschild radius.}
  \label{fig:vainshtein}
\end{figure}
%%%%%%%%%%%%%%%%%%%%%%%%%%%%%%%%%%%%%%%%%%%%%%%%%%%%%%%%%%%%%%%%%%%%%%%%%%%%%%%%%

Again following \cite{2015-rev-Joyce-et-al}, we describe how this profile encodes the functioning of the Vainshtein screening mechanism:
\begin{itemize}

\item $r \gg r_{\rm V}$: Far away from the source, the profile goes approximately as $1/r^2$,
\be
\phi'(r\gg r_{\rm V}) \simeq \frac{g}{3} \cdot \frac{M}{8\pi M_{\rm Pl} r^2}\,.
\ee
The ratio of the Galileon scalar force to the gravitational force is given by
\be
\left. \frac{F_{\phi}}{F_{\rm gravity}}\right\vert_{r\gg r_{\rm V}}\simeq \frac{g^2}{3}\,.
\ee
So the gravitational force is enhanced as well-known for DGP models for example. 

\item $r \ll r_{\rm V}$: close to the source, the profile ~\eqref{galfieldprofile} is given by
\be
\phi'(r\ll r_{\rm V})  \simeq \frac{\Lambda^3r_{\rm V}}{2} \sqrt{\frac{r_{\rm V}}{r}}\sim \frac{1}{\sqrt{r}} \,.
\label{ErpprV}
\ee
The ratio of the galilean force to the gravitational force is given by
\be
\left. \frac{F_{\phi}}{F_{\rm gravity}}\right\vert_{r\ll r_{\rm V}} \sim \left(\frac{r}{r_{\rm V}}\right)^{3/2}\ll 1\,,
\label{galforcesup}
\ee
so the fifth force is strongly suppressed at distances much smaller than the Vainshtein radius.  We provide a schematic picture of the Vainshtein mechanism in Fig.~\ref{fig:vainshtein}.

\end{itemize}

It is worth ending this section by pointing out a number of papers that have used astrophysical constraints on screening mechanisms some practically ruling out chameleon mechanisms or putting constraints on Vainshtein mechanisms, e.g., \cite{Salzano2017}. Studies also provided forecasts on such constraints from future surveys and space-missions including \citep{Sakstein2017}.

%%%%%%%%%%%%%%%%%%%%%%%%%%%%%%%%%%%%%%%%%%%%%%%%%%%%%%%%%%%%%%%%%%%%%%%%%%%%%%%%%%%
%%%%%%%%%%%%%%%%%%%%%%%%%%%%%%%%%%%%%%%%%%%%%%%%%%%%%%%%%%%%%%%%%%%%%%%%%%%%%%%%%%%
%%%%%%%%%%%%%%%%%%%%%%%%%%%%%%%%%%%%%%%%%%%%%%%%%%%%%%%%%%%%%%%%%%%%%%%%%%%%%%%%%%%
%%%%%%%%%%%%%              CONSTRAINTS ON MG MODELS                      %%%%%%%%%%
%%%%%%%%%%%%%%%%%%%%%%%%%%%%%%%%%%%%%%%%%%%%%%%%%%%%%%%%%%%%%%%%%%%%%%%%%%%%%%%%%%%
%%%%%%%%%%%%%%%%%%%%%%%%%%%%%%%%%%%%%%%%%%%%%%%%%%%%%%%%%%%%%%%%%%%%%%%%%%%%%%%%%%%
%%%%%%%%%%%%%%%%%%%%%%%%%%%%%%%%%%%%%%%%%%%%%%%%%%%%%%%%%%%%%%%%%%%%%%%%%%%%%%%%%%%

\section{Constraints on MG models from current cosmological data sets}

%%%%%%%%%%%%%%%%%%%%%%%%%%%%%%%%%%%%%%%%%%%%%%%%%%%%%%%%%%%%%%%%%%%%%%%%%%%%%%%%%%%
%%%%%%%%%%%%%%%%%%%%%%%                                         %%%%%%%%%%%%%%%%%%%
%%%%%%%%%%%%%%%%%%%%%%%          SUB-SECTION                    %%%%%%%%%%%%%%%%%%%
%%%%%                Constraints on Horndeski and beyond Horndeski models      %%%%
%%%%%                                                                          %%%%
%%%%%%%%%%%%%%%%%%%%%%%%%%%%%%%%%%%%%%%%%%%%%%%%%%%%%%%%%%%%%%%%%%%%%%%%%%%%%%%%%%%
%%%%%%%%%%%%%%%%%%%%%%%%%%%%%%%%%%%%%%%%%%%%%%%%%%%%%%%%%%%%%%%%%%%%%%%%%%%%%%%%%%%

\subsection{Constraints on Horndeski and beyond Horndeski models}

Planck2015MG used the $\alpha_x$-parameterization of Sect.~\ref{sec:horndeski} in order to put constraints on Horndeski models with a number of restrictions to reduce the number of parameters. This is necessary in view of the relatively limited constraining power of current data sets. They used EFTCAMB \citep{EFTCAMB1} so they adapted the parameterization accordingly. 

They considered Horndeski models with $\alpha_M = - \alpha_B$, $\alpha_T = \alpha_H = 0$, and $\alpha_K$ fixed by setting $M_2=0$ in equations \eqref{eq:M_ast}-\eqref{eq:alpha_T}. So they simply considered non-minimally coupled K-essence type models as in \cite{BelliniSawicki2014} with the only free function being $\alpha_M$. As discussed in Sect.~\ref{sec:horndeski}, a non-zero $\alpha_M$ parameter represents a non-zero anisotropic stress and a modification of the lensing potential. Additionally, Planck2015MG used the ansatz,
\be
\alpha_M = \alpha_M^{today} a^\beta
\ee
where $\alpha_M^{today}$ is a constant and $\beta>0$ determines its backward time evolution.  In the minimal model they considered, $\alpha_M$ is related to the EFT function $\Omega(a)$ 
which after integration gives \citep{Planck2015MG}
\begin{equation}
\Omega(a) = \exp\left\{\frac{\alpha_M^{today}}{\beta} a^\beta \right\} - 1.
\end{equation}
They called this the exponential EFT model while they called a second model with 
$\beta=1$ and $\Omega(a)=\Omega_0\,a$ the linear EFT model. In their Table~5, they give their results from where we extract the following representative constraints: 
\begin{itemize}
\item For the linear EFT case: 
$\alpha_M^{today} < 0.052$ (95\% confidence level) for the Planck TT+lowP data set combination and $\alpha_M^{today} < 0.043$ (95\% confidence level) for the Planck TT+TE+EE+BSH data set combination (BSH standing again for BAO, SN and local Hubble constraints). 
\item For the exponential EFT case:
$\alpha_M^{today} < 0.063$ and $\beta=0.87^{0.57}_{0.27}$ (95\% confidence level) for the TT+lowP data set combination  
and $\alpha_M^{today} < 0.062$ and $\beta=0.92^{0.53}_{0.24}$ (95\% confidence level) for the TT+TE+EE+BSH data set combination. 
\end{itemize}
with stringent bounds on the $\alpha_M$ and its time evolution index $\beta$ with the $\Lambda$CDM values of $0$ and $1$, respectively, within those constraints.

We discuss in Sect.~\ref{sec:GW170817} how the gravitational-wave (GW) event GW170817 and its electromagnetic counterpart GRB170817A constrained the speed of propagation of GW
to be practically equal to the speed of light and thus strongly reduced the number of viable Horndeski models to generalized Brans--Dicke theories and cubic Galileons although the latter are ruled out by ISW observations (see Sect.~\ref{sec:constraints_on_galileons}). However, it was commented in \cite{Simone2017} that Horndeski models with non-trivial modifications to GR remain possible at the level of linear perturbations as they explored it using MG parameters. Furthermore, models that are beyond Horndeski and in particular the more general class of degenerate higher order scalar-tensor theories (DHOST) (see Sect.~\ref{sec:horndeski}) provide a much more general class to look for further viable self-accelerating models \citep{Crisostomi2017DHOST}. The latter study found DHOST models with late-time self-acceleration. They performed perturbations in the quasi-static limit and showed that the models can satisfy constraints from solar interior structure \citep{Saito2015MG,Sakstein2015MT,Babichev2016} and the GW orbital decay of the Hulse--Taylor pulsar \citep{Beltran2016ET}.  
\cite{Sakstein2017b} studied how the interior of astrophysical bodies and pulsations of stars can be used to test beyond horndeski models. They found that brown dwarfs and Cepheid stars are particularly sensitive to such tests.   
These beyond Horndeski models will be subject to full cosmological analyses once full CMB analysis tools will become available.

\cite{Kreisch2017} performed a cosmological constraint analysis on Horndeski models that were not ruled out by the implication of the gravitational-wave event GW170817 and its electromagnetic counterpart GRB170817A (i.e., $c_{_{GW}=c}$ so $\alpha_T=0$). They used  
CMB data from Planck and the joint analysis of the BICEP2/Keck Array and Planck, galaxy clustering data SDSS LRGs, BOSS BAO data, and RSD measurements to constrain the remaining  parameters. They modeled the evolution of each parameter with an amplitude (the parameter value today) and an index parameter for a scale factor power law as follows (see Sect.~\ref{sec:horndeski} for further discussion of the $\alpha_x$-parameterization):

\begin{itemize}
\item{
Running rate of the effective Planck mass, $M_*^2$,
\be
\label{eq:alpha_M}
\alpha_{\rm M}=\frac{d\ln \left(M_*^2\right)}{d\ln a},
\ee
which they evolved as 
\be 
\widetilde{M}  = \widetilde{M}_0 a^{\beta} \implies \alpha_{\rm M} =\widetilde{M}_0 a^{\beta} \frac{\beta}{1+\widetilde{M}_0 a^{\beta}},
\ee
where $M^2_*/m_0^2 = 1 + \widetilde{M}$ as implemented in the software \texttt{EFTCAMB} they used. $m_0^2$ is the Planck mass so $\widetilde{M}_0$ is the fractional deviation of $M_*^2$ from $m_0^2$ today.
}
\item{
Kinecity of the scalar field due to the presence of its kinetic terms in the action
\be
\alpha_{\rm K}  =\alpha^{\rm K}_0 a^{\kappa}
\ee
}
\item{
Braiding/mixing of scalar and tensor terms
\be
\alpha_{\rm B}  =\alpha^{\rm B}_0 a^{\xi} 
\ee
}
\end{itemize}

They obtained the following results from using CMB and all the LSS data when fixing the kinecity to $\alpha_{K} = 0.1 a^3$: The friction $\alpha_0^{M}$ has an upper limit of $0.38$ when $\alpha_0^{B} \neq 0$  and $0.41$ when $\alpha_0^{B} = 0$ (all at the 95\% confidence level). They found in the case $\alpha_0^{B} \neq 0$ that the bound excludes GR but in the 
  $\alpha_0^{B} = 0$ case they attributed this to stability constraints imposed by  
  fixing $\alpha_0^{K}=0.1$. They also conclude that the effects of Horndeski theory on primordial B-modes are constrained by CMB and LSS data to be insignificant with 95\% confidence.

They caution though that making assumptions on some parameters in Horndeski models can cause dramatic changes in the results on other parameters, and fixing the kinecity is one case of this. 

They use the Akaike information criterion (AIC) \citep{AIC1974} to compare the two models $\alpha_0^{B} \neq 0$ versus $\alpha_0^{B} = 0$. They found that all the data sets prefer the model with $\alpha_0^{B} = 0$ where the data is then consistent with GR. 

It is worth ending this sub-section with some useful remarks about constraining Horndeski models. Before the measurement of $\alpha_T \approx 0$ from double Neutron star merger event (i.e., the GW signal GRB170817 and its electromagnetic counterpart GRB170817A, see Sect.~\ref{sec:GW170817}), there was too much freedom in the parameter space of the Horndeski models. They can produce a $\lcdm$ background and large-scale observables while providing self-acceleration with no need for a cosmological constant or dark energy component; see for example \cite{Lombriser2015a}. 
This degeneracy has now been broken by the GW170817/GRB170817A event as was anticipated in \cite{Lombriser2016a} (see Fig.~4 therein). With the constraint $a_T=0$, a MG cosmic self-acceleration effect now must manifest itself in LSS observables. Indeed, \cite{Lombriser2017CT} showed that a minimal signature such a model must produce in LSS provides a 3-$\sigma$ worse fit than $\lcdm$ model. They concluded that $\alpha_T=0$ will challenge the self-acceleration from a genuine scalar-tensor modification to GR (i.e., breaking the strong equivalence principle). In beyond-Hordeski models, other free functions are introduced and bring back further degeneracies between, for example, cosmic acceleration and LSS from $\alpha_M$. However, \cite{Lombriser2016a} discussed how Standard Sirens are not affected by this degeneracy from $\alpha_M$ and should be able to test a self-acceleration at the 5-$\sigma$ level for both Horndeski and beyond-horndeski. 

%%%%%%%%%%%%%%%%%%%%%%%%%%%%%%%%%%%%%%%%%%%%%%%%%%%%%%%%%%%%%%%%%%%%%%%%%%%%%%%%%%%
%%%%%%%%%%%%%%%%%%%%%%%                                         %%%%%%%%%%%%%%%%%%%
%%%%%%%%%%%%%%%%%%%%%%%          SUB-SECTION                    %%%%%%%%%%%%%%%%%%%
%%%%%                    Constraints on Brans-Dicke theory                     %%%%
%%%%%                                                                          %%%%
%%%%%%%%%%%%%%%%%%%%%%%%%%%%%%%%%%%%%%%%%%%%%%%%%%%%%%%%%%%%%%%%%%%%%%%%%%%%%%%%%%%
%%%%%%%%%%%%%%%%%%%%%%%%%%%%%%%%%%%%%%%%%%%%%%%%%%%%%%%%%%%%%%%%%%%%%%%%%%%%%%%%%%%

\subsection{Constraints on Brans--Dicke theory}

The most stringent constraint on Brans--Dicke (BD) theory comes from solar system tests where the Cassini mission put the bound $\omega_{\rm BD}>40,000$ \citep{Cassini2003,WillReview2014}. However, as argued for example in \cite{Avilez2014}, the theory can be a sub-category of a more general theory (e.g., Horndeski) that has a screening mechanism that makes it very close to GR at small scales and departs from it at cosmological scales. It is also interesting to obtain independent constraints on the theory at very different scales and times. \cite{Avilez2014} used CMB data from Planck, WMAP and SPT and ACT \citep{Planck2013,WMAP7,SPT2011,ACT2013}, and constraints from Big-Bang Nucleosynthesis (BBN) light element abundances \citep{Iocco2009BBN}. They use initial conditions on the scalar field such that the gravitational constant today on Earth is Newton's constant, $G_N$. They find then $\omega_{\rm BD} > 692$ at the 99\% CL.  
They also consider the case where the scalar is free and allowed to vary as a parameter. They find in this case, $\omega_{\rm BD} > 890$ and $0.981 < G_{\rm eff}/G_N < 1.285$ at the 99\% CL. This provided an order of magnitude improvement on previous measurements that were reported in the following analyses,  \citep{Acquaviva2005BD,Wu2010BD1,Wu2010BD}

%%%%%%%%%%%%%%%%%%%%%%%%%%%%%%%%%%%%%%%%%%%%%%%%%%%%%%%%%%%%%%%%%%%%%%%%%%%%%%%%%%%
%%%%%%%%%%%%%%%%%%%%%%%                                         %%%%%%%%%%%%%%%%%%%
%%%%%%%%%%%%%%%%%%%%%%%          SUB-SECTION                    %%%%%%%%%%%%%%%%%%%
%%%%%   Constraints on Vector-Tensor and Generalized Einstein Aether theories   %%%%
%%%%%                                                                          %%%%
%%%%%%%%%%%%%%%%%%%%%%%%%%%%%%%%%%%%%%%%%%%%%%%%%%%%%%%%%%%%%%%%%%%%%%%%%%%%%%%%%%%
%%%%%%%%%%%%%%%%%%%%%%%%%%%%%%%%%%%%%%%%%%%%%%%%%%%%%%%%%%%%%%%%%%%%%%%%%%%%%%%%%%%

\subsection{Constraints on Vector-Tensor and Generalized Einstein Aether theories}

\cite{Zuntz2010EA} conducted a thorough investigation of Einstein Aether theory finding that while in principle the vector field can source a dark matter component and also cause late-time cosmic acceleration, only the latter was found as a viable possibility. Indeed, they found that using the vector field effective effect as a substitute to dark matter does not fit large scale structure from SDSS DR6 \citep{SDSS6} and CMB WMAP7 data \citep{WMAP7}. They also found physical problems with such a possibility. On the other hand, an effective dark energy based on the vector field was found consistent with the data.

Some time earlier, \cite{Carroll2004LV} expressed the gravitational constant 
appearing in the effective Friedmann equation 
of a Lorentz-Violating Vector Field theory as 
(we follow here the notation of \citealt{Oost2018})

\begin{equation}
G_{\rm cos}= \frac{G_{\rm ae}}{1+\frac{1}{2}(c_1+c_3+3 c_2)}
\end{equation}

where 

\begin{equation}
G_{\rm ae}=G_N \left( 1-\frac{1}{2}(c_1+c_4)\right)
\end{equation}

This results in a decrease in the expansion rate with 
consequences during the big bang nucleosynthesis and 
will result in weak interactions freezing-out later. 
This leads to a lower freeze-out temperature and 
a decrease in the production of the primordial $^4 He$ and
a lower $^4 He$-to-hydrogen mass ratio \citep{Carroll2004LV}. 
This modifies the abundance of the primordial helium.
In order to be consistent with current observations (see, e.g., \citealt{Izotov2014,Aver2015}), the cosmological gravitational 
constant must satisfy the constraint \citep{Oost2018}
\begin{equation}
\left|\frac{G_{cos}}{G_N} -1   \right| \lesssim \frac{1}{8}. 
\end{equation}

As we will discuss in Sect.~\ref{sec:GW170817}, stringent constraints have been obtained on Einstein-Aether theories from the binary neutron star merger event GW170817 and GRB170817A by constraining $c_1=c_3$. \cite{Oost2018} used higher order expansion of the  $c_i$ parameter and combined this GW constraint with other theoretical and observational constraints in order to plot allowed regions in the $c_i$ parameter spaces. 
The constraints from GW170818/GRB170817B can be summarized as (see \citealt{Oost2018b})
\begin{equation}
|c_1+c_3| < 10^{-15}
\end{equation}
and 
\begin{equation}
0 \lesssim c_1+c_4 \lesssim 2.5 \times 10^{-5},\,\,\,\,\,c_4 \lesssim 0,\,\,\,\,\, 0 \lesssim c_2 \lesssim 0.095.
\end{equation}
Other additional constraints from astrophysics and theory within these bounds can be found in \cite{Oost2018b}. 

\cite{DeFelice2017Proca3} worked out perturbations for some specific Proca vector-tensor models and compared them to current CMB distance data, BAO, SN, and RSD growth rate data. They found the models to be consistent with the data used and equally (or slightly more) competitive than the $\Lambda$CDM. They found that models fit the expansion data with an effective dark energy equation of state of $w_{\rm DE}=-1-s$ with $s=0.254^{{}+ 0.118}_{{}-0.097}$ at 95\,\% confidence level (CL). When the growth data is added to the fit, they obtain, $s=0.16^{+0.08}_{-0.08}$ (95\,\% CL). It remains interesting to perform a full CMB and large scale structure analysis of the models. 

%%%%%%%%%%%%%%%%%%%%%%%%%%%%%%%%%%%%%%%%%%%%%%%%%%%%%%%%%%%%%%%%%%%%%%%%%%%%%%%%%%%
%%%%%%%%%%%%%%%%%%%%%%%                                         %%%%%%%%%%%%%%%%%%%
%%%%%%%%%%%%%%%%%%%%%%%          SUB-SECTION                    %%%%%%%%%%%%%%%%%%%
%%%%%                 Constraints on massive gravity and bigravity             %%%%
%%%%%                                                                          %%%%
%%%%%%%%%%%%%%%%%%%%%%%%%%%%%%%%%%%%%%%%%%%%%%%%%%%%%%%%%%%%%%%%%%%%%%%%%%%%%%%%%%%
%%%%%%%%%%%%%%%%%%%%%%%%%%%%%%%%%%%%%%%%%%%%%%%%%%%%%%%%%%%%%%%%%%%%%%%%%%%%%%%%%%%

\subsection{Constraints on massive gravity and bigravity}
\label{sec:constraints_massive}

\cite{Koennig2014b} considered perturbations of bimetric massive gravity and identified a self-accelerating branch that is consistent with the expansion history and stable to linear perturbations. They call this the infinite-branch of bimetric gravity (IBB) based on the behavior of the ratio of the scale factor in the two metrics. 
They found that the only models with a stable cosmological evolution are the ones with non vanishing $\beta_0$, $\beta_1$ and $\beta_4$ parameters. Since $\beta_0$ is equivalent to a cosmological constant and they were rather interested in self-accelerating stable models, they restricted the analysis to the $\beta_1$ and $\beta_4$ IBB models.  
They compared the models, in the quasi-static approximation, to available growth rate data in the form of $f\sigma_8$  from 6dFGS \citep{BeutlerEtAl2012T6}, LRG200, LRG60 \citep{SamushiaEtAl2012IL},  BOSS \citep{Tojeiro2012TC}, WiggleZ \citep{BlakeEtAl2012}, and VIPERS \citep{delaTorreEtAl2013TV} surveys, as well as the Union 2.1 Compilation of SNe Ia data  \citep{SuzukiEtAl2012}. They found that the IBB model fits the data with $\Omega_m^0=0.18$ and an effective dark energy equation of state $w(z)=-0.79+0.21z/(1+z)$. They also found that growth rate of structure in IBB is well-approximated at late times by $f(z)\approx \Omega_m^{.47}[1+0.21z/(1+z)]$. 
They found that the combination of growth and supernova data gives the IBB parameter constraints $\beta_1 = 0.48^{+0.05}_{-0.16}$ and $\beta_4 = 0.94^{+0.11}_{-0.51}$, although  the strongest constraints come from the supernova data. They also find that the anisotropic stress MG parameter (or slip) in this model tends to $1/2$ and the gravitational coupling parameter, as they defined it, tends to $4/3$ at early times and they are different from the GR unity values. These MG parameters will be then a route to test and distinguish between these models and $\Lambda$CDM. Finally, they also found for these models that the usual ansatz used in GR for $f\approx\Omega_m^{\gamma}$ does not work. It is rather a two-parameter form given by  
\begin{equation}
f\approx\Omega_{m}^{\gamma_{0}}\left(1+\alpha\frac{z}{1+z}\right),\label{eq:IBB_f_ansatz_LR}
\end{equation}
that provides good fit to the growth rate with $\gamma_{0}=0.47$ and $\alpha=0.21$ as best-fit values.

\cite{GengEtAl2017CP} studied background and linear perturbation evolution for a minimum nontrivial case by setting only $\beta_0$ and $\beta_1$ non-zero so the models are not self-accelerating, see Sect.~\ref{sec:bigravity}.  As we reported earlier, they found an effective dark energy equation of state of a phantom type. They also compared the matter power obtained to SDSS LRG DR7 finding that it puts stringent constraints on $\bar{\beta}_1$ from Eq.~\eqref{eq:modelpar_LR} to be $\lesssim \mathcal{O}(10^{-2})$ at the linear perturbation level.
Other seminal papers that compared bimetric theory to observations include \cite{Akrami2012} and \cite{Enander2015}.

Bigravity models such the IBB above, and others, remain interesting to test using full CMB and large scale data analysis, however, it is worth mentioning that most of these solutions have been found to suffer from  gradient or Higuchi instabilities in for example \cite{Konning2015H} which compromises their viability. 

%%%%%%%%%%%%%%%%%%%%%%%%%%%%%%%%%%%%%%%%%%%%%%%%%%%%%%%%%%%%%%%%%%%%%%%%%%%%%%%%%%%
%%%%%%%%%%%%%%%%%%%%%%%                                         %%%%%%%%%%%%%%%%%%%
%%%%%%%%%%%%%%%%%%%%%%%          SUB-SECTION                    %%%%%%%%%%%%%%%%%%%
%%%%%                  Constraints on $f(R)$ models                            %%%%
%%%%%                                                                          %%%%
%%%%%%%%%%%%%%%%%%%%%%%%%%%%%%%%%%%%%%%%%%%%%%%%%%%%%%%%%%%%%%%%%%%%%%%%%%%%%%%%%%%
%%%%%%%%%%%%%%%%%%%%%%%%%%%%%%%%%%%%%%%%%%%%%%%%%%%%%%%%%%%%%%%%%%%%%%%%%%%%%%%%%%%

\subsection{Constraints on $f(R)$ models}

Planck2015MG also constrained $f(R)$ models in terms of the scalaron Compton wavelength today $B_0$ (see Eq.~\eqref{eq:Compton_FR}). When using Planck TT+lowP+BSH, they noted a degeneracy between $B_0$ and the optical depth $\tau$. This is removed when adding CMB lensing. They find at 95 \% CL: $B_0< 0.12$ for the Planck TT+lowP+CMB Lensing and a very tight bound of $B_0 < 0.79 \times 10^{-4}$ when Planck TT+lowP+CMB Lensing+BAO+WL+RSD is used. The result are thus close to that of a $\Lambda$CDM model and put very stringent limit on a departure toward $f(R)$ gravity.  

Although not among the most recent papers on the subject, \cite{Giannantonio2010} provided one of the most stringent and clear analysis on constraining $f(R)$ models using WMAP5 CMB data \citep{WMAP5}, ISW data from cross-correlating WMAP maps with six galaxy data sets in different bands (i.e., 2MASS, SDSS main galaxies, LRGs, and QSOs, NVSS, and HEAO) \citep{Giannantonio2008}. The data covers a redshift ranging from $\bar z= 0.1$ to $\bar z= 1.5$ and thus probes variations of the gravitational potentials over a large redshift range. They also added the union compilation of SN data from \cite{SNUnion2008}
They used the parameterization $(\mu(a,k), \gamma(a,k))$ as in \eqref{eq:MGmu} and \eqref{eq:MGeta} with time and scale dependencies given by a refined version of  \eqref{eq:BZparamLR} \citep{ZhaoEtAl2009}. For f(R) models mimicking $\Lambda$CDM expansion, they obtained an upper bound of $B_0<0.4$ at the 95\% C.L.  

\cite{2013-Okada-etal-RSD-MG} conducted an analysis using $f \sigma_8$ RSD data for redshift range $z = 0.06$--$0.8$ from WiggleZ, SDSS LRG, BOSS, and 6dFGRS. They tested the Hu--Sawicki's $f(R)$ model finding that only the parameter space that makes the model practically indistinguishable from $\Lambda$CDM is allowed at 95\% CL. 
\cite{Dossett2014FR} combined large scale data from WiggleZ, BAO (from 6dF, SDSS DR7 and BOSS DR9), and Planck-2013 CMB data (and WMAP Polarization data) to find $\log_{10} < -4.07$ at 95\% C.L. thus also putting a tight low bound on the Compton wavelength parameter $B_0$. They also found that $f(R)$ models cannot explain the tension in the lens amplitude ``parameter'' of the CMB spectrum. Again, this reduces the allowed $f(R)$ parameter space to be very close to $\Lambda$CDM. Further recent cosmological constraints on $f(R)$ models can be found in \cite{Nunes2017,2018-Perez-Romero-Nesseris,2018-Li-Shirasaki}.  

Other very stringent limits on $f(R)$ and other Chameleon theories came from astrophysical constraints using distance measurements in the nearby universe \citep{2013-Jain-Vikram-Sakstein}. For example, this screening mechanism affects to different levels the enhanced gravitational force when using Cepheid stars versus when using tip of red giant branch stars to estimated distances. The screening mechanism leads to opposite effects on the inferred distances and offers the possibility to test such theories. The authors found no evidence for an enhancement of the gravitational force and put a constraint of $f_{_{R0}}\le 5\times 10^{-7}$ at 95\% C.L. \citep{2013-Jain-Vikram-Sakstein}.  Finally, \cite{2014-Sakstein-Jain-Vikram} made the point that while our galaxy and similar ones are screened but less dense galaxies may be subject to less or no screening. In that case, stars in such dwarf galaxies must be hotter, brighter and pulsate with a shorter period. They used a samples of 25 unscreened  galaxies and showed that the chameleon mechanism is practically ruled out. We note that using star interior physics has become a promising and effective probe of modification to gravity and associated screening mechanisms and we refer the reader to further works in  \cite{Sakstein2013a,Sakstein2015TG,Sakstein2015MT,Saltas2018} and references therein. 

An excellent review of cosmological and astrophysical constraints on Chameleon fields and in particular $f(R)$ models can be found in \cite{Lombriser2014}. The author provides a thorough compilation of bounds on $|f_{R0}|$ including relevant redshifts and scale as well as the measurements and probes used (see Table~I there).  Another very useful compilation (compendium) of constraints on Chameleon models including astrophysical and laboratory bounds can be found in \cite{Burrage2016}.

Most recently, \cite{Battye2018} used the designer approach to $f(R)$ models to compare them to Planck CMB temperature anisotropy, polarisation and lensing data as well as the BAO data from SDSS and WiggleZ. They showed that such approach based on the equation of state to the dark sector perturbations is numerically stable and provides analytical insights of the dynamics of such perturbations in the designer approach to $f(R)$. 
They put stringent constraints on $B_0$ finding $B_0 < 0.006$ (95\%CL) for the designer models with $w = -1$, $B_0 < 0.0045$ and $|w + 1| < 0.002$ (95\%CL) for the designer models with $w \ne -1$. The authors discuss that their results indicate that 
for these models, $w$ is strongly constrained to be $-1$, due to the strong dependence 
of $\sigma_8$ on $w$. They state that similar results were found in previous works of  \cite{Raveri2014,Hu2016} for the designer and Hu-Sawicki models using the Effective Field Theory (EFT) approach.  They conclude that this hints for the fact that generic $f(R)$ models with $w \ne -1$ can be ruled out from current cosmological data. 

In sum, combining stringent cosmological constraints, astrophysical bounds and no-go theorems on Chameleon mechanism practically rules out $f(R)$ models and in particular as a possible explanation for cosmic acceleration. 

%%%%%%%%%%%%%%%%%%%%%%%%%%%%%%%%%%%%%%%%%%%%%%%%%%%%%%%%%%%%%%%%%%%%%%%%%%%%%%%%%%%
%%%%%%%%%%%%%%%%%%%%%%%                                         %%%%%%%%%%%%%%%%%%%
%%%%%%%%%%%%%%%%%%%%%%%          SUB-SECTION                    %%%%%%%%%%%%%%%%%%%
%%%%%                   Constraints on DGP models                              %%%%
%%%%%                                                                          %%%%
%%%%%%%%%%%%%%%%%%%%%%%%%%%%%%%%%%%%%%%%%%%%%%%%%%%%%%%%%%%%%%%%%%%%%%%%%%%%%%%%%%%
%%%%%%%%%%%%%%%%%%%%%%%%%%%%%%%%%%%%%%%%%%%%%%%%%%%%%%%%%%%%%%%%%%%%%%%%%%%%%%%%%%%

\subsection{Constraints on DGP models}
\label{sec:constraints_DGP}

Despite being plagued by the presence of ghost fields, the self-accelerating branch of the Dvali--Gabadadze--Porrati (sDGP) has been compared extensively to various astrophysical and cosmological observations. Using various distance measurements, \cite{Alcaniz2002SO,Jain2002BW,Deffayet2002SC,Fairbairn2006SL} found that the characteristic scale does verify $r_c \sim H_0^{-1}$. \cite{Maartens2006OC} used SN-Ia from \cite{Riess2004TI,Astier2006TS}, CMB shift parameter from \cite{Wang2006RD} and BAO data from \cite{Eisenstein2005DO} to constrain the self-accelerating branch finding it consistent with the data at the 2-$\sigma$ level but the $\Lambda$CDM provided a better fit to the data. Some stringent constraints came from \cite{Song2007GI} using the angular diameter distance to surface of last scattering from WMAP Y3 \citep{Spergel2007TW}, SN data from \cite{Riess2004TI,Astier2006TS} and local measurements of Hubble to show that the flat self-accelerating DGP model is inconsistent with the data at the 3-$\sigma$ level. They also found that the curved self-accelerating models remained consistent with the data but with a poorer fit than the $\Lambda$CDM. However, by using BAO data, growth data from the ISW and ISW-galaxy cross-correlations, they showed that any models with the same self-acceleration history as a wCDM model are strongly disfavored by such data.    
\cite{Fang2008CT,Lombriser2009CC} used CMB data, galaxy-ISW cross-correlations data, and distance measurements, to show that both flat and curved self-accelerating DGP models are much disfavored by the data in comparison to $\Lambda$CDM.  

While less appealing due to the lack of self-acceleration, the normal branch (nDGP) continues to be used to derive benchmark constraints, to run simulation and build mock data for MG studies in the nonlinear regime; see for example \cite{Barreira2016c,Hellwing2017,Bose2018DGP}. 

%%%%%%%%%%%%%%%%%%%%%%%%%%%%%%%%%%%%%%%%%%%%%%%%%%%%%%%%%%%%%%%%%%%%%%%%%%%%%%%%%%%
%%%%%%%%%%%%%%%%%%%%%%%                                         %%%%%%%%%%%%%%%%%%%
%%%%%%%%%%%%%%%%%%%%%%%          SUB-SECTION                    %%%%%%%%%%%%%%%%%%%
%%%%%                  Constraints on Galileon models                          %%%%
%%%%%                                                                          %%%%
%%%%%%%%%%%%%%%%%%%%%%%%%%%%%%%%%%%%%%%%%%%%%%%%%%%%%%%%%%%%%%%%%%%%%%%%%%%%%%%%%%%
%%%%%%%%%%%%%%%%%%%%%%%%%%%%%%%%%%%%%%%%%%%%%%%%%%%%%%%%%%%%%%%%%%%%%%%%%%%%%%%%%%%

\subsection{Constraints on Galileon models}
\label{sec:constraints_on_galileons}

\cite{2013-Okada-etal-RSD-MG} used $f \sigma_8$ RSD data for redshift range $z = 0.06-0.8$ from WiggleZ, SDSS LRG, BOSS, and 6dFGRS. 
They tested covariant Galileon models with late-time acceleration and found that the
model parameter space consistent with the observed background expansion is excluded by  RSD data at more than $8-\sigma$ level. The models they considered have too strong of a growth rate and do not fit the data. However, they found that the extended Galileons of \cite{DeFelice2012EG} have solutions that are consistent with the RSD data within a 2-$\sigma$ level. As mentioned in Sect.~\ref{sec:ClusteringRSD}, we recall that one needs to keep in mind that $f\sigma_8$ data points should be corrected for any assumptions of the $\lcdm$ model when reducing/calibrating the data. Most recently, \cite{2016-Okumura-etal-RSD-MG}  used  their high redshift RSD data point at $z\sim1.4$ from the FastSound survey using the Subaru telescope as well as lower-z previous measurements in order to constrain deviations from GR. They used covariant Galileon models as well as extended covariant Galileons but considered models with growth less strong than that of models used in \cite{2013-Okada-etal-RSD-MG}. While the models were found to be within the 1-$\sigma$ level at low redshifts, they deviate significantly from the GR-$\Lambda$CDM model at high redshift where they fall outside the 1-$\sigma$ bounds and possibly outside the 2-$\sigma$ for the covariant Galileons, see Fig.~\ref{fig:RSDMG} from \cite{2013-Okada-etal-RSD-MG}. However, even more stringent constraints have been put now in the way of Galileons models from ISW measurements and from the gravitational-wave event GW170817 and its electromagnetic counterpart as we discuss below and further in Sect.~\ref{sec:GW170817}. 

\cite{BarreiraEtAl2014} analyzed cubic Galileon models and found that in the presence of massive neutrinos, the models provide a very good fit to CMB temperature, CMB lensing and BAO data. The authors used Planck-2013 \citep{Planck2013} temperature data, WMAP9 polarization data \citep{WMAP9}, and Planck-2013 CMB Lensing \citep{Planck2013WL}. They noted these as the PL data set. They added to these data sets, BAO measurements from the 6dF, SDSS DR7 and BOSS DR9 which they noted at the PLB data sets. They dubbed the models as $\nucubic$ and $\nulcdm$ that each having seven cosmological parameters.  $\nulcdm$ cosmic acceleration is due to a cosmological constant while that of $\nucubic$ is due to a different coupling between curvature and sources. They found that while in the absence of massive neutrino $\Lambda$CDM is clearly favored by the data, the two models have close $\chi^2$ when massive neutrinos are added to the analysis. That is $(\chi^2_{P} ; \chi^2_{L} ; \chi^2_{B})=(9813.5\ ; 4.5\ ; 1.0)$ and $(9805.4\ ;8.7\ ; 1.4)$ for $\nucubic$ and $\nulcdm$ respectively and with close total $\chi^2$'s. They noted that the $\nucubic$ best-fit model is also consistent with the local measurements of the Hubble constant, unlike the $\Lambda$CDM model. However, they noted that the $\nucubic$  shows a negative ISW effect that is hard to reconcile with current observations. The models they considered are plotted against CMB, CMB Lensing and BAO data available at the time of their analysis in Fig.~\ref{fig:bfs_lr} that we reproduce here.

In a most recent analysis, \cite{Renk-etal-2017} (including some of the same authors as above) performed a further thorough analysis of self-accelerating Galileon models using CMB data from Planck-2015 temperature and polarization \citep{Planck2015} plus CMB lensing \citep{Planck2015XV}, BAO (same as above), $H_0$ \citep{RiessEtAl2016} and ISW data. For ISW, they used CMB temperature maps cross-correlated with foreground galaxies from the Wide-field Infrared Survey Explorer (WISE) survey \citep{WrightEtAl2010}. They found again that the Cubic Galileon models predict a negative ISW effect and thus is in a 7.8$\sigma$ tension with available data which rules the cubic models out. They also found that the ISW data constrain significantly the parameter spaces for the quartic and quintic Galileon models but leave regions of the parameter space where the models provide fits to the data comparable to the $\Lambda$CDM. However, this time the Galileon models are found in some 2-$\sigma$ tension with the BAO data. They concluded that the models are likely to be decisively constrained by future ISW and BAO data. In sum, while the cubic Galileons have been excluded by the ISW effect here, the quartic and quintic have been excluded by the 
the gravitational-wave event GW170817 and its electromagnetic counterpart GRB170817A, as we discuss in Sect.~\ref{sec:GW170817}. 

Finally, \cite{Peirone2017CG} investigated further the effect of neutrino masses and different mass hierarchies on fitting covariant Galileons to current data. They use the Planck 2015 temperature and polarization data, BAO from BOSS, local measurements of $H_0$, weak lensing from KiDS, and supernova JLA compilation. This analyses found that even with neutrinos and considering different mass hierarchies, the data considered rule out all covariant Galileons including the cubic, the quartic, and the quintic, in agreement with other previous results from ISW for the cubic and GW170817/GRB170817A for the quartic and quintic models. 

%%%%%%%%%%%%%%%%%%%%%%%%%%%%%%%%%%%%%%%%%%%%%%%%%%%%%%%%%%%%%%%%%%%%%%%%%%%%%%%%%%%
%%%%%%%%%%%%%%%%%%%%%%%                                         %%%%%%%%%%%%%%%%%%%
%%%%%%%%%%%%%%%%%%%%%%%          SUB-SECTION                    %%%%%%%%%%%%%%%%%%%
%%%%%                        Constraints on TeVeS                              %%%%
%%%%%                                                                          %%%%
%%%%%%%%%%%%%%%%%%%%%%%%%%%%%%%%%%%%%%%%%%%%%%%%%%%%%%%%%%%%%%%%%%%%%%%%%%%%%%%%%%%
%%%%%%%%%%%%%%%%%%%%%%%%%%%%%%%%%%%%%%%%%%%%%%%%%%%%%%%%%%%%%%%%%%%%%%%%%%%%%%%%%%%

\subsection{Constraints on TeVeS}

\cite{Xu2015TEVES} used the galaxy velocity power spectrum from 6dF survey  and the kinetic Sunayev Zel'dovich (kSZ) power spectrum from ACT/SPT \citep{ACTSZ2013,SPT2011} to put constraints on TeVeS theory (see Sect.~\ref{sec:TEVES}). They used these two particular probes in order to provide complementary constraints to those of $E_G$ from \cite{ReyesEtAl2010} (see  Sect.~\ref{sec:E_G_constraints}) since the latter is insensitive to the amplitude of perturbations. For the TeVeS cosmology they also added one sterile neutrino and 3 massless neutrinos following the suggestion of \cite{Angus2009} in order to fit observations of the CMB temperature power spectrum. They found that the linear kSZ power spectrum is consistent with upper limits of the ACT/SPT data. However, they found that the nonlinear kSZ TeVeS spectrum is ruled out by SPT observations and the ACT data put stringent constraints on the model parameters. They also constrained the models using \cite{Planck2013}  data and allowed for the three parameters $K_B,l_B$ and $\mu_0$ of the TeVeS model to vary as well as the neutrino physical energy density. The best fit cosmological parameter for the TeVeS models were found to be difficult to reconcile with other observations. Namely, the model gives a very small optical depth indicating that re-ionization would have ended at z=1.2 and an inferred value of the Hubble constant $H_0 < 50.8$\Hu  which is hard to reconcile with any other measurement of this constant. 
They also performed $\chi^2$ goodness of the fit test and found that the TeVeS models have an excess of $\delta \chi^2=501.36$ compared to the $\Lambda$CDM model and concluded that these results from Planck data rule out the TeVeS models.

This is a good example of how cosmological data can be used to rule out models that have evaded so far a number of tests such as solar system constraints and galaxy rotation curves without dark matter. This is a good example to show the promise of cosmological tests in constraining gravity theories and departures from GR at large scales. 

However, the TeVeS theory structure remains an example of a complex theory that may have not said its last words as some other developments continue. For example, there is a general version of TeVeS in \cite{Skordis2008GT} that has not been constrained in \cite{Xu2015TEVES}. Also, the theory has been combined with a Galilean scalar field \citep{gTEVES}, although with some continuing but less stringent challenges \citep{gTEVESCosmo}. 

%%%%%%%%%%%%%%%%%%%%%%%%%%%%%%%%%%%%%%%%%%%%%%%%%%%%%%%%%%%%%%%%%%%%%%%%%%%%%%%%%%%
%%%%%%%%%%%%%%%%%%%%%%%                                         %%%%%%%%%%%%%%%%%%%
%%%%%%%%%%%%%%%%%%%%%%%          SUB-SECTION                    %%%%%%%%%%%%%%%%%%%
%%%%%                 Constraints on Non-Local gravity models                  %%%%
%%%%%                                                                          %%%%
%%%%%%%%%%%%%%%%%%%%%%%%%%%%%%%%%%%%%%%%%%%%%%%%%%%%%%%%%%%%%%%%%%%%%%%%%%%%%%%%%%%
%%%%%%%%%%%%%%%%%%%%%%%%%%%%%%%%%%%%%%%%%%%%%%%%%%%%%%%%%%%%%%%%%%%%%%%%%%%%%%%%%%%

\subsection{Constraints on Non-Local gravity models}
\label{sec:constraintsonNL}

Full comparison of the ${\mathit{RR}}$ Non-Local gravity model to CMB and other cosmological data has been performed in \cite{Dirian2014CP,Dirian2015NG,Dirian2016NG,Dirian2017CT} with further model exploration in \cite{Belgacem2017NL}. Cosmological background and perturbation equations have been put into the Boltzmann-Einstein code CLASS by \cite{Dirian2016NG} allowing for a full comparison to CMB and matter power spectra data.  

We report here results from constraining ${\mathit{RR}}$ non-local model from \cite{Dirian2016NG,Dirian2017CT,Belgacem2017NL}. The authors used the following data sets. CMB from Planck-2015 \citep{Planck2015I} including: lowTEB data ($\ell \leq 29$) and the high-$\ell$ TT,TE,EE ($\ell > 29$) of temperature and polarization spectra \citep{Planck2015MG}; temperature plus polarization lensing data in the conservative range $\ell =40-400$ \citep{Planck2015CMB,Planck2015XV}. They also used Type Ia SN from the JLA data of SDSS-II/SNLS3 from the JLA data of SDSS-II/SNLS3 \citep{BetouleEtAl2014}; and BAO data from \citep{BeutlerEtAl2011,RossEtAl2014,AndersonEtAl2014}. 

First, \cite{Belgacem2017NL} explained that the results found in \cite{Dirian2016NG} favoring $\lcdm$ to ${\mathit{RR}}$ minimal models is mainly due to fixing the $\sum_{\nu}m_{\nu}=0.06$~eV. When letting this parameter vary, the two models fit the data with practically equal $\chi$-squares and Bayes' factors \citep{Dirian2017CT}. Next, they stressed two particular results from their use of CMB+BAO+SN analysis. They find constraints on the Hubble constant of $H_0= 69.49\pm 0.80$ which is higher than the one from using the $\lcdm$. Compared to the local measurement of  $H_0 = 73.24 \pm 1.74$ of  \cite{RiessEtAl2016}, this is only in $2.0\sigma$ tension compared to that which they find for the $\lcdm$, i.e., $3.1\sigma$. Second they find neutrino masses with the constraints $\sum_{\nu}m_{\nu}=0.219^{+0.083}_{-0.084}\hspace{3mm}{\rm eV}$, which they remark falls within the window $0.06~{\rm eV}\,\lsim \,\sum_{\nu}m_{\nu}\, \lsim\, 6.6~{\rm eV}$ provided by oscillation and beta-decay experiments and is more consistent than the lower limit in the $\lcdm$. We refer the reader to \cite{Belgacem2017NL} for result summary tables and more discussions. 

Next, since the $H_0$ is not in significant tension with the ${\mathit{RR}}$ minimal model, the authors of \cite{Belgacem2017NL} added the local measurements to use CMB+BAO+JLA+$H_0$. They found then $H_0=70.13_{-0.72}^{+0.76}$ and $\sum_{\nu}m_{\nu}=0.168_{-0.084}^{+0.078}\,\,\,{\rm eV}$ with a slightly better $\chi^2$ for the ${\mathit{RR}}$ model compared to $\lcdm$, although not statistically significant. However, the authors finish their analysis by considering comparison of the ${\mathit{RR}}$ model and the $\nu\lcdm$ models to current data of the growth factor, $f\sigma_8$, data from 6dF GRS \citep{BeutlerEtAl2012T6}, SDSS LRG \citep{Oka2014SC}, BOSS CMASS \citep{SamushiaEtAl2014}, WiggleZ \citep{BlakeEtAl2012}, VIPERS \citep{delaTorreEtAl2013TV} and BOSS DR12 \citep{AlamEtAl2016}. They found that $\chi^2$ is lower in $\nu\lcdm$, compared to the minimal ${\mathit{RR}}$ model with $\delta\chi^2\simeq 2.01$. They state that when this is combined with the $\delta\chi^2=-1.0$ from comparison with CMB+BAO+JLA+$H_0$ the models are then statistically equivalent. However, they also reported that when $H_0$ is not considered then overall the difference rises to $\delta\chi_{\rm tot}^2\simeq 4.95 $ which favors weakly $\nu\Lambda$CDM over the ${\mathit{RR}}$ models. This is certainly to be followed closely with incoming growth data. 

Finally, the authors concluded their comparison of the ${\mathit{RR}}$ minimal non-local gravity models by discussing the effect of the recent results from the GW event from the neutron star merger GW170817 and its electromagnetic counterpart GRB170817A. They showed that that gravitational waves in the ${\mathit{RR}}$ model propagate at the speed of light and thus comply with the limit $c_T\approx 0$. However, they pointed out to the possibility of using standard sirens to distinguish between $\lcdm$ and the ${\mathit{RR}}$ model using third-generation GW interferometers which they discussed in a companion paper \cite{Belgacem2}. As they stress there, one can define a ``GW luminosity distance'' which is different from the standard luminosity distance for electromagnetic signal. 
They take advantage of the predictivity of their ${\mathit{RR}}$ model and provide a concrete prediction for the ratio of the GW and EM luminosity distances.
They found that the effect due to modified GW propagation is more easily detectable, at  future GW interferometers,  than the effect from the dark energy equation of state \citep{Belgacem2}. Furthermore, the authors give a much more detailed discussion of how their model can be tested with modified GW propagation in \cite{Belgacem2018b}. The discussion is more general where they propose a parametrization of the effect of modified GW propagation that could be used for any modified gravity theory. They obtain some limits already by comparing the LIGO/Virgo measurement of $H_0$ using standard sirens with that from standard candles, and they compute in detail the sensitivity of the Einstein Telescope to the parameter related to modified GW propagation, in generic modified gravity theories. This will be very relevant to future GW detectors such as LISA and ET. We refer the reader to their papers for more on this new avenue.

%%%%%%%%%%%%%%%%%%%%%%%%%%%%%%%%%%%%%%%%%%%%%%%%%%%%%%%%%%%%%%%%%%%%%%%%%%%%%%%%%%%
%%%%%%%%%%%%%%%%%%%%%%%%%%%%%%%%%%%%%%%%%%%%%%%%%%%%%%%%%%%%%%%%%%%%%%%%%%%%%%%%%%%
%%%%%%%%%%%%%%%%%%%%%%%%                                  %%%%%%%%%%%%%%%%%%%%%%%%%
%%%%%%%%%%%%%%%%%%%%%%%%   GW170817/GRB170817A            %%%%%%%%%%%%%%%%%%%%%%%%%
%%%%%%%%%%%%%%%%%%%%%%%%                                  %%%%%%%%%%%%%%%%%%%%%%%%%
%%%%%%%%%%%%%%%%%%%%%%%%%%%%%%%%%%%%%%%%%%%%%%%%%%%%%%%%%%%%%%%%%%%%%%%%%%%%%%%%%%%
%%%%%%%%%%%%%%%%%%%%%%%%%%%%%%%%%%%%%%%%%%%%%%%%%%%%%%%%%%%%%%%%%%%%%%%%%%%%%%%%%%%

\section{Constraints on deviations from GR and MG models from neutron star merger  event GW170817/GRB170817A}
\label{sec:GW170817}

The beginning of the 21st century will be remembered for the first detection of gravitational waves (GW) from compact objects. It all started when the Laser Interferometer Gravitational Observatory (LIGO) detected GW signals from the merger of black hole event which confirmed the existence of black holes and the prediction of GW \cite{GWEvent1}. Almost two years later, LIGO and the VIRGO interferometer made the detection of GW from a merger of two neutron stars (GW170817) \citep{GW170817}. Incidentally, the Fermi Gamma-ray Burst Monitor, and the Anti-Coincidence Shield for the Spectrometer for the International Gamma-Ray Astrophysics Laboratory observed a gamma ray burst (GRB170817A) event within the following 1.7 seconds and in a close location to GW170817 \citep{Goldstein2017AO,Savchenko2017ID}. There were no doubts that GRB170817A was the electromagnetic counterpart of GW170817 \citep{Abbott2017GW}. This was a consequential event to test some aspects of gravity at cosmological scales as one can confront the two completely different types of \textit{astrophysical messengers}.  That exactly what was done immediately after the announcement of the event, see for example  \cite{Tessa2017GW,Crem2017GW,Ezq2017GW,Sak2017GW,Langlois2017GW,Amandola2017GW}.  

Indeed, the scientific community was well-prepared to exploit such an event since a number of papers had already studied the implications that can be drawn from comparing the propagation of gravitational and electromagnetic waves \citep{Amendola2013OA,Nishizawa2014,Amendola2014EO,Linder2014AS,Raveri2015MT,Saltas2014,
Beltran2016ET,Bettoni2017SO,Sawicki2017NG,Lombriser2016a,Lombriser2017CT}

In particular, \cite{Lombriser2016a} explicitly studied the implications for scalar-tensor gravity from an electromagnetic counterpart measurement to a LIGO/VIRGO gravitational wave emitted by a neutron star merger. In their Fig.~4, they predicted a constraint that closely matches that of GW170817/GRB170817A further below. They also discussed the implications of such a measurement for Horndeski scalar-tensor gravity (and beyond), and estimated that such a simultaneous measurement should be anticipated within a few years from writing their paper. Their paper followed a previous analysis by \cite{Nishizawa2014} which also made predictions close to the constraint below from GW170817/GRB170817A.

In GR, GWs travel at the speed of light, however in MG models, this is not always the case. As we discussed in Sect.~\ref{sec:horndeski}, it is common to parametrize deviations of the speed of GW, $c_T$, from $c=1$ (keeping our notation convention) by using the tensor speed excess parameter $\alpha_T$ \citep{BelliniSawicki2014} 
\be
\alpha_T=c_T^2-1.
\label{eq:alpha_TGW}
\ee
Note that the first term on the RHS is actually $(c_{_T}/c)^2$ but we kept the notation convention of setting $c=1$. 

\cite{Tessa2017GW} discussed the implication of this event using the $\alpha_x$ (Sect.~\ref{sec:horndeski}) parameterization while \cite{Crem2017GW}  used directly the EFT formulation of dark energy and modified gravity theories (see Sect.~\ref{sec:EFT}). Following \cite{Tessa2017GW}, let us note the travel time of GW from GW170817 event to the GW detectors as 
\be
t_D-t_e= \frac{d_s}{c_T},
\ee
where $t_D$ is the merger time identified in the GW detectors, $t_e$ is the time of emission of GW and light from the event, and $d_s\simeq 40$ Mpc is the distance to the source event. It is worth mentioning that an Euclidean treatment of the distances is used here because of the relatively short distances involved. Similarly, we note the travel time of light from GRB170817A to the GRB light detectors as 
\be
t_L-t_e= d_{s},
\ee
where $t_L$ is the time of arrival (or peak brightness) measured at the GRB detectors. 
Taking the difference of the two above equations gives
\be
t_G-t_L=d_s\left(\frac{1}{c_T}-1\right).
\label{eq:diffrence}
\ee
Using the arrival time difference of $t_D-t_L \simeq 1.7$ seconds and the value of $d_s$ into the above equation and translating the results to \eqref{eq:alpha_TGW} gives the stringent bound
\begin{eqnarray}
\left|\alpha_T\right|\lesssim 1\times10^{-15}. 
\label{eq:GWGRBbound}
\end{eqnarray}
It is worth mentioning here that one assumed here that the gamma-ray photons and GWs are released simultaneously. In reality, there could be a delay of order a few hours between these two events. Therefore, taking into account such a possible delay weakens the bound by a few orders of magnitude. However, $10^{-15}$ or $10^{-12}$ are both very tight constraints, leading to practically the same outcome. 

Incidentally, the bound in \eqref{eq:GWGRBbound} is consistent with the bound derived in \cite{Moore2001,Kimura2012} from gravitational Cherenkov radiation which constrains GW speed to not exceed the speed of light, assuming a galactic origin for the high energy cosmic rays. 

The bound \eqref{eq:GWGRBbound} suggests that $\alpha_T\simeq 0$ so a number of papers studied the same consequences of assuming this is the case in order to constrain deviations from GR and MG models or, in other cases, propagating the stringent bound to constrain departures from GR.

%%%%%%%%%%%%%%%%%%%%%%%%%%%%%%%%%%%%%%%%%%%%%%%%%%%%%%%%%%%%%%%%%%%%%%%%%%%%%%%%%%%
%%%%%%%%%%%%%%%%%%%%%%%                                         %%%%%%%%%%%%%%%%%%%
%%%%%%%%%%%%%%%%%%%%%%%          SUB-SECTION                    %%%%%%%%%%%%%%%%%%%
%%%%%              Implications for scalar-tensor theories                     %%%%
%%%%%                                                                          %%%%
%%%%%%%%%%%%%%%%%%%%%%%%%%%%%%%%%%%%%%%%%%%%%%%%%%%%%%%%%%%%%%%%%%%%%%%%%%%%%%%%%%%
%%%%%%%%%%%%%%%%%%%%%%%%%%%%%%%%%%%%%%%%%%%%%%%%%%%%%%%%%%%%%%%%%%%%%%%%%%%%%%%%%%%

\subsection{Implications for scalar-tensor theories}

%%%%%%%%%%%%%%%%%%%%%%%%%%%%%%%%%%%%%%%%%%%%%%%%%%%%%%%%%%%%%%%%%%%%%%%%%%%%%%%%%%%
%%%%%%%%%%%%%%%%%%%%%%%%%%%%%%%%%%%%%%%%%%%%%%%%%%%%%%%%%%%%%%%%%%%%%%%%%%%%%%%%%%%
%%%%%%%%%%%%%%%%%%%%%%%%%%%%%%%%%%%%%%%%%%%%%%%%%%%%%%%%%%%%%%%%%%%%%%%%%%%%%%%%%%%
\subsubsection{Implications for Horndeski models}

The Horndeski class of MG models is a large class of scalar-tensor theories that was discussed in Sect.~\ref{sec:horndeski} and for which the gravitational action was given by \eqref{Horndeski_Action}. \cite{Tessa2017GW} discussed that the constraint \eqref{eq:GWGRBbound} can be realized by a highly tuned cancellation between the Horndenski action terms $G_{4,X}$, $G_{5,\phi}$ and $G_{5,X}$ that can all contribute to $\alpha_T$. However, as they stated, a more logical implication of $\alpha_T\simeq 0$ is that each of the three terms vanishes identically. Furthermore using the Bianchi identity, the Horndeski action then reduces, besides the potential term and the cubic term, to \cite{Tessa2017GW}
\be 
{\cal L}_4=f(\phi)R
\label{eq:JDBL}
\ee 
leaving only conformally coupled theories of the Jordan--Brans--Dicke (JBD) type. 
This then eliminates the quartic and quintic Galileons theories. These consequences on the Hordeski terms were also given prior to this event in \cite{McManus2016}. See also \cite{Crem2017GW} for the same conclusions from GW170817/GRB170817A. \cite{Sak2017GW} combined the constraint  \eqref{eq:GWGRBbound} with the lack of violation of the strong equivalence principle in the supermassive black hole in M87 in order to exclude the quartic Galileon model. 

The JDB like models \eqref{eq:JDBL} can be divided in two sub-classes as discussed in \cite{Tessa2017GW}. The first sub-class is the generalized JDB where the scalar field does not evolve significantly on cosmic timescales. However, this sub-class of models requires the chameleon screening mechanism to pass solar-system tests of gravity and thus cannot be self-accelerating due to the no-go theorems discussed in Sect.~\ref{sec:Large-mass} \citep{WangEtALNG}.  In the second sub-class, the scalar field evolves significantly on cosmic timescales, for example as caused by terms in $G_2$ and $G_3$, producing self-acceleration. Thus the event does not exclude cubic Galileons, kinetic gravity braiding models \citep{Deffayet2010} and k-essence models \citep{kessence1,kessence2}. See more discussion in  \cite{Tessa2017GW}  

\cite{Ezq2017GW} considered the implications of the GW170817 and GRB170817A on MG models by starting from the covariant Galileon models and then moving to their generalizations to Horndeski and beyond Horndeski models. They translated the stringent bound \eqref{eq:GWGRBbound} into bounds on the Galileon model coefficients and their generalizations. They arrived at similar conclusions as in \cite{Crem2017GW,Tessa2017GW,Langlois2017GW}. They tabulated models explicitly indicating that, in the Horndeski general class, Brans-Dicke, f(R), kinetic gravity braiding \citep{Deffayet2010} are not affected, while quartic and quintic Galileons \citep{NicolisEtAL2009Galileons,DeffayetEtAl2009}, Fab Four \citep{Charmousis2012GS}, de Sitter Horndeski \citep{Martin-Moruno2015HT}, and $f(\phi)$Gauss-Bonnet \citep{Nojiri2005GD} are all excluded. 

%%%%%%%%%%%%%%%%%%%%%%%%%%%%%%%%%%%%%%%%%%%%%%%%%%%%%%%%%%%%%%%%%%%%%%%%%%%%%%%%%%%
%%%%%%%%%%%%%%%%%%%%%%%%%%%%%%%%%%%%%%%%%%%%%%%%%%%%%%%%%%%%%%%%%%%%%%%%%%%%%%%%%%%
%%%%%%%%%%%%%%%%%%%%%%%%%%%%%%%%%%%%%%%%%%%%%%%%%%%%%%%%%%%%%%%%%%%%%%%%%%%%%%%%%%%
\subsubsection{Implications for Beyond Horndeski models}

The beyond-Horndeski models of \cite{BeyondHorndeski1,BeyondHorndeski2} receive almost the same consequences as Horndeski models except for a specific combination of terms in the beyond-Horndeski action which can realize $\alpha_T=0$. The cosmology and motivation for such a specific combination remains to be explored and it is not clear if such models have any particular motivation \citep{Tessa2017GW}. \cite{Sak2017GW} also excluded the quartic beyond horndeski models. \cite{Ezq2017GW} with  their approach above found that beyond-Horndeski models with disformal tuning and the $A_1=0$ class of quadratic Degenerate Higher-Order Scalar-Tensor (DHOST) theories \citep{Langlois2016D} are not excluded, while quartic/quintic beyond-Horndeski models \citep{Langlois2016D}, quadratic (with $A_1\ne 0$) \citep{Langlois2016D} and cubic DHOST models \citep{BenAchour2016DH} are all excluded. They showed that only three alternatives (or their combination) are possible for scalar-tensor theories: 1) restricting Horndeski models to their minimum simplest terms that keep $c_T=1$; 2) applying a conformal transformation to these minimal Horndeski models which preserves the causal structure; 3) using Horndeski models but compensating the terms that modify the speed of GW to keep it luminal. This is done by a specific disformal factor to tune away the departure of the speed of GW from light speed. \cite{Langlois2017GW} presented an analysis of the implications of the event GW170817/GRB170817A using the DHOST framework. For Horndeski and Beyond Horndeski theories, they came to the same conclusions discussed above from \cite{Crem2017GW,Tessa2017GW,Ezq2017GW}.  

%%%%%%%%%%%%%%%%%%%%%%%%%%%%%%%%%%%%%%%%%%%%%%%%%%%%%%%%%%%%%%%%%%%%%%%%%%%%%%%%%%%
%%%%%%%%%%%%%%%%%%%%%%%                                         %%%%%%%%%%%%%%%%%%%
%%%%%%%%%%%%%%%%%%%%%%%          SUB-SECTION                    %%%%%%%%%%%%%%%%%%%
%%%%%                 Implications for Vector-Tensor Theories                %%%%
%%%%%                                                                          %%%%
%%%%%%%%%%%%%%%%%%%%%%%%%%%%%%%%%%%%%%%%%%%%%%%%%%%%%%%%%%%%%%%%%%%%%%%%%%%%%%%%%%%
%%%%%%%%%%%%%%%%%%%%%%%%%%%%%%%%%%%%%%%%%%%%%%%%%%%%%%%%%%%%%%%%%%%%%%%%%%%%%%%%%%%

\subsection{Implications for Vector-Tensor Theories}

The constraints $\alpha_T=0$ imposes on Generalized Einstein-Aether theories \citep{Jacobson2001EA,Zlosnik1} the condition $c_1=-c_3$ (see Sect.~\ref{sec:EAther}) which makes the effective Planck mass reduce to the GR value, while the cosmological background evolution remains different from GR \citep{Tessa2017GW}. For the Generalized Proca theories \citep{Tasinato2014CA,Heisenberg2014PROCA}, the condition $\alpha_T=0$ imposes either a fine tuned cancellation of terms in the action or imposes that the terms related to $\alpha_T$ be all identically zero. The latter natural interpretation gives a branch with a cosmological evolution different from GR  with a rescaled Planck mass in the modified Friedmann equation \citep{Tessa2017GW}. 
Similar results were obtained in \cite{Oost2018} about the Einstein-Aether theories.
Since the Generalised Proca theory has a similar structure to Horndeski, the effects of $\alpha_T = 0$ on Generalised Proca is similar to the effects on Horndeski, i.e., the quartic and quintic terms are (effectively) ruled out. The same implication applies to the beyond Generalized Proca models of \cite{Heisenberg2016BG}. 

%%%%%%%%%%%%%%%%%%%%%%%%%%%%%%%%%%%%%%%%%%%%%%%%%%%%%%%%%%%%%%%%%%%%%%%%%%%%%%%%%%%
%%%%%%%%%%%%%%%%%%%%%%%                                         %%%%%%%%%%%%%%%%%%%
%%%%%%%%%%%%%%%%%%%%%%%          SUB-SECTION                    %%%%%%%%%%%%%%%%%%%
%%%%%            Implications for Massive gravity and bigravity theories       %%%%
%%%%%                                                                          %%%%
%%%%%%%%%%%%%%%%%%%%%%%%%%%%%%%%%%%%%%%%%%%%%%%%%%%%%%%%%%%%%%%%%%%%%%%%%%%%%%%%%%%
%%%%%%%%%%%%%%%%%%%%%%%%%%%%%%%%%%%%%%%%%%%%%%%%%%%%%%%%%%%%%%%%%%%%%%%%%%%%%%%%%%%

\subsection{Implications for Massive gravity and bigravity theories}

For massive gravity and bimetric gravity \citep{deRhamEtAl2011RO,deRhamEtAl2010GO,HassanEtAl2012BG}, the new results from GW170817 and GRB170817A have no significant cosmological consequences. For massive gravity, one can just obtain further weak constraints on the mass of the graviton. Similar to bounds obtained from previous Black Hole merger events \citep{Abbott2016TO}, \cite{Tessa2017GW} used the time delay of the GRB170817A electromagnetic counterpart to find $m \lesssim 10^{-22}$eV for the graviton mass. This is again much weaker than the solar system bound of the 
order of $m \lesssim 10^{-33}$eV \citep{deRham2017GM} 
or galaxy cluster bound of order $m \lesssim 10^{-29}$eV
(see e.g. \cite{Desai2018}.
This is an independent bound though. More relevant to our review, the local bounds obtained from GW propagation and the electromagnetic counterpart have no consequence on the cosmology of massive gravity and bigravity theory  \citep{Lagos2014,Cusin2015,DeFelice2014PE,Narikawa2015DO,Max2017GW}.
However, see \cite{Brax2017GW} and \cite{Akrami2018} for constraints on doubly
coupled metrics to matter models.

%%%%%%%%%%%%%%%%%%%%%%%%%%%%%%%%%%%%%%%%%%%%%%%%%%%%%%%%%%%%%%%%%%%%%%%%%%%%%%%%%%%
%%%%%%%%%%%%%%%%%%%%%%%                                         %%%%%%%%%%%%%%%%%%%
%%%%%%%%%%%%%%%%%%%%%%%          SUB-SECTION                    %%%%%%%%%%%%%%%%%%%
%%%%%      Implications for Ghost condensates and Horava-Lifshitz Gravity      %%%%
%%%%%                                                                          %%%%
%%%%%%%%%%%%%%%%%%%%%%%%%%%%%%%%%%%%%%%%%%%%%%%%%%%%%%%%%%%%%%%%%%%%%%%%%%%%%%%%%%%
%%%%%%%%%%%%%%%%%%%%%%%%%%%%%%%%%%%%%%%%%%%%%%%%%%%%%%%%%%%%%%%%%%%%%%%%%%%%%%%%%%%

\subsection{Implications for Ghost condensates and Ho\v{r}ava--Lifshitz Gravity}

For ghost condensates \citep{GCTheory2004}, the modification of the GW speed is given by 
$c_T^2-1 \sim M^2_{GC}/M^2_{Pl}$, where $M_{GC}$ is the typical scale of the model.
Now, experimental bounds on modifications of Newton law give $M_{GC}\leq 10$ MeV. So it is not expected to see any significant changes in the speed of GW and the constraint \eqref{eq:GWGRBbound} does not affect this theory \citep{Crem2017GW}. This is not the case for Ho\v{r}ava--Lifshitz theory as stated in \cite{Crem2017GW} where $C_T$ is expected to deviate from the speed of light.

\cite{Gumrukcuoglu2017} argued that the implications of the bound \eqref{eq:GWGRBbound} are more subtle for HL parameters. As they explain, the theory has 3 independent IR parameters ($\alpha$, $\beta$ and $\gamma$ in their paper). Before the constraint on the speed of GW, the tightest constraints on HL in IR had come from ppN constraints where  one assumes $\alpha = 2 \beta$. Papers then normally considered the 2-dimensional sub-region determined by ($\alpha=2\beta$ versus $\gamma$) in the parameter space. But the recent result from GW170817/GRB170817B set a constraints of $|\beta| < 10^{-15})$ so it is not justified to set $\alpha =2 \beta$ since it would require $\alpha$ and $\beta$ to be highly fine-tuned to the $10^-{15}$ level. Accordingly, they motivated in their paper to look at the ($\alpha$, $\gamma$) parameter sub-space, as the current constraints on $\alpha$ and $\gamma$ are orders of magnitudes looser compared to $\beta$ and focus on the $\beta=0$ plane. In the limit where HL and GR becomes indistinguishable (from IR perspective), HL becomes strongly coupled and loses its use as a perturbative alternative theory for GR. As they discussed in the paper, this puts a lower bound on $\alpha$ and $\gamma$ parameters set by experiments testing the validity of perturbative GR. This means that future tighter constraint on HL in IR regime combined with upper energy bounds on the validity of perturbative GR from future experiments could rule out HL as  perturbatively renormalizable theory of gravity and make HL absolute. See  \cite{Gumrukcuoglu2017} for a more detailed discussion. 

%%%%%%%%%%%%%%%%%%%%%%%%%%%%%%%%%%%%%%%%%%%%%%%%%%%%%%%%%%%%%%%%%%%%%%%%%%%%%%%%%%%
%%%%%%%%%%%%%%%%%%%%%%%                                         %%%%%%%%%%%%%%%%%%%
%%%%%%%%%%%%%%%%%%%%%%%          SUB-SECTION                    %%%%%%%%%%%%%%%%%%%
%%%%%           Implications for higher dimension models                       %%%%
%%%%%                                                                          %%%%
%%%%%%%%%%%%%%%%%%%%%%%%%%%%%%%%%%%%%%%%%%%%%%%%%%%%%%%%%%%%%%%%%%%%%%%%%%%%%%%%%%%
%%%%%%%%%%%%%%%%%%%%%%%%%%%%%%%%%%%%%%%%%%%%%%%%%%%%%%%%%%%%%%%%%%%%%%%%%%%%%%%%%%%

\subsection{Implications for higher dimension models}

As we discussed further above, in these models our universe is a 3+1 brane embedded in a higher dimensional space, for example 4+1 dimensional anti-de-Sitter space (see Sect.~\ref{sec:HigherDimensions}). In such a universe, gravity is the only force that propagates in the extra dimension (or the bulk space), while other forces are constrained to the brane hypersurface. As a consequence, GW and EM signals follow different paths leading to a time lag between the two signals propagating from a given point to another. 
GW170817 and GRB170817A can thus be used to put constraints on such models.
 
\cite{Visinelli2018} considered the setting where the GW and EM signals travel at the same speed but where the GW can take a shortcut in the bulk space and thus arrive ahead of the EM signal. This can be used to put a constraint on the radius of curvature, $\ell$, of the AdS5 bulk space. They used a $\lcdm$ model and performed a likelihood analysis to set an upper limit of $\ell \lesssim 0.535$ Mpc (68\% CL). 
As the authors mention, this bound is not competitive with current Solar
System constraints (e.g., \citealt{Long2003,Tan2016}), but is the first constraints from multi-messenger measurements. 

\cite{Pardo2018} used the GW170817/GRB170817B result in a different way to put a constraint on the possible number of spacetime dimensions. 
They used the fact that in these higher dimension models, there is gravitational leakage into extra dimension leading to dumping of the amplitude of GW that reflects on the inferred distance to gravitational source. 
They used GW as standard sirens and extracted directly the luminosity distance, $d^{GW}_L$, to GW170817. They compared this distance with the inferred luminosity distance to the EM counterpart, $d^{EM}_L$. The latter is determined using the Hubble law at the small redshift from the source, i.e., $v_H=c z= H_0 d^{EM}_L$ (but taking into account the peculiar velocity of the host galaxy with respect of its galaxy-group precessing velocity). Following \cite{Deffayet2007}, they used a dumping parameter $\gamma$ to write $d^{GW}_L = (d^{EM}_L)^{\gamma}$. This parameter is related to the number of dimension, $D$, by $\gamma=\frac{D-2}{2}$.  
From the two distances as inferred above, they find $\gamma = 1.01^{+0.04}_{-0.05}$ at the 68\% CL (using the local value of $H_0$) or $\gamma = 0.99^{+0.03}_{-0.05}$ (using Planck value of $H_0$). This in turn allowed them to put constraints on the spacetime dimension number as $D = 4.02^{+0.07}_{-0.10}$ (using local $H_0$) and $D = 3.98^{+0.07}_{-0.09}$ (using Planck $H_0$). They concluded that their results are in favor of the 3+1 dimensions of GR.  

%%%%%%%%%%%%%%%%%%%%%%%%%%%%%%%%%%%%%%%%%%%%%%%%%%%%%%%%%%%%%%%%%%%%%%%%%%%%%%%%%%%
%%%%%%%%%%%%%%%%%%%%%%%                                         %%%%%%%%%%%%%%%%%%%
%%%%%%%%%%%%%%%%%%%%%%%          SUB-SECTION                    %%%%%%%%%%%%%%%%%%%
%%%%% Implications for Results on MG parameters and large-scale-structure from GW170817 and GRB170817A  %%%%
%%%%%                                                                          %%%%
%%%%%%%%%%%%%%%%%%%%%%%%%%%%%%%%%%%%%%%%%%%%%%%%%%%%%%%%%%%%%%%%%%%%%%%%%%%%%%%%%%%
%%%%%%%%%%%%%%%%%%%%%%%%%%%%%%%%%%%%%%%%%%%%%%%%%%%%%%%%%%%%%%%%%%%%%%%%%%%%%%%%%%%

\subsection{Implications for Results on MG parameters and large-scale-structure from GW170817 and GRB170817A}

Interestingly, \cite{Saltas2014,Sawicki2017NG} showed that there is a one-to-one relationship between modification to the propagation of GW and the gravitational slip parameter  when the source is a perfect fluid matter. 
\cite{Amandola2017GW} noted that this result in combination with the constraint $c_T=c$ from GW170817/GRB170817A implies that the presence of a slip MG parameter in scalar-tensor theories can be attributed to only a conformal coupling to gravity.  
They also showed that the surviving vector-tensor theories cannot have any slip at all so detecting any slip parameter will rule all of them out.  
They demonstrated  then that the growth rate in the surviving models must be at least as fast as that of GR except possibly for beyond Horndeski theories. Finally, they showed that if the slip parameter is to have any scale dependence at all then it should be in a way that the parameter reduces to the GR unity value at large scales with no-slip and so the model cannot be distinguished from GR at large scales.

In light of the implication of GW170817/GRB170817A (i.e., $\alpha_T=0$) and its consequences  for the slip parameter \citep{Saltas2014,Sawicki2017NG}, the study \cite{Linder2018NSG} considered scalar-tensor models where the slip parameter is identically zero.
\cite{Linder2018NSG} noted that with the vanishing of $\alpha_T$, the no slip 
criterion is simply given by $\alpha_B=-2 \alpha_M$. It was then shown that 
stability conditions for absence of ghosts and a positive sound speed squared for perturbations impose further restrictions on $(\alpha_B,\alpha_K)$ reducing the independent parameters to only one. So this no slip gravity can be characterized by one MG parameter. Using some guidance from cosmological evolution and stability requirements, some forms on the time evolution of the one parameter (e.g. $\alpha_M$)  was proposed and studied. \cite{Linder2018NSG} then compared the growth rate data of $f\sigma_8$ to some of these models and found them to fit the data better than $\Lambda$CDM as they have a 
lower growth. It was noted that, unlike many other scalar-tensor theories, no-slip gravity predicts a weaker gravity than GR which explains the growth fit and, as the author states, could potentially inform the tension in the low amplitude found in weak lensing studies. 
 The study concludes with forecasts of constraints from the DESI galaxy redshift survey showing that it could be distinguished from GR at the 3-$\sigma$ level.

\cite{Simone2017} Performed an extensive analytical and numerical analysis of the MG parameters $\Sigma$ and $\mu$ or equivalently $G_{light}/G$ and $G_{matter}/G$. They considered Horndeski models that are consistent with tests of gravity and the cosmic expansion history with late time acceleration. They also take into account the recent result from from GW170817 and its counterpart GRB170817A, setting $c_T=c$. They confirmed a conjecture they made in their earlier work \citep{PogosianEtAl2016} about MG parameters in Horndeski models. That is $(\sigma-1)(\mu-1)\ge0$ must hold in viable Horndeski models. They also test the validity of the quasi-static approximation in Horndeski models finding that it holds well at small and intermediate scales but fails at $k\le 0.001h$/Mpc. They concluded in their analysis that despite the stringent result from GW170817/GRB170817A, there remain Horndeski models with non-trivial modifications to gravity at the level of linear perturbations and large scale structure. They stressed the complementarity of different approaches to modifications to GR and the practicality of using the phenomenological $\Sigma$ and $\mu$ parameterization and their consistency relations, see also \cite{PogosianEtAl2016}. 

Finally, \cite{Battye2018b} explored the results from GW170817/GRB170817A using an equation of state approach to modified gravity models. They confirmed the strong constraints found for Hordeski and Einstein-Aether models. They discuss how it is possible to construct MG models that evade GW170817/GRB170817A constraint but still provide cosmologically interesting modifications to gravity. These include $f(R)$, non-local, and higher order derivative models. 

%%%%%%%%%%%%%%%%%%%%%%%%%%%%%%%%%%%%%%%%%%%%%%%%%%%%%%%%%%%%%%%%%%%%%%%%%%%%%%%%%%%
%%%%%%%%%%%%%%%%%%%%%%%                                         %%%%%%%%%%%%%%%%%%%
%%%%%%%%%%%%%%%%%%%%%%%          SUB-SECTION                    %%%%%%%%%%%%%%%%%%%
%%%%%  Implications for Vainshtein screening mechanism after GW170810 and GRB170817A  %%%%
%%%%%                                                                          %%%%
%%%%%%%%%%%%%%%%%%%%%%%%%%%%%%%%%%%%%%%%%%%%%%%%%%%%%%%%%%%%%%%%%%%%%%%%%%%%%%%%%%%
%%%%%%%%%%%%%%%%%%%%%%%%%%%%%%%%%%%%%%%%%%%%%%%%%%%%%%%%%%%%%%%%%%%%%%%%%%%%%%%%%%%

\subsection{Implications for Vainshtein screening mechanism after GW170810 and GRB170817A}

\cite{CK2017} applied the implication of GW170817 and its counterpart GRB170817A to study the Vainshtein screening mechanism in the very general class of Degenerate Higher-Order Scalar-Tensor (DHOST) theories (including Horndeski and beyond-Horndeski models). They set $c_T=c$ and find that the Vainshtein mechanism generally works outside a matter source but it fails inside the matter. This then opens the door to test these theories using astrophysical observations inside matter sources such as stars, galaxies and clusters of galaxies and large scale structure. The formalism for such structures in this context depends on 3 parameters and some astrophysical constraints have already been derived on them \citep{Koyama2015AP,Saito2015MG,Sakstein2015MT,Sakstein2015TG,Sakstein2016TG,
Sakstein2017TS}. \cite{Dima2017GWS} found further implications and results on the Vainshtein screening mechanism from GW170817 and its counterpart that are consistent with the results above about the breaking of the Vainshtein screening inside astrophysical bodies. Finally, \cite{Langlois2017GW} study the Vainshtein mechanism in the Degenerate Higher-Order Scalar-Tensor (DHOST) framework. They derive, for the DHOST theories satisfying $c_T=c$, the gravitational equations for inside and around a non-relativistic spherical object. Unlike outside the object, they found that gravity inside the object deviates from standard gravity. They also found that the deviation from standard gravity inside the object can be described by 3 parameters that satisfy consistency relations and can be constrained using present and future astrophysical data \citep{Langlois2017GW}. This concurs with the findings above. It is also worth noting that the breaking of the Vainshtein screening mechanism inside matter has been discussed prior to GW170817 and GRB 170817B; see, for example \cite{Beltran2016ET}.

%%%%%%%%%%%%%%%%%%%%%%%%%%%%%%%%%%%%%%%%%%%%%%%%%%%%%%%%%%%%%%%%%%%%%%%%%%%%%%%%%%%
%%%%%%%%%%%%%%%%%%%%%%%%%%%%%%%%%%%%%%%%%%%%%%%%%%%%%%%%%%%%%%%%%%%%%%%%%%%%%%%%%%%

\begin{table}[h!]
%\footnotesize 
\caption{{\textbf{Status of some selected MG models}. Self-acceleration means that the models can have acceleration 
without a cosmological constant or other equivalent constant coming from the theory. BH stands for Beyond Horndeski models. 
For screening, C stands for Chameleon, V for Vainshtein, and W for weak coupling. NS stands for Neutron Stars. 
The reasoning here is that if a theory has specific models or branches that pass the constraint then we put a check. A question mark means that the point is still under debate or is unclear in the current literature.}}
\label{tab:model_summary}
\begin{tabular}{|l|c|c|c|c|c|}
\hline
                     &Has            &Passes Ghost  &Screening    &Passes current            &Other       \\ 
                     &self-          &and other     &mechanism    &cosmological,             &features,   \\ 
                     &acceleration   &instability   &type         &astrophysical,            &limitations,\\ 
                     &               &constraints   &             &\small{and GW170817}      &or constraints\\ 
                     &               &              &             &                          &             \\
                     
\hline
$f(R)$               &\acmarklr      & \acmarklr    &Chameleon    &\axmarklr\,(self-acc)     &no-go theorem\\
                     &               &              &             &\acmarklr\,(non self-acc) &for self-acc.;\\
                     &               &              &            &                           &some models \\
                     &               &              &            &                           &inst. inside NS\\
\hline
sDGP                 &\acmarklr      & \axmarklr    &Vainshtein    &\axmarklr                 &              \\
                     &               &              &             &                          &               \\
\hline
nDGP                 &\axmarklr      & \acmarklr    &Vainshtein    & \acmarklr?              &              \\
                     &               &              &             &                          &               \\
\hline
\textbf{Cubic} (Galileons,&\acmarklr & \acmarklr    &Vainshtein   & \axmarklr                &              \\
Horndeski, B-H)      &               &              &             &                          &               \\
\hline
\textbf{Quartic} (Galileons,&\acmarklr& \acmarklr   &Vainshtein   & \axmarklr                &              \\
Horndeski, B-H)      &               &              &             &                          &               \\
\hline
\textbf{Quintic} (Galileons,&\acmarklr& \acmarklr   &Vainshtein   & \axmarklr                &              \\
Horndeski, B-H) &               &              &                  &                          &               \\
\hline
Brans-Dicke (BD),     &\acmarklr      & \acmarklr   &             & \acmarklr ?              &\small{BD very tightly}\\
Generalized BD       &               &              &   ?         & \acmarklr                &constrained\\
                     &               &              &             &                          &by solar syst.\\
\hline
Einstein-Aether,     &\axmarklr?     & \acmarklr    &C?, V         & \acmarklr?                & Lorentz-       \\
Generalized EA       &\acmarklr      &              &             &                          & violation        \\
\hline
Proca and gene-      &\acmarklr      & \acmarklr    &Vainshtein   &Full CMB and              &             \\
ralized Proca        &               &              &             &LSS analysis              &             \\
                     &               &              &             &not completed             &              \\
\hline
TeVeS                &\axmarklr      & \acmarklr    &\small{Reduces to MOND}&\axmarklr \,but debated&              \\
                     &               &              &at small scales& \small{for some models}&               \\
\hline
Massive gravity      &\acmarklr      & \acmarklr?    &Vainshtein   & \acmarklr?               &No-FLRW sols.\\
                     &               &              &             &                          &Higuchi inst.\\
\hline
Bimetric massive     &\acmarklr      & \axmarklr?    &Vainshtein   & \acmarklr                &Higuchi or\\
gravity              &               &              &             &                          &gradient inst.\\
\hline
Horava-Lifshitz      &\axmarklr      &  \acmarklr   &close to GR at& \acmarklr\,but narrow   & Lorentz-   \\
                     &               &              &small scales& param. space             & violation  \\
\hline
Non-local $RR$       &\acmarklr      & \acmarklr    &Reduces to GR& \acmarklr                &              \\
                     &               &              &            &                          &               \\
\hline
Beyond-Horndeski,    &\acmarklr      & \acmarklr    & C, V, W     &Full CMB and              &some models\\
DHOST                &               &              &             &LSS analysis              &have gradient\\
                     &               &              &             &not completed             &inst.?       \\
\hline
\end{tabular}
%\footnotesize
\end{table}

%%%%%%%%%%%%%%%%%%%%%%%%%%%%%%%%%%%%%%%%%%%%%%%%%%%%%%%%%%%%%%%%%%%%%%%%%%%%%%%%%%%

%%%%%%%%%%%%%%%%%%%%%%%%%%%%%%%%%%%%%%%%%%%%%%%%%%%%%%%%%%%%%%%%%%%%%%%%%%%%%%%%%%%
%%%%%%%%%%%%%%%%%%%%%%%                                         %%%%%%%%%%%%%%%%%%%
%%%%%%%%%%%%%%%%%%%%%%%          SUB-SECTION                    %%%%%%%%%%%%%%%%%%%
%%%%% Further notes or caveats on the implications of GW170817 and GRB170817   %%%%
%%%%%                                                                          %%%%
%%%%%%%%%%%%%%%%%%%%%%%%%%%%%%%%%%%%%%%%%%%%%%%%%%%%%%%%%%%%%%%%%%%%%%%%%%%%%%%%%%%
%%%%%%%%%%%%%%%%%%%%%%%%%%%%%%%%%%%%%%%%%%%%%%%%%%%%%%%%%%%%%%%%%%%%%%%%%%%%%%%%%%%

\subsection{Further notes or caveats on the implications of GW170817 and GRB170817A}

Some caveats were raised in \cite{Tessa2017GW} about the fact that the result $\alpha_T\simeq 0$ is based on a measurement at very low redshift ($z_s=0.01$) corresponding to practically the present time in cosmic history so it is possible, in principle, that this was not always the case. Another possible caveat is the limitation that can come from noting that cosmological gravitational waves have long wavelengths and propagate in a higher cosmological average density, while GW170817 has short wavelength and propagated to us in almost empty space. So it will be interesting to see/confirm if gravitational waves at cosmological scales would travel at the speed of light. See further discussions in \cite{Tessa2017GW}. 

%%%%%%%%%%%%%%%%%%%%%%%%%%%%%%%%%%%%%%%%%%%%%%%%%%%%%%%%%%%%%%%%%%%%%%%%%%%%%%%%%%%

%%%%%%%%%%%%%%%%%%%%%%%%%%%%%%%%%%%%%%%%%%%%%%%%%%%%%%%%%%%%%%%%%%%%%%%%%%%%%%%%%%%
%%%%%%%%%%%%%%%%%%%%%%%%                                  %%%%%%%%%%%%%%%%%%%%%%%%%
%%%%%%%%%%%%%%%%%%%%%%%%   MG COMPUTER CODES              %%%%%%%%%%%%%%%%%%%%%%%%%
%%%%%%%%%%%%%%%%%%%%%%%%                                  %%%%%%%%%%%%%%%%%%%%%%%%%
%%%%%%%%%%%%%%%%%%%%%%%%%%%%%%%%%%%%%%%%%%%%%%%%%%%%%%%%%%%%%%%%%%%%%%%%%%%%%%%%%%%
%%%%%%%%%%%%%%%%%%%%%%%%%%%%%%%%%%%%%%%%%%%%%%%%%%%%%%%%%%%%%%%%%%%%%%%%%%%%%%%%%%%

\section{Computer codes and packages for testing gravity at cosmological scales}
\label{sec:codes}
A number of codes and software packages have been developed following the rapid development of the subject of testing GR and MG models at cosmological scales. 
Similar to the theoretical developments, codes have been developed according to two types. The first type is where a generic parametrization of deviations from GR is  implemented using one of the generic parametrizations of Sect.~\eqref{sec:MGparameters}. The second type is where the codes have focused on implementing a specific MG model or  a broad class of models such as those described in Sect.~\ref{sec:MGtheories}.   

It is worth noting that most codes that solve Einstein--Boltzmann equations are  based on a modification of two popular codes that solve the Boltzmann and gravitational field equations to calculate CMB temperature and polarization power spectra as well as the matter power spectrum. The first is \texttt{CAMB} (Code for Anisotropies in the Microwave Background) and is available at \url{http://camb.info/}, see also \cite{CAMB}. The second code is \texttt{CLASS} (Cosmic Linear Anisotropy Solving System) and is available at \url{http://class-code.net/}, see also \cite{CLASS-I,CLASS-II}. There are however other codes that are not based on these two systems such as for example DASh \citep{DASH} and COOP \citep{COOP} (available at \url{http://cita.utoronto.ca/~zqhuang/coop/}). 

We describe further below two examples of codes of the first type for generic deviation from GR, i.e., \texttt{ISitGR} \citep{ISITGR} and \texttt{MGCAMB} \citep{ZhaoEtAl2009,MGCAMB2}. We also describe two examples of codes of the second type, i.e., \texttt{hi\_class} \citep{HICLASS} and \texttt{EFTCAMB} \citep{EFTCAMB1,EFTCAMB2,Raveri2014} that both deal with broad classes of scalar-tensor MG models, and we refer the reader to the comparative study of \cite{BelliniEtAl2017AC} for a detailed list and description of other codes. 

Codes of the second type include: Cosmology Object Oriented Package (COOP) \citep{COOP,COOPII}  which implements an EFT approach to dark energy and modified gravity theories including the Horndeski broad class of scalar-tensor theories;  
Davis Anisotropy Shortcut Code (DASh) \citep{DASH}; 
CLASSig \citep{CLASSIG}; a code used in \cite{Avilez2014} for Jordan--Bran--Dicke gravity; a modified version of CMBEASY \citep{CMBEASY} for Einstein-Aether gravity \citep{AETHER};  
modified versions of CAMB \citep{CAMB} for $f(R)$ models \citep{Dossett2014FR,BEAN2007,Battye2018,EOS1,EOS2}; 
a modified version of CAMB \citep{CAMB} for covariant Galileons \citep{BarreiraEtAl2012}; 
CLASS-LVDM for Ho\v{r}ava--Lifshitz gravity  \citep{LVDM2}; 
and modified versions of CAMB and CLASS for models of nonlinear gravity with respective references \citep{NLCAMB} and \citep{NLCLASS2}.  

We reproduce Table~I from \cite{BelliniEtAl2017AC} (as Table~\ref{tab:code_theory_overview} here) that provides a good list of such codes with tested models, to which we added the corresponding references.   

Finally, we do not cover here N-Body simulation codes for MG models or implementation of semi-analytical models but we refer the reader to \cite{COMPARENL} (and references therein) for a recent comparative analysis of MG N-Body codes. See also other recent works using  the Comoving Lagrangian Acceleration
(COLA) approach in \cite{Cola2017MG,Cola2017MG2}.  
The presence of screening mechanisms in MG models makes the implementation of MG simulations more complicated. A parameterization for modified gravity on nonlinear cosmological scales was proposed in \cite{Lombriser2016NL} and a fitting formula for $f(R)$ Hu--Sawicki model has been derived in \cite{Zhao2014Halot}.
\begin{table*}
\caption{The codes used in the comparison by \cite{BelliniEtAl2017AC} along with the models tested. As they note, the table shows only the models used in their paper, not all the models that each code can test.
We added here to each line the references to corresponding papers. Reproduced with permission from Table~I of \cite{BelliniEtAl2017AC}, copyright by APS.}
\label{tab:code_theory_overview}
\footnotesize 
\begin{tabular}{|l|c|c|c|c|c|c|c|}
\hline
\hline
&  $\alpha$ Param-& EFT Para- & JBD & Covariant & f(R)&Ho\v rava &Non-Local \\ 
& etrization & metrization &  & Galileon & designer & Lifshitz& Gravity\\ 
\hline
\texttt{EFTCAMB}&\acmarklr&\acmarklr&\acmarklr&\acmarklr&\acmarklr&\acmarklr&\axmarklr\\
\citep{EFTCAMB1}&&&&&&&\\
\hline
\texttt{hi\_class}&\acmarklr&\acmarklr&\acmarklr&\acmarklr& \axmarklr&\axmarklr &\axmarklr\\
\citep{HICLASS}&&&&&&&\\
\hline
COOP&\acmarklr&\axmarklr&\axmarklr&\axmarklr&\axmarklr&\axmarklr&\axmarklr\\
\citep{COOP}&&&&&&&\\
\hline
GalCAMB&\axmarklr&\axmarklr&\axmarklr&\acmarklr&\axmarklr&\axmarklr&\axmarklr\\
\citep{BarreiraEtAl2012}&&&&&&&\\
\hline
BD-CAMB&\axmarklr&\axmarklr&\acmarklr&\axmarklr&\axmarklr&\axmarklr&\axmarklr\\
\citep{Avilez2014}&&&&&&&\\
\hline
DashBD&\axmarklr&\axmarklr&\acmarklr&\axmarklr&\axmarklr&\axmarklr&\axmarklr\\
\citep{DASH}&&&&&&&\\
\hline
CLASSig&\axmarklr&\axmarklr&\acmarklr&\axmarklr&\axmarklr&\axmarklr&\axmarklr\\
\citep{CLASSIG}&&&&&&&\\
\hline
{\tt CLASS\_EOS\_fR}&\axmarklr&\axmarklr&\axmarklr&\axmarklr&\acmarklr&\axmarklr&\axmarklr\\
\citep{Battye2018}&&&&&&&\\
\hline
CLASS-LVDM&\axmarklr&\axmarklr&\axmarklr&\axmarklr&\axmarklr&\acmarklr&\axmarklr\\
\citep{LVDM2}&&&&&&&\\
\hline
NL-CLASS&\axmarklr&\axmarklr&\axmarklr&\axmarklr&\axmarklr&\axmarklr&\acmarklr\\
\citep{NLCLASS2}&&&&&&&\\
\hline
NL-CAMB&\axmarklr&\axmarklr&\axmarklr&\axmarklr&\axmarklr&\axmarklr&\acmarklr\\
\citep{NLCAMB}&&&&&&&\\
\hline
\hline
\end{tabular}
\end{table*}
%

%%%%%%%%%%%%%%%%%%%%%%%%%%%%%%%%%%%%%%%%%%%%%%%%%%%%%%%%%%%%%%%%%%%%%%%%%%%%%%%%%%%
%%%%%%%%%%%%%%%%%%%%%%%                                         %%%%%%%%%%%%%%%%%%%
%%%%%%%%%%%%%%%%%%%%%%%          SUB-SECTION                    %%%%%%%%%%%%%%%%%%%
%%%%%                           ISITGR                                         %%%%
%%%%%                                                                          %%%%
%%%%%%%%%%%%%%%%%%%%%%%%%%%%%%%%%%%%%%%%%%%%%%%%%%%%%%%%%%%%%%%%%%%%%%%%%%%%%%%%%%%
%%%%%%%%%%%%%%%%%%%%%%%%%%%%%%%%%%%%%%%%%%%%%%%%%%%%%%%%%%%%%%%%%%%%%%%%%%%%%%%%%%%

\subsection{{\it \textbf{I}ntegrated \textbf{S}oftware \textbf{i}n \textbf{T}esting \textbf{G}eneral \textbf{R}elativity} (\texttt{ISiTGR})}
\label{sec:ISITGR}
We start with \texttt{ISiTGR} (pronounced \textit{Is it GR?}) that is publically available at \url{http://www.utdallas.edu/~jdossett/isitgr/}) and described in   \cite{ISITGR}. 
\texttt{ISiTGR} is an integrated set of modified modules for the publicly available packages \texttt{CosmoMC} (Cosmological Monte Carlo) \citep{COSMOMC} and \texttt{CAMB} \citep{CAMB}. 
\texttt{CosmoMC} software uses a Markov-Chain Monte-Carlo (MCMC) approach to explore cosmological parameter spaces (see more information at \url{http://cosmologist.info/cosmomc/}). 

\texttt{ISiTGR} introduces all the MG modifications to those two packages and combines them to a modified version of the Integrated Sachs--Wolfe (ISW)-galaxy cross correlations module of \cite{ISWHo,ISWHirata} to test GR. It also includes a modified weak-lensing likelihood module for the refined Hubble Space Telescope (HST) Cosmic Evolution Survey (COSMOS) lensing tomography analysis as described in \cite{Schrabback2010} which has also been modified to test GR. It also includes a new baryon acoustic oscillation (BAO) likelihood module for the WiggleZ Dark Energy Survey BAO measurement data \citep{2011WiggleZ-BAO}. 
\texttt{ISiTGR} also has a version tailored specially to constrain $f(R)$ models and is for example described and used in \cite{Dossett2014FR} and available at the same website above.

\texttt{ISiTGR} uses the modified growth parameters as described in Eqs.~\eqref{eq:PoissonModLR}, \eqref{eq:Mod2ndEinLR}, \eqref{eq:PoissonModSumLR}, and \eqref{eq:SigmaMG} as well as their time and scale evolution given by Eqs.\,\eqref{eq:BeanEvo}, \eqref{eq:GBmuEvo1_LR}, \eqref{eq:GBetaEvo_LR}, \eqref{eq:ZBinEvo}, \eqref{eq:kBin}, and \eqref{eq:kHybridQ}, see also Table~\ref{tab:Grid}. 

For \texttt{ISiTGR} and other codes discussed further below, it is worth noting that \texttt{CAMB} is written in the synchronous gauge and uses the metric potentials $h$ and $\eta$ as described in \cite{Ma-Bertschinger1995} instead of the potentials $\Phi$ and $\Psi$ of the conformal Newtonian gauge used in Sect.~\ref{sec:MGparameters}. In order to give a brief description of the implementation of \texttt{ISiTGR} (and other software further below), we will outline some common conversion and implementation steps using \texttt{CAMB}. The metric potentials in the two gauges are related to one another by, e.g., \cite{Ma-Bertschinger1995}
\bea
{\Phi} & = & \eta -\mathcal{H}\alpha,
\label{eq:PhiSG_LR} \\
\Psi &= & \dot{\alpha}+\mathcal{H} \alpha,
\label{eq:PsiSG_LR}
\eea
where 
\be
k^2\alpha = \frac{\dot{h}}{2} +3\dot{\eta}.
\label{eq:alphaSG_LR}
\ee 
Now, \texttt{CAMB} evolves the metric potential $\eta$ (or $k\eta$) as well as the matter perturbations, $\delta_i$, heat flux, $q_i$, and the shear stress $\sigma_i$ for each matter species in the synchronous gauge according to the evolution equations given in \cite{Ma-Bertschinger1995}.  Furthermore, \texttt{CAMB} uses two other variables noted $\sigma_{\rm CAMB}$ and $\mathcal{Z}$ that are defined and evaluated at each time step as follows 
\bea
\sigma_{CAMB} \equiv k\alpha = \frac{k(\eta-\Phi)}{\mathcal{H}},
\label{eq:sigcamb}\\
\mathcal{Z} \equiv \frac{\dot{h}}{2k} = \sigma_{CAMB}- 3\frac{\dot{\eta}}{k}.
\eea
The idea is that these variables allow \texttt{CAMB} to be written in such a way that the evolution of all other variables is changed simply by adjusting the evolution of the metric potential $\eta$.  Thus it is important that one derives an equation for the evolution of $\eta$ consistent with the modified growth equations \eqref{eq:PoissonModLR} and \eqref{eq:PoissonModSumLR}.  
As described in \cite{ISITGR}, after some steps, one obtains
\be
\dot{\eta}=\frac{-1}{2f_Q}\left \{2(\mathcal{H}^2-\dot{\mathcal{H}})k^2\alpha +\sum_i\bar{\rho_i}(a)\left[ \left(2\mathcal{H}\left[\mathcal{D}-Q\right]+\dot{Q}\right)\delta_i -Q(1+w_i)k^2\alpha -  Q f_1\frac{q_i}{k} \right]\right\},
\label{eq:etadotfin_LR}
\ee
with
\be
f_Q  =  k^2 + \frac{3}{2} Q\sum_i\bar{\rho_i}(1+w_i).
\ee
Finally, the next necessary change is to redefine the derivatives of the Newtonian metric potentials, $\dot{\Phi}+\dot{\Psi}$, which go into evaluating the ISW effect in the CMB temperature anisotropy spectrum.  This can be done quickly by observing that the quantities $\delta_i$ and $\sigma_i$ are invariant in transformations between the synchronous and conformal Newtonian gauges. Thus one can simply take the time derivative of \eqref{eq:PoissonModSumLR} and sub in for $\dot{\delta}$ and $\dot{\bar{\rho}}_i$ to get:
\bea
\dot{\Phi}+\dot{\Psi} = \frac{1}{k^2}\sum_i\bar{\rho_i}(a)\Bigg{\{}\left[((1+3w_i)Q+2\mathcal{D})\mathcal{H}-\dot{Q}\right]\frac{3(1+w_i)\sigma_i}{2} -\frac{3Q(1+w_i)\dot{\sigma}_i}{2} \\ \nonumber
+(\mathcal{D}\mathcal{H}-\dot{\mathcal{D}})\delta_i+\mathcal{D}(1+w_i)\left(k^2\alpha-3\dot{\eta}\right)+ \mathcal{D} f_1\frac{q_i}{k} \Bigg{\}}.
\eea
Beside these changes other small adaptations, modifications and additions to both 
\texttt{CAMB} and \texttt{CosmoMC} are necessary to assure a smooth running and accurate output of modified CMB spectra according to the MG parameters. These can be found in   \cite{ISITGR}.    

\texttt{ISiTGR} has been used or cited in over 50 papers. \texttt{ISiTGR} was used in the recent KiDS survey MG analyses \citep{JoudakiEtAl2017} and KiDS+2dFLenS \citep{Joudaki2018} as well as CFHTLenS+Planck data analysis including intrinsic alignment of galaxies as a systematic effect in \cite{Dossett2015}. 

%%%%%%%%%%%%%%%%%%%%%%%%%%%%%%%%%%%%%%%%%%%%%%%%%%%%%%%%%%%%%%%%%%%%%%%%%%%%%%%%%%%
%%%%%%%%%%%%%%%%%%%%%%%                                         %%%%%%%%%%%%%%%%%%%
%%%%%%%%%%%%%%%%%%%%%%%          SUB-SECTION                    %%%%%%%%%%%%%%%%%%%
%%%%%                            MGCAMB                                        %%%%
%%%%%                                                                          %%%%
%%%%%%%%%%%%%%%%%%%%%%%%%%%%%%%%%%%%%%%%%%%%%%%%%%%%%%%%%%%%%%%%%%%%%%%%%%%%%%%%%%%
%%%%%%%%%%%%%%%%%%%%%%%%%%%%%%%%%%%%%%%%%%%%%%%%%%%%%%%%%%%%%%%%%%%%%%%%%%%%%%%%%%%

\subsection{{\it \textbf{M}odification of \textbf{G}rowth with} \texttt{CAMB} (\texttt{MGCAMB}) and \texttt{MGCosmoMC}}

\texttt{MGCAMB} provides a set of patches to the code \texttt{CAMB} in which the linearized Einstein equations were modified according to MG equations \eqref{eq:MGmu} and \eqref{eq:MGeta}. The software is publically available at \url{http://aliojjati.github.io/MGCAMB/mgcamb.html} and described in \cite{MGCAMB2,ZhaoEtAl2009}. As described on its website, there was a major upgrade to MGCAMB in \cite{MGCAMB2} from the original version of \cite{ZhaoEtAl2009}, making it easier to use with \texttt{CosmoMC} and working for the entire redshift range. Similarly, \textbf{M}odified \textbf{G}ravity models with \texttt{CosmoMC} (\texttt{MGCosmoMC}) is a modified version of CosmoMC that allows one to fit modified gravity parameters to data sets in addition to other cosmological parameters. 

The most recent versions of \texttt{MGCAMB} and \texttt{MGCosmoMC} include  a wide range of parametrizations to accommodate MG models such as screened scalar-tensor theories as described in \cite{BraxEtAl2012}, Symmetron parameterization, generalized Dilaton parametrization, Hu-Sawicki f(R) gravity, as well as the time and scale evolution parametrizations of MG parameters \eqref{eq:MGmu} and \eqref{eq:MGeta}. \texttt{MGCAMB} has been used or cited in over 100 papers and has been used, in for example, \cite{Planck2015MG}.

%%%%%%%%%%%%%%%%%%%%%%%%%%%%%%%%%%%%%%%%%%%%%%%%%%%%%%%%%%%%%%%%%%%%%%%%%%%%%%%%%%%
%%%%%%%%%%%%%%%%%%%%%%%                                         %%%%%%%%%%%%%%%%%%%
%%%%%%%%%%%%%%%%%%%%%%%          SUB-SECTION                    %%%%%%%%%%%%%%%%%%%
%%%%%                           HiCLASS                                        %%%%
%%%%%                                                                          %%%%
%%%%%%%%%%%%%%%%%%%%%%%%%%%%%%%%%%%%%%%%%%%%%%%%%%%%%%%%%%%%%%%%%%%%%%%%%%%%%%%%%%%
%%%%%%%%%%%%%%%%%%%%%%%%%%%%%%%%%%%%%%%%%%%%%%%%%%%%%%%%%%%%%%%%%%%%%%%%%%%%%%%%%%%

\subsection{{Horndeski in CLASS} (\texttt{hi\_class})}

\texttt{hi\_class} \citep{HICLASS} is an extension to the Boltzmann solver code \texttt{CLASS} \citep{CLASS-I,CLASS-II} to include modification to GR based on Horndeski models. \texttt{hi\_class} inherits all the functionality of \texttt{CLASS} and can calculate cosmological distances, CMB, matter, and galaxy number count power spectra for this class of models. A publicly available version noted as hi\_class teaser can be cloned or downloaded from the repository \url{https://github.com/miguelzuma/hi_class_public} or from the webpage \url{http://miguelzuma.github.io/hi_class.html}. This version is described in \cite{HICLASS} and the latter websites.  

The implementation of Horndeski in \texttt{hi\_class} code is based on the EFT parameterization (see Sect.~\ref{sec:EFT}). \texttt{CLASS} and the \texttt{hi\_class} extension are written in C programming language but use a class-structure and modularity similar to that of object-oriented languages such as C++ or Java in order to make the code more readable while easier to parallelize (see  \url{http://class-code.net/} for a discussion).    

Since it encompasses a large class of models, \texttt{hi\_class} has been used in a number of recent analyses including  \cite{Bellini-Zumalacarregui-2015,Bellini-et-al-2016,Renk-2016-CMB-LSS,Alonso2016,Renk-etal-2017,BelliniEtAl2017AC,Lorenz-etal-2017-cosmo-param,Ezq2017GW} 

\subsection{{Effective Field Theory  with \texttt{CAMB}} (\texttt{EFTCAMB}) and (\texttt{EFTCosmoMC})}

\texttt{EFTCAMB} \citep{EFTCAMB1} and \citep{EFTCAMB2} is a set of patches to the code CAMB which implements the EFT approach to dark energy and modified gravity models of cosmic acceleration as described in Sect.~\ref{sec:EFT}. The package comes along with a modified version of \texttt{CosmoMC}, called \texttt{EFTCosmoMC}, that allows one to use the software with cosmological data sets. The code description and download are available at \url{http://eftcamb.org/index.html} and the corresponding  papers \citep{EFTCAMB1,EFTCAMB2}. A useful flowchart of the code and models covered is also accessible at \url{http://eftcamb.org/images/EFTCAMB_structure.pdf}. 

\texttt{EFTCAMB} implements the evolution of scalar and tensor perturbation equations including all the second order EFT operators. The implementation takes into account a consistent inclusion of more than one second order operator at a time and allows the use of a wide range of equation of state of dark energy for the background evolution. A number of options are made available to the user and can be found on the website and the  flowchart above. 

\texttt{EFTCAMB} has been used in a number of recent cosmological analyses including 
\cite{BelliniEtAl2017AC,LiuEtAt2017AN,RaveriEtAl2017PO,PeironeEtAl2017IO,Hu2016,
FruscianteEtAl2016HG,Planck2015MG,HuEtAl2015EM,EFTCAMB1,Raveri2014,EFTCAMB2}

%%%%%%%%%%%%%%%%%%%%%%%%%%%%%%%%%%%%%%%%%%%%%%%%%%%%%%%%%%%%%%%%%%%%%%%%%%%%%%%%%%%
%%%%%%%%%%%%%%%%%%%%%%%                                         %%%%%%%%%%%%%%%%%%%
%%%%%%%%%%%%%%%%%%%%%%%          SUB-SECTION                    %%%%%%%%%%%%%%%%%%%
%%%%%                  COMPARISON OF BOLTZMAN CODES                            %%%%
%%%%%                                                                          %%%%
%%%%%%%%%%%%%%%%%%%%%%%%%%%%%%%%%%%%%%%%%%%%%%%%%%%%%%%%%%%%%%%%%%%%%%%%%%%%%%%%%%%
%%%%%%%%%%%%%%%%%%%%%%%%%%%%%%%%%%%%%%%%%%%%%%%%%%%%%%%%%%%%%%%%%%%%%%%%%%%%%%%%%%%

\subsection{Comparison of Einstein--Boltzmann solver codes for testing General Relativity}

A recent careful comparative study of codes that solve Einstein--Boltzmann equations can be found in  \cite{BelliniEtAl2017AC}. Motivated by the high precision requirements from upcoming surveys such as LSST, WFIRST, Euclid, SKA, 
and Stage IV CMB experiments, the study aimed at finding at what level of accuracy such codes would agree with each other in calculating various CMB and matter power spectra.

The study  compared codes of the second type as discussed in Sect.~\ref{sec:codes}. They compared \texttt{EFTCAMB}, \texttt{hi\_class} and \texttt{COOP} for general scalar-tensor theories. They found that CMB and matter power spectra from \texttt{EFTCAMB} and  \texttt{hi\_class} agree with one with another to a sub-percent level. They also found that \texttt{COOP} has the required accuracy and agrees with the two other at large scales but needs calibration to remain in agreement at scales below Mpc. 
Then they compared these three codes to the following six codes and found them in good agreement: DASh \citep{DASH}, BD-CAMB \citep{Avilez2014} and CLASSig \citep{CLASSIG} that model Jordan--Brans--Dicke (JBD) gravity; GalCAMB \citep{BarreiraEtAl2012} for Galileon models; CLASS\_EOS\_fR \citep{Battye2018,EOS1,EOS2} for f(R) models; and CLASS-LVDM for Ho\v{r}ava--Lifshitz gravity \citep{LVDM2}.
Finally, they also compared the two codes \texttt{NLCAMB} \citep{NLCAMB} and  \texttt{NLCLASS} \citep{NLCLASS2} for non-local gravity and found them in good agreement. 

While the comparison was done for some specific points in the cosmological parameter space, the authors stated that they expect that their comparison should hold for other models and parameters in view of the stability found for these codes. However, the authors clarify that future code comparisons should include more models, the nonlinear regime and the effect of screening mechanisms. 

The authors conclude their analysis with a set of steps and warnings that a user should take into account when using these codes with various MG models to avoid any common possible sources of errors due to code versions, untested models, parameter conversion, initial conditions, and model-dependent precision requirements. We refer the interested reader to the full paper \citep{BelliniEtAl2017AC} for detailed discussions and comparisons.     

%%%%%%%%%%%%%%%%%%%%%%%%%%%%%%%%%%%%%%%%%%%%%%%%%%%%%%%%%%%%%%%%%%%%%%%%%%%%%%%%%%%
%%%%%%%%%%%%%%%%%%%%%%%%%%%%%%%%%%%%%%%%%%%%%%%%%%%%%%%%%%%%%%%%%%%%%%%%%%%%%%%%%%%
%%%%%%%%%%%%%%%%%%%%%%%%                                  %%%%%%%%%%%%%%%%%%%%%%%%%
%%%%%%%%%%%%%%%%%%%%%%%%   SYSTEMATIC EFFECTS  VS MG      %%%%%%%%%%%%%%%%%%%%%%%%%
%%%%%%%%%%%%%%%%%%%%%%%%                                  %%%%%%%%%%%%%%%%%%%%%%%%%
%%%%%%%%%%%%%%%%%%%%%%%%%%%%%%%%%%%%%%%%%%%%%%%%%%%%%%%%%%%%%%%%%%%%%%%%%%%%%%%%%%%
%%%%%%%%%%%%%%%%%%%%%%%%%%%%%%%%%%%%%%%%%%%%%%%%%%%%%%%%%%%%%%%%%%%%%%%%%%%%%%%%%%%

\section{Systematic effects in cosmological probes and degeneracies with modifications to GR}
\label{sec:systematics}

As we review in Sect.~\ref{sec:forecasts}, constraining decisively modifications to GR will depend on how well ongoing and future surveys and experiments can control and mitigate systematic effects in the data. First, uncertainties on MG parameters will become soon systematic-error dominated rather than statistical-error dominated. So the precision needed to distinguish between MG and GR will depend on how well systematic uncertainties will be mitigated down. Second, some systematic effects can mimic physical effects on observables and therefore introduce a bias (shift) in the corresponding cosmological parameters including MG parameters, causing them to deviate from their GR values. We describe below some of these systematic effects taking weak gravitational lensing and intrinsic alignments as an illustrative example and refer the reader to corresponding reviews and papers for other probes and effects.  

Weak gravitational lensing is a promising probe for measuring MG parameters to a one-percent precision level as forecast studies show in the next section. However, in order to reach this potential, one needs to get rid of some systematic effects such as galaxy intrinsic alignments, baryonic effects, and photometric redshift uncertainties, see for example the reviews \cite{Hoekstra2008,2015-Troxel-Ishak-lensing,2015-KirK-etal-Galaxy-aligments,2015-Eifler-etal-baryonic-effects-WL,Mandelbaum2017}.

For example, Intrinsic alignments (IA) of galaxies have been recognized as one of the most serious contaminants to weak gravitational lensing and the cosmological constraints obtained from it. For example, \cite{Bridle2007} found a 50\% bias due to IA on determining the dark energy equation of state from weak lensing. There are two types of IA correlations. The first is the intrinsic ellipticity correlation, also known as the II signal, and is due to the fact that two physically close galaxies could be aligned by the tidal force field of the same dark matter structure surrounding them. The second type of alignment has been pointed out more recently by \cite{Hirata2004} and is due to the fact that if a matter structure causes the alignment of a nearby galaxy and also contributes to the lensing signal of a background galaxy, then it produces an (anti-)correlation between gravitational lensing and intrinsic ellipticities, also known as the GI signal. The GI 2-point signal has been measured in SDSS, MegaZ-LRG and other samples by various groups including \cite{Mandelbaum2006,Hirata2007,Okumura2009,Faltenbacher2009}. The 3-point IA correlations follow the same mechanisms and are known as III, GGI, and GII correlations.  While the II and III correlations of IA can be, in principle, greatly reduced with photo-z's by using cross-spectra of galaxies in two different redshift bins, so that the galaxies are separated by large enough distances to assure that the tidal effect is weak, this does not work for the GI, GGI, and GII types which happen between galaxies at different redshifts and large separations. Proposed mitigation methods for IA include parametrization-marginalization   \cite{2013CFHTlens,Krause2016}, nulling techniques \citep{Joachimi2008,Joachimi2009}, or self-calibration methods\citep{Zhang2010a,Zhang2010b,Troxel2012SCa,Troxel2012b,Troxel2012c}.

\cite{Laszlo2012}  conducted a forecast analysis to study the disentanglement of cosmic tests of gravity from weak lensing systematics. They considered ongoing and upcoming photometric stage III surveys such as DES and stage IV such as Euclid, LSST and WFIRST.
They found that using galaxy bias and intrinsic alignment models that depend on the redshift give figures of merit in constraining modifications to gravity that are a factor of four weaker than when no redshift dependence is assumed. This reflects the fact that not accounting for systematics or not properly modeling them can lead to overestimating the constraints on MG. They also found that adding Planck CMB data helps in adding a number of parameters to model systematic effect in lensing without loss of constraining power. 

\cite{Ferte2017} constrained MG parameters $\mu$ and $\Sigma$ using weak lensing data from CFHTLenS and DES-SV, RSD data from BOSS DR 12 and the 6dF galaxy survey, and CMB data from Planck (see Sect.~\ref{sec:constraints}). They included three lensing systematics in their analysis. First, the shape measurement error that they model with a multiplicative factor. Second, the calibration bias of the photometric redshift distribution that they model with another parameter. Third, the intrinsic alignments that they use with a one amplitude parameter for the IA nonlinear model of \cite{Hirata2004,Bridle2007}. They marginalized in their analysis over these three systematic parameters and compared the effect of ignoring one systematic at a time. They found that ignoring the effect of calibration bias or photometric redshift bias does affect significantly the constraints on MG parameters. 
However, ignoring intrinsic alignments  shifts the constraints toward lower values of $\Sigma$. They found thus a degeneracy between the amplitude of IA and the $\Sigma$ MG parameter leading to higher values of $\Sigma$ when IA is included. 
Similar shifts in the dark energy equation of state parameters as caused by including or not including IA systematics have been studied in \cite{Krause2016,Yao2017}. 
Furthermore, \cite{Ferte2017} also found when forecasting constraints on MG, using 5 years data of DES, that including IA increases the uncertainties on MG parameters as shown in their figure 11 (right panel of Fig.~\ref{fig:mg_desy5_LR} here). This shows that ignoring IA leads to overestimating MG parameter constraints from lensing. 

\cite{Dossett2015} performed a constraint analysis on MG parameters using binned, functional and hybrid parameterizations including intrinsic alignment systematic effect. 
They used data from Planck temperature anisotropies, the galaxy power spectrum from
WiggleZ survey, weak lensing tomography shear-shear cross correlations from the CFHTLenS survey, Integrated Sachs Wolfe-galaxy cross correlations, and baryon acoustic oscillation data. They found that the constraints on the amplitude of intrinsic alignment depend on the MG parametrization used but the correlation parameters between MG parameters and IA amplitude are weak to moderate. 

The lesson to take from this illustrative example is that systematic effects in cosmological probes of gravity can be degenerate with MG parameters and also limit the precision that one can reach in constraining these parameters. This is the case also for other systematics such as baryonic feedback effects that can  enhance growth of structure and be degenerate with some modifications to gravity as reflected on the matter power spectrum at smaller scales \citep{Puchwein2013}. The scale dependence of the $\beta$  parameter in redshift space distortion measurements, if ignored, can also introduce bias on determining the growth factor of structure leading to incorrect constraints on MG theories \citep{Okumura2011}. We refer the reader to the following review articles including systematic effects in cosmological probes and their effect on dark energy or modified gravity models: e.g., \cite{Weinberg2013,Mandelbaum2017,Nishizawa2014SW} and references therein. 

%%%%%%%%%%%%%%%%%%%%%%%%%%%%%%%%%%%%%%%%%%%%%%%%%%%%%%%%%%%%%%%%%%%%%%%%%%%%%%%%%%%
%%%%%%%%%%%%%%%%%%%%%%%%%%%%%%%%%%%%%%%%%%%%%%%%%%%%%%%%%%%%%%%%%%%%%%%%%%%%%%%%%%%
%%%%%%%%%%%%%%%%%%%%%%%%                                  %%%%%%%%%%%%%%%%%%%%%%%%%
%%%%%%%%%%%%%%%%%%%%%%%%   FORECASTS ON MG                %%%%%%%%%%%%%%%%%%%%%%%%%
%%%%%%%%%%%%%%%%%%%%%%%%                                  %%%%%%%%%%%%%%%%%%%%%%%%%
%%%%%%%%%%%%%%%%%%%%%%%%%%%%%%%%%%%%%%%%%%%%%%%%%%%%%%%%%%%%%%%%%%%%%%%%%%%%%%%%%%%
%%%%%%%%%%%%%%%%%%%%%%%%%%%%%%%%%%%%%%%%%%%%%%%%%%%%%%%%%%%%%%%%%%%%%%%%%%%%%%%%%%%

\section{Future cosmological constraints on GR and MG parameter forecasts}
\label{sec:forecasts}

There are a number of promising cosmological surveys of large scale structure, CMB and distance probe experiments that are being built or planned  such as (AdvACT, eBOSS, DESI, Euclid, HSC/PFS, LSST, POLARBEAR, SPT-3G, WFIRST and others \dots). These will provide an overwhelming large amount of data with high precision. As we discussed above, huge efforts are also being made to develop and advance the mitigation of systematic effects to allow these surveys to reach their full constraining potential. 

We will here provide a brief overview of some parameter forecast analyses that examined how well we will be able to constrain MG parameters using these future surveys. The commonly used formalism for such forecasts is the Fisher formalism \citep{Fisher1935} or the Markov Chain Monte-Carlo (MCMC) simulated spectra and likelihoods \citep{Metropolis1953}. 

The fisher matrix $\mathbf{F}$ can be determined from the theoretical observable functions and specifications of a survey, e.g., \cite{Vogeley1996,Tegmark1997}. It can provide a forecasted covariance matrix $\mathbf{C}$  since $\mathbf{C}=\mathbf{F^{-1}}$. This allows one to forecast uncertainties on individual cosmological parameters $\sigma(p^i)=\sqrt{C_{ii}}$. This also allows one to calculate the correlations between parameters as $P_{ij}=C_{ij}/\sqrt{C_{ii}C_{jj}}$. Although, the Fisher formalism has shortcomings, as for example, it does not cover non-Gaussian constraints, it has been used extensively in the literature for cosmological parameter forecasts. Another quantity that is worth mentioning here is the Figure of Merit (FoM) that can be used to determine the constraining power of probes or combinations of probes, e.g., \cite{Albrecht2006,Albrecht2009}. FoM is often defined to be proportional to the reciprocal of the square root of the determinant of the covariance matrix, i.e., $\sqrt{\rm det(\mathbf{C})}$ since the latter is proportional to the super-volume of the super-ellipsoides in the parameter hyperspace. Various constants of proportionality have been used including unity. As the constraints get tighter, the ellipsoid volumes get smaller and the FoM get stronger: $\mathrm{FoM}= (\det\, C)^{-1/2}$ or  $\mathrm{FoM} = -\frac{1}{2} \ln(\det(\mathbf{C}))$. FoM has been used for the dark energy equation of state constraints in, e.g., \cite{Albrecht2006,Albrecht2009,Acquaviva2010,2010Mortonson-Huterer-Hu-FOM-DE,Wang2010} and for MG parameter constraints in, e.g., \cite{DossettEtAl2011,Laszlo2012,Casas2017}.

A second approach to parameter forecasting is to use simulated likelihoods using Markov-Chain-Monte-Carlo methods. This allows one to go beyond the Gaussian assumptions in the Fisher formalism. While the Fisher analyses can in principle provide accurate estimates in the vicinity of the best fit points in parameter space, it becomes less accurate away from such regions and in particular in higher dimensional spaces where systematic effect parameters are added to the analysis. MCMC simulation methods can be computationally intensive and have been used for dark energy equation of state forecasts with or without systematic effects, see for example \cite{Upadhye2005,Krause2016}.   

\begin{figure}[t]
\begin{center}
\includegraphics[scale=0.4]{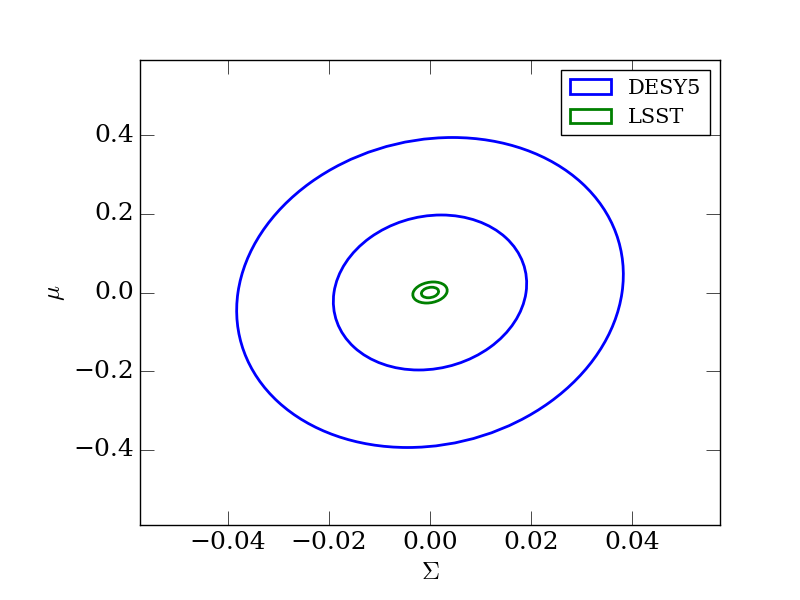}
\includegraphics[scale=0.4]{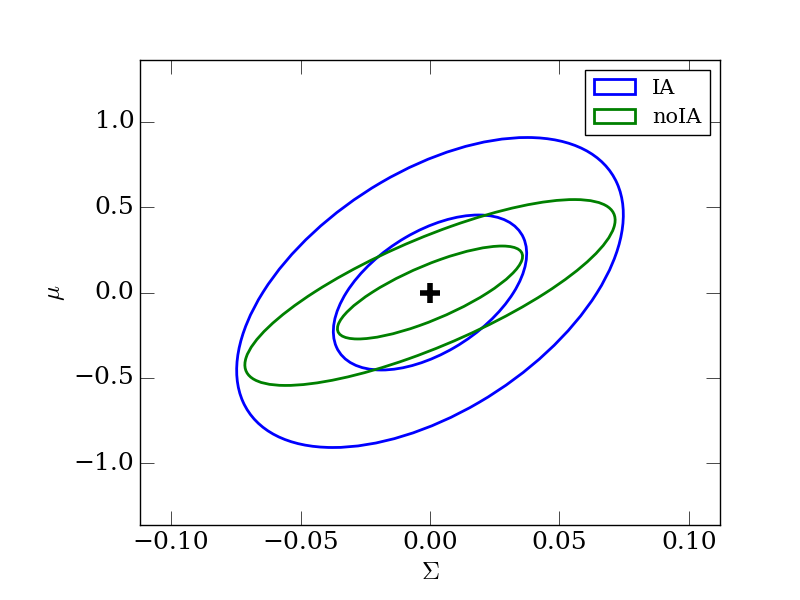}
\caption{Left: Forecasted 68\% and 95\% confidence contours on MG parameters $\Sigma$ and $\mu$ for a future DES-Y5 survey in blue and for an LSST-like survey in green. No intrinsic alignment (IA) systematics are assumed here. While DES-Y5 promises to make tight constrains on $\Sigma$ as expected from a weak lensing survey, LSST will provide an order magnitude further improvement on the two parameters. Right:  A comparison between uncertainties for DES-Y5 survey when ignoring IA in green and taking them into account in blue. This means that ignoring IA leads to underestimating the uncertainties on the parameters. This also shows that these uncertainties will be systematic-error dominated. Figure reproduced with permission from \cite{Ferte2017}.}
\label{fig:mg_desy5_LR}
\end{center}
\end{figure}

\cite{Ferte2017} added to their paper a parameter forecast analysis including the two MG parameters  $[1 + \mu (a)]$ and $[1 + \Sigma (a)]$ in the Poisson and weak lensing equations taking 0 values in GR. They used the full five-year DES survey and an LSST-like survey. They marginalized over five other cosmological parameters as defined previously $\{A_S, n_s, \Omega_m, \Omega_b, h_0\}$ and assumed a $\Lambda$CDM fiducial model. They used a Fisher analysis and used specifications for DES-5Y and LSST-like in the respective order: 5,000 and 18,000 square degrees of sky coverage; 5 and 10 redshift bins; 10 and 55 galaxies per arc-minutes squared; 0.25 and 0.20 for the intrinsic ellipticity standard deviation; $0.05(1+z)$ and $0.05(1+z)$ for the standard deviation of the photo-z estimation as a function of the redshift $z$. They first derive results for DES-Y5 and LSST-like without any use of intrinsic alignment systematics. We display their  
Fig.~10 (left panel of Fig.~\ref{fig:mg_desy5_LR} here) showing forecasted 
68\% and 95\% confidence contours on ($\Sigma$,$\mu$) around their GR values for both surveys. They give the projected uncertainties  as 
\begin{eqnarray}
\sigma_{\Sigma}  =  0.019, \;\;\;\;\;\;\; 
\sigma_{\mu}  =   0.20, 
\end{eqnarray} 
for DES-Y5, and
\begin{eqnarray}
\sigma_{\Sigma}  =  0.0017, \;\;\;\;\;\;\; 
\sigma_{\mu}  =   0.013,
\end{eqnarray} 
for an LSST-like survey. 

An ideal case without IA systematics, the uncertainties on $\Sigma$ from DES seem already significant while more uncertainty will persist on $\mu$. However, LSST would provide a further order of magnitude improvement on the errors and constraining even $\mu$ down to a decisive bound. However, as the authors show, including IA systematics changes significantly these forecasts. They show the effect only for constraints from DES-Y5 using an IA nonlinear model that they represent by adding an IA amplitude parameter in each of the 5 bins, so bringing the number of parameters to 12. Their Fig.~11 (right panel of Fig.~\ref{fig:mg_desy5_LR} here) shows a comparison between constraints with and without IA systematics. It indicates that ignoring IA leads to under-estimating uncertainties. It could also be pointed out that ignoring IA can also falsely shift the best fit MG parameters in a real-data analysis. Here, the analysis did not even include complementary probes such as RSD, CMB and BAO and yet the statistical errors are of the order of a percent or sub-percent level. Clearly, the future of constraining MG models to such a level will be systematic error dominated. The analysis showed the promise of future weak lensing surveys in providing decisive answers on any modification of GR at cosmological scales.

\cite{Casas2017} performed an extended forecast analysis of MG parameters for the weak lensing and galaxy clustering surveys of Euclid, Square Kilometer Array 1 (SKA1), SKA2, and the dark energy spectroscopic instrument (DESI). They also combined the above surveys to a prior covariance matrix from Planck CMB constraints. They used the MG parameters $\mu(a)$, $\eta(a)$ and $\Sigma(a)$ as defined in equations \eqref{eq:MGmu}, \eqref{eq:MGeta} and \eqref{eq:Sigma2} of Sect.~\ref{sec:MGparameters}. 

First, they employed a binning method where they divided the redshift range $0 \le z \le 3$ into six bins with some smoothed transitions. They did not consider any scale dependence and assumed that the two parameters take their GR values of unity at redshift above 3. They assumed that the background has a flat $\lcdm$ evolution and considered 15 parameters $\theta=\{\Omega_{m},\Omega_{b},h,\ln10^{10} A_{s},n_{s},\{\mu_{i}\},\{\eta_{i}\}\}$, with $i=1..5$. All $\eta_i$ and $\mu_i$ are equal to 1 for GR. 

Second, they considered a functional time parametrization for the 2 MG parameters following \cite{Planck2015MG}, with no scale dependence again. They considered 
two sub-cases:

\begin{itemize}

\item case-1 where the time evolution of MG parameters is parameterized via the effective dark energy density parameter. They call this the late-time parameterization since it reduces to GR at early times
\begin{eqnarray}
\mu(a,k)\equiv1+E_{{\rm 11}}\Omega_{{\rm
DE}}(a)\,\,\,,\label{eq:DE-mu-parametrization}\\
\eta(a,k)\equiv1+E_{{\rm 22}}\Omega_{{\rm
DE}}(a)\,\,\,;\label{eq:DE-eta-parametrization_LR}
\end{eqnarray}

\item case-2 where the time dependence is parameterized directly using a Taylor series in the scale factor. They call this the early-time parameterization since it allows for modifications to GR even at early times:  
\begin{eqnarray}
\mu(a,k)\equiv1+E_{{\rm 11}}+E_{{\rm
12}}(1-a)\,\,\,,\label{eq:TR-mu-parametrization_LR}\\
\eta(a,k)\equiv1+E_{{\rm 21}}+E_{{\rm
22}}(1-a)\,\,\,.\label{eq:TR-eta-parametrization_LR}
\end{eqnarray}
See \cite{Casas2017,Planck2015MG} for more discussion about these parametrizations. 
\end{itemize}
In addition to the five standard cosmological parameters, they added $E_{11}$ and $E_{22}$ for the late-time parametrization and $E_{11},E_{12},E_{21},E_{22}$ for the early-time case. The $E_{ij}$s are then used to reconstruct $\mu(a)$, $\eta(a)$ and $\Sigma(a)$.  
Here all $E_{ij}$s are zero in GR. 

The fiducial model values used for the binned and the two functional parametrizations were taken as the best fit values to the data \textit{Planck}+BAO+SNe+H$_{0}$ (BSH) as used in \cite{Planck2015MG} (unlike the analysis discussed right above where the fiducial values were taken as the GR ones).

The results of \cite{Casas2017} are summarized in three tables (\ref{tab:errors-all-MGBin3_LR}, \ref{tab:errors-GC-SKAcompare-MG-DE-mu-eta-sigma_LR}, and \ref{tab:errors-GC-SKAcompare-MG-TR-mu-eta-sigma-Zhao-1_LR}) that we reproduce from \cite{Casas2017}.
They give results for weak lensing, galaxy clustering for Euclid, SKA1 and SKA2 separating  constraints when using linear and nonlinear scales as well as with and without the combination to Planck CMB constraint covariance matrix. They used two semi-analytical approximations to include nonlinear regimes for lensing and clustering. They find that using nonlinear scales reduces and even breaks degeneracies between MG parameters in different bins and also with the overall amplitude of the matter power spectrum. They also show in the tables how much gain is obtained when adding constraints from nonlinear scales. In sum, they find that future surveys Euclid, SKA1, SKA2 and DESI (in combination with Planck priors) can constrain the present values (i.e., at $z=0$) of the parameters $\eta$, $\mu$, and $\Sigma$ to 2--3\% when only linear scales are used but these reduce to about 1\% or less when nonlinear scales are included. Clearly again, the determining factor for these future surveys in obtaining decisive answers on deviation from GR will be the successful mitigation of  systematic effects.  

\begin{table}[htbp]
%\begin{sidewaystable}[htbp]
\caption{Forecasted uncertainties for the MG binned parameterization from \cite{Casas2017}: 1-$\sigma$ marginalized errors expressed as percentage of the corresponding fiducial value on  parameters for Euclid Galaxy Clustering (GC) and Weak Lensing (WL) surveys used separately and combined. Results with the addition of the Planck-2015 prior covariance are also presented. Results are also presented using linear spectra (lin) and nonlinear prescription (nl-HS). For the GC survey, they set the cutoff to $k_{\max}=0.15$ h/Mpc in the linear case and $k_{\max}=0.5$ h/Mpc in the nonlinear case.
For WL, they set the maximum cutoff in the linear case at $\ell_{\max} = 1000$, 
while setting it in the nl-HS case at $\ell_{\max} = 5000$.
The last row is for combined GC+WL+\textit{Planck} using the Halofit nonlinear fitting formula for power spectra. 
The last column shows the increase of Figure of Merit (FoM) (see top of Sect.~\ref{sec:forecasts}) with respect to the reference GC linear case. Just as the FoM, a larger relative increase indicates more constraining power for the probe. 
One can see that each time nonlinearities are added, considerable improvements are  obtained. As the authors pointed out, showing errors on $\mu$ and $\eta$ make WL look unfairly poor at constraining parameters. But when these errors are converted into errors on $\Sigma$, the constraints on $\Sigma_{1,2,3}$ are slightly better, of the order of 40\% for WL(nl-HS). This is due to the fact that WL is more sensitive to the parameter $\Sigma$. The FoM itself is almost unaffected by the choice of the parameter pair because the area of a 2D ellipse is invariant under rotation. Table reproduced with permission from \cite{Casas2017}, copyright by Elsevier.}
\label{tab:errors-all-MGBin3_LR}
\footnotesize 
\centering
\begin{tabular}{|l|c|c|c|c|c|c|c|c|c|c|c|}
\hline 
\aTstrutlr \textbf{Euclid}& $\mu_{1}$  & $\mu_{2}$  & $\mu_{3}$  & $\mu_{4}$  & $\mu_{5}$  & $\eta_{1}$  
& $\eta_{2}$  & $\eta_{3}$  & $\eta_{4}$ & $\eta_{5}$ &Rel.
\\
(Redbook)&   &  &   &  &   &   &   &  & &  &MG
\\
\hline 
\aTstrutlr Fiducial& 1.108  & 1.027  & 0.973  & 0.952  & 0.962  & 1.135  & 1.160  & 1.219  & 1.226 & 1.164 &FoM\\
\hline 
\aTstrutlr  {GC (lin)}\aTstrutlr& 119\% & 159\% & 183\% & 450\% & 1470\% & 509\% & 570\% & 586\% & 728\% & 3390\% & 0 \\
\aTstrutlr  {GC (nl-HS)}\aTstrutlr& 7.0\% & 6.7\% & 10.9\% & 27.4\% & 41.1\% & 20\% & 24.3\% & 19.9\% & 38.2\% & 930\% & 19  \\
\hline
\hline
\aTstrutlr  {WL (lin)}& 165\% & 2210\% & 4150\% & 13100\% & 22500\% & 2840\% & 3140\% & 8020\% & 29300\% & 39000\% 
& -27\\
\aTstrutlr  {WL (nl-HS)}& 188\% & 255\% & 419\% & 222\% & 206\% & 330\% & 488\% & 775\% & 8300\% & 9380\%
& -10  \\
\hline
\hline
\aTstrutlr  {GC+WL (lin)}& 5.8\% & 10\% & 19.2\% & 282\% & 469\% & 7.9\% & 9.6\% & 16.1\% & 276\% & 2520\% & 12
 \\
\aTstrutlr  {GC+WL+}&   &  &   &  &   &   &  &   &  & & 
\\
\aTstrutlr  {\textit{Planck} (lin)}& 3.4\% & 4.8\% & 7.8\% & 9.3\% & 13.1\% & 6.2\% & 7.7\% & 9.1\% & 12.7\% & 23.6\%
& 27 \\
\hline
\hline
\aTstrutlr  {GC+WL (nl-HS)}& 2.2\% & 3.3\% & 8.2\% & 24.8\% & 34.1\% & 3.6\% & 5.1\% & 8.1\% & 25.4\% & 812\%
& 24 \\
\aTstrutlr  {GC+WL+\textit{Planck}}&   &  &   &  &   &   &  &   &  & & 
\\  {(nl-HS)}  
& 1.8\% & 2.5\% & 5.8\% & 7.8\% & 10.3\% & 3.2\% & 4.1\% & 5.9\% & 9.6\% & 19.5\%
& 33 \\
\aTstrutlr  {GC+WL+\textit{Planck}}&   &  &   &  &   &   &  &   &  & & 
\\  {(nl-Halofit)}  
& 2.0\% & 2.4\% & 5.1\% & 7.4\% & 10.2\% & 3.5\% & 4.1\% & 5.8\% & 9.2\% & 18.9\% 
& 33 \\
\hline
\end{tabular}
%\end{sidewaystable}
\end{table}

\newpage

\begin{table}[htbp]
\caption{Forecasted uncertainties for the MG late-time functional parameterization from \cite{Casas2017}:  
1$\sigma$ marginalized errors expressed as percentage of the corresponding fiducial value on  parameters for Galaxy Clustering (GC) and Weak Lensing (WL) surveys used separately and combined for Euclid, SKA1 and SKA2. Results with and without Planck-2015 prior covariance. Results are also presented using linear spectra (lin) and nonlinear prescription (nl-HS).
Last column shows the FoM for each probe for MG parameters relative to the Euclid GC linear base case (not shown here).
One can see that in general, SKA2 is the most powerful survey, followed by Euclid and SKA1.
The authors note that in the case of GC alone, DESI-ELG is more constraining than SKA1-SUR. As expected, the GC survey would only constrain $\mu$ with
a high accuracy, while a WL survey would constrain $\Sigma$ with
a very good accuracy. The combination of both breaks the degeneracy and provides much more powerful constraints than each probe alone. 
However, as the authors noted, adding \planck$\,$  priors in the last row improves considerably the constraints on the base $\lcdm$ parameters but has almost no effect on the MG parameters. This is also indicated by the almost constant MG FoM. Table reproduced with permission from \cite{Casas2017}, copyright by Elsevier.}
\label{tab:errors-GC-SKAcompare-MG-DE-mu-eta-sigma_LR}
\footnotesize 
\centering{}%
\begin{tabular}{|l|c|c|c|c|c||c|c|c|c|}
	\hline
	 & $\Omega_{c}$ & $\Omega_{b}$ & $n_{s}$ & $\ell\mathcal{A}_{s}$ & $h$ & $\mu$ & $\eta$ & $\Sigma$ & MG FoM \aTBstrutlr\\
\hline 
Fiducial & {0.254 } & {0.048 } & {0.969 } & {3.060 } & {0.682 } & {1.042 } & {1.719 } & {1.416 } & relative\aTBstrutlr\\
\hline 
\hline 
 {GC(nl-HS)} &  &  &  &  &  &  &  & & \aTBstrutlr\\
Euclid   
& 0.9\% & 2.5\% & 1.3\% & 0.8\% & 1.7\% & 1.7\% & 475\% & 291\%
& 2.9
\\
SKA1-SUR  
& 5\% & 15.3\% & 8.7\% & 3.8\% & 10.8\% & 18.1\% & 165\% & 108\%
& 1.7
\\
SKA2  
& 0.5\% & 1.3\% & 0.4\% & 0.4\% & 0.8\% & 0.7\% & 86.8\% & 53.2\%
& 5.5
\\
DESI-ELG   
& 1.6\% & 4.1\% & 2.3\% & 1.3\% & 2.9\% & 3.3\% & 899\% & 552\%
& 1.8
\\
\hline 
\hline 
 {WL(nl-HS)} &  &  &  &  &  &  &  &  & \aTBstrutlr\\
Euclid   
& 6.3\% & 20.7\% & 4.6\% & 5.8\% & 13.8\% & 23.3\% & 40.9\% & 4.6\%
& 4.5
\\
SKA1  
& 30.8\% & 109\% & 35\% & 36.5\% & 77.6\% & 220\% & 405\% & 36.8\%
& 0.5
\\
SKA2  
& 6\% & 22.5\% & 5.9\% & 6.8\% & 15.9\% & 19\% & 33.2\% & 3.7\%
& 4.9
\\
\hline 
\hline 
 {GC+WL(lin)} &  &  &  &  &  &  &  & & \aTstrutlr\\
Euclid  
& 1.8\% & 5.9\% & 2.8\% & 2.3\% & 4.2\% & 7.1\% & 10.6\% & 2\%
& 6.6
\\
SKA1  
& 10.1\% & 47.6\% & 25.4\% & 21.7\% & 40.4\% & 26.4\% & 28.8\% & 13.6\%
& 3.7
\\
SKA2 
& 1.2\% & 4.5\% & 2.2\% & 1.9\% & 3.3\% & 4.1\% & 5.5\% & 1.6\%
& 7.5
\\
\hline 
\hline 
 {GC+WL(lin)+\textit{Planck}} &  &  &  &  &  &  &  & & \aTstrutlr\\
Euclid  
& 1.0\% & 0.7\% & 0.4\% & 0.4\% & 0.4\% & 6.2\% & 9.8\% & 1.5\%
& 6.9
\\
SKA1  
& 2.4\% & 1.2\% & 0.4\% & 1.2\% & 0.7\% & 12\% & 19.8\% & 3.8\%
& 5.3
\\
SKA2 
& 0.7\% & 0.6\% & 0.3\% & 0.4\% & 0.3\% & 3.6\% & 5.2\% & 1.2\%
& 7.8
\\
\hline 
\hline 
 {GC+WL(nl-HS)} &  &  &  &  &  &  &  & & \aTstrutlr\\
Euclid  
& 0.8\% & 2.2\% & 0.8\% & 0.7\% & 1.5\% & 1.6\% & 2.4\% & 1.0\%
& 8.7
\\
SKA1  
& 4.7\% & 14.3\% & 6.2\% & 3.6\% & 9.6\% & 12.8\% & 11\% & 7.3\%
& 5.5
\\
SKA2 
& 0.4\% & 1.3\% & 0.3\% & 0.4\% & 0.8\% & 0.7\% & 0.9\% & 0.6\%
& 10.3
\\
\hline
\hline
 {GC+WL(nl-HS)+\textit{Planck}} &  &  &  &  &  &  &  & & \aTstrutlr\\
Euclid  
& 0.7\% & 0.6\% & 0.2\% & 0.2\% & 0.3\% & 1.6\% & 2.4\% & 0.9\%
& 8.9
\\ 
SKA1  
& 2.0\% & 1.0\% & 0.4\% & 0.8\% & 0.6\% & 3.5\% & 6\% & 2.7\%
& 6.9
\\ 
SKA2 
& 0.4\% & 0.5\% & 0.2\% & 0.1\% & 0.2\% & 0.6\% & 0.9\% & 0.5\%
& 10.3
\\
\hline
\end{tabular}
\end{table}

\newpage

\begin{table}[htbp]
\caption{Forecasted uncertainties for the MG early-time functional parameterization from \cite{Casas2017}:
1$\sigma$ marginalized errors expressed as percentage of the corresponding fiducial value on  parameters for Galaxy Clustering (GC) and Weak Lensing (WL) surveys used separately and combined for Euclid, SKA1 and SKA2. Results with the addition of the Planck-2015 prior covariance are also presented. Results are also presented using linear spectra (lin) and nonlinear prescription (nl-HS).
Last column shows the FoM for each probe for MG parameters relative to the Euclid GC linear base case (not shown here).
Also, adding Planck to the last combination does not provide any additional improvements in MG parameters.  
The authors note that in this parameterization, a GC survey alone is able to constrain both $\mu$ and $\Sigma$ to a good level for all surveys, better than with the late time parameterization, more often used in literature. Of course, WL still does better on $\Sigma$. The combination of GC+WL is however less constraining in the early time parametrization than in the late time parameterization one. 
The nonlinear forecast for GC+WL+\textit{Planck} would yield, for Euclid and SKA2, contraints at the 1--2\% accuracy level on $\mu$, $\Sigma$, while
for SKA1 the contraints would be at the 8\% level. Table reproduced with permission from \cite{Casas2017}, copyright by Elsevier.}
\label{tab:errors-GC-SKAcompare-MG-TR-mu-eta-sigma-Zhao-1_LR}
\footnotesize 
\centering
\begin{tabular}{|l|c|c|c|c|c||c|c|c|c|}
\hline
	 & $\Omega_{c}$ & $\Omega_{b}$ & $n_{s}$ & $\ell\mathcal{A}_{s}$ & $h$ & $\mu$ & $\eta$ & $\Sigma$  & MG FoM \aTBstrutlr\\
\hline 
Fiducial & {0.256} & {0.0485} & {0.969} & {3.091} & {0.682} & {0.902} & {1.939} & {1.326} & relative \aTBstrutlr\\ 
\hline 
\hline 
 {GC(nl-HS)} &  &  &  &  &  &  &  & & \aTBstrutlr\\
Euclid   
& 1.1\% & 2.3\% & 1.3\% & 0.7\% & 1.6\% & 1.8\% & 7.9\% & 4.8\%
& 6.6
\\
SKA1-SUR  
& 7.9\% & 14.2\% & 13.4\% & 4.2\% & 11\% & 12.6\% & 82.7\% & 52.6\%
& 2.2
\\
SKA2  
& 0.6\% & 1.3\% & 0.7\% & 0.4\% & 0.9\% & 0.9\% & 3.4\% & 1.8\%
& 8.3
\\
DESI-ELG   
& 2.0\% & 4.3\% & 2.7\% & 1.4\% & 3.0\% & 8.2\% & 32\% & 28.6\%
& 4.3
\\
\hline 
\hline 
 {WL(nl-HS)} &  &  &  &  &  &  &  &  & \aTBstrutlr\\
Euclid   
& 6.5\% & 21.9\% & 6.6\% & 5.9\% & 15.8\% & 2.8\% & 8.0\% & 3.4\%
& 6.6
\\
SKA1  
& 32\% & 106\% & 37.2\% & 33\% & 79.3\% & 13.1\% & 37.1\% & 16.4\%
& 3.4
\\
SKA2  
& 5.9\% & 22.1\% & 6.7\% & 6.1\% & 16.1\% & 2.4\% & 7.0\% & 2.9\%
& 6.9
\\
\hline 
\hline 
 {GC+WL(lin)} &  &  &  &  &  &  &  & & \aTstrutlr\\
Euclid  
& 1.8\% & 6.6\% & 3.4\% & 5.6\% & 5.2\% & 3.0\% & 6.8\% & 3.4\%
& 6.4
\\
SKA1  
& 10.3\% & 46.4\% & 24.2\% & 33.6\% & 40.2\% & 14.4\% & 29.6\% & 15.5\%
& 3.3
\\
SKA2 
& 1.3\% & 4.9\% & 2.5\% & 4.2\% & 3.9\% & 2.5\% & 5.7\% & 2.7\%
& 6.8
\\
\hline 
\hline 
 {GC+WL(lin)+{\it Planck}} &  &  &  &  &  &  &  & & \aTstrutlr\\
Euclid  
& 1.2\% & 0.9\% & 0.6\% & 2.3\% & 0.4\% & 2.4\% & 6.5\% & 2.8\%
& 6.8
\\
SKA1  
& 2.5\% & 1.5\% & 0.8\% & 2.9\% & 0.8\% & 8.8\% & 22.2\% & 8.5\%
& 4.5
\\
SKA2 
& 0.9\% & 0.7\% & 0.6\% & 2.1\% & 0.3\% & 2.1\% & 5.4\% & 2.3\%
& 7.2
\\
\hline 
\hline 
 {GC+WL(nl-HS)} &  &  &  &  &  &  &  & & \aTstrutlr\\
Euclid  
& 1.0\% & 2.2\% & 1.2\% & 0.7\% & 1.6\% & 1.3\% & 4.4\% & 1.9\%
& 8.1
\\
SKA1  
& 7.1\% & 13.4\% & 10.7\% & 4\% & 10\% & 8.2\% & 24.4\% & 10.5\%
& 4.4
\\
SKA2 
& 0.6\% & 1.3\% & 0.7\% & 0.4\% & 0.9\% & 0.8\% & 2.7\% & 1.3\%
& 8.8
\\
\hline
\hline
 {GC+WL(nl-HS)+{\it Planck}} &  &  &  &  &  &  &  & & \aTstrutlr\\
Euclid  
& 0.8\% & 0.7\% & 0.3\% & 0.3\% & 0.3\% & 1.3\% & 4.4\% & 1.9\%
& 8.1
\\ 
SKA1  
& 2.1\% & 1.3\% & 0.5\% & 0.9\% & 0.7\% & 7\% & 20.8\% & 8.2\%
& 4.9
\\ 
SKA2 
& 0.5\% & 0.5\% & 0.3\% & 0.2\% & 0.2\% & 0.8\% & 2.7\% & 1.3\%
& 8.8
\\
\hline
\end{tabular}
\end{table}

\newpage

Other relatively recent MG parameter constraint forecast studies include \cite{Alonso2016}, where the authors considered scalar-tensor theories and the $\alpha_x$ parameterization. They focused on Stage IV CMB-S4 and photometric surveys such as LSST, and SKA1. They also used a Fisher analysis and the FoM metric. The analysis was not restricted to the quasi-static approximation and included relativistic effects. 
They showed how combinations of probes can constrain redshift and scale evolution.
They found that combination of probes can constrain the MG parameters down to a few percent level as well. It is even more optimistic to note that these should only improve now that the event GW170817/GRB170817A has constrained $\alpha_T$ to practically zero. 
\cite{Harrison2016} determined dark energy and MG parameter constraint forecasts for weak lensing surveys from SKA1 and SKA2. They find that SKA1 can provide constraints similar to stage-II experiments such as DES while SKA2 can provide tighter constraints than  stage-IV experiments such as LSST, WFIRST and Euclid. Further MG parameters or $f\sigma_8$ forecast studies can be found in \cite{Spurio2018,BeutlerEtAl2012T6,Majerotto2012}.

%%%%%%%%%%%%%%%%%%%%%%%%%%%%%%%%%%%%%%%%%%%%%%%%%%%%%%%%%%%%%%%%%%%%%%%%%%%%%%%%%%%
%%%%%%%%%%%%%%%%%%%%%%%%%%%%%%%%%%%%%%%%%%%%%%%%%%%%%%%%%%%%%%%%%%%%%%%%%%%%%%%%%%%
%%%%%%%%%%%%%%%%%%%%%%%%                                  %%%%%%%%%%%%%%%%%%%%%%%%%
%%%%%%%%%%%%%%%%%%%%%%%%   CONCLUDING REMARKS             %%%%%%%%%%%%%%%%%%%%%%%%%
%%%%%%%%%%%%%%%%%%%%%%%%                                  %%%%%%%%%%%%%%%%%%%%%%%%%
%%%%%%%%%%%%%%%%%%%%%%%%%%%%%%%%%%%%%%%%%%%%%%%%%%%%%%%%%%%%%%%%%%%%%%%%%%%%%%%%%%%
%%%%%%%%%%%%%%%%%%%%%%%%%%%%%%%%%%%%%%%%%%%%%%%%%%%%%%%%%%%%%%%%%%%%%%%%%%%%%%%%%%%

\section{Concluding remarks and outlook} 

Cosmological surveys and experiments are increasing in number and sophistication. Interesting ideas with new theoretical developments in gravity theories continue to emerge. In the midst of this buildup, general relativity continues to be so far prosperous and consistent with various cosmological tests and observations. It is worth noting though that while relativity is found to be consistent with all current data sets, the constraints are still too large to exclude some other possible theories. 

There are some small tensions that appear between different data sets when the $\lcdm$ model of general relativity is being used as an underlying theoretical model. While these tensions are likely due to systematic effects in various data sets, it is worth following closely how they will evolve with upcoming and future more precise data. 

Constraints on modified gravity parameters are quickly tightening up due to increasing statistical power in the data. However, this shows that for upcoming and planned  surveys, the uncertainty in testing general relativity at cosmological scales will be rather systematic-error dominated. Therefore, understanding and mitigating systematic effects in cosmological probes of gravity will play a major role in obtaining decisive answers from observations.

Progress is also needed in working on modified gravity numerical simulations in order to exploit nonlinear regimes where probes such as weak lensing and galaxy clustering can reach more constraining power.  

Astrophysical tests at galactic and stellar levels are found to be complementary to cosmological tests of gravity and will prove to be very useful in testing screening mechanisms of modified gravity models. 

There are some interesting proposed viable theories of gravity that are still consistent with some cosmological observations and have luminal speed of gravitational waves (see Table~\ref{tab:model_summary}). It will be useful to develop frameworks to test them against full large-scale structure and CMB data. 

Finally, in the next decade or so, upcoming and future surveys or experiments (e.g. AdvACT, DES, DESI, Euclid, HSC/PFS, LiteBIRD, LSST, PIXIE, SKA, SPT-3G, WFIRST and others) along with ongoing efforts in mitigating systematic effects promise to tighten the constraints on MG parameters and provide conclusive answers on gravity physics at cosmological scales.

 %%%%%%%%%%%%%%%%%%%%%%%%%%%%%%%%%%%%%%%%%%%%%%%%%%%%%%%%%%%%%%%%%%%%%%%%%%%%%%%%%%
%%%%%%%%%%%%%%%%%%%%%%%%%%%%%%%%%%%%%%%%%%%%%%%%%%%%%%%%%%%%%%%%%%%%%%%%%%%%%%%%%%%
%%%%%%%%%%%%%%%%%%%%%%%%                                  %%%%%%%%%%%%%%%%%%%%%%%%%
%%%%%%%%%%%%%%%%%%%%%%%%       Acknowledgments            %%%%%%%%%%%%%%%%%%%%%%%%%
%%%%%%%%%%%%%%%%%%%%%%%%                                  %%%%%%%%%%%%%%%%%%%%%%%%%
%%%%%%%%%%%%%%%%%%%%%%%%%%%%%%%%%%%%%%%%%%%%%%%%%%%%%%%%%%%%%%%%%%%%%%%%%%%%%%%%%%%
%%%%%%%%%%%%%%%%%%%%%%%%%%%%%%%%%%%%%%%%%%%%%%%%%%%%%%%%%%%%%%%%%%%%%%%%%%%%%%%%%%%

\begin{acknowledgements}
We thank Tessa Baker, Alex Barreira, Clare Burrage, Yves Dirian, Logan Fox, Shahab Joudaki, Austin Joyce, Kazuya Koyama, Lucas Lombriser, Michele Maggiore, Francesco Pace, Austin Peel, Levon Pogosian, Bharat Ratra, Jeremy Sakstein, Mehdi Saravani, Alessandra Silvestri, Constantinos Skordis,  Mark Trodden, Michael Troxel, Amol Upadhye, and Anzhong Wang for proofreading parts of the manuscript and providing useful comments. Particular thanks go to Weikang Lin for proofreading the whole manuscript and providing a number of useful comments. Thanks also go to Ji Yao for helping with formatting some of the references. 
The author acknowledges that this material is based upon work supported in part by the National Science Foundation under grant AST-1517768 and the U.S.\ Department of Energy, Office of Science, under Award Number DE-SC0019206.
\begin{flushleft}
\textbf{Note:} Comments and references are welcome for future versions. 
\end{flushleft}
\end{acknowledgements}

%%%%%%%%%%%%%%%%%%%%%%%%%%%%%%%%%%%%%%%%%%%%%%%%%%%%%%%%%%%%%%%%%%%%%%%%%%%%%%%%%%%
%%%%%%%%%%%%%%%%%%%%%%%%%%%%%%%%%%%%%%%%%%%%%%%%%%%%%%%%%%%%%%%%%%%%%%%%%%%%%%%%%%%
%%%%%%%%%%%%%%%%%%%%%%%%                                  %%%%%%%%%%%%%%%%%%%%%%%%%
%%%%%%%%%%%%%%%%%%%%%%%%           REFERENCES             %%%%%%%%%%%%%%%%%%%%%%%%%
%%%%%%%%%%%%%%%%%%%%%%%%                                  %%%%%%%%%%%%%%%%%%%%%%%%%
%%%%%%%%%%%%%%%%%%%%%%%%%%%%%%%%%%%%%%%%%%%%%%%%%%%%%%%%%%%%%%%%%%%%%%%%%%%%%%%%%%%
%%%%%%%%%%%%%%%%%%%%%%%%%%%%%%%%%%%%%%%%%%%%%%%%%%%%%%%%%%%%%%%%%%%%%%%%%%%%%%%%%%%

\bibliographystyle{spbasic}      % basic style, author-year citations
%\bibliography{CosmoTestsGR_merged_BIBV8}   % name your BibTeX data base
\bibliography{CosmoTestsGR_merged_BIBV8F}

\end{document}